\documentclass[]{jfm}
\usepackage{lipsum}
\usepackage{graphicx}
\usepackage{dcolumn}
\usepackage{bm}
\usepackage{epsfig}
\usepackage{float}
\usepackage{amsmath}
\usepackage{times}
\usepackage{wrapfig}
\usepackage{float}
\usepackage{xcolor}
\usepackage{comment}

\usepackage{newtxtext}
\usepackage{newtxmath}
\usepackage{changepage}
\usepackage{epstopdf}
\usepackage{natbib}
\usepackage{hyperref}
\hypersetup{
    colorlinks = true,
    urlcolor   = blue,
    citecolor=blue,
    linkcolor=blue,
}

\newcommand{\RomanNumeralCaps}[1]

\usepackage[utf8]{inputenc}
\usepackage{amsfonts}
\usepackage{amssymb}
\usepackage{mathtools}
\usepackage{subfigure}

\usepackage{caption}
\usepackage{subcaption}
\DeclareGraphicsExtensions{.pdf,.png,.jpg,.eps,.emf}

\usepackage{epstopdf}
\usepackage{relsize}
\usepackage{varioref}
\usepackage{booktabs}
\long\def\comment#1{}

\usepackage{fancyhdr}
\title{Transition in elastic Dean flow: the centre-mode versus hoop-stress pathways}

\author{P.\,S.\,D.\,Surya Phani Tej \aff{1},
   Ganesh Subramanian\aff{2}\corresp{\email{sganesh@jncasr.ac.in}},
 \and V.\,Shankar\aff{1}\corresp{\email{vshankar@iitk.ac.in}}}

\affiliation{\aff{1}Department of Chemical Engineering, Indian Institute of Technology, Kanpur-208016, India 
\aff{2}Engineering Mechanics Unit, Jawaharlal Nehru centre for Advanced Scientific Research, Bangalore - 560064, India 
}

\begin{document}
\maketitle
\begin{abstract}
We analyse the stability of viscoelastic Dean flow (flow of an elastic fluid through a curved two-dimensional channel, driven by an azimuthal pressure gradient) in the absence of fluid inertia. This configuration is well known to exhibit a hoop-stress-driven `purely elastic' instability (referred to henceforth as the hoop-stress mode -- `HSM') on account of the base-flow streamline curvature. The objective of this study is to demonstrate the existence and importance of a distinct elastic instability in this flow configuration, which is not driven by hoop-stresses, but instead is a continuation of a novel `centre-mode' (CM) instability recently identified in rectilinear shear flows. 
On account of its origins, the CM instability in Dean flow is expected for two-dimensional (azimuthally varying) disturbances with no axial variation, but continues to exist for three-dimensional disturbances. In contrast, the HSM instability is expected primarily in the axisymmetric limit, and again continues to exist for three-dimensional disturbances.

We use both the Oldroyd-B and FENE-P models to map out parameter regimes in the $W\!i$--$\epsilon$--$\beta$ space where the aforementioned instabilities are present. Here, $W\!i$ is a suitably defined Weissenberg number that characterizes fluid elasticity, $\beta$ is the ratio of solvent to total solution viscosity, and $\epsilon$ is the ratio of the gap (channel) width to the radius of curvature. While its origin in rectilinear shearing flows might lead one to expect the CM instability to only be present for small $\epsilon$’s (the `narrow-gap’ limit), we show that it exists even for $O(1)$ values of $\epsilon$, and over a larger range of $W\!i$. Nevertheless, within the Oldroyd-B framework, the HSM determines the stability threshold for experimentally relevant gap-width ratios, corresponding to
$0.1 \leq \epsilon \leq 1$, with the CM becoming the most unstable mode only for $\epsilon < 0.001$. For the more accurate FENE-P model, however, decreasing the finite extensibility parameter $L$ has opposing effects on the HSM and CM instabilities -- stabilising the former, but destabilising the latter. In the dilute solution regime ($\beta > 0.95$), and for realistic values of $L \sim O(100)$, corresponding to polymer molecular weights of $O(10^{5-6})$g/mol, the CM remains the most unstable mode for $\epsilon \leq 0.25$, rendering it potentially relevant to the onset of elastic turbulence in the flow of such polymer solutions through curved channels.
\end{abstract}

\section{Introduction}
\label{sec:intro}
Viscoelastic fluids subject to a shearing flow often exhibit strikingly novel instabilities engendered by their elastic nature, and that are absent in their Newtonian counterparts. For instance, it is well known, since the pioneering work of \cite{larson_shaqfeh_muller_1990}, that Taylor-Couette flow of a viscoelastic fluid exhibits a linear instability even when fluid inertia is negligible. Analogous instabilities have been demonstrated for the flow of polymer solutions in other curvilinear geometries such as cone-and-plate flow \citep{McKinley1991}, torsional flow between parallel circular discs \citep{byars1994}, and the Dean and Taylor-Dean flow configurations \citep{joo_shaqfeh_1991,joo1992purely,joo_shaqfeh_1994}. These instabilities, referred to as `purely elastic' instabilities \citep[see][]{Shaqfeh1996}, rely on two key ingredients in the base state -- streamline curvature and a streamwise normal stress that has a tensile character (owing to stretched polymers). The latter results in a `hoop stress' in the aforesaid curvilinear flows that then drives an instability even in the absence of inertial forces; note that the latter forces come into play even in a Newtonian fluid, leading to inertial instabilities beyond a threshold Reynolds number ($Re$) in the aforementioned configurations. The Pakdel-McKinley (PM) scaling arguments \citep{Pakdel_McKinley,PakdelMcKinley1996} provide a unifying expression for the onset of hoop-stress-driven instabilities, given by
\begin{equation}
\frac{\lambda V}{{\cal R}} \frac{N_1}{|\tau|} \geq M_{crit}^2 \, ,
\label{eq:Pakdelcriterion}
\end{equation}
where $\lambda$ is the (longest) relaxation time of the fluid, $V$ is a characteristic velocity scale, ${\cal R}$ is the pertinent radius of curvature, $N_1$ is the first normal stress difference, $\tau$ is the total shear stress in the base-state, and $M_{crit}$ is an $O(1)$ number sensitive to the actual geometry.
The PM criterion cannot be satisfied for rectilinear shearing flows (with ${\cal R} \rightarrow \infty$),  implying that hoop-stress-driven linear instabilities are absent in such flows. We note, however, that hoop stresses in these flows can become relevant at a nonlinear order, owing to curvature of 
the perturbed streamlines, as has been demonstrated within the framework of a weakly nonlinear stability analysis \citep{morozov_saarloos2005,morozov_saarloos2007,morozov_saarloos2019}. The aforesaid absence of a hoop-stress-driven linear instability had led to the viewpoint that  purely elastic instabilities are restricted to curvilinear configurations, even if they be more complicated than the canonical geometries above \citep{Haward_McKinley_Shen_SciRep_2016}, and that rectilinear viscoelastic shear flows are therefore linearly stable \citep{bertola2003experimental,Pan_2013_PRL}.

In a significant departure from this paradigm, recent work by \cite{khalid_creepingflow_2021} demonstrated that inertialess plane Poiseuille flow of an Oldroyd-B fluid is linearly unstable in the limit of high Weissenberg numbers ($W\!i = \lambda V/H \sim 10^3$, $V$ being the base-state maximum speed, and $H$ the channel half-width) and a near-unity solvent to solution viscosity ratio\,($\beta > 0.99)$; this corresponds to ultra-dilute highly elastic polymer solutions. The unstable mode has a phase speed close to the base-state maximum, belonging therefore to a class of `centre modes', and the instability arises due to a critical-layer mechanism that does not rely on streamline curvature. 
\cite{khalid_creepingflow_2021} further showed that, with increasing $R\!e$, the elastic centre-mode instability smoothly crosses over to the elastoinertial centre-mode instability, predicted in an earlier effort by \cite{khalid2021centermode}.  
  \cite{Kerswell2024asymptotics} have recently carried out a matched asymptotic expansions analysis to capture this unstable centre mode. Further, \cite{buza_page_kerswell_2022} and subsequently \cite{KhalidFENEP2025}, have shown using the FENE-P model, that accounts for finite extensibility of the polymer molecules, that the elastic centre-mode instability is present even
  for $W\!i \sim O(100)$ and $\beta \sim 0.97$. Thus, use of the more realistic FENE-P model extends the unstable domain to experimentally accessible parts of the parameter space.

  Recent efforts have identified an unstable centre mode in other rectilinear shearing flows as well. \cite{Yadav2024} have shown that the elastic centre-mode instability is present in the Couette-Poiseuille family\,(CPF), but only when the base-state velocity profile is non-monotonic, with a maximum within the flow domain; the instability being absent in plane Couette flow, and more generally, in shearing flows with monotonic velocity profiles. This intimate connection between the presence of a base-state maximum, and the existence of a centre-mode instability, is further corroborated by the presence of the latter in viscoelastic Kolmogorov flow \citep{Kerswell2024asymptotics,Lewy_Kerswell_2025}; although the significance of this finding was not recognized in the original analysis of \cite{BOFFETTA_2005}. An analogue of the centre-mode instability was also shown to be present in viscoelastic film flow down an inclined plane \citep{Mamta2023}, with a semi-parabolic base-state velocity profile.

Thus, there are two qualitatively different elastic instabilities that operate in viscoelastic shearing flows: (i) the hoop-stress mode (`HSM' henceforth) which requires the presence of streamline curvature, but exists even for monotonic velocity profiles (such as Taylor-Couette flow), and (ii) the elastic centre mode (`CM' henceforth) which does not rely on streamline curvature, but requires a non-monotonic base-state velocity profile. This raises the intriguing possibility that  these modes might co-exist in flow configurations which possess
both the aforementioned ingredients. That is to say, one has the potential for elastic instabilities of different physical origins in the same flow, a possibility that has not yet been explored in the literature. One such configuration is Dean flow -- the flow in a curved channel driven by an azimuthal pressure gradient. In the Newtonian context, this configuration was first analysed by \cite{Dean1928}, who showed the presence of a centrifugal instability. The threshold for this instability, in the narrow-gap limit, can be obtained from an inertial analogue of the PM criterion above, with the Reynolds number $Re$ replacing $W\!i$; this was indeed pointed out by  
\cite{PakdelMcKinley1996} while formulating their criterion for elastic instabilities.
As alluded to above, \cite{joo_shaqfeh_1991,joo1992purely,joo_shaqfeh_1994} analysed the viscoelastic Dean flow problem, albeit in the inertialess limit ($Re = 0$), and showed the presence of an HSM instability in this configuration. One could also have a combination of an azimuthal pressure gradient and wall motion driving the flow, the resulting configuration being dubbed the `Taylor--Dean' flow \citep{joo_shaqfeh_1991}, and being more easily realized in experiments. 

In this study, we carry out a linear stability analysis of viscoelastic Dean flow to explore the relevance of the centre- and hoop-stress modes in a viscoelastic parameter space comprising   $W\!i$, $\beta$, and the gap-width ratio $\epsilon = d/{\mathcal R}$, $d$ being the channel width. For the case where the curved channel is  the gap between concentric rotating cylinders with $R_2$ and $R_1$ being the outer and inner cylinder radii, as is typically the case in experiments, $d= (R_2 - R_1)$ may be identified with the gap width, and the radius of curvature with $R_1$. We consider the inertialess regime in this effort, and therefore, set $Re = 0$. Traditionally, the HSM has been first analysed in the narrow-gap limit \citep{larson_shaqfeh_muller_1990,joo_shaqfeh_1991}, corresponding to $\epsilon \ll 1$, with this assumption being relaxed in subsequent numerical investigations \citep{joo1992purely,joo_shaqfeh_1994}. In the narrow-gap limit, in order for the hoop stress to remain relevant in the leading order radial momentum balance, one requires $\epsilon^{1/2} W\!i \sim O(1)$.  Here, $W\!i = \lambda U_m/d$ is the Weissenberg number based on the gap width $d$ and a velocity scale $U_m$ associated with the 
Dean flow profile (defined below). Thus, in the narrow-gap limit, the critical $W\!i$ for the HSM instability must diverge as $\epsilon^{-1/2}$. In the same limit, the critical $W\!i$ for the CM instability is independent of $\epsilon$, and approaches that of plane Poiseuille flow \citep{khalid2021centermode}. At a scaling level, therefore, as $\epsilon \rightarrow 0$,  the critical $W\!i$ for HSM should eventually become larger than that for the CM. However, owing to the relatively large threshold, $W\!i \sim O(10^3)$ for the CM (within the Oldroyd-B framework), it is necessary to carry out the stability analysis to determine the precise value of $\epsilon$ corresponding to this crossover, and thereby assess its relevance to typical configurations used in experiments and applications. To this end, we present here a comprehensive account of the parameter regimes in which either of the aforementioned modes is the most unstable in viscoelastic Dean flow, by considering arbitrary gap-width ratios ($\epsilon$), and by including both axisymmetric and non-axisymmetric disturbances. 

While the existence of two types of elastic instabilities in the Dean flow configuration is of interest from the fundamental viewpoint, it is also significant from the applications perspective. Viscoelastic Dean flow is often used for hydrodynamic focusing of particles in curved \citep{Nikdoost_Dean_application3}, spiral \citep{Bai_Dean_application1}, and serpentine  \citep[][]{Chen_Dean_application2} microchannels. Although such experiments involve finite Reynolds numbers, the secondary flows emerging from the elastic instabilities reported in this study could certainly impact particle migration, and more generally, the efficiency of focussing protocols on microfluidic platforms. Polymer solutions are also used as displacing agents in enhanced oil recovery, in order to remove capillary entrapments of oil in the pores. In this context, flow in serpentine microfluidic channels has been used to mimic the tortuous nature of the flow in porous media \citep{Shakeri_etal_PoF_2021}, wherein elastic instabilities are harnessed in triggering the breakup of large capillary entrapments and in the eventual  removal of the same. Importantly, the elastic turbulence regime accessed in previous experiments involving serpentine microchannels \citep[also regarded as an important means of achieving efficient mixing in microfluidic devices; see][]{Groisman2001,Groisman2004} is either explicitly or implicitly understood as the eventual consequence of an initial hoop-stress-driven instability. Our findings suggest that there might exist two distinct pathways\,(viz., an HSM-based route, and a CM-based route, as suggested by the title) to the eventual turbulent regime, and perhaps, even two different kinds of turbulent states, depending on the wavelength of modulation of the serpentine channel.

We consider both the Oldroyd-B and FENE-P constitutive equations to model the viscoelastic fluid. The Oldroyd-B model has the necessary ingredients to capture both HSM and CM, but owing to its prediction of a shear-rate-independent viscosity and first normal stress coefficient, it may be not accurate for fast shearing flows of polymer solutions\,(translating to a large threshold $W\!i$ for the CM in the present context) where shear thinning becomes important. This lacuna is due to the restrictive assumption of the polymer molecule being an infinitely extensible bead-spring dumbbell. The FENE-P model \citep{Bird1980} overcomes this shortcoming by accounting for finite extensibility via the parameter $L$, the ratio of the fully-extended length of the polymer (dumbbell) to its equilibrium size; the Oldroyd-B model being realized in the limit $L \rightarrow \infty$. The FENE-P model predicts shear thinning of both the viscosity and the first normal stress coefficient, and hence is a more realistic model at higher $W\!i$. As will be shown below, the consideration of finite extensibility is particularly important in the present scenario, on account of its contrasting effects on the HSM and CM instabilities. Well, why then should one consider the Oldroyd-B model at all? As noted by \cite{ShaqfehKhomami2021}, in their review article commemorating the birth centenary of James Oldroyd, the results obtained using the Oldroyd-B fluid have their own value in terms of providing a base reference, and (we quote) ``it is perilous...to ignore the analysis based on the Oldroyd-B model for any elastic flow problem''!  Before we proceed with the present work, we provide a brief review of the relevant literature on HSM and CM instabilities.

\subsection{The hoop-stress instability in viscoelastic Dean flow}
\label{subsec:litreviewHSM}
 Since the focus of this work is on Dean flow, we do not review here the significant progress made in understanding the HSM in viscoelastic Taylor-Couette flow, since its discovery by \cite{larson_shaqfeh_muller_1990}. Despite sharing some similarities, there are many differences vis-a-vis the HSM instabilities in these two flows, and the reader is referred to recent reviews \citep{Datta_etal2022,ShaqfehKhomami2021,CastilloSanchez2022} for the latest developments in this area.  Suffice it to say that this flow configuration still commands significant interest among researchers, as can be seen from the recent experiments of \cite{YiBaoZhang_etal}, who showed that there is a continuous transition between the elastic and elastoinertial turbulent states in Taylor-Couette flow, by probing elasticity numbers ($E = W\!i/Re$) ranging over two orders of magnitude.  Although a connection between the elastoinertial and elastic turbulent states  was first speculated by  \cite{Samanta2013} in their observations concerning  the onset of elastoinertial turbulence in pipe flow of polymer solutions, a demonstration of a  connected linearly unstable region, across the entire range of $E$, was provided by 
  \cite{khalid2021centermode}, in the context of plane Poiseuille flow.
 
 Following the identification of a purely elastic instability in the Taylor-Couette set-up by \cite{larson_shaqfeh_muller_1990}, \cite{joo_shaqfeh_1991,joo_shaqfeh_1994} first demonstrated, again  using the Oldroyd-B model, the presence of an HSM instability in the inertialess flow through a curved planar channel, driven by an azimuthal pressure gradient (the Dean flow configuration), as well as in the Taylor-Dean configuration where there is the added tangential motion of one of the boundaries. The authors assumed axisymmetric disturbances to begin with, and used the narrow-gap limit to show the presence of the hoop-stress instability when $W\!i \gtrsim  \epsilon^{-1/2}$, $\epsilon$ being the gap-width ratio mentioned above; subsequently, the authors explored finite $\epsilon$'s and non-axisymmetric disturbances. They also carried out experiments, wherein a Taylor-Dean flow (with zero net azimuthal flow rate) was realized in the gap between concentric rotating cylinders with an azimuthal obstruction, and found a qualitative (but not quantitative) agreement between the experimental and theoretical thresholds. In contrast to the Taylor-Couette configuration \citep{Avgousti1993,joo_shaqfeh_1994}, the unstable mode in the Dean configuration is stationary and axisymmetric. Later, \cite{Ramanan_1999} demonstrated the stabilizing effect of an imposed axial flow on the Dean flow HSM, again focusing on stationary, axisymmetric modes.

 The above conclusions for the HSM in Dean flow were arrived at within the Oldroyd-B framework. With the narrow-gap limit in mind, it is natural to enquire as to how these conclusions are modified at higher $W\!i$, using a more accurate nonlinear constitutive model. 
 Over the past decade, several experimental studies
\citep{dutcher2013effects,schaefer2018geometric,lacassagne2021shear,Steinberg2021,more2024elasto} have explored elastic instabilities in various curvilinear geometries, in regimes where shear thinning is important, thereby making the case for a nonlinear constitutive model even more compelling.
 The general expectation is that 
 shear thinning of the first normal stress coefficient, that arises in any nonlinear model, will weaken the elastic stresses that drive the HSM instability.  Indeed, the earlier work of \cite{byars1994} and \cite{McKinley1995} used the FENE-CR model to demonstrate the stabilizing role of finite extensibility in torsional flows and  in the cone-and-plate configuration. 
  While the finite-extensibility-induced stabilising argument can been made more precise \citep{CastilloSanchez2022} by accounting for shear thinning within the Pakdel-McKinley framework, rather surprisingly, we are not aware of the Dean-flow HSM being analysed quantitatively using any of the well-known nonlinear constitutive equations, such as the FENE-P/FENE-CR or Giesekus models. For this reason, and owing to the contrasting effect of finite extensibility on the HSM and CM demonstrated below, we revisit the HSM in viscoelastic Dean flow using both Oldroyd-B and FENE-P models. In doing so, we examine both axisymmetric and non-axisymmetric disturbances in the narrow-gap limit ($\epsilon \ll 1$), and for finite gap-width ratios  ($\epsilon \sim O(1)$).  The consideration of non-axisymmetric disturbances gains relevance because, while the HSM is present for axisymmetric disturbances, a non-zero azimuthal wavenumber can lead to a slight destabilization in certain cases, as has been shown by  \cite{Avgousti1993} for viscoelastic Taylor-Couette flow.

\subsection{The centre-mode instability in viscoelastic shearing flows}
\label{subsec:litreviewCM}

In the narrow-gap limit ($\epsilon \rightarrow 0$), $W\!i$ has to increase as $\epsilon^{-1/2}$ in order for the hoop stress to remain in the leading order momentum balance. Thus, the HSM instability disappears, for a fixed $W\!i$, as $\epsilon \rightarrow 0$, with the Dean flow 
configuration approaching plane Poiseuille flow. As mentioned above, the latter flow of an Oldroyd-B fluid was shown, by \cite{khalid_creepingflow_2021}, to be susceptible to the CM instability in the highly elastic ultra-dilute limit. The authors showed that the unstable mode has a phase speed close to the base-state maximum at the channel centreline (thence, the `centre mode' terminology), with the critical $W\!i$ being $O(10^3)$. Furthermore, the most unstable mode is two-dimensional with no spanwise variation, owing to the existence of a Squire's theorem for rectilinear shearing flows of an Oldroyd-B fluid \citep{BISTAGNINO2007}, which stipulates that two-dimensional disturbances are more unstable than three-dimensional ones. However, at such high $W\!i$'s, one expects shear thinning to become important \citep{KhalidFENEP2025}, potentially rendering the Oldroyd-B predictions less accurate. It is then appropriate to use, for instance, the FENE-P model which accounts for shear thinning of both the viscosity and first normal stress coefficient. 

Rather surprisingly, and contrary to the aforementioned  expectation of a stabilising influence of finite extensibility, \cite{buza_page_kerswell_2022} used the FENE-P model to show that the critical $W\!i$ decreased to $O(100)$.
\cite{KhalidFENEP2025} have recently built on this finding, and presented a more comprehensive picture of the parameter regimes in which the elastic CM instability is present in channel flow of a FENE-P fluid. The authors showed that, while finite extensibility has the expected stabilising influence at lower elasticity numbers, it has a destabilising influence at higher $E$'s, for $L \sim O(100)$.
Such $L$'s correspond to polyacrylamide solutions with molecular weights of $O(10^5)$, used in earlier experimental studies \citep{choueiri2021experimental}. As indicated above, the most unstable CM in plane Poiseuille flow has a non-zero streamwise wavenumber and zero spanwise wavenumber, which translates, in the Dean flow context, to two-dimensional perturbations varying in the azimuthal direction with no axial variations. In other words, contrary to the HSM above, the CM instability can arise even in the two-dimensional limit with axially invariant  perturbations. 
Nevertheless, the absence of Squire's theorem for any non-zero $\epsilon$ motivates us 
to also consider three-dimensional perturbations to arrive at the critical $W\!i$ as a function of $\epsilon$. 
We note that, in light of \cite{Yadav2024}'s demonstration, of the existence of the CM instability for members of the Couette-Poiseuille family with non-monotonic velocity profiles, the present Dean-flow results should be relevant to the Taylor-Dean configuration as well, and thence, amenable to experiments \citep{joo_shaqfeh_1994}. \\

The rest of this paper is organized as follows. The governing equations for both Oldroyd-B and FENE-P models are discussed in Sec.\,\ref{subsec:goveqns}. The base state velocity and stresses are discussed in Sec.\,\ref{subsec:BaseOldB} for the Oldroyd-B model, and in Sec.\,\ref{subsec:BaseFENEP} for the FENE-P model. The linearization procedure is briefly introduced in Sec.\,\ref{subsec:linstab}, with Appendices \ref{app:Old_b_Linearized_eq} and \ref{app:FENE_P_Linearized_eq} providing the full system of linearized equations governing small amplitude disturbances in the respective cases. Since earlier efforts examining the Dean flow configuration have solely focused on the unstable modes, we begin here by discussing the nature of the complete elastic eigenspectrum for Dean flow of both Oldroyd-B and FENE-P fluids in Sec.\,\ref{sec:Nature_of_Spectrum}, with additional details of representative eigenspectra provided in Appendix\,\ref{sec:Appendixspectra}; sections\,\ref{subsec:spectrum_axisym} and \ref{subsec:nonaxy_spectrum} discuss the eigenspectra for the axisymmetric and non-axisymmetric cases, respectively.
We then discuss the CM instability in viscoelastic Dean flow in Sec.\,\ref{sec:CMinDeanflow}. Herein, we first use the Oldroyd-B model in Sec.\,\ref{subsec:CMOldroydB} to map out the unstable parameter regimes for arbitrary $\epsilon$, both for two- and three-dimensional perturbations, and then proceed to examine the role of finite extensibility\,(using the FENE-P model) in Sec.\,\ref{subsec:LonCM}.  Section\,\ref{sec:HSMinDeanflow} begins with a  brief recap of the key results for HSM within the Oldroyd-B framework, following which we examine the role of finite extensibility on the HSM for both axisymmetric (Sec.\,\ref{subsec:Lonaxisym}) and non-axisymmetric (Sec.\,\ref{subsec:Lonnonaxy}) disturbances. Appendix\,\ref{sec:Appendixeigenfunctions} provides a brief compilation of eigenfunctions of the relevant unstable modes. Finally, Sec.\,\ref{sec:concl} provides a brief summary of the salient conclusions emerging from this effort, along with a discussion on which of the two modes (CM or HSM) is more critical in the viscoelastic parameter space.

\begin{figure}
  \centering
  \includegraphics[width = 0.8\textwidth]{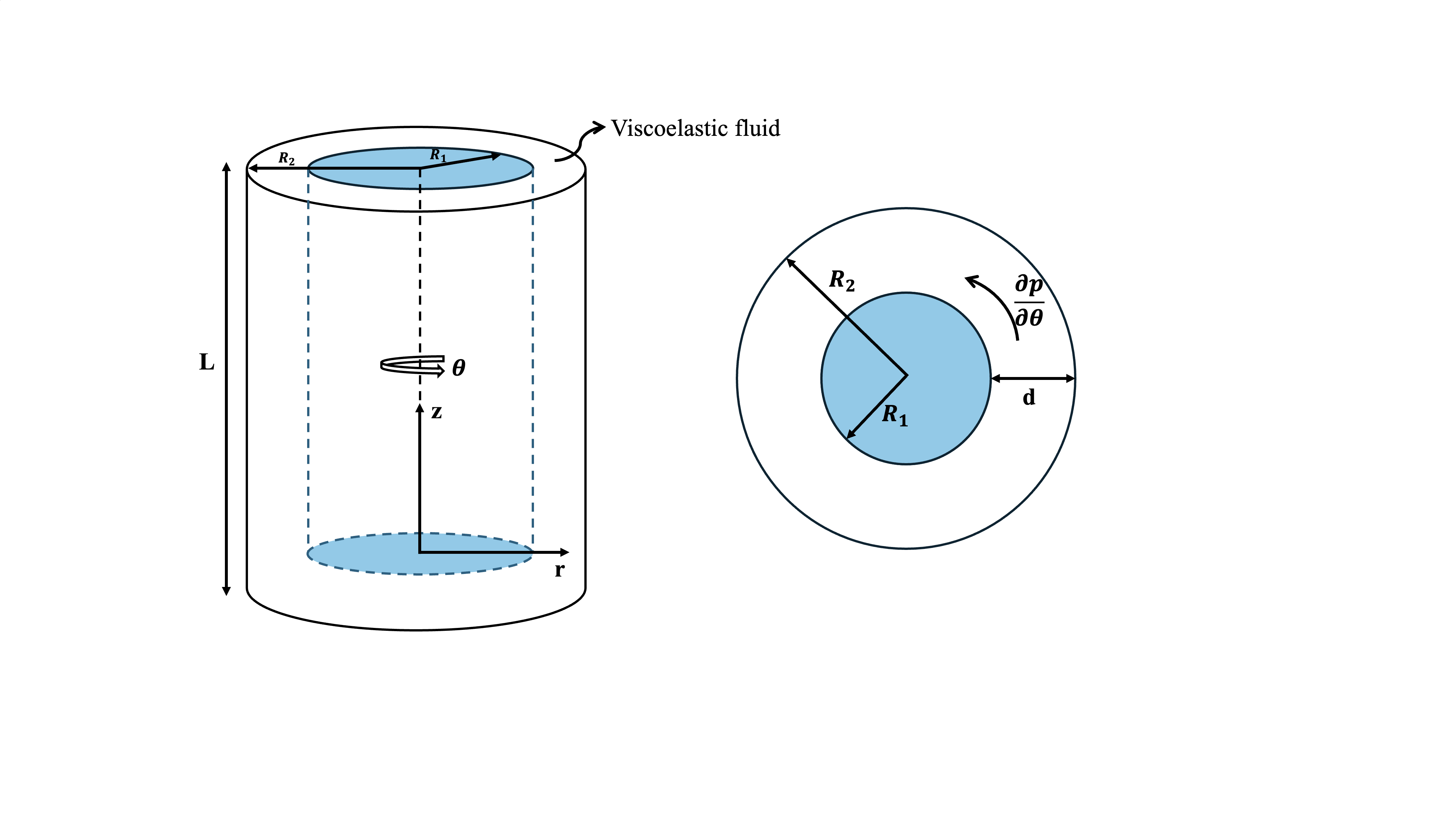}
  \caption{Schematic of the geometry and the coordinate system considered.}
  \label{fig:deanpic}
\end{figure}
\section{Problem Formulation and numerical method}
\label{sec:problem_formulation}
\subsection {Governing Equations}
\label{subsec:goveqns}
We consider steady, fully developed pressure-driven flow of a viscoelastic fluid between concentric cylindrical surfaces 
with radii $R_1$ and $R_2$ ($R_1 < R_2$), corresponding to a gap width $d = R_2 - R_1$; the gap-width ratio being defined as $\epsilon = d/R_1$.
A cylindrical coordinate system with $r$, $\theta$, $z$ denoting the radial, azimuthal and axial coordinates, respectively, shown in Fig.\,\ref{fig:deanpic}, is used with the pressure gradient imposed along the azimuth.
The scales used for rendering dimensionless the physical quantities, associated with both Oldroyd-B and FENE-P models,  are summarized in Table \ref{Scaling}. 
The dimensionless continuity and inertialess Cauchy momentum equations take the form
\begin{equation}
 \label{eq:fluid-continuity}
\nabla \cdot \boldsymbol{v}=0 ,   
\end{equation}
\begin{equation}
\label{eq:fluid-momentum}
0 =-\nabla {p}+ {\beta} \,\nabla^2 \, \boldsymbol{v} +  {(1-\beta)} \, \nabla \cdot \boldsymbol{\tau} \, ,
\end{equation}
with $\boldsymbol{v}$ and $p$ being the velocity and pressure fields, and $\boldsymbol{\tau}$ the polymeric stress tensor. 
In Eq.\,\ref{eq:fluid-momentum}, $\beta = \eta_s/\eta$ is the solvent to solution viscosity ratio, where $\eta = \eta_s + \eta_p$ is the total (zero-shear) solution viscosity with $\eta_p$ and $\eta_s$ being the zero-shear polymer and solvent viscosities, respectively.

\begin{table}
  \begin{center}
\def~{\hphantom{0}}
  \begin{tabular}{lccc}
  \vspace{5pt}
      Parameter  & Dimensional &  Scale & Dimensionless  \\
                 & Variable    &        &  variable \\
      \hline
      \vspace{5pt}
       Length & $r^*$ & $d$ (gap width) & $r$ \\
       Velocity & $\boldsymbol{v}^*$ & $U_m$ (average velocity) & $\boldsymbol{v}$ \\
       \vspace{5pt}
      Time & $t^*$ &  $\frac{d}{U_m}$  & $t$ \\
      \vspace{5pt}
       Pressure  & $p^*$ & $ \frac{\eta U_m}{d}$ & $p$ \\
       \vspace{5pt}
       Polymeric stress & $\boldsymbol{\tau}^*$ & $\frac{\eta_p U_m}{d}$ & $\boldsymbol{\tau}$ \\
       \vspace{5pt}
       Disturbance frequency & $\omega^*$ & $\lambda$ (relaxation time)  & $\omega$  \\
       \vspace{5pt}
       Conformation tensor & $\boldsymbol{C}^*$ & $k_B T/H_{s\!p}$  & $\boldsymbol{C}$  \\
  \end{tabular}
  \caption{Non-dimensionalisation scheme adopted in this study. Most symbols are defined in the text. Here, $H_{s\!p}$ is the equilibrium spring constant in the Oldroyd-B and  FENE-P models, $k_B$ is the Boltzmann constant and $T$ is the absolute temperature.}
  \label{Scaling}
  \end{center}
\end{table}

In the Oldroyd-B model
\citep{larson1988constitutive}, the polymer solution is modelled as a non-interacting suspension of Hookean dumbbells, and the polymeric stress $\boldsymbol{\tau}$ is governed by:
\begin{eqnarray}
\boldsymbol{\tau}+{W\!i} \,\Big(\frac{\partial \boldsymbol{\tau}}{\partial t}+ \boldsymbol{v}\cdot\nabla 
\boldsymbol{\tau}-\nabla \boldsymbol{v}^\intercal\cdot\boldsymbol{\tau}- \,\boldsymbol{\tau}\cdot\nabla \boldsymbol{v}\Big)= \Big(\nabla \boldsymbol{v}+\nabla \boldsymbol{v}^\intercal \Big).
\label{eq:stress constitutive eq}
\end{eqnarray}
Here, $W\!i = \lambda U_m/d$ is the Weissenberg number.
In the FENE-P model, the dumbbells are finitely extensible, and the polymeric contribution to the stress tensor is given by
\begin{equation}
    \boldsymbol{\tau} =  \frac{f\boldsymbol{C} - \boldsymbol{I}}{W\!i}.
    \label{eq:Stress constitutive eq}
\end{equation}
Here, $\boldsymbol{C}$ denotes the dimensionless conformation tensor, while $\boldsymbol{I}$ is the identity tensor representing the isotropic  distribution of dumbbell conformations in the absence of an imposed flow. The Peterlin closure function $f$ \citep{Bird1980,Herrchen_Ottinger_1997} is defined as
\begin{equation}
    f = \frac{L^2 - 3}{L^2 -tr(\boldsymbol{C})}\,,
    \label{eq:f constitutive eq}
\end{equation}
where $L$ represents the dimensionless extensibility parameter, defined as the ratio of the polymer's maximum extension to its equilibrium root-mean-square value; the Oldroyd-B model being attained in the limit of infinite extensibility viz., $L \rightarrow \infty$, when $f \rightarrow 1$.
The conformation tensor is governed by the following evolution equation:
\begin{equation}
    \frac{\partial \boldsymbol{C}}{\partial t}+ \boldsymbol{v}\cdot\nabla 
\boldsymbol{C}-\nabla \boldsymbol{v}^\intercal \cdot \boldsymbol{C}- \,\boldsymbol{C}\cdot \nabla \boldsymbol{v} = -\boldsymbol{\tau}.
    \label{eq:Conformation Tensor constitutive eq}
\end{equation}
 \subsection {Base State for the Oldroyd-B Model}
 \label{subsec:BaseOldB}
 The base state velocity profile for Dean flow of an Oldroyd-B fluid is identical to its Newtonian counterpart
 \citep{Diprima_1959,joo1992purely}, and therefore, independent of $W\!i$. The non-dimensional velocity profile, after using the coordinate transformation $r^{*} = R_1(1 + \epsilon  y)$, is given by:
\begin{equation}
\setlength{\arraycolsep}{0pt}
\renewcommand{\arraystretch}{1.5}
\overline{\boldsymbol{v}} = \left[
\begin{array}{ccccc}
  \overline v_{r} \\
    \overline v_{\theta} \\
    \overline v_{z} \\
\end{array}  \right]  = \left[
\begin{array}{ccccc}
  0 \\
    U(y) \\
    0 \\
\end{array}  \right] ,
\label{Base_State_Velocity}
\end{equation}
\begin{equation}
U(y) = \frac{ -K \left[ (1 + \epsilon y) - \frac{1}{(1 + \epsilon y)}\right] +  \left[(1 + \epsilon y) \ln (1 + \epsilon y) \right] }{K \left[\frac{ \ln (1 + \epsilon )}{\epsilon}\right] - \frac{1}{4} (2 + \epsilon)} ,
\label{eq:OldBbasevel}
\end{equation}
where $K = \frac{(1 + \epsilon)^2 \ln(1+ \epsilon)}{(1 + \epsilon)^2  - 1}  $, and the average velocity that has been used as the scale in Eq.\,\ref{eq:OldBbasevel}, is given by: $U_m = \frac{\frac{\partial p}{\partial \theta} R_1}{2 \eta} \left (K \left[\frac{ \ln (1 + \epsilon )}{\epsilon}\right] - \frac{1}{4} (2 + \epsilon) \right)$ .
The polymer contribution to the  base state  stress tensor is given by 
\begin{equation}
\setlength{\arraycolsep}{3pt}
\renewcommand{\arraystretch}{2}
\boldsymbol{\overline\tau} = \left[
\begin{array}{ccccc}
  \overline\tau_{rr} & \overline\tau_{r \theta} & \overline\tau_{rz} & \\
    \overline\tau_{\theta r} & \overline\tau_{\theta \theta} & \overline\tau_{\theta z} & \\
    \overline\tau_{zr } & \overline\tau_{z \theta} & \overline\tau_{zz} & \\
\end{array}  \right]  = \left[
\begin{array}{ccccc}
  0 & U' -  \frac{\epsilon}{1 + \epsilon y}U & 0 & \\
    U' -  \frac{\epsilon}{1 + \epsilon y}U & 2 W\!i (U' -  \frac{\epsilon}{1 + \epsilon y}U)^2 & 0 & \\
    0 & 0 & 0 & \\
\end{array}  \right] ,
\label{Base_State_Polymeric_Stress}
\end{equation}
where  $ \zeta' \equiv D\zeta \equiv \frac{d\zeta}{dy}$.
\begin{figure}
    \centering
    \subfigure[Data collapse]{
        \includegraphics[width=0.45\textwidth]{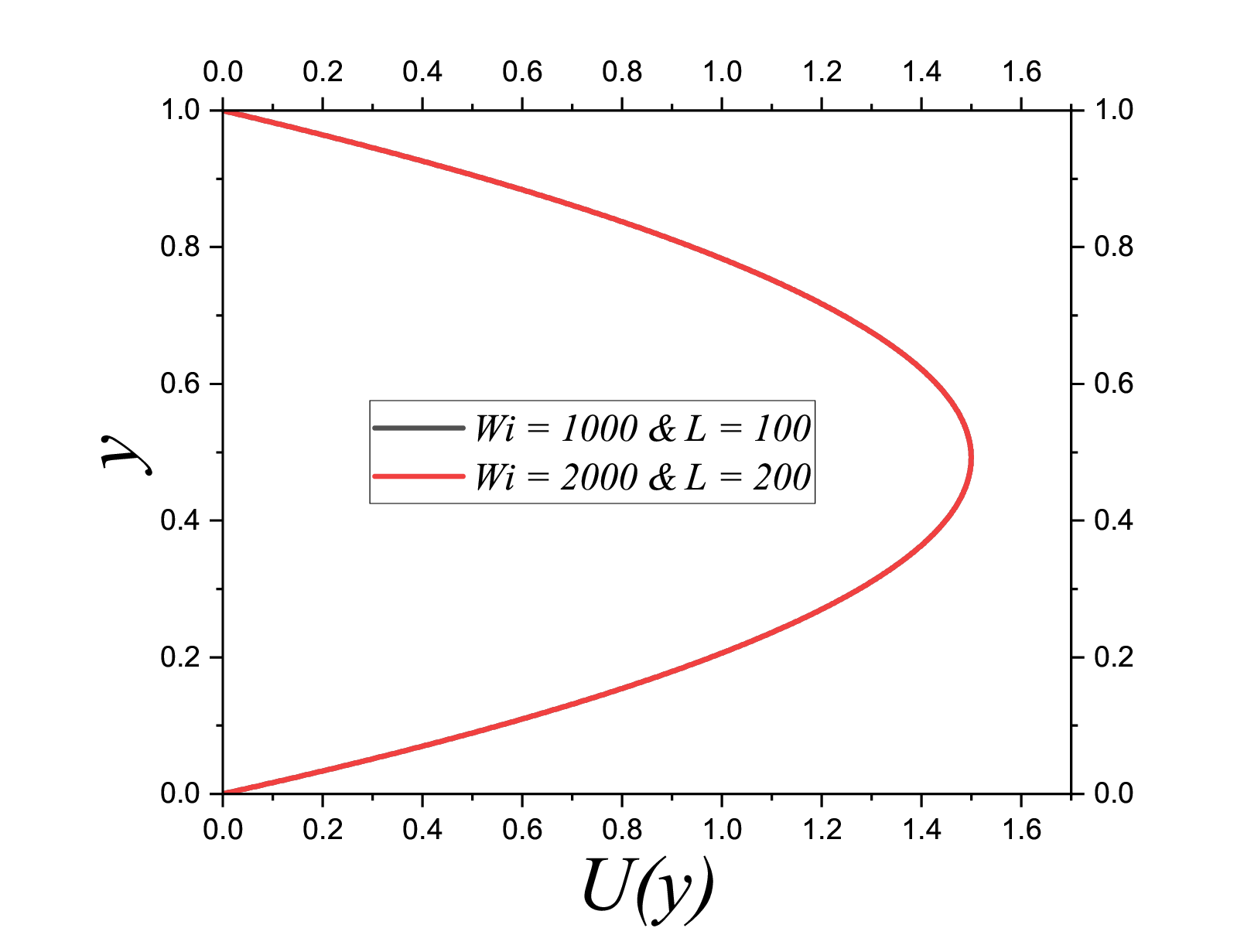} \label{fig:Deandiff_W_by_L_10_beta_0.98}
    }
    \subfigure[Effect of $Wi/L$]{
        \includegraphics[width=0.45\textwidth]{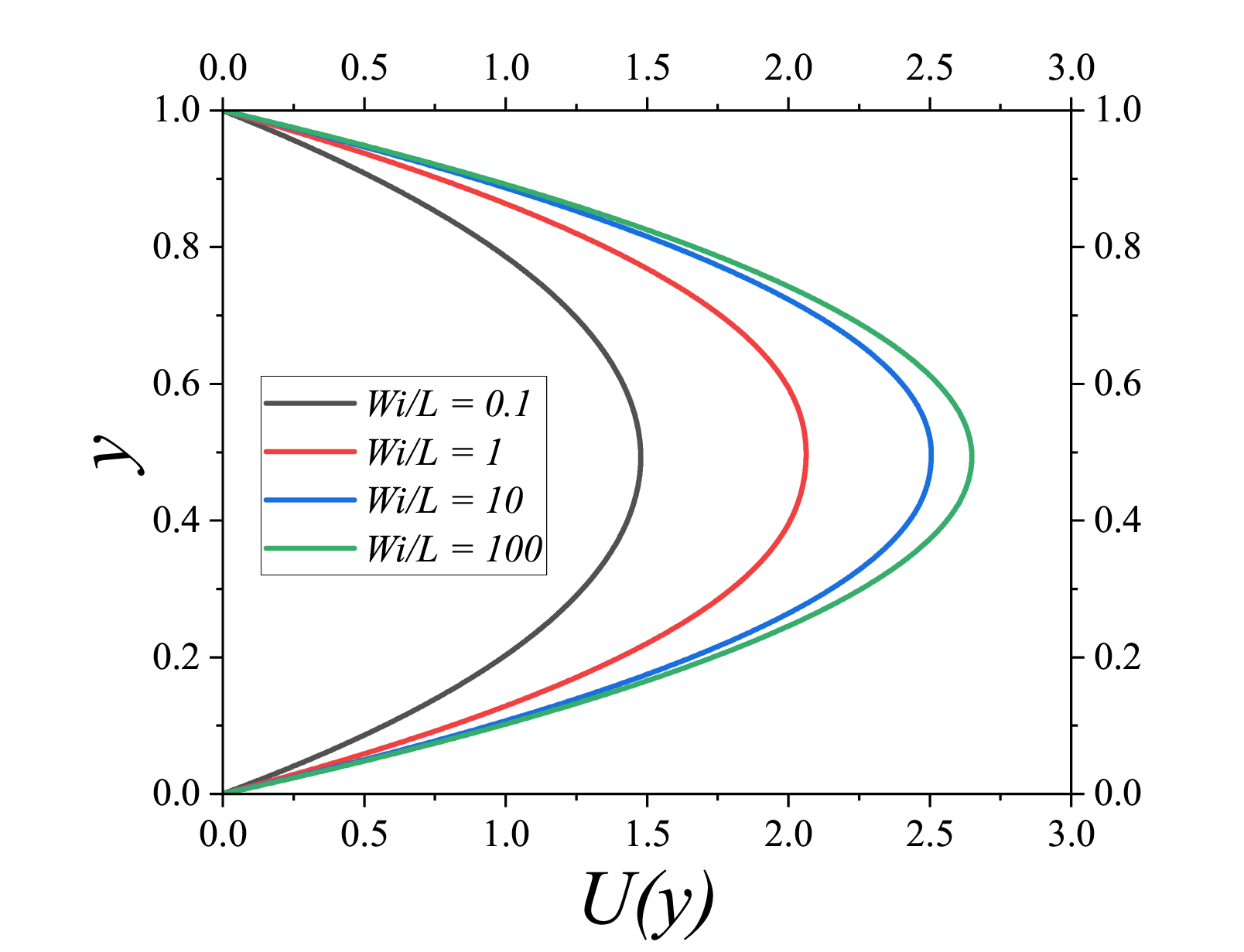}\label{fig:Deandiff_W_by_L_beta_0.5}
    }
    \subfigure[Effect of $\beta$]{
        \includegraphics[width=0.45\textwidth]{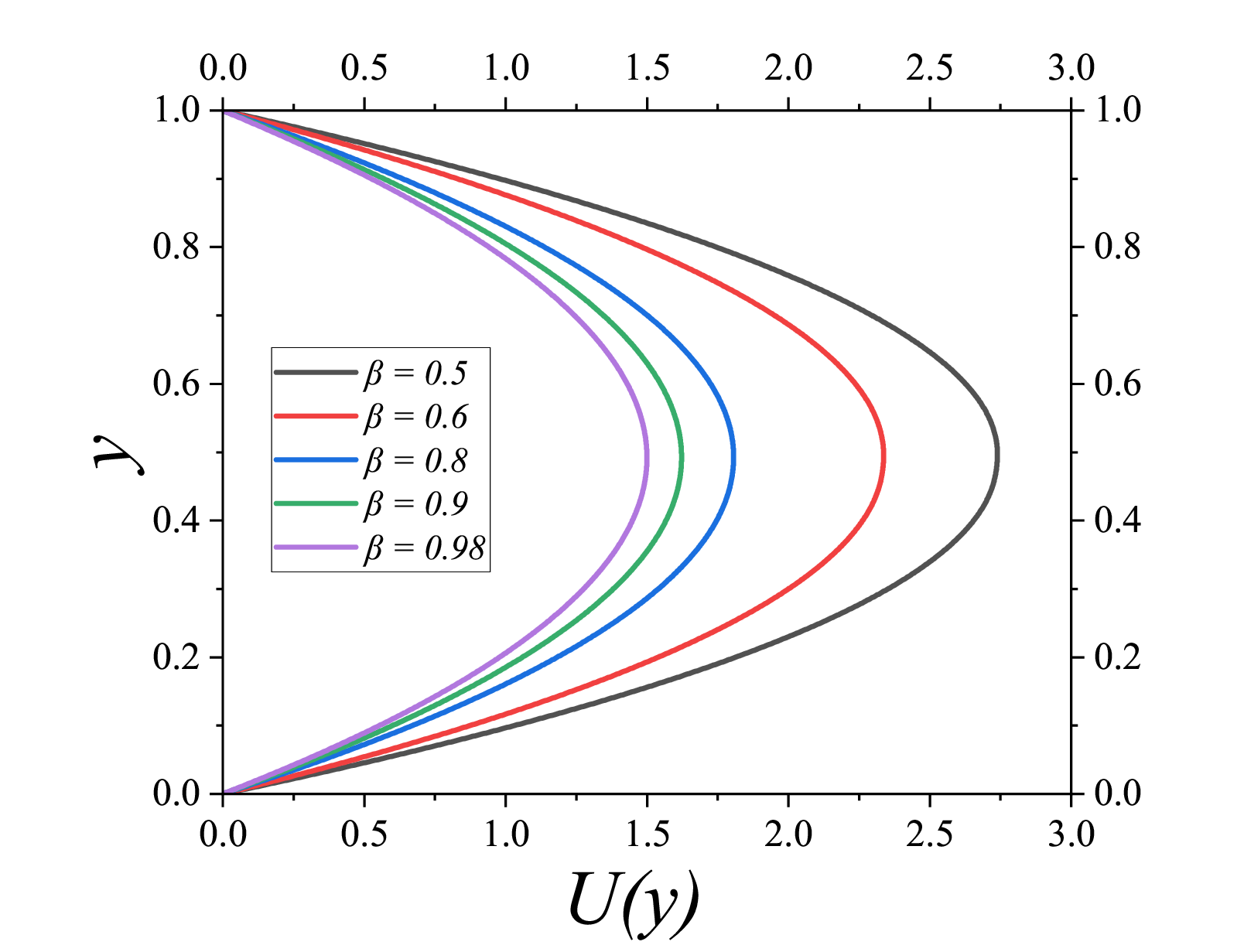} \label{fig:Deandiffbetas}
    }
    \subfigure[Effect of $\epsilon$]{
        \includegraphics[width=0.45\textwidth]{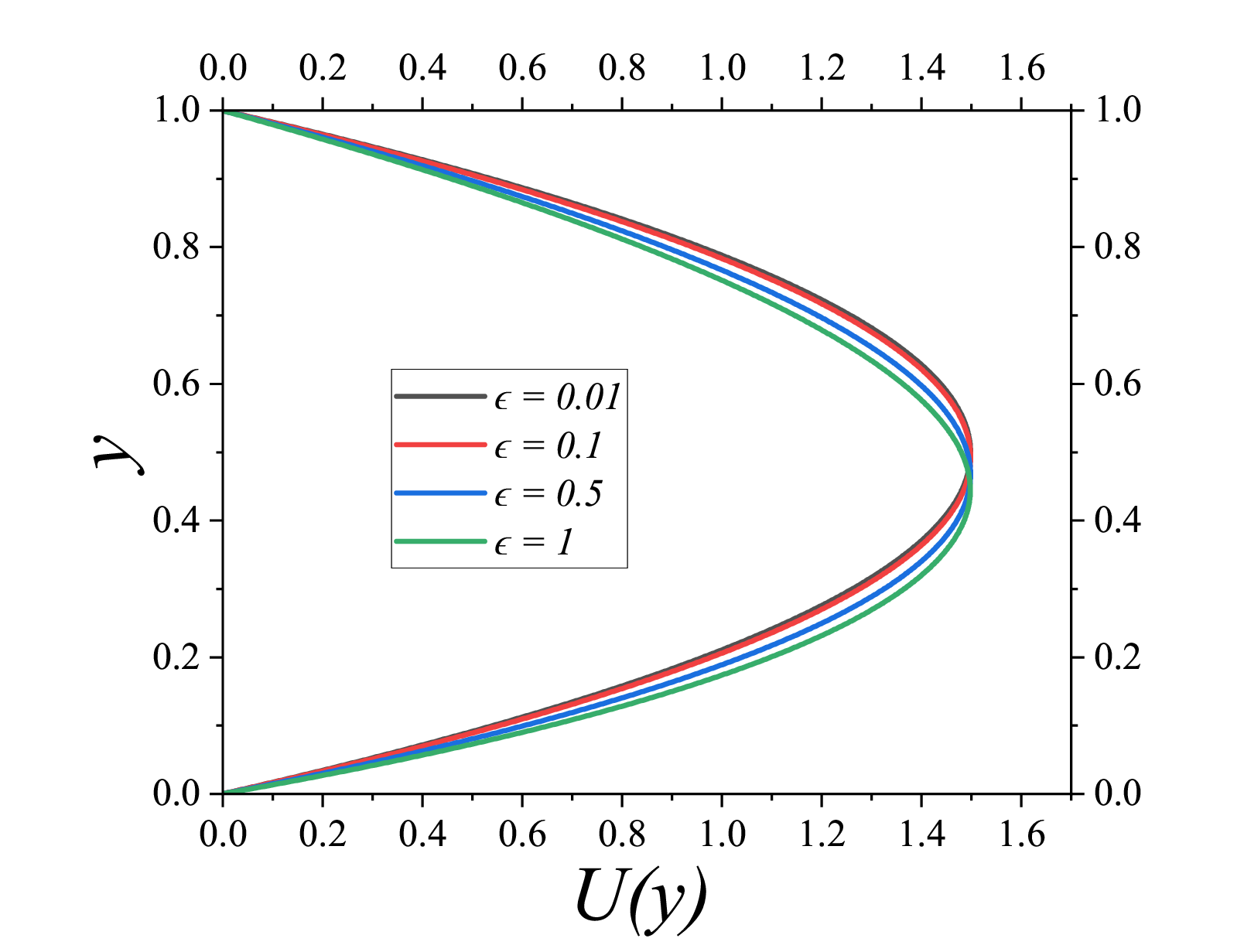}\label{fig:Deandiffeps}
    }
    \caption{Dean flow velocity profiles for a FENE-P fluid: (a) $\beta = 0.98$ and $\epsilon = 0.1$ for two $(Wi, L)$ pairs, with $W\!i/L = 10$; (b) $\beta = 0.5$ and  $\epsilon = 0.1$   with varying $Wi/L$; (c) master curves for $Wi/L = 10$ and $\epsilon = 0.1$ for varying $\beta$; (d) master curves for $Wi/L = 10$ and $\beta = 0.9$ for varying $\epsilon$.
 }
    \label{fig:Deandiff_W_by_L_same_beta_0.98}
\end{figure}

\subsection {Base State for the FENE-P model}
\label{subsec:BaseFENEP}
The base state velocity field, conformation tensor, and the polymer stress tensor in this case are given by
 \begin{equation}
\setlength{\arraycolsep}{0pt}
\renewcommand{\arraystretch}{1.5}
\overline{\boldsymbol{v}} = \left[
\begin{array}{ccccc}
  \overline v_{r} \\
    \overline v_{\theta} \\
    \overline v_{z} \\
\end{array}  \right]  = \left[
\begin{array}{ccccc}
  0 \\
    U(y) \\
    0 \\
\end{array}  \right] ,
\label{Base_State_Velocity_fene_p}
\end{equation}
\begin{equation}
\setlength{\arraycolsep}{3pt}
\renewcommand{\arraystretch}{2}
\boldsymbol{\overline C} = \left[
\begin{array}{ccccc}
    \overline C_{rr} &  \overline C_{r\theta} &  \overline C_{rz} & \\
     \overline C_{\theta r} &  \overline C_{\theta \theta} &   \overline C_{\theta z} & \\
      \overline C_{zr} &  \overline C_{z\theta} &   \overline C_{zz} & \\
\end{array}  \right]  =  \left[
\begin{array}{ccccc}
  \frac{1}{\bar f} & \frac{W\!i~ \overline C_{rr} (U'- \frac{U \epsilon}{1+ \epsilon y})}{\bar f} & 0  & \\
    \frac{W\!i~ \overline C_{rr} (U'- \frac{U \epsilon}{1+ \epsilon y})}{\bar f} &  \frac{1  +  2 W\!i~ \overline C_{r\theta}( U' -\frac{U \epsilon}{1+ \epsilon y})}{\bar f} & 0 & \\
    0 & 0 &  \frac{1}{\bar f} \\
\end{array}  \right],
\label{Base_State_Conformation_Tensor_DF}
\end{equation}
\begin{equation}
\setlength{\arraycolsep}{3pt}
\renewcommand{\arraystretch}{2}
\boldsymbol{\overline \tau} = \left[
\begin{array}{ccccc}
  \overline \tau_{rr} & \overline \tau_{r \theta} &  \overline \tau_{rz} & \\
\overline \tau_{\theta r} & \overline \tau_{\theta \theta} & \overline \tau_{\theta z} & \\
    \overline \tau_{zr } & \overline \tau_{z \theta} &  \overline \tau_{zz} & \\
\end{array}  \right]  =  \left[
\begin{array}{ccccc}
  0 & \overline C_{rr} (U'- \frac{U \epsilon}{1+ \epsilon y}) & 0  & \\
   \overline C_{rr} (U'- \frac{U \epsilon}{1+ \epsilon y})  & 2 \overline C_{r\theta} (U'- \frac{U \epsilon}{1+ \epsilon y})  & 0 & \\
    0 & 0 &  0 \\
\end{array}  \right].
\label{Base_State_Polymeric_Stress_DF}
\end{equation}
By combining Eq.\,\ref{eq:f constitutive eq} in the base state with $\overline \tau_{r \theta} = \overline C_{rr} (U'- \frac{U \epsilon}{1+ \epsilon y}) = \frac{U' - \frac{U \epsilon}{1 + \epsilon y}}{\bar f}$ and eliminating $\bar f$, we obtain
\begin{equation}
\tau_{r \theta}^3 + \frac{L^2}{2 W\!i^2} \tau_{r \theta} - \frac{L^2}{2 W\!i^2} \left(U' - \frac{U \epsilon}{1+ \epsilon y}\right) = 0\, .
\label{Dean_Flow_Tau_r0_Cubic_eq}
\end{equation}
This equation for $\tau_{r \theta}$ is coupled to $U(y)$ through the $\theta$-component of Eq.\,\ref{eq:fluid-momentum}, which provides a second relation between $\tau_{r \theta}$ and $U(y)$. The two equations are simultaneously solved, numerically, to obtain the base state fields; more details regarding the numerical procedure can be found in \cite{Tej2024}. The base state velocity profile does depend on $W\!i$ and $L$, although only via the combination $W\!i/L$ for $L \gg 1$. Figure \ref{fig:Deandiff_W_by_L_10_beta_0.98} illustrates the collapse of the profiles for  different ($W\!i$, $L$) pairs, with $W\!i/L$ fixed\,(and for $L\gg 1$). This collapse, first demonstrated by
\cite{Yamani_McKinley2023} for simple shear flow, was later generalized by \cite{Tej2024} to steady unidirectional rectilinear and curvilinear shear flows of a FENE-P fluid.  Figure\,\ref{fig:Deandiff_W_by_L_beta_0.5} shows the velocity profiles for different $W\!i/L$, for a fixed $\beta$, where the maximum velocity $U_{max}$ is seen to increase with increasing $W\!i/L$, due to shear thinning; when normalized by $U_{max}$, expectedly, the profiles would be flatter for larger $W\!i/L$. The effect of changing $\beta$ at a fixed $Wi/L$ and $\epsilon$ is shown in Fig.\,\ref{fig:Deandiffbetas}, where shear thinning again leads to an increased maximum, relative to the Oldroyd-B limit, for the smaller $\beta$ values; the profile reverts to the Newtonian form for $\beta \rightarrow 1$. Finally, the role of varying $\epsilon$ is shown in Fig.\,\ref{fig:Deandiffeps} where, for $\epsilon \ll 1$ (narrow-gap limit), the maximum velocity occurs at the centreline; this location shifting towards the inner cylinder for finite gap-width ratios.

\subsection {Linear stability analysis}
\label{subsec:linstab}
A temporal linear stability analysis is carried out wherein the aforementioned base states (Eqs.\,\ref{Base_State_Velocity}-\ref{Base_State_Polymeric_Stress} for the Oldroyd-B model and Eqs.\,\ref{Base_State_Velocity_fene_p}-\ref{Base_State_Polymeric_Stress_DF} for the FENE-P model)  are subjected to small-amplitude  perturbations. We consider both axisymmetric and non-axisymmetric disturbances in this study. The total velocity, pressure and stress are each expressed as a sum of a base-state contribution ($\overline{\chi}$) and an imposed perturbation ($\hat{\chi}$):
\begin{equation}
    \chi = \overline{\chi} + \hat \chi ,
\end{equation}
where the variable $\chi$ represents any of the field variables. Next, the perturbation fields are assumed to be of the Fourier mode form:  
\begin{equation}
      \hat{\chi}(y,z,\theta,t)
 = \skew2\tilde{\chi}(y) \exp \left[\mathrm{i} \left(\alpha z + n \theta - \frac{\omega}{W\!i} t \right)\right].
 \label{Normal mode form}
\end{equation}
Here $\alpha$  and  $n$ are the axial and azimuthal wavenumbers respectively and $\omega = \omega_r + i \omega_i$ is the complex frequency. 
Note that axisymmetric disturbances correspond to $n = 0$, while $\alpha = 0$ corresponds to two-dimensional perturbations in the $r$-$\theta$ plane; three-dimensional disturbances correspond to $\alpha \neq 0$ and $n \neq 0$.
The flow is temporally unstable if $\omega_i > 0$. Substituting Eq.\,\ref{Normal mode form}  in the linearized versions of \eqref{eq:fluid-continuity}, \eqref{eq:fluid-momentum} and \eqref{eq:stress constitutive eq} we obtain the set of governing equations, given in Appendices\,\ref{app:Old_b_Linearized_eq} and \ref{app:FENE_P_Linearized_eq}  for the Oldroyd-B and FENE-P models, respectively.
\subsection {Numerical methods}
\label{subsec:nummethods}
\begin{figure}
  \centering
  \includegraphics[width=0.45\textwidth]{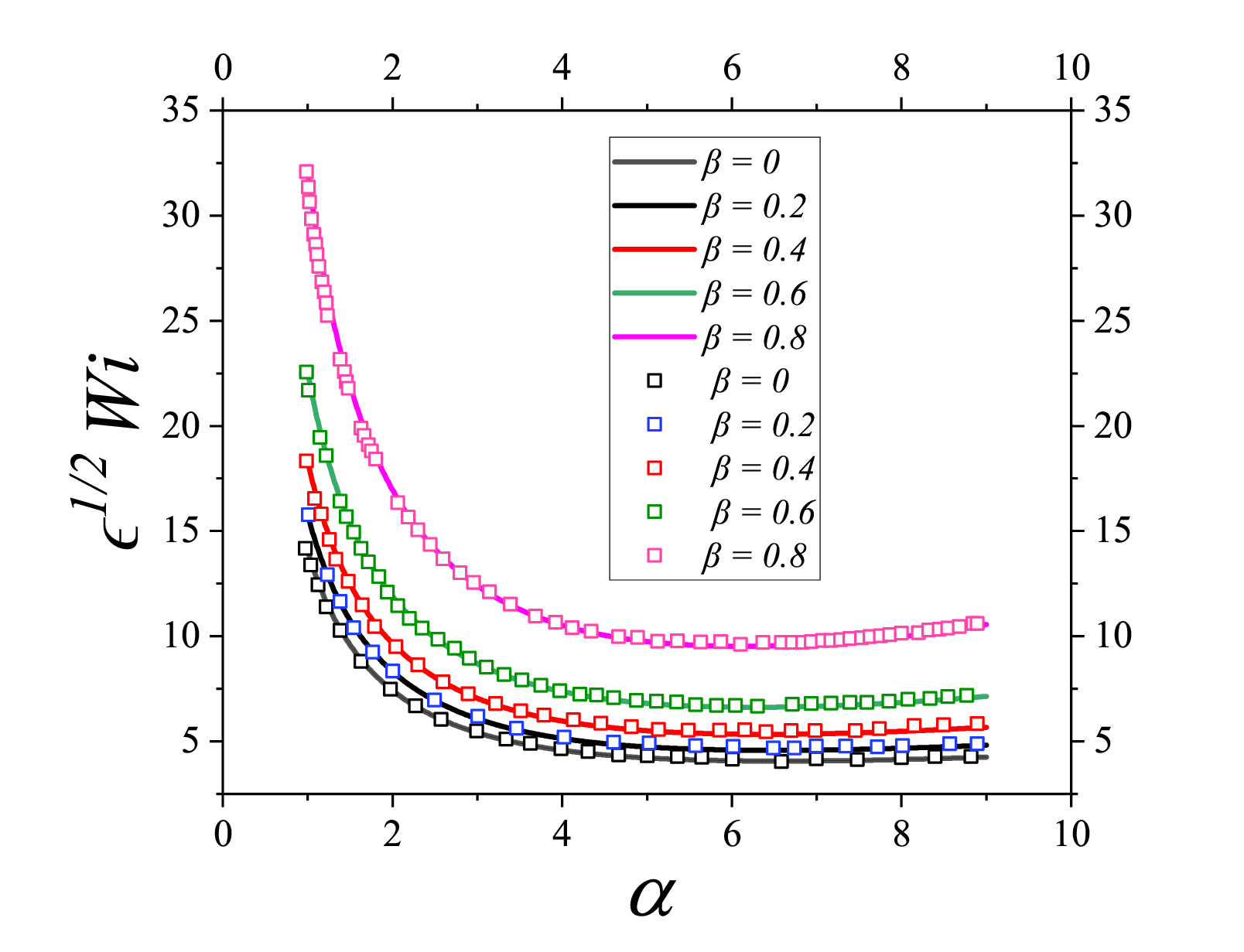}
  \caption{Benchmarking of neutral curves obtained from our numerical (spectral-cum-shooting) procedure (for $Re = 0$, $ n = 0$) with those of \cite{joo1992purely}, for Dean flow of an Oldroyd-B fluid, for different $\beta$. Continuous lines show results from the present work, while discrete points represent the data of \cite{joo1992purely}.}
  \label{fig:HSM_NeutralCurves_in_narrowgaplimit}
\end{figure}

We solved the linearized governing equations using a spectral method \citep{trefethen2000spectral, weideman2000matlab} and a shooting procedure. In the spectral method, we expand the dynamical variables as a finite sum of Chebyshev polynomials, and substitute this expansion in the linearized governing equations -- (\ref{Normal mode form continuity}) – (\ref{Normal mode form ZZ stress}), for the Oldroyd-B model or (\ref{FENE_P_Normal_mode_form_continuity}) – (\ref{FENE_P_Normal_mode_form_Z_Momemtum}) for the FENE-P model -- to obtain a generalized eigenvalue problem as follows
\begin{equation}
   \boldsymbol{A} \boldsymbol{x} = \omega \boldsymbol{B} \boldsymbol{x} \, .
\end{equation} 
Here $\boldsymbol{A}$ and $\boldsymbol{B}$ are coefficient matrices, with  $\boldsymbol{x} = (\Tilde v _r , \Tilde v _\theta, \Tilde v _z,  \Tilde p,  \Tilde \tau _{rr}, \Tilde \tau _{r\theta}, \Tilde \tau _{rz}, \Tilde \tau _{\theta\theta}, \Tilde \tau _{\theta z}, \Tilde \tau _{zz} )^T$ for the Oldroyd-B model and  $\boldsymbol{x} = (\Tilde v _r , \Tilde v _\theta, \Tilde v _z,  \Tilde p,  \Tilde C _{rr}, \Tilde C_{r\theta}, \Tilde C_{rz}, \Tilde C_{\theta\theta}, \Tilde C _{\theta z}, \Tilde C_{zz} )^T$ for the FENE-P model.
The size of the $\boldsymbol{A}$ matrix is $10N \times 10N $, where $N$ is the number of Gauss–Lobatto collocation points. The generalized eigenvalue problem is solved using the ‘eig’ solver of \emph{Matlab}. Depending on the parameter regime, the number of collocation points $N$ required to capture physically genuine eigenvalues differs. Convergence in this method is achieved by comparing the eigenspectra for two different values of $N$, and filtering them using a prescribed tolerance. The shooting method is based on a Runge-Kutta integrator with a Gram-Schmidt orthonormalization procedure \citep{ho1977stability}, and a Newton-Raphson iteration for the eigenvalue. This procedure is implemented only for the Oldroyd-B model. It is known that spectral method can give unphysical (spurious) eigenvalues \citep{SchmidHenningson2001}, but the shooting method gives genuine eigenvalues, given a good initial guess. Thus, the genuineness of the eigenvalues, obtained from the spectral method, is confirmed by providing them as initial guesses for the shooting method. To benchmark the implementation of our numerical methodology in the narrow-gap limit, we compare (Table\,\ref{Scaling1} and Figure\,\ref{fig:HSM_NeutralCurves_in_narrowgaplimit})  results from our procedure with those of \cite{joo1992purely}, finding excellent agreement.
\begin{table}
  \begin{center}
\def~{\hphantom{0}}
  \begin{tabular}{lcc}
  \vspace{5pt}
   Dean Flow  & $\alpha_c$ & $ \epsilon^{1/2} W\!i_c $  \\
   Joo and Shaqfeh & 6.6 $\pm$ 0.1 &  4.06 $\pm$ 0.02    \\ 
   Present &  6.56 & 4.06     \\
  \end{tabular}
  \caption{Validation of UCM $(\beta = 0)$ results with those of \cite{joo1992purely}.  }
  \label{Scaling1}
  \end{center}
\end{table}

\begin{figure}
  \centering
  \subfigure [$ \alpha = 7 $]{\includegraphics[width=0.45\textwidth]{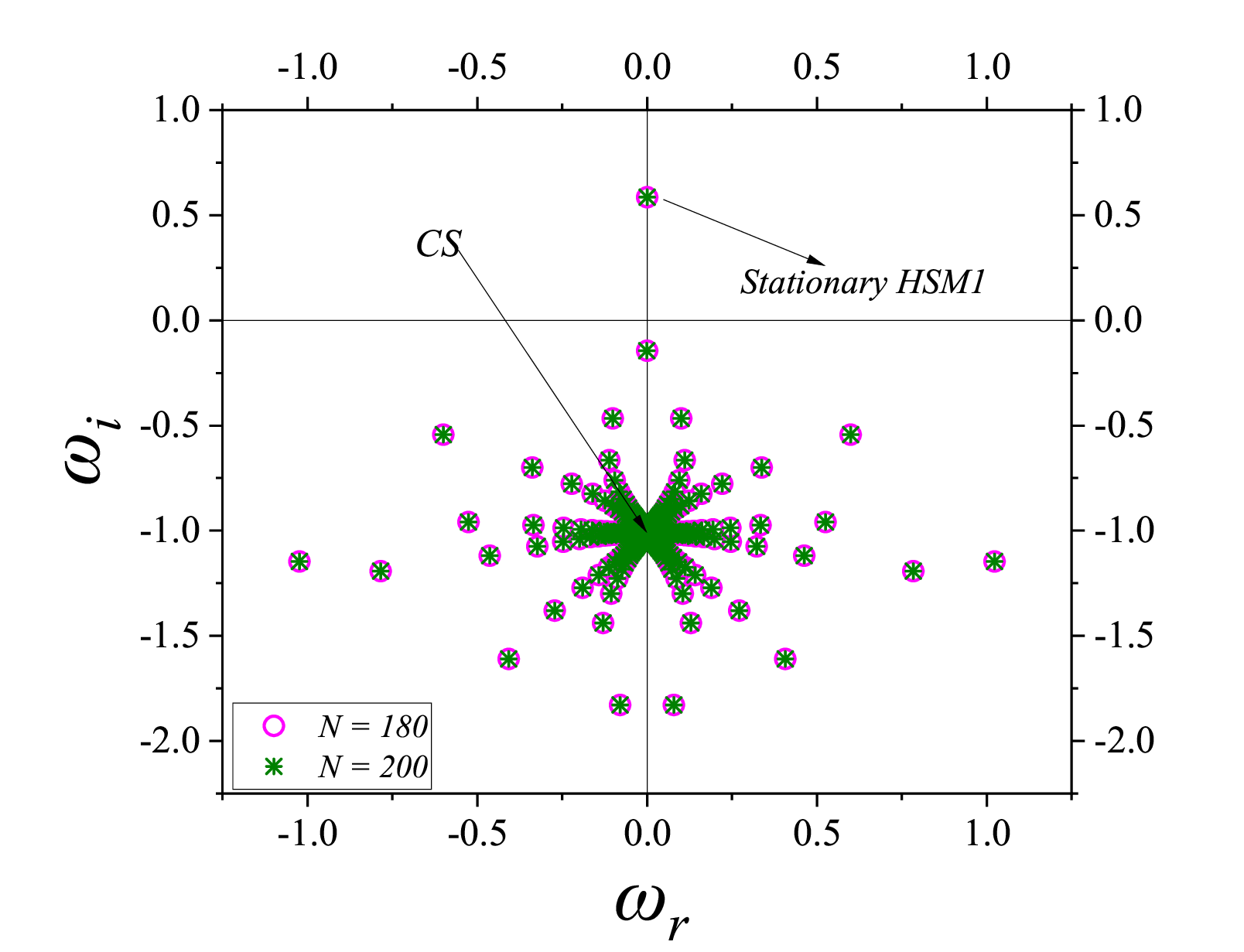}\label{fig:Axisymmetric_Stationary_alpha_7}}
   \subfigure[$ \alpha = 14 $]{\includegraphics[width=0.45\textwidth]{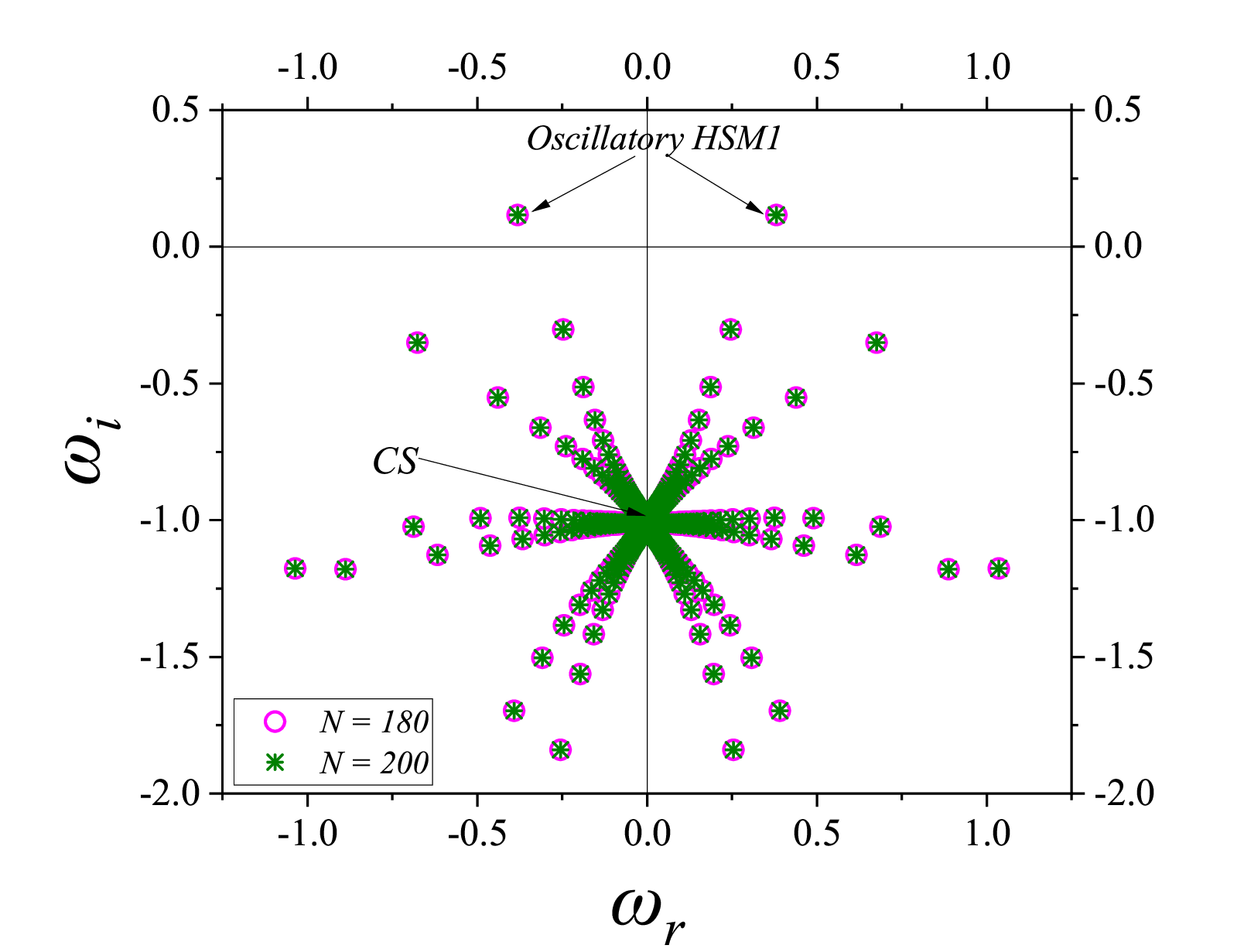}\label{fig:Axisymmetric_Oscillatory_alpha_14}}
  \caption{Eigenspectrum showing stationary ($\alpha = 7$) and propagating  ($\alpha = 14$) modes of instability via the HSM1 mode in Dean flow of an Oldroyd-B fluid at $Re = 0$, $n = 0$, $\beta = 0.98$, $\epsilon = 0.1$, and $W\!i = 25$.} \label{Axisymmetric_Stationary_vs_Oscillatory}
\end{figure}

\section {Nature of the elastic spectrum in Dean flow}
\label{sec:Nature_of_Spectrum}
Herein, we provide an overview of the elastic eigenspectrum for Dean flow, 
before proceeding to discuss the two classes of unstable modes (viz., CM and HSM) in subsequent sections. A clear understanding of the structure of the elastic spectrum is necessary to interpret the origin of different unstable modes, and  to also enable us to distinguish them from one another.
\subsection{Axisymmetric disturbances ($n = 0$)}
\label{subsec:spectrum_axisym}
Figure\,\ref{Axisymmetric_Stationary_vs_Oscillatory} shows the axisymmetric elastic spectrum for Dean flow of an Oldroyd-B fluid at two different axial wavenumbers ($\alpha$). For both $\alpha$'s, there appear to be a large number of discrete modes, perhaps even an infinite number of them, accumulating towards the point  $(\omega_r = 0, \omega_i = -1)$; this point is shown below to be part of the continuous spectrum for $n = 0$.  In both spectra, there are also unstable modes, which could either be stationary ($\omega_r = 0$; see Fig.\,\ref{fig:Axisymmetric_Stationary_alpha_7}) or propagating ($\omega_r \neq 0$; see Fig.\,\ref{fig:Axisymmetric_Oscillatory_alpha_14}) ones; in the latter case they appear symmetrically on either side of the imaginary axis. 
For the stationary scenario, the least stable/most unstable of the discrete modes is the classical axisymmetric HSM, originally identified by \cite{joo_shaqfeh_1991,joo_shaqfeh_1994}, and is labelled as `HSM1' henceforth. A more detailed discussion of the parameter regimes in which HSM1 is stationary or propagating is provided in Appendix\,\ref{sec:Appendixspectra}.  It will be seen therein that the stationary-to-propagating modal transition can happen with an increase in $\alpha$ for a fixed $W\!i$, or for a fixed $\alpha$ and with decreasing $W\!i$.

\begin{figure}
  \centering
  \subfigure [$ n = 0 $]{\includegraphics[width=0.35\textwidth]{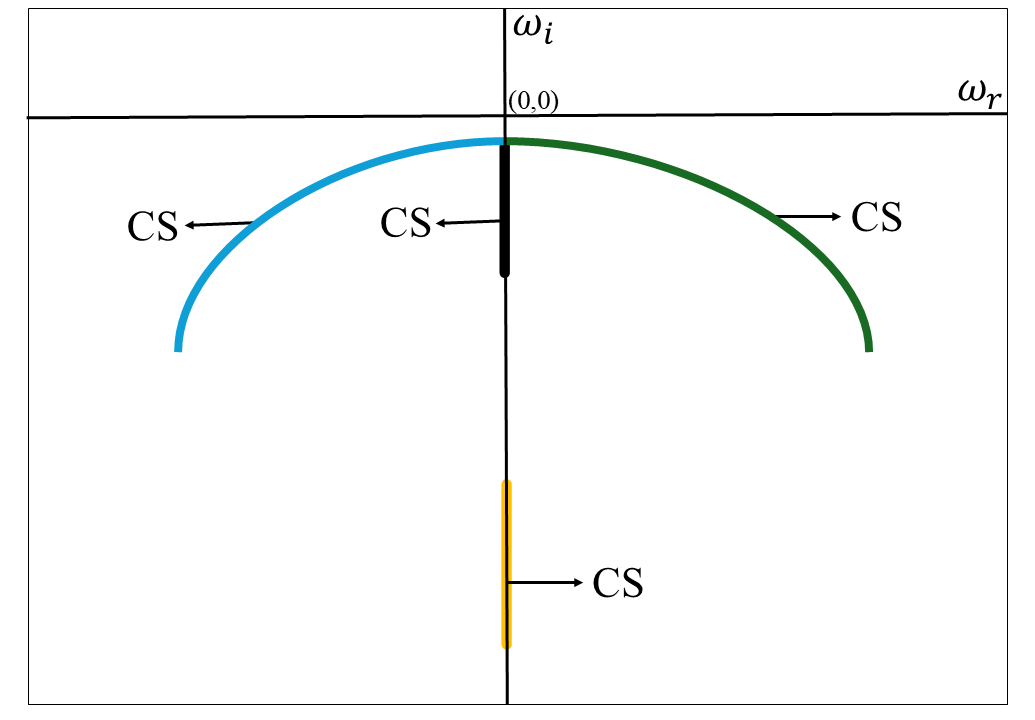}\label{fig:schematic_axi}}
  \quad \quad
   \subfigure[$ n \neq 0 $]{\includegraphics[width=0.35\textwidth]{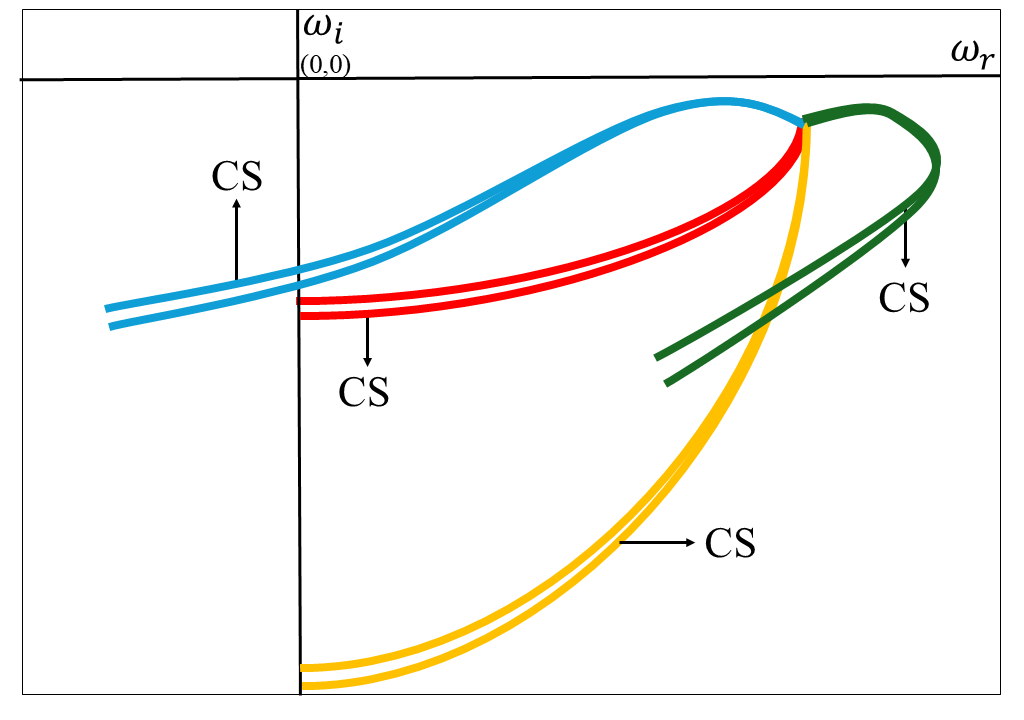}\label{fig:schematic_nonaxi}}
  \caption{Schematic of the continuous spectra branches in Dean flow of a FENE-P fluid with $W\!i/L \sim O(1)$ and $\beta \rightarrow 1$.} \label{fig:CS_schematic_Dean}
\end{figure}

\begin{figure}
  \centering
   \subfigure [$ L = 10^4 $ (Oldroyd-B)]{\includegraphics[width=0.35\textwidth]{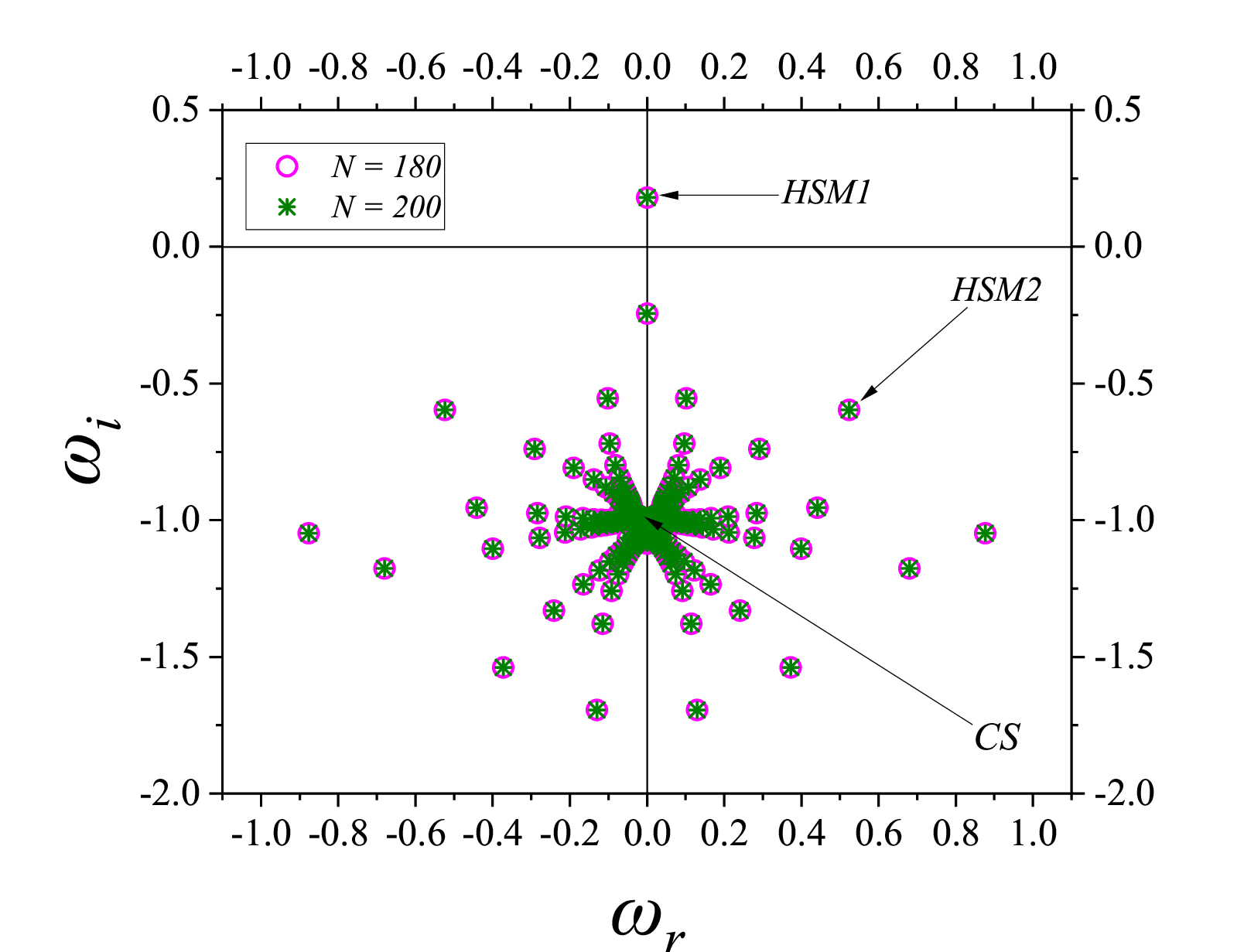}\label{fig:Axisymm_HSM_dean_e_0.1_beta_0.98_L_10000}}
   \subfigure[$ L = 5000 $]{\includegraphics[width=0.35\textwidth]{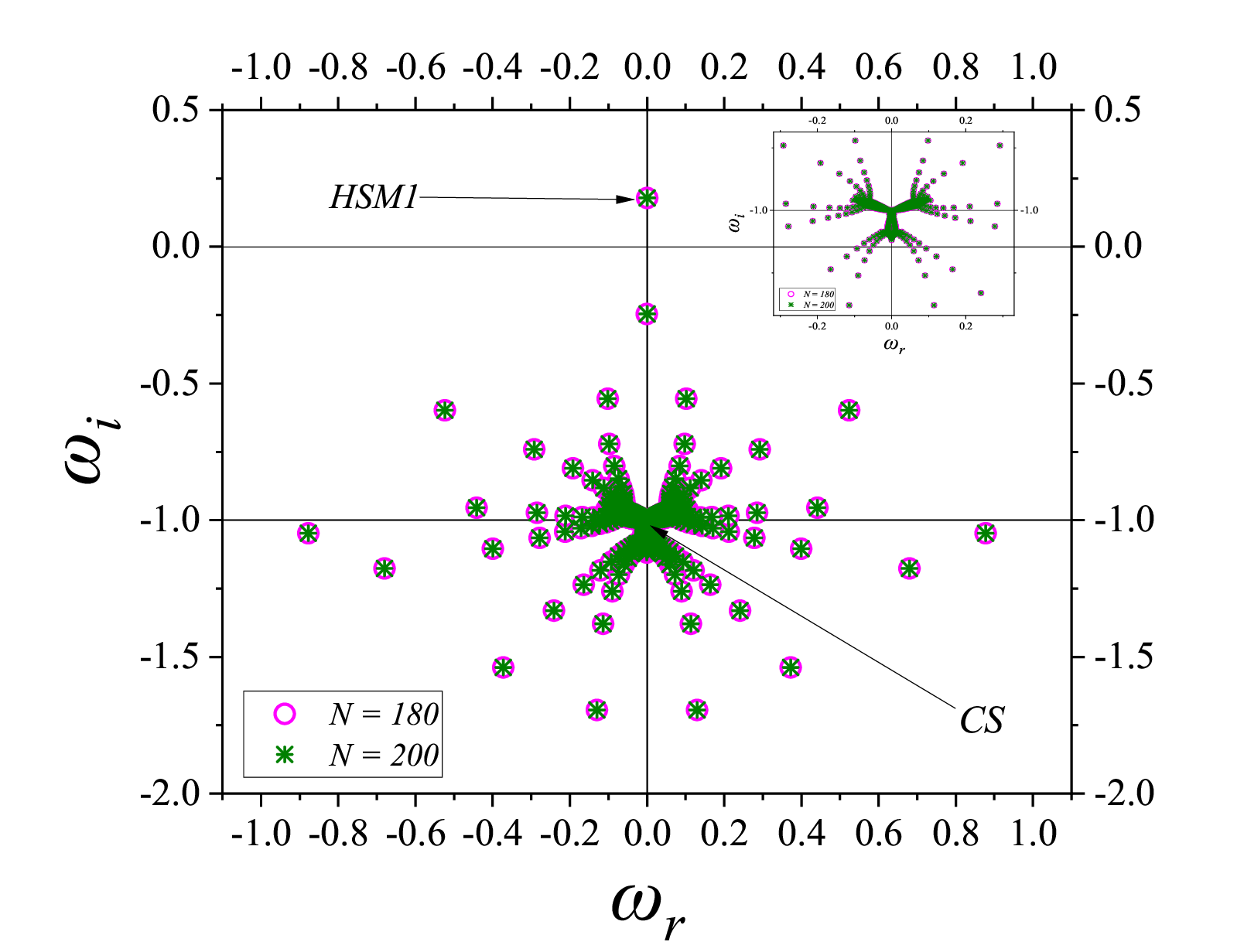}\label{fig:Axisymm_HSM_dean_e_0.1_beta_0.98_L_5000}}
   \subfigure[$ L = 1000 $]{\includegraphics[width=0.35\textwidth]{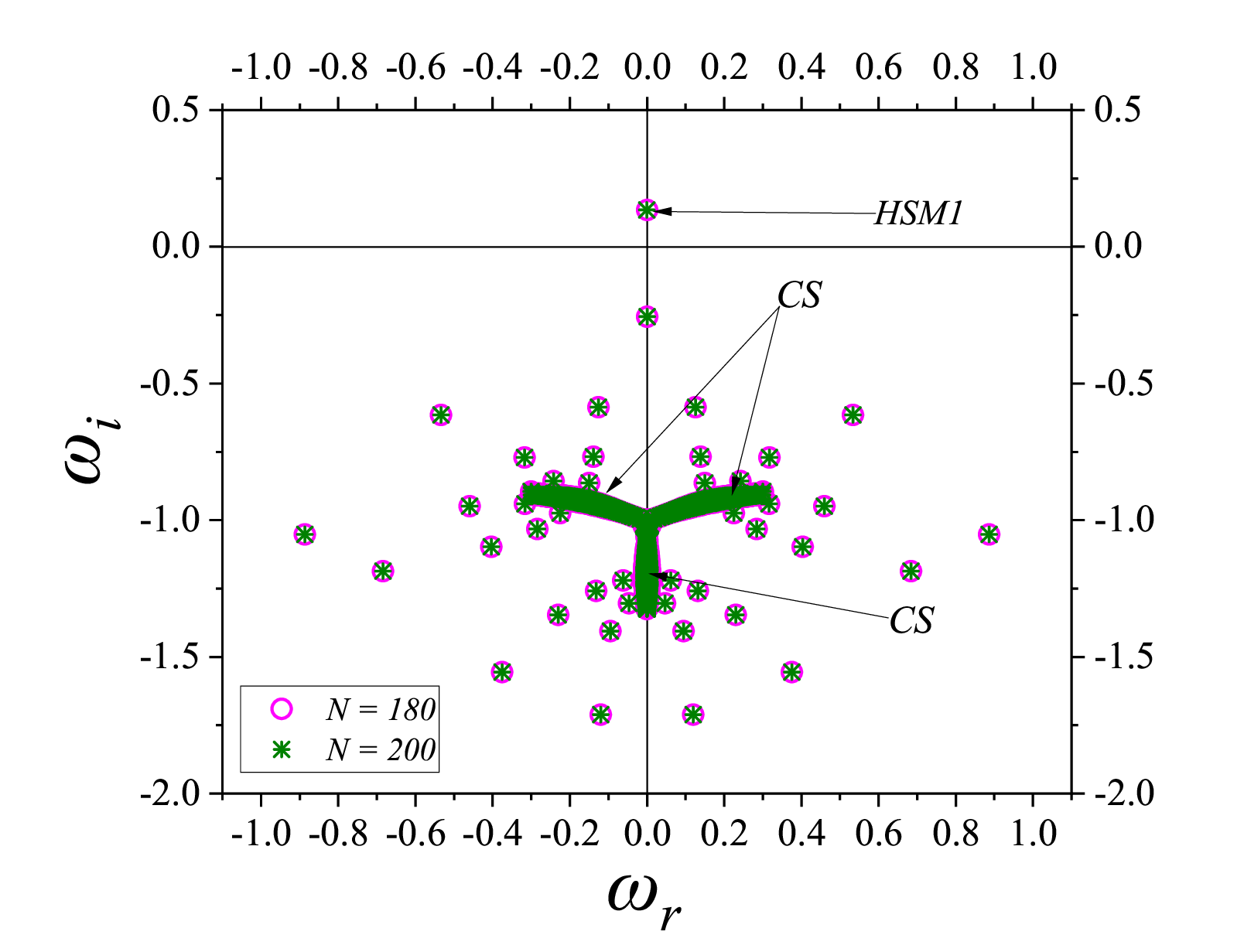}\label{fig:Axisymm_HSM_dean_e_0.1_beta_0.98_3}}
  \subfigure[$ L = 500 $]{\includegraphics[width=0.35\textwidth]{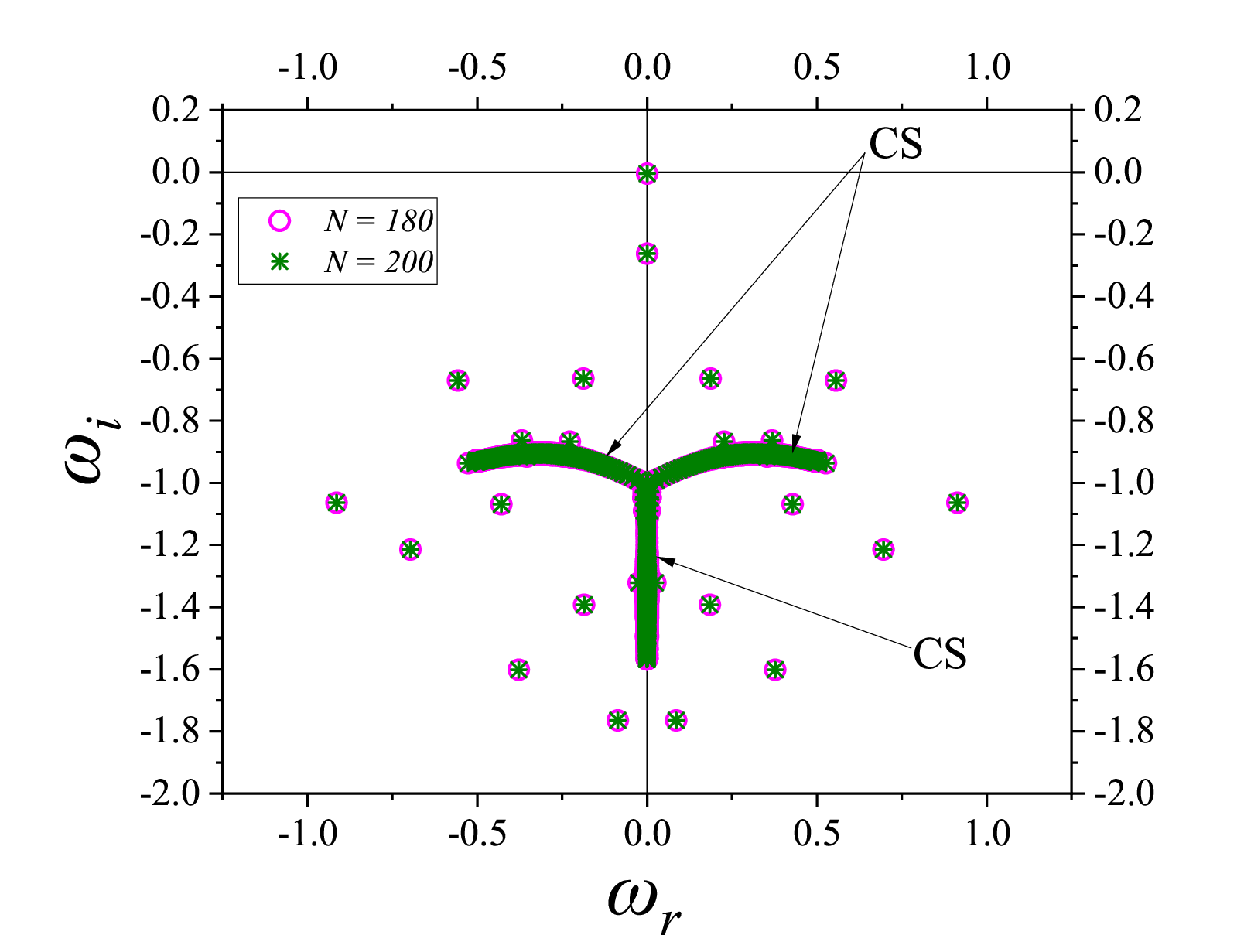}\label{fig:Axisymm_HSM_dean_e_0.1_beta_0.98_L_500}}
   \subfigure [$ L = 398 $]{\includegraphics[width=0.35\textwidth]{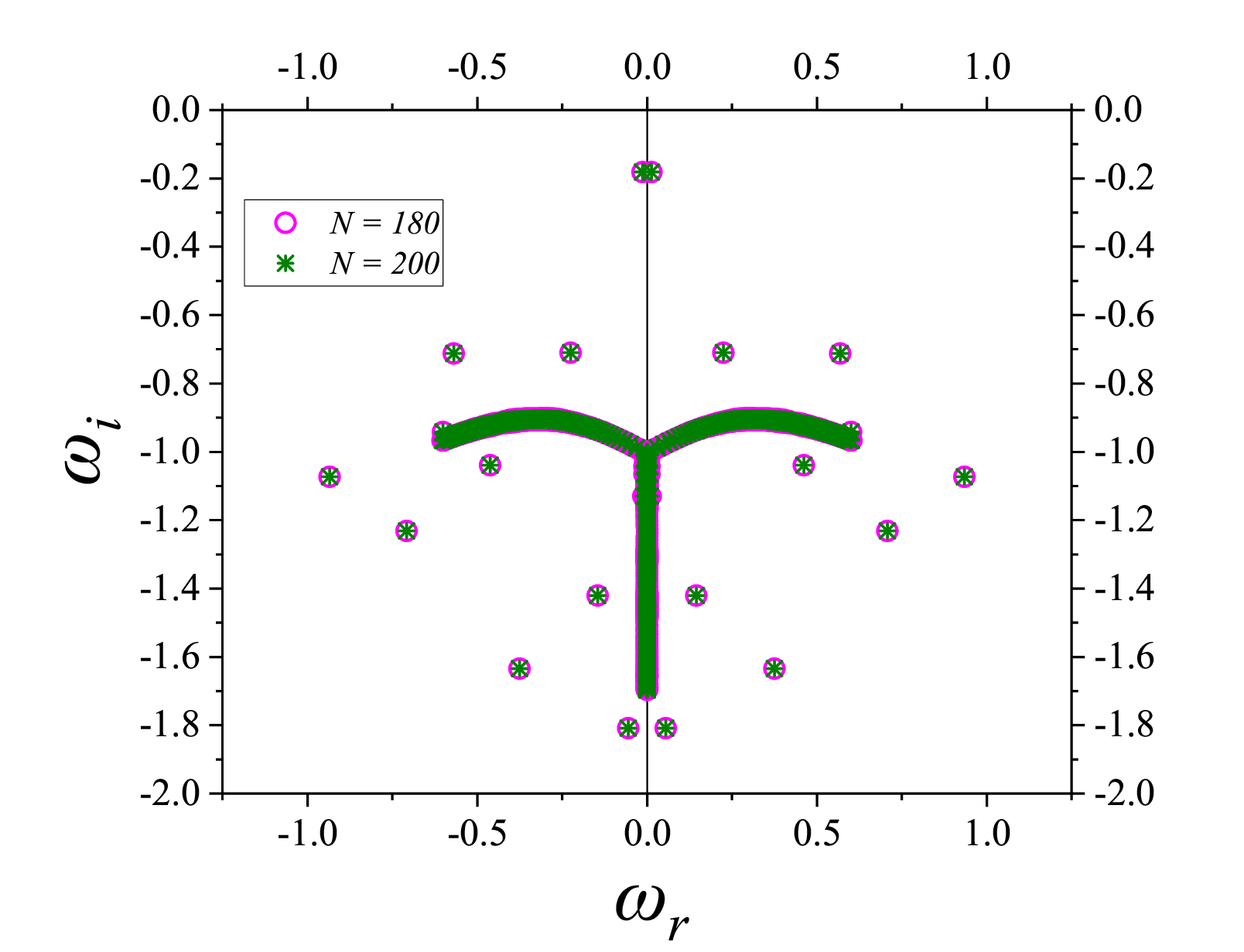}\label{fig:Axisymm_HSM_dean_e_0.1_beta_0.98_L_395}}
  \subfigure[$ L = 390 $]{\includegraphics[width=0.35\textwidth]{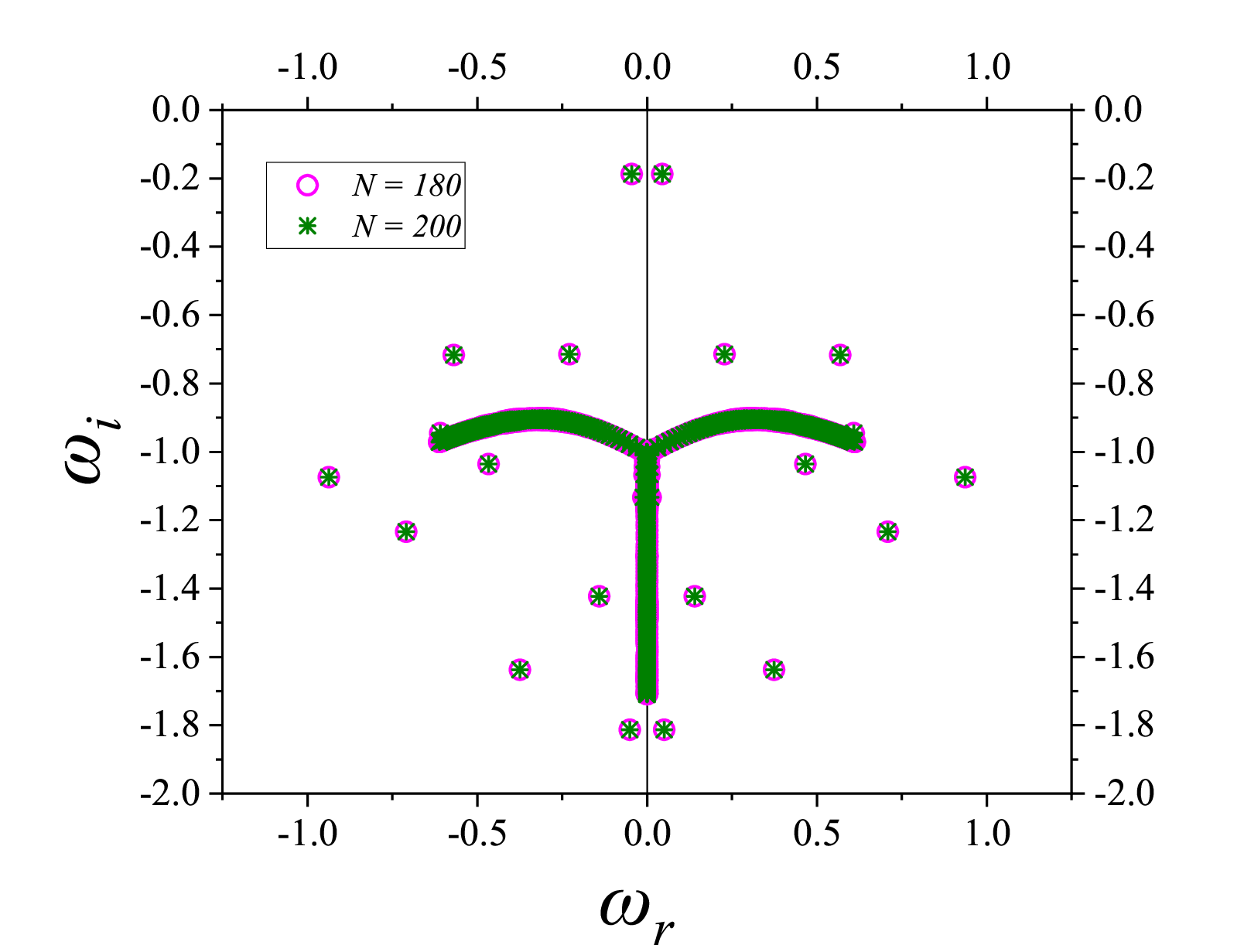}\label{fig:Axisymm_HSM_dean_e_0.1_beta_0.98_L_390}}
   \subfigure [$ L = 150 $]{\includegraphics[width=0.35\textwidth]{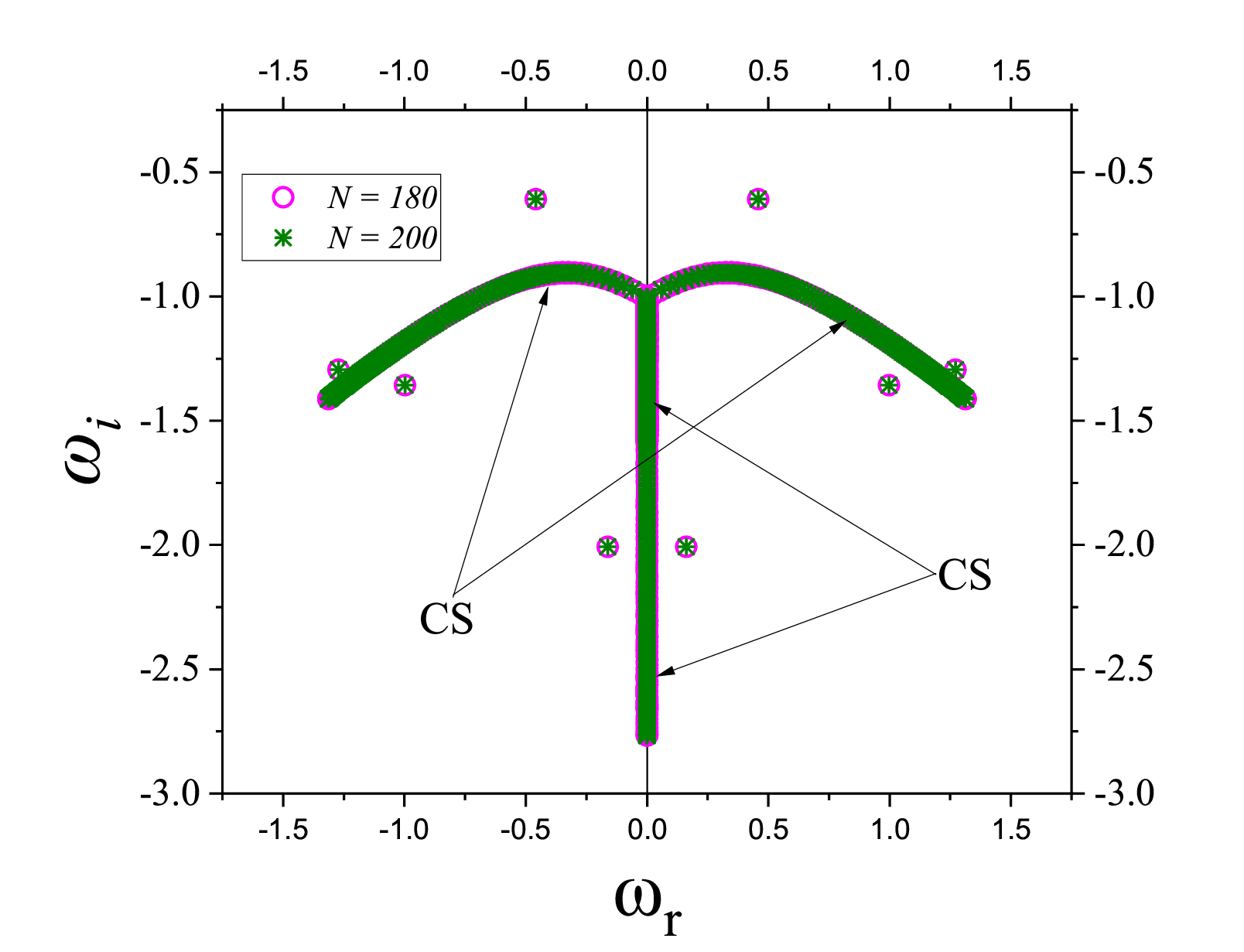}\label{fig:Axisymm_HSM_dean_e_0.1_beta_0.98_L_150}}
  \subfigure[$ L = 100 $]{\includegraphics[width=0.35\textwidth]{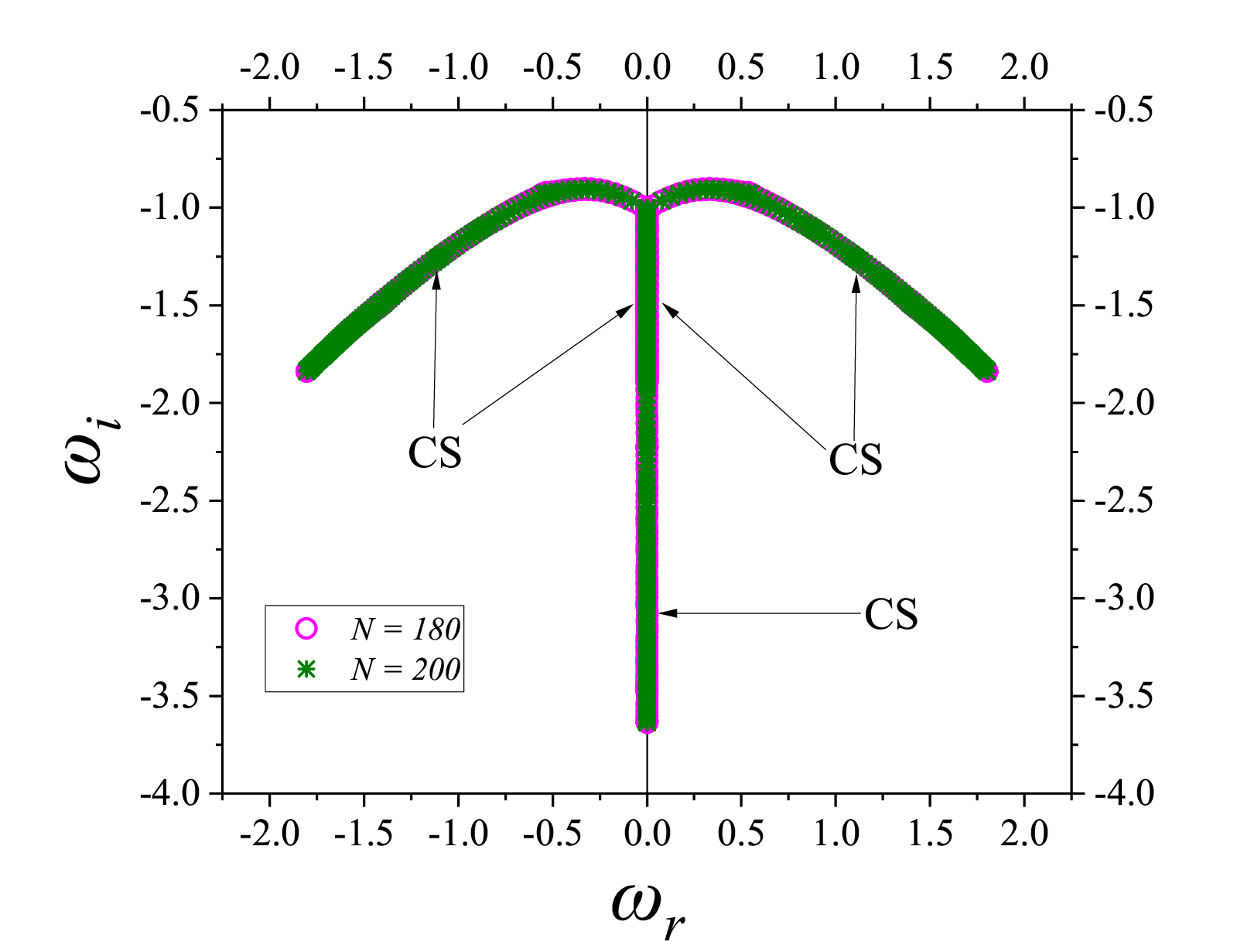}\label{fig:Axisymm_HSM_dean_e_0.1_beta_0.98_L_100}}
  \caption{Eigenspectra for Dean flow of FENE-P fluids at different $L$'s; data for $n = 0$, $\epsilon = 0.1$, $\alpha = 7$, $\beta = 0.98$, $Wi = 20$, and $Re = 0$.}
\label{fig:multi_subfigures_FENE_P_Axisymmetric_same_Wi_diff_L}
\end{figure}

Now, it is well known that the eigenspectra for viscoelastic flows also contain (stable) continuous spectra (CS)  of eigenvalues \citep{Wilson1999,Roy_Garg_Reddy_Subramanian_2022}. These are obtained by setting the coefficient of the highest order derivative of the governing linearized equation for the wall-normal velocity perturbation to zero.
For rectilinear flows such as plane Couette and Poiseuille flows of an Oldroyd-B fluid, this leads to two distinct CS with decay rates $\omega_i = -1$ and $\omega_i = -1/\beta$. In general, these CS are horizontal line segments in the $\omega_r$-$\omega_i$ plane whose length is proportional to  $k \Delta V$, with $\Delta V$ being the  base range of velocities, and $k$ the streamwise wavenumber. Along similar lines, for Dean flow, the horizontal extent of the CS is proportional to $n \Delta{V}$.
For Dean flow subjected to axisymmetric disturbances ($n = 0$) as in Fig.\,\ref{Axisymmetric_Stationary_vs_Oscillatory},  the CS,
within the Oldroyd-B framework, therefore degenerate to a pair of points on the imaginary axis with $\omega_i = -1$ and $-1/\beta$; the
the two points approaching each other in the limit $\beta \rightarrow 1$ ($\beta = 0.98$ in Fig.\,\ref{Axisymmetric_Stationary_vs_Oscillatory}).

We now turn to the effect of $L$ on the spectra. As $L$ is decreased, owing to shear thinning of rheological properties, the nature of CS for both rectilinear and curvilinear shearing flows, including Dean flow in particular, is significantly more complex compared to that for the Oldroyd-B fluid mentioned above. In the interests of brevity, we provide here only the key features of the CS that are relevant to the present work; a detailed discussion is provided in \cite{Mohanty_etal_CS}. A schematic of the CS for Dean flow, pertinent to the dilute-solution limit ($\beta \gtrsim 0.95$), and for $W\!i/L \sim O(1)$ (so shear thinning effects remain significant), is shown in Fig.\,\ref{fig:schematic_axi} for $n = 0$.
As $L$ is decreased from the Oldroyd-B limit, the `point CS' on the imaginary axis, in the Oldroyd-B limit mentioned above, turn into vertical line segments. The two line segments are shown to be distinct in the schematic, but may overlap depending on the particular choice of parameters.
The reason underlying the extended CS, even for $n = 0$, is due to the variation of the base-state Peterlin function $\bar{f}$ as a function of the wall-normal coordinate, which leads to a range of relaxation times, and thence, decay rates.
Concurrently, a pair of symmetrically placed `wing'-like CS emerge on either side of the imaginary axis \citep{Mohanty_etal_CS}. As $L$ is decreased, the horizontal extent of the wings and the vertical line segments (belonging to the CS) increases. The computed numerical spectra illustrating these features are shown in Fig.\,\ref{fig:multi_subfigures_FENE_P_Axisymmetric_same_Wi_diff_L}, for different $L$'s, at a fixed $W\!i$.

Apart from the CS, the spectra in Fig.\,\ref{fig:multi_subfigures_FENE_P_Axisymmetric_same_Wi_diff_L} also contain discrete modes,
which are not depicted in the Fig.\,\ref{fig:schematic_axi} schematic.
In fact, Figs.\,\ref{fig:Axisymm_HSM_dean_e_0.1_beta_0.98_L_10000}--\ref{fig:Axisymm_HSM_dean_e_0.1_beta_0.98_L_100} show that decreasing $L$ has a strong influence on the configuration of discrete modes. 
 In contrast to the Oldroyd-B case above, where there were (apparently) an infinite number of discrete modes accumulating at the (point) CS,
 a decrease in $L$ leads to a progressive reduction in the number of modes. 
 When $L$ is decreased to $398$, the most unstable mode changes from a stationary one, to a pair of propagating ones; this change is due to a coalescence of HSM1 and the next (least stable) mode, followed by a bifurcation. The propagating pair moves down towards the CS as $L$ is decreased further, this shear-thinning-induced stabilization being discussed further in Sec.\,\ref{subsec:Lonaxisym} below.
 Eventually, for $L = 100$, there are no discrete modes, either stable or unstable, and the  spectrum comprises solely of modes belonging to the CS for $W\!i \gtrsim 16$ (see Fig.\,\ref{fig:Axisymm_HSM_dean_e_0.1_beta_0.98_L_100}).

\begin{figure}
  \centering
   \subfigure[Oldroyd-B, $Wi = 20$]{\includegraphics[width=0.45\textwidth]{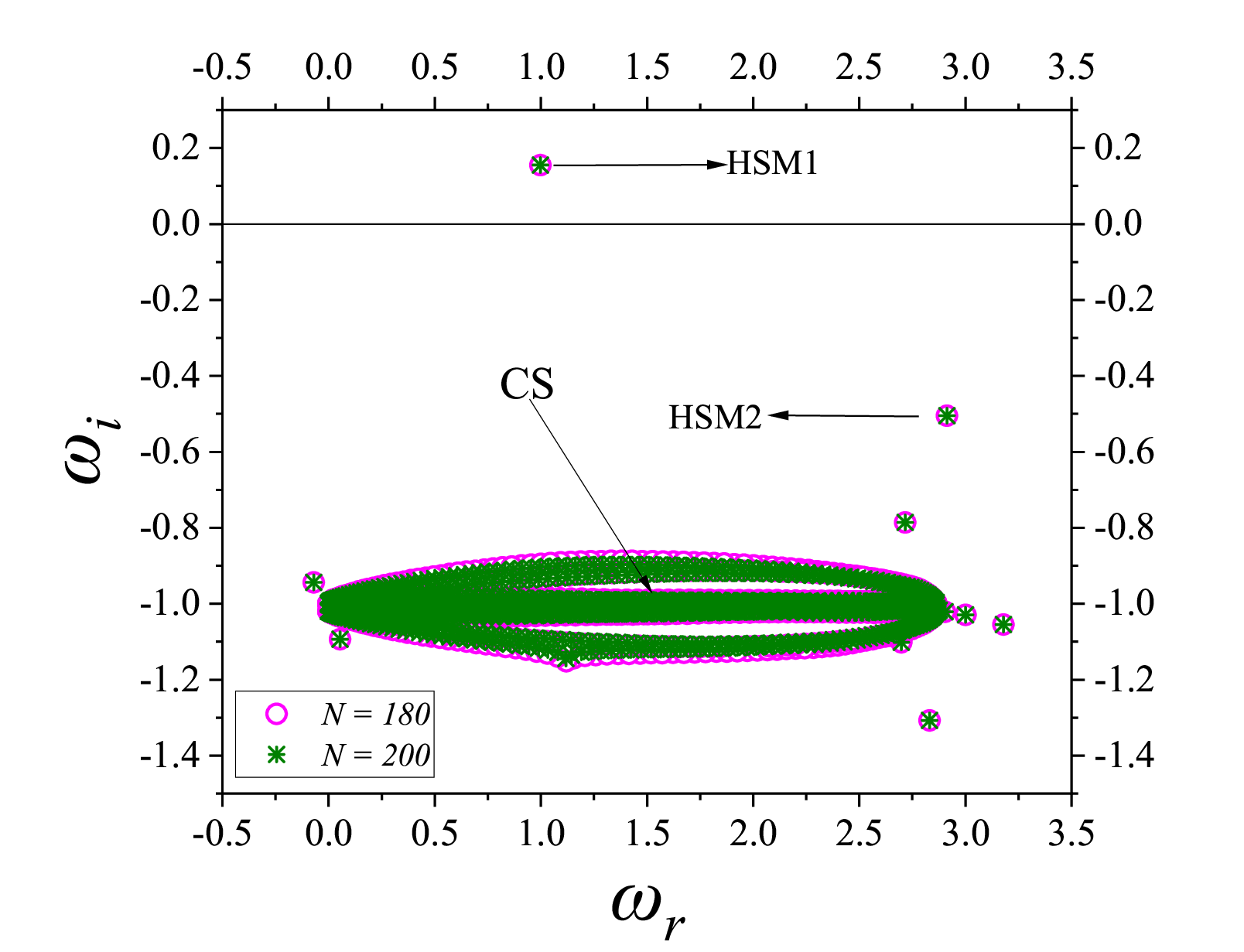}\label{fig:Non_Axisymmetric_Spectrum_mode_1_and_mode_2}}
     \subfigure[$ L = 200 $, $W\!i = 400$ ]{\includegraphics[width=0.45\textwidth]{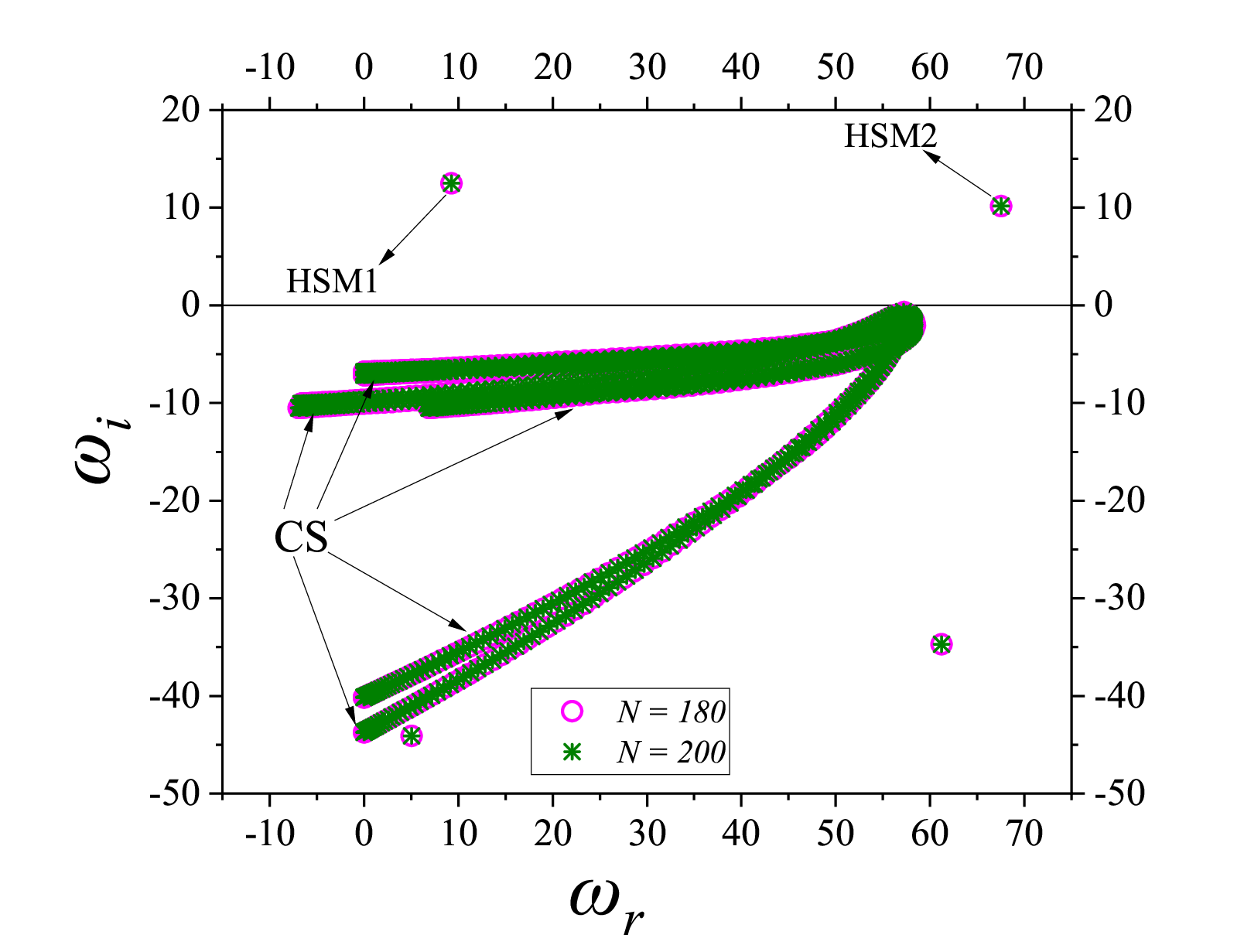}\label{Showing_HSM2_L_200} }
  \caption{Eigenspectrum for non-axisymmetric ($n = 1$) disturbances showing both HSM1 and HSM2 in Dean flow of an Oldroyd-B and FENE-P ($L = 200$) fluids. Data for $\alpha = 7$, $\beta = 0.98$, and $ \epsilon = 0.1$.}
   \label{fig:Spectrum_mode_1_and_mode_2}
\end{figure}

\subsection{Non-axisymmetric disturbances ($n \neq 0$)}
\label{subsec:nonaxy_spectrum}
  The schematic of the CS for $n \neq 0$ is shown in Fig.\,\ref{fig:schematic_nonaxi} for FENE-P fluids with $W\!i/L \sim O(1)$; again, the details of how these are obtained theoretically are discussed in \cite{Mohanty_etal_CS}.  
  There are four distinct CS (for $L \gtrsim 50$) in Fig.\,\ref{fig:schematic_nonaxi}, and each CS has two branches. These branches arise due to the lack of a mid-plane symmetry in Dean flow, and degenerate in the case of  plane-Poiseuille flow.
  The actual numerical spectra for a nonzero $n$ ($n = 1$), for both the Oldroyd-B and FENE-P cases, are shown in Figs.\,\ref{fig:Non_Axisymmetric_Spectrum_mode_1_and_mode_2} and \ref{Showing_HSM2_L_200}.
  
  In Fig.\,\ref{fig:Non_Axisymmetric_Spectrum_mode_1_and_mode_2}, we present the spectra for the same set of parameters as for the axisymmetric case ($n = 0$) in Fig.\,\ref{fig:Axisymm_HSM_dean_e_0.1_beta_0.98_L_10000}. For $n = 1$, the theoretical CS are no longer points even in the Oldroyd-B fluid,  but are horizontal  line segments  
  located again at $\omega_i = -1$ and $-1/\beta$; these are only resolved as balloons in Fig.\,\ref{fig:Non_Axisymmetric_Spectrum_mode_1_and_mode_2}.
 
 The constitution of the discrete modes in the spectra for $n = 1$ also differs qualitatively from those for $n = 0$. In particular, the discrete modes in Fig.\,\ref{fig:Axisymm_HSM_dean_e_0.1_beta_0.98_L_10000} (for $n = 0$) with negative $\omega_r$ are virtually absent in the spectrum for $n = 1$ (see Fig.\,\ref{fig:Non_Axisymmetric_Spectrum_mode_1_and_mode_2}). The absence is understandable since a negative $\omega_i$ implies a propagation in the direction opposite to the base state shear flow, which is an unlikely possibility in the absence of inertia. We have verified (data not shown) that as $n$ is increased continuously from $0$ to $1$, 
 there is also an overall reduction in the number of discrete modes, this reduction arising because  many of these discrete modes disappear into the CS.
 In Fig.\,\ref{fig:Non_Axisymmetric_Spectrum_mode_1_and_mode_2}, the classical hoop stress mode (HSM1) is the mode  that dictates the instability in the Oldroyd-B limit. This mode could be stationary or propagating for axisymmetric disturbances, but is propagating  for non-axisymmetric disturbances. 
 
\begin{figure}
  \centering
  \subfigure[$ \epsilon = 0.1 , n = 1, \alpha = 7, W\!i = 20 $]{\includegraphics[width=0.45\textwidth]{HSM1_and_HSM2_Oldroyd_B_nonaxisymmetric_n_1_e_0.1_alpha_7_beta_0.98_Wi_20.eps}\label{Showing_HSM2_Wi_20}}
   \subfigure[$ \epsilon = 0.1 , n = 1, \alpha = 7 , W\!i = 60 $]{\includegraphics[width=0.45\textwidth]{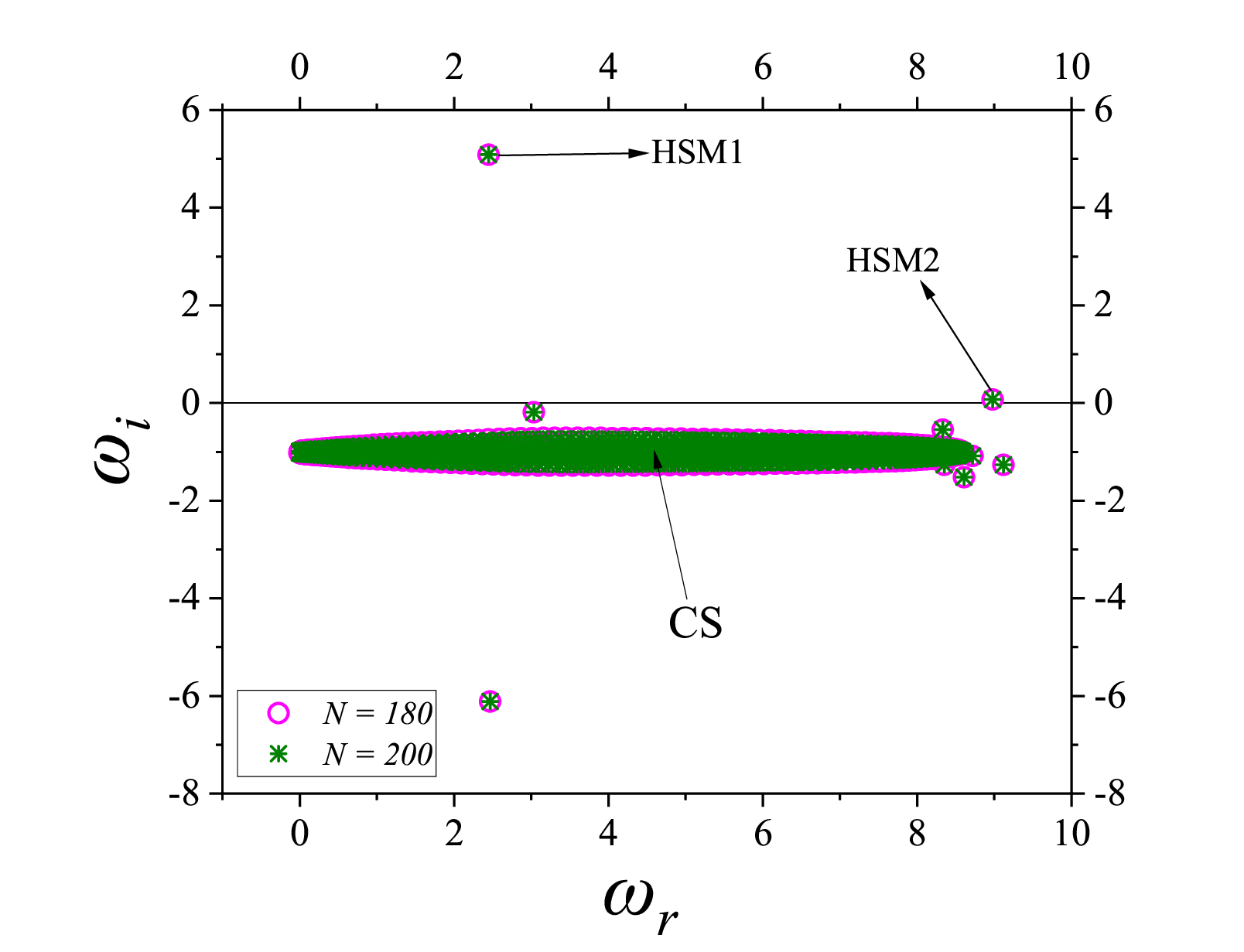}\label{Showing_HSM2_Wi_60}}
    \subfigure[$ \epsilon = 0.1 , n = 1, \alpha = 7, W\!i = 200 $]{\includegraphics[width=0.45\textwidth]{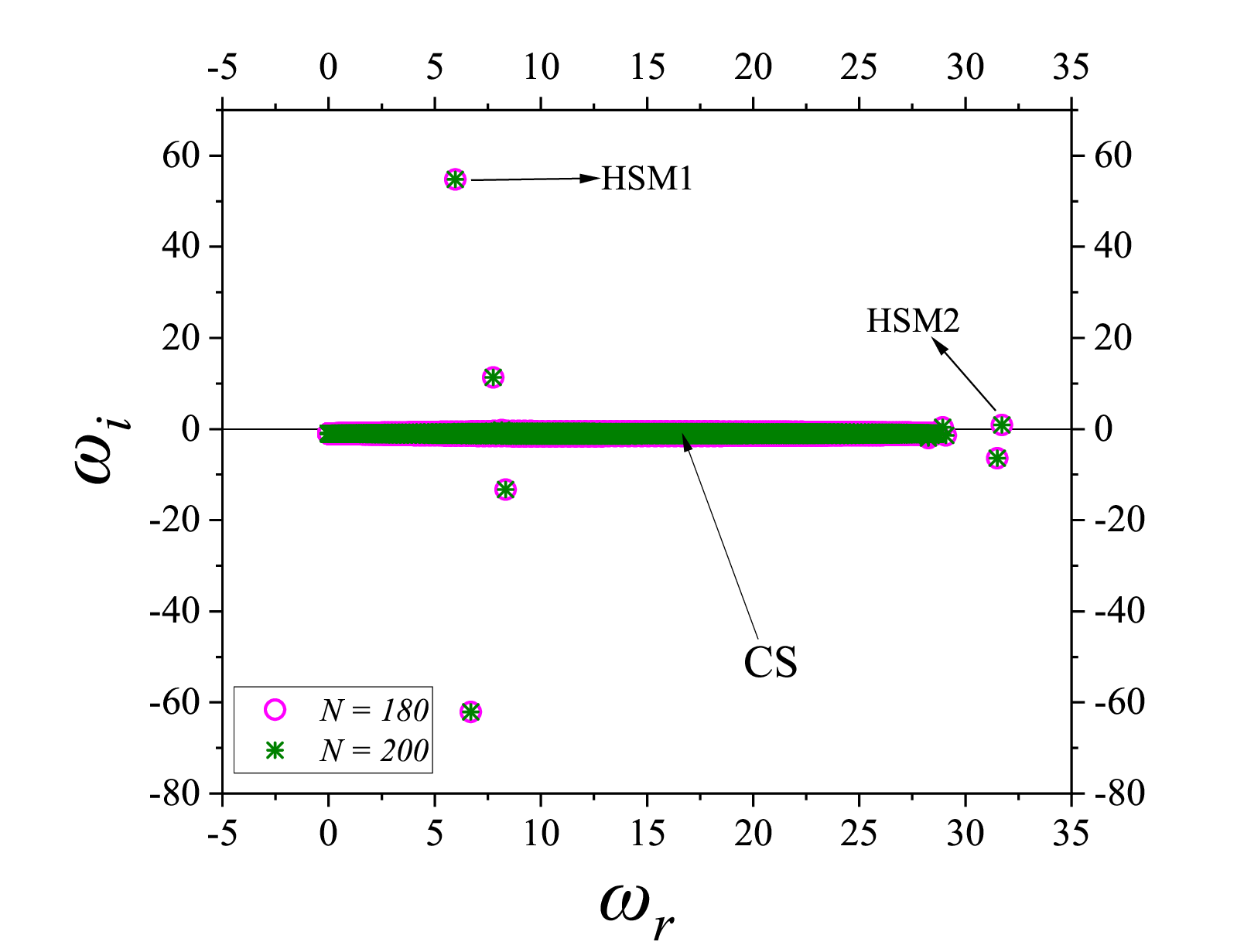}\label{Showing_HSM2_Wi_200}}
   \subfigure[$ \epsilon = 0.1 , n = 1, \alpha = 7 , W\!i = 400 $]{\includegraphics[width=0.45\textwidth]{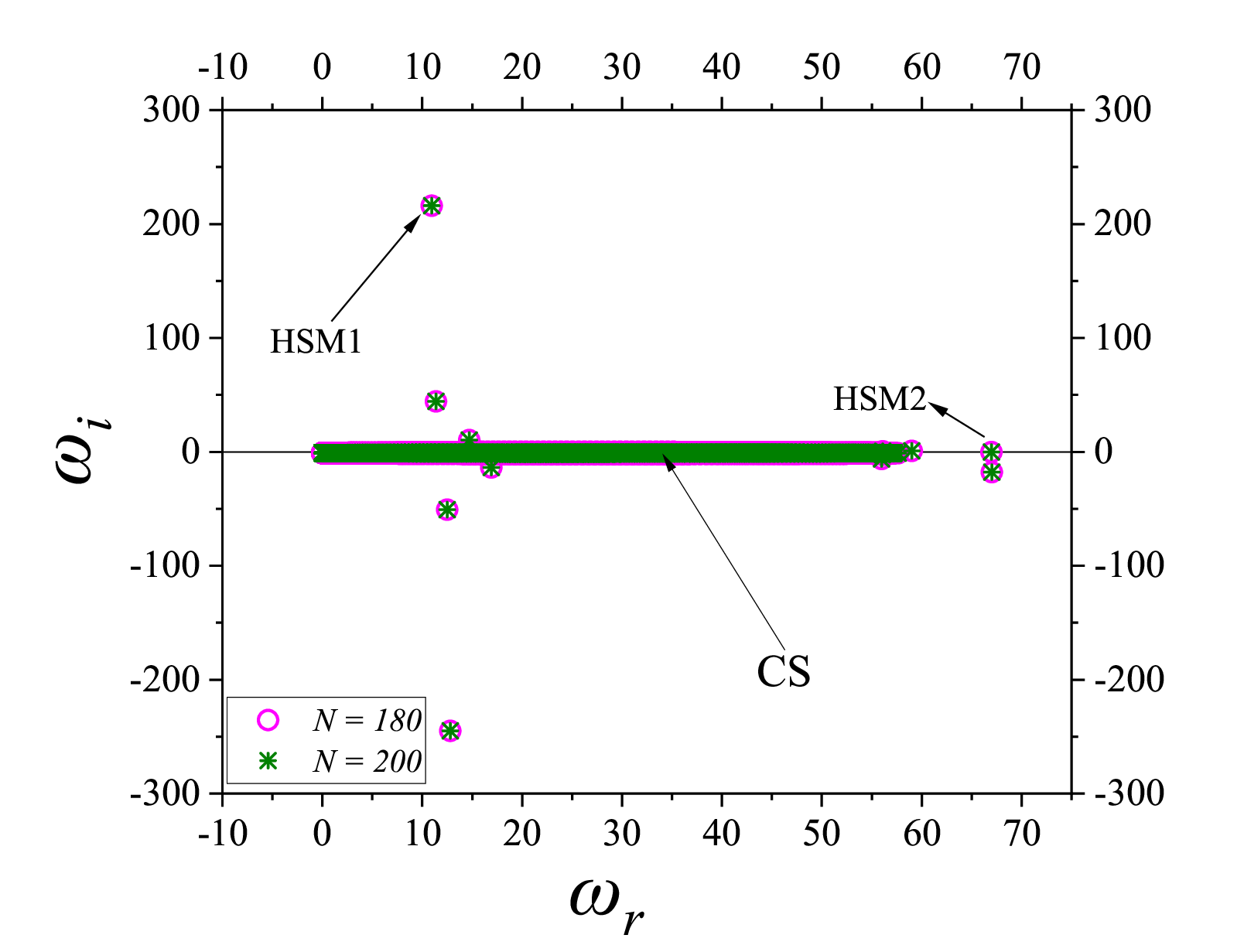}\label{Showing_HSM2_Wi_400}}
  \caption{Eigenspectra showing HSM1 and HSM2 in Dean flow of an Oldroyd-B fluid at $Re = 0$, $\alpha = 7$, $\beta = 0.98$, $ \epsilon = 0.1, n = 1$ for different $Wi$'s.}
   \label{Showing_HSM2_Unstable}
\end{figure}

\begin{figure}
  \centering
  \includegraphics[width=0.6\textwidth]{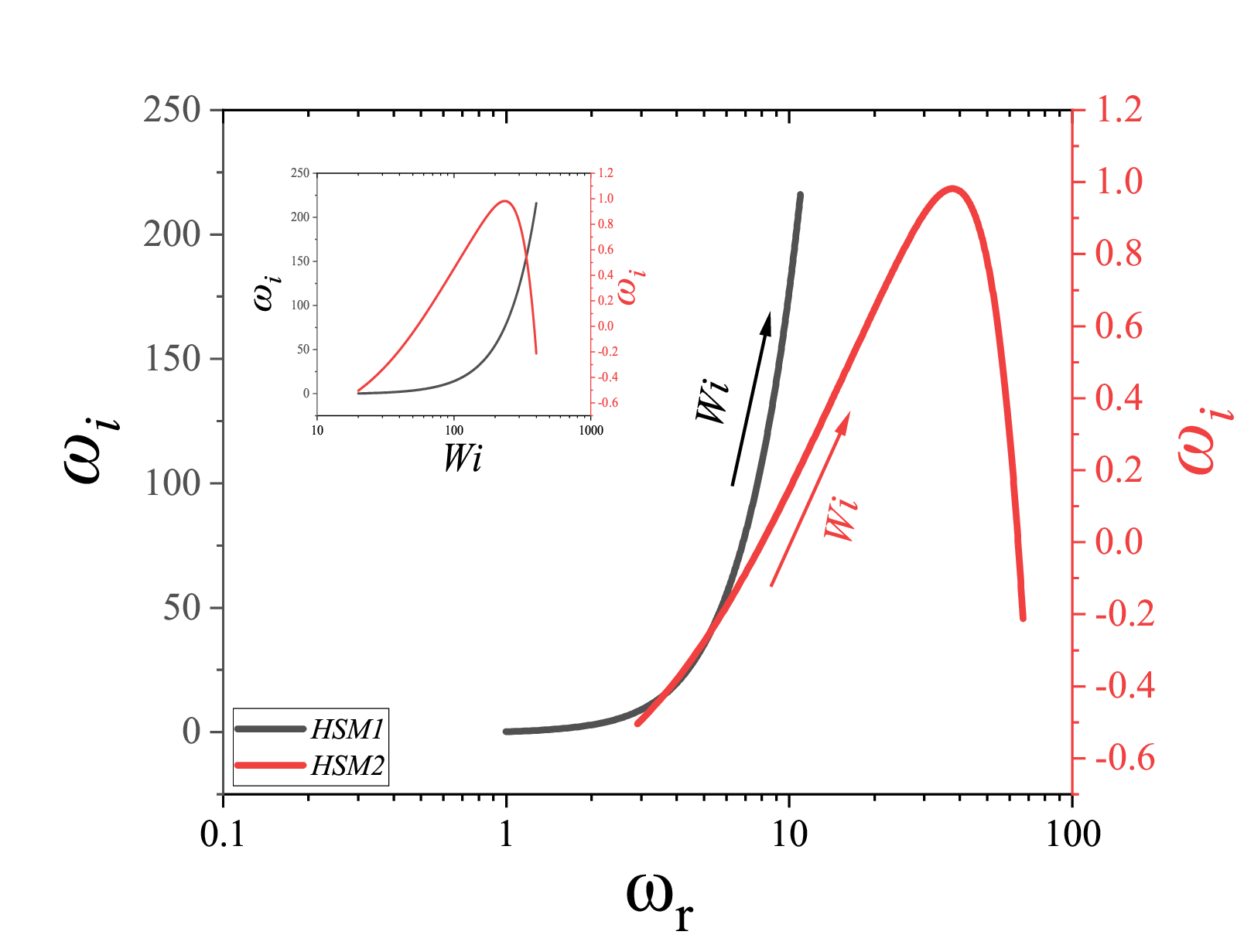}
  \caption{Variation of HSM1 and HSM2 growth rates with increasing $W\!i$. Note the difference in scales of the left and right $y$-axes (for HSM1 and HSM2, respectively) on account of the vast difference in the growth rate magnitudes. The arrows indicate the direction of increasing $W\!i$. The inset shows the variation of $\omega_i$ with $W\!i$ for both the modes.  Data for $Re = 0$, $\alpha = 7$, $\beta = 0.98$, $ \epsilon = 0.1, n = 1$.}
  \label{fig:tracking_mode_with_Wi_in_old_B}
\end{figure}

 Amidst the collection  of other discrete modes, we also identify another mode (labelled `HSM2'), which remains stable at the $Wi$ chosen in Fig.\,\ref{fig:Non_Axisymmetric_Spectrum_mode_1_and_mode_2}. One reason behind singling out HSM2, amongst other discrete modes, maybe seen in Fig.\,\ref{Showing_HSM2_L_200}, where it is unstable with a growth rate that is nearly equal to that of HSM1, for $L = 200$, albeit at a higher $W\!i  = 400$.
Figures\,\ref{Showing_HSM2_Wi_20}--\ref{Showing_HSM2_Wi_400} show the complete spectra for the Oldroyd-B fluid at different $W\!i$, with both HSM1 and HSM2 identified, while Fig.\,\ref{fig:tracking_mode_with_Wi_in_old_B} tracks the loci of HSM1 and HSM2 in the complex-$\omega$ plane with increasing $W\!i$.  The inset in this figure shows the variation of the growth rate with $W\!i$ for both modes.
While an increase in $W\!i$ has a monotonic destabilizing influence on HSM1, it has a non-monotonic effect on HSM2, with this mode being unstable only over an intermediate range of $W\!i$, and eventually stabilizing at sufficiently high $W\!i$; note that the growth rates of HSM2 are much smaller compared to those of HSM1.

  Returning to Fig.\,\ref{Showing_HSM2_L_200}, the spectrum (for $L = 200$) also demonstrates the presence of multiple non-trivial CS branches, for $n \neq 0$,  which nevertheless conform to the  schematic shown in Fig.\,\ref{fig:schematic_nonaxi} above.
 While Fig.\,\ref{Showing_HSM2_L_200} is for $L = 200$, HSM2 is further destabilized with decreasing $L$, leading to it becoming the critical mode for sufficiently small $L$, in certain regions of the parameter space. This is in contrast to the Oldroyd-B limit, where HSM1 is always the critical mode.
 The detailed behaviour of HSM1 and 2 growth rates, for the FENE-P case,  is discussed in Sec.\,\ref{subsec:Lonnonaxy}. It is nevertheless worth a brief mention here, since the unexpected destabilizing effect of finite elasticity first seen for the CM by \cite{buza_page_kerswell_2022}, is also present for HSM2.

\begin{figure}
  \centering
  \includegraphics[width= 0.45\textwidth]{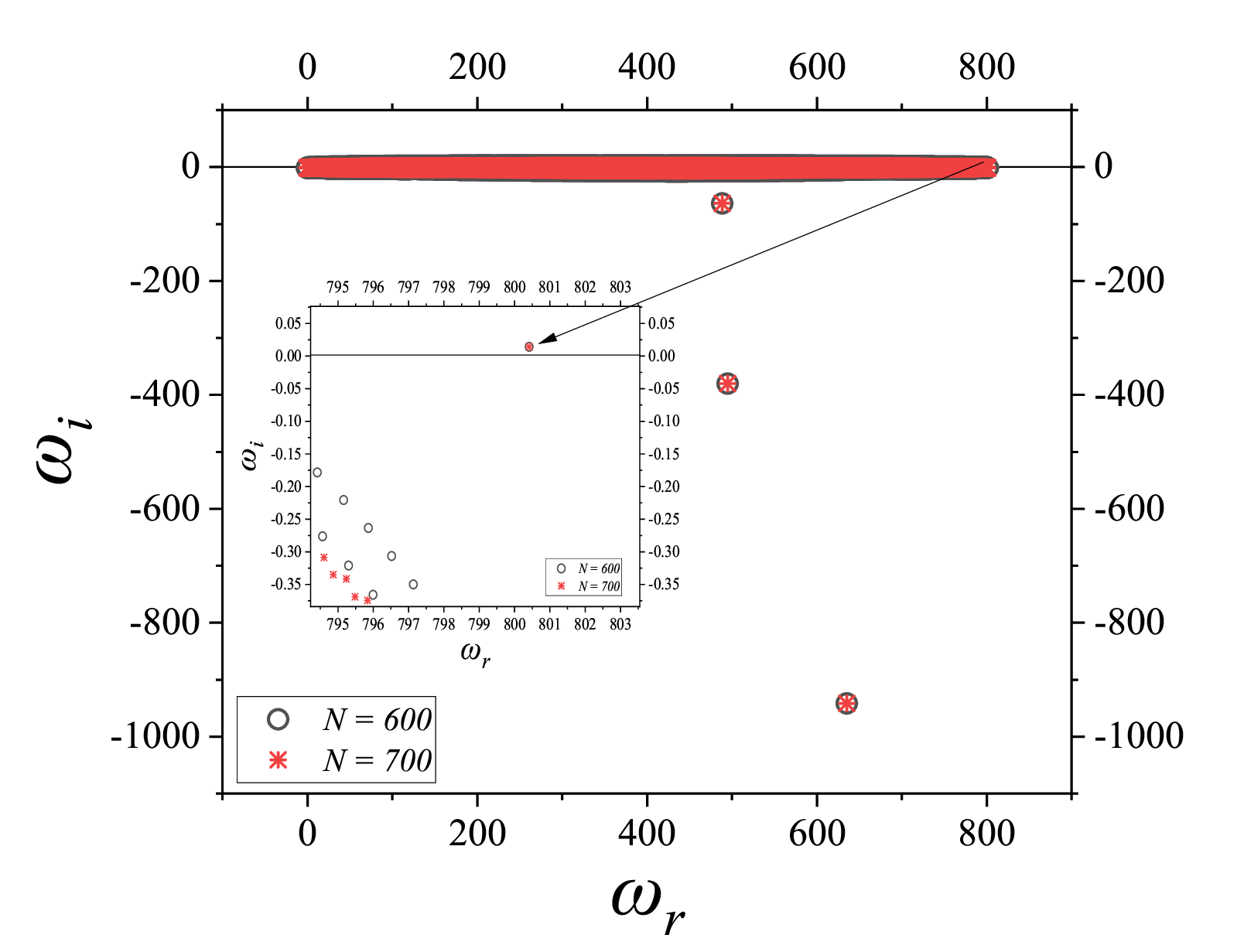}
  \caption{Eigenspectrum showing the presence of the centre mode in Dean flow of an Oldroyd-B fluid (with $n = 16$, $\epsilon = 0.1$, $\alpha = 0$, $\beta = 0.994$ and $W\!i = 350$). The inset shows zoomed-in view highlighting the region near the unstable centre mode; the discrete eigenvalues obtained for $N = 600$ and $N = 700$ show good convergence.}
   \label{fig:CM_EigenSpectrum_at_beta_0.994_e_0.1}
\end{figure}

We end our discussion of the elastic spectrum with  Fig.\,\ref{fig:CM_EigenSpectrum_at_beta_0.994_e_0.1}, which presents the elastic spectrum for $\beta = 0.994$ and $\epsilon = 0.1$; the spectrum includes an unstable CM.
The contrast in the choice of parameters, relative to the spectra in Figs.\,\ref{Axisymmetric_Stationary_vs_Oscillatory} and \ref{fig:multi_subfigures_FENE_P_Axisymmetric_same_Wi_diff_L} above,
is on account of the restricted nature of the unstable portion of the parameter space, in the limit of $\epsilon \rightarrow 0$, corresponding to plane Poiseuille flow -- the centre-mode instability only exists for $\beta > 0.99052$ in this limit, and for $W\!i \sim O(300)$ \citep[within the Oldroyd-B model; see][]{khalid_creepingflow_2021}. It is pertinent to note here that the Weissenberg number used in the present study and the one used in the earlier effort of \cite{khalid_creepingflow_2021} are related by a factor of three, as we elaborate in Sec.\,\ref{subsec:CMOldroydB} below.
The role of decreasing $L$ on the CM is discussed subsequently in Sec.\,\ref{subsec:LonCM}.

\section {Elastic centre-mode instability in Dean flow}
\label{sec:CMinDeanflow}
Having described the overall structure of the spectrum in the regions of parameter space relevant to the two elastic instabilities, we first focus on the results for the CM instability, using the Oldroyd-B model, both in the narrow- and finite-gap regimes. We then proceed to explore the role of finite extensibility. Owing to the difficulty in the experimental realization of pure Dean flow, the existing observations \citep[by][]{joo_shaqfeh_1994}
are for a subset of the Taylor-Dean family with zero azimuthal flow rate. As noted in the Introduction, the latter was realised in the gap between concentric cylinders, with the flow driven by rotation of either cylinder, and with the (adverse) pressure gradient arising from a vertical obstruction in the gap at a particular azimuthal location. For this setup, the unidirectional approximation of the base Taylor-Dean velocity profile is strictly valid only for $\epsilon \ll 1$, when `end effects' due to the azimuthal obstruction are negligible. For $\epsilon \sim O(1)$, the gap width is comparable to the radius of the inner cylinder, and the unidirectional velocity profile given by Eq.\,\ref{eq:OldBbasevel}
will no longer be representative of the actual flow (which deviates from uni-directionality in a region whose size becomes comparable to $R_1$). Therefore, keeping in mind relevance to experiments, we restrict our results below to $\epsilon \leq 1$ .

\subsection {Centre-mode instability within the Oldroyd-B framework}
\label{subsec:CMOldroydB}
In the narrow-gap limit, $\epsilon \ll 1$, the base state Dean profile approaches plane Poiseuille flow, and  we therefore begin with a validation of our numerical procedure by recovering the expected results  \citep{khalid_creepingflow_2021} in the aforementioned limit. 
To establish the connection between the two flows, note that the angular displacement $\theta$ in the Dean (cylindrical) geometry 
is the ratio of the arc length $x^*$  to the radial distance $r^*$, 
that is, $\theta  = \frac{x^{*}}{r^{*}}$.
In the narrow-gap limit, $r^* = R_1(1 + \epsilon y) \approx R_1$, and 
the above relation reduces to $\theta = x^*/R_1$, leading to $n\theta = n x^*/R_1$. Thus, 
$n/R_1$ in cylindrical coordinates, for Dean flow, corresponds to the dimensional streamwise wavenumber $k^*$ for plane Poiseuille flow; 
using $k^* = 2k/d$ gives $n = 2k/\epsilon$.
\begin{figure}
  \centering
 \includegraphics[width=0.45\textwidth]{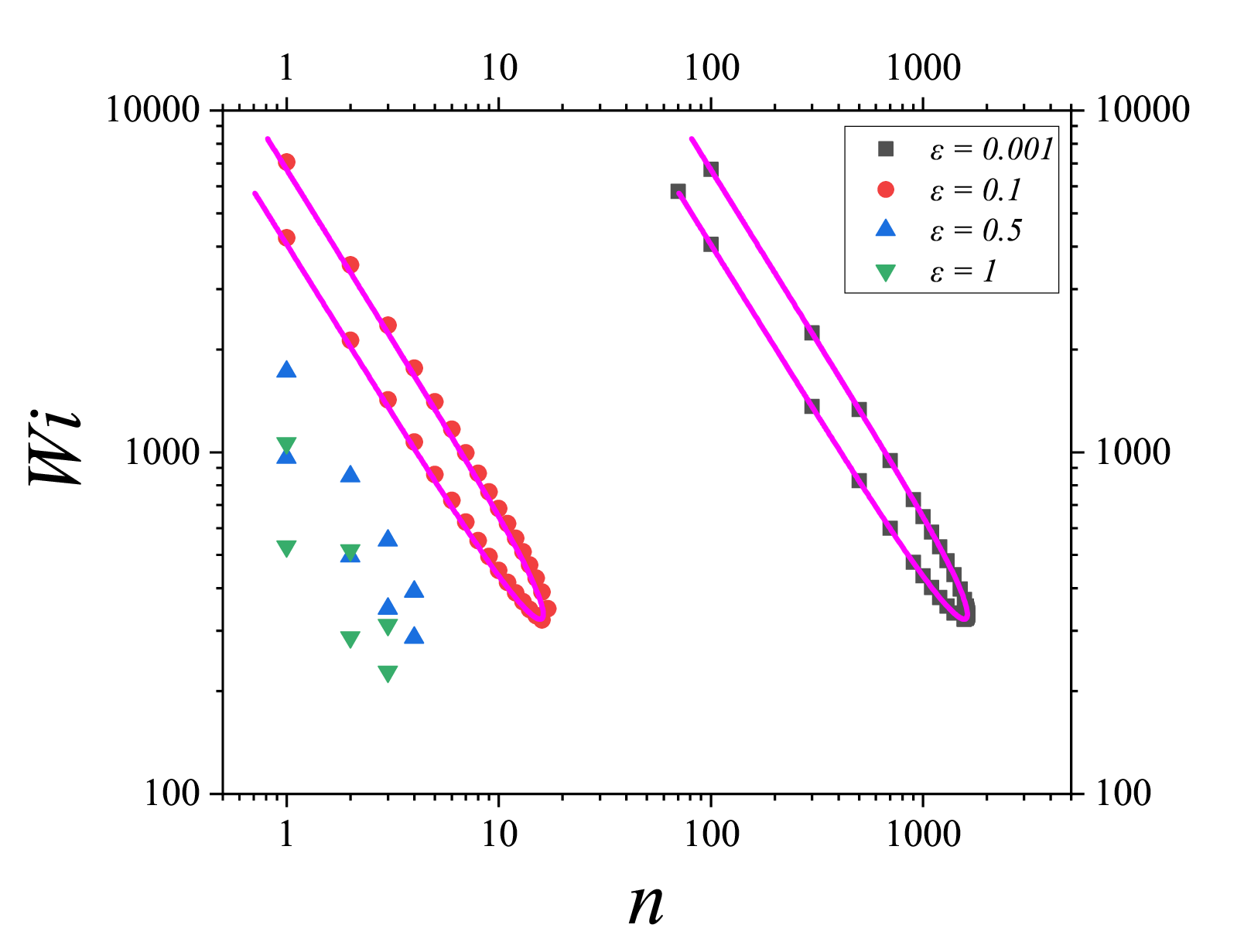}%
  \caption{The centre-mode instability in Dean flow of an Oldroyd-B fluid at $Re = 0$, with $\alpha = 0$: Neutral curves in the $W\!i$--$n$ plane at different $\epsilon$'s  for $\beta = 0.994$. For $\epsilon = 0.001$ and $0.1$, we also plot  the neutral curve (continuous lines) obtained for pressure-driven flow in a rectilinear channel \citep{khalid_creepingflow_2021}; incipient departure from the $n \sim \epsilon^{-1}$ scaling is evident in the comparison for $\epsilon = 0.1$.}
  \label{fig:CM_NeutralCurve_at_beta_0.994}
\end{figure}
Now, \cite{khalid_creepingflow_2021} report the minimum critical Weissenberg number ($W\!i_c = 973.8$) and critical wavenumber ($k_c = 0.783$), for $\beta = 0.994$, for the elastic CM in plane Poiseuille flow, and these values arise from the authors using the maximum speed ($U_{max}$) and the half gap-width ($d/2$) as the characteristic velocity and length scales; the Weissenberg number and non-dimensional wavenumber being defined as  $ W\!i_K = \frac{\lambda U_{max}}{(d/2)}$ and $k = k^{*}d/2$ respectively. In the present study, the Weissenberg number is defined as  $W\!i = \lambda U_m/d$. Noting that the ratio of maximum to mean speed is $3/2$ for plane Poiseuille flow, yields the following relation between the critical parameters: $W\!i/W\!i_K = 1/3$, $n_c = 2k_c/\epsilon$, in turn leading to $W\!i_c = 324.6$ and $n_c = 1566$ in the narrow-gap limit. 
Figure\,\ref{fig:CM_NeutralCurve_at_beta_0.994} shows the neutral curves in the $W\!i$-$n$ plane for $\epsilon$'s ranging from $0.001$ to $1$. 
The figure also shows the neutral curves obtained by \cite{khalid_creepingflow_2021}  for pressure-driven (rectilinear) channel flow,
highlighting the excellent agreement (for $\epsilon = 0.001$) between the Dean  and channel flow cases.
Although it was found convenient to vary $n$ continuously, when tracking the neutral points in the $W\!i$-$n$ plane using the shooting procedure, it is worth keeping in mind that the discreteness of $n$ implies that the critical $W\!i$ corresponds to the integer $n$ closest to the minimum of the tongues in Fig.\,\ref{fig:CM_NeutralCurve_at_beta_0.994}. 
There is a leftward movement of the neutral curves with increasing $\epsilon$, as is expected on account of the $n \sim \epsilon^{-1}$ scaling for $\epsilon \ll 1$; although, this movement slows down for finite $\epsilon$. There is, in addition, an overall lowering of the unstable tongue, resulting in a decrease in the critical Weissenberg number $W\!i_c$ with an increase in $\epsilon$.

\begin{figure}
  \centering
  \subfigure [$\alpha = 0$]{\includegraphics[width=0.4\textwidth]{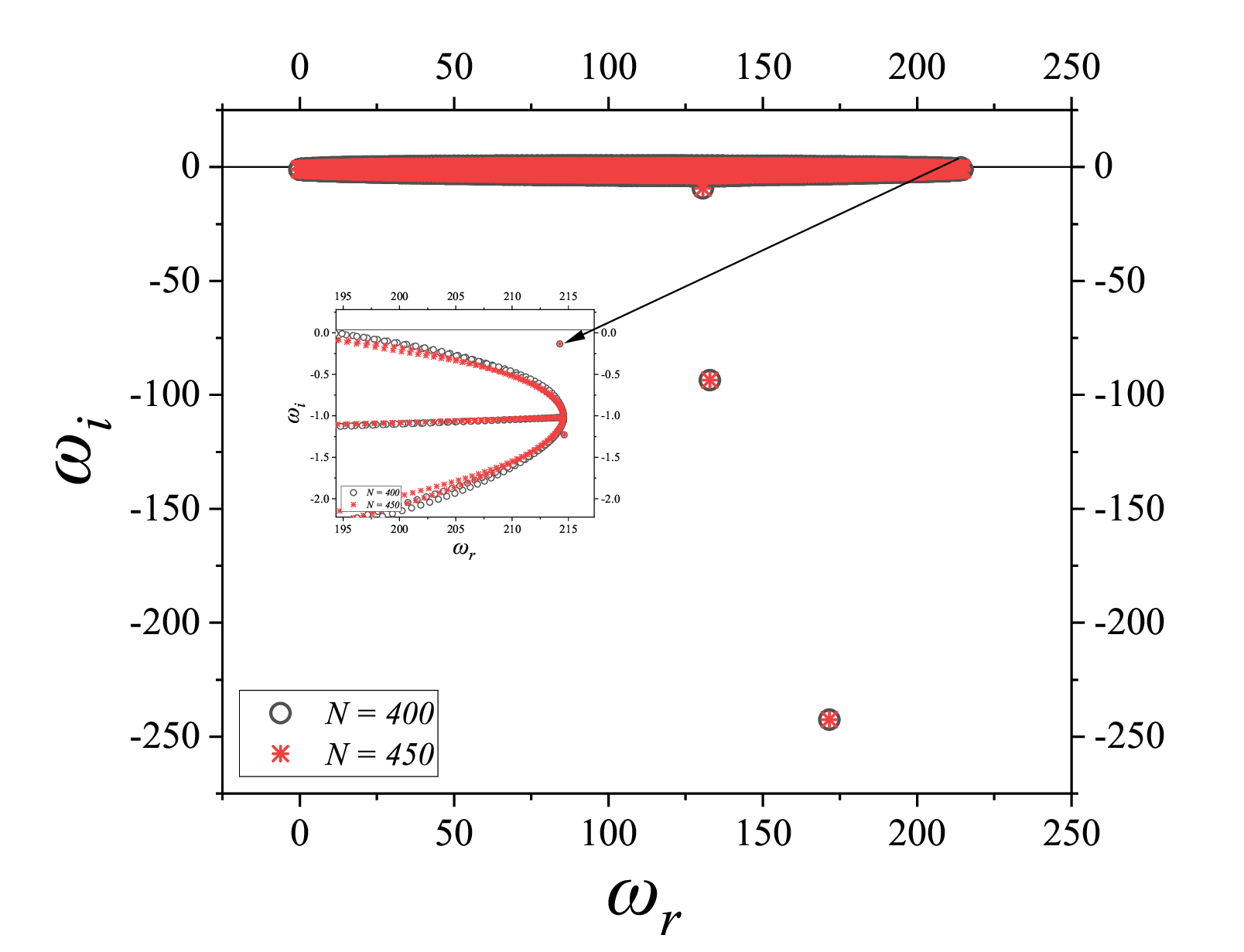}\label{fig:CM_ES_e_0.1_beta_0.98_alpha_0}}
   \subfigure[$\alpha = 0.5$]{\includegraphics[width=0.4\textwidth]{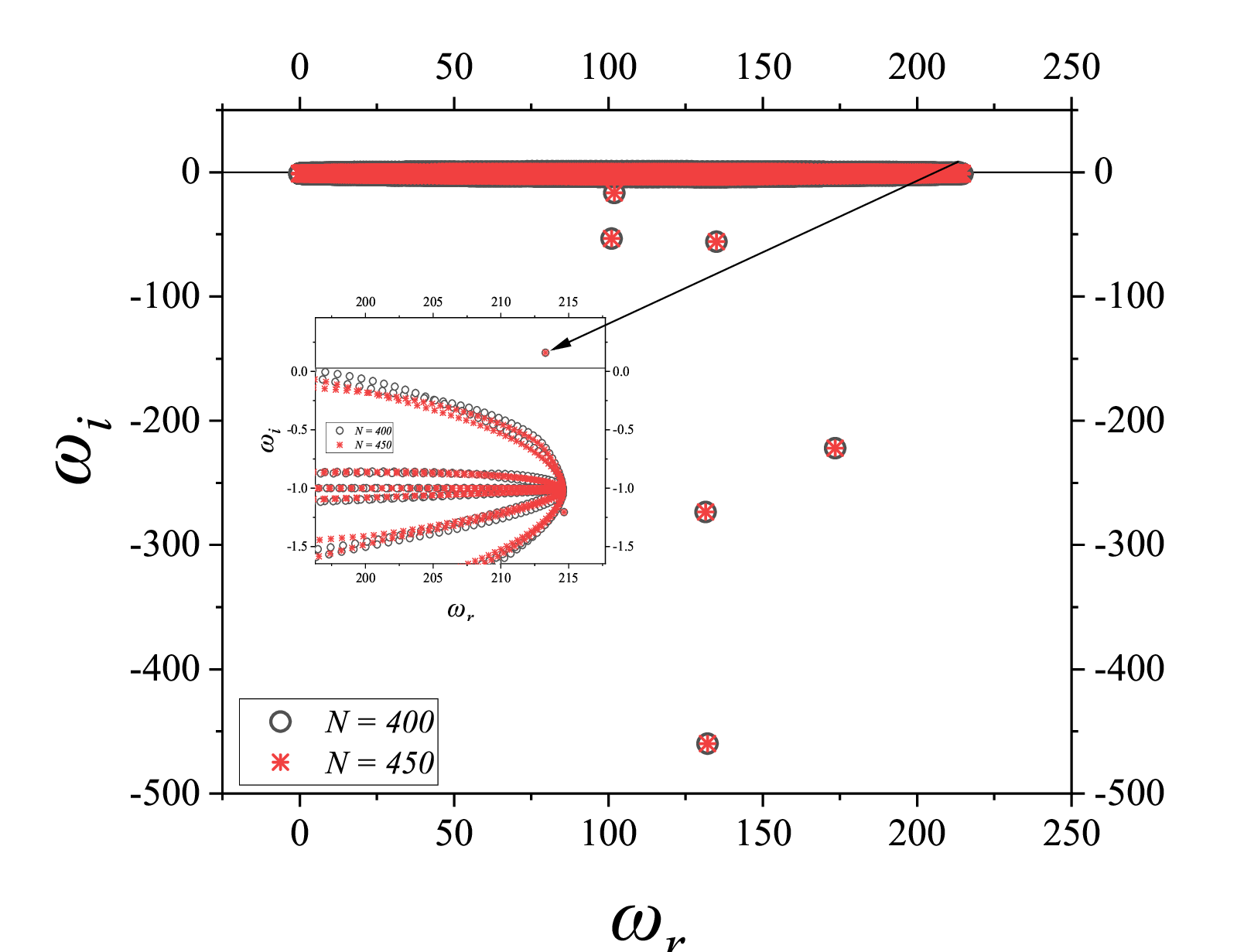}\label{fig:CM_ES_e_0.1_beta_0.98_alpha_0.5}}
  \caption{Eigenspectra showing the destabilising role of axially varying perturbations on the centre mode in Dean flow of an Oldroyd-B fluid. Data for $n = 1$, $\epsilon = 0.1$, $\beta = 0.98$ and $W\!i = 1500$. The insets show the enlarged region of the spectra near the centre mode.}
   \label{fig:CM_EigenSpectrum_at_beta_0.98_diff_alpha_e_0.1}
\end{figure}

\begin{figure}
  \centering
  \subfigure[]{\includegraphics[width=0.45\textwidth]{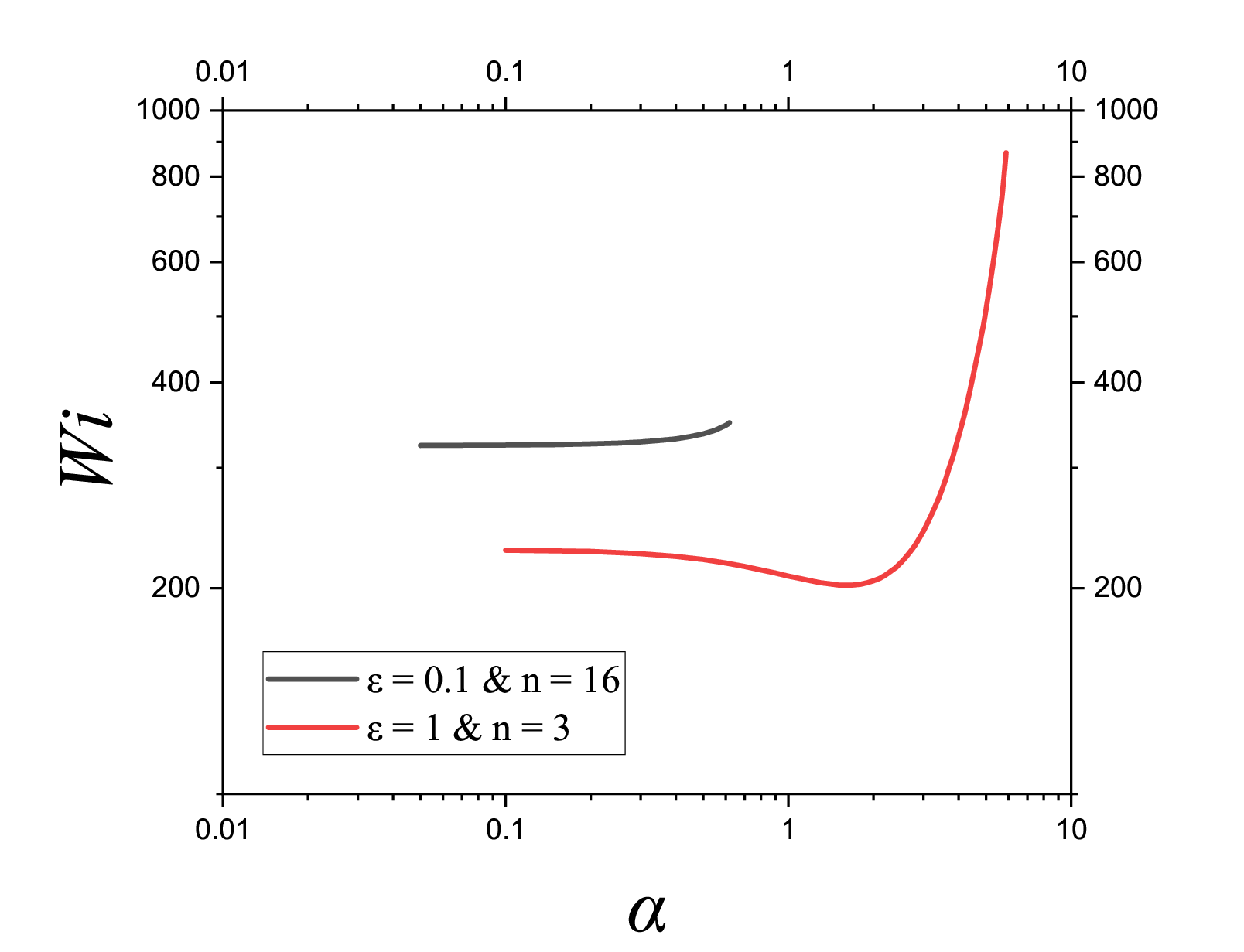}%
  \label{fig:CM_effect_of_alpha}}
  \subfigure[]{\includegraphics[width=0.45\textwidth]{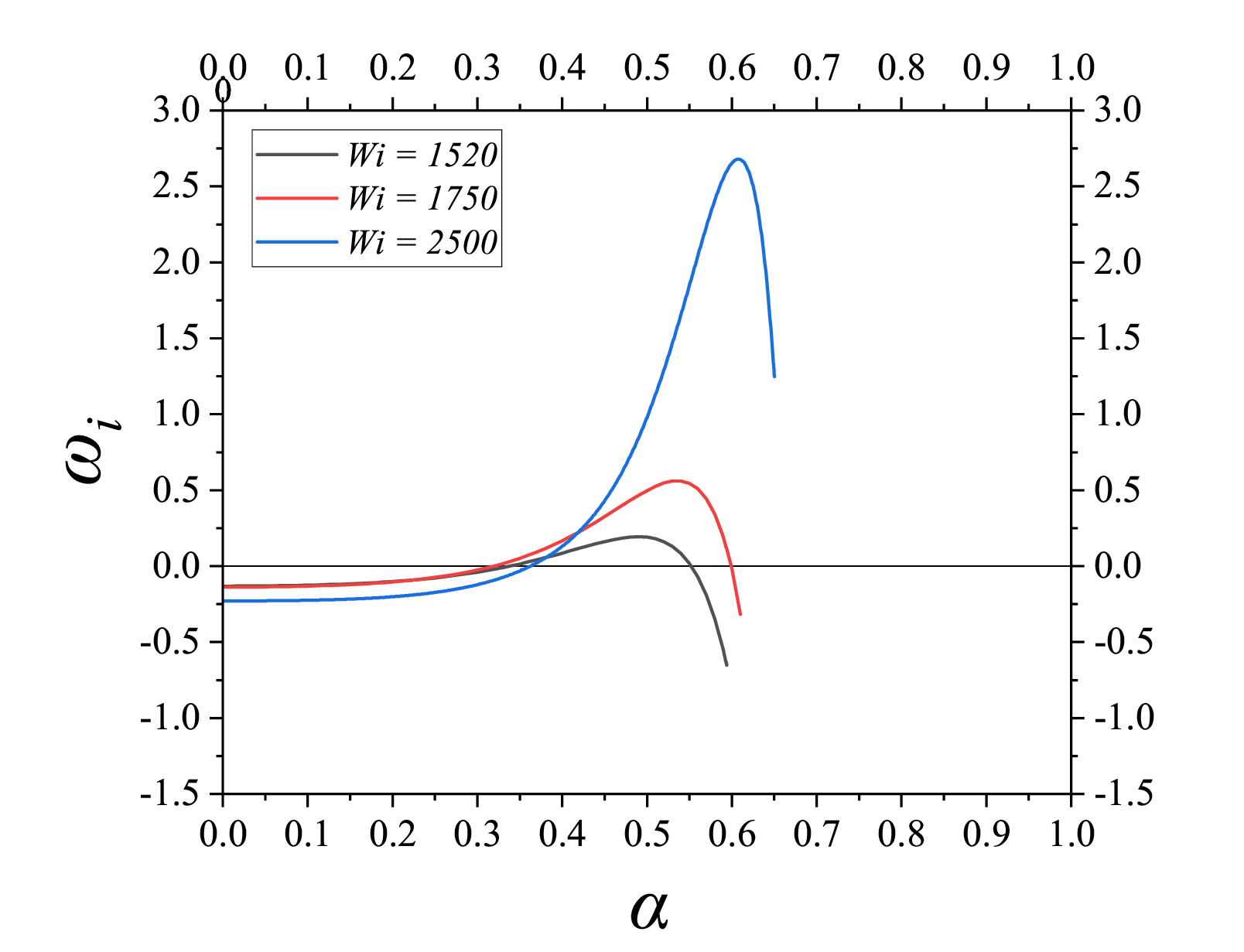}\label{fig:CM_omega_i_vs_alpha_diff_Wi}}
  \caption{(a) Neutral curves in the $W\!i$--$\alpha$ plane  for two ($\epsilon$, $n$) pairs for $\beta = 0.994$. (b) Variation of $\omega_i$ with $\alpha$ showing the destabilisation of the centre mode in Dean flow for non-zero $\alpha$. Data for the Oldroyd-B fluid with $n = 1$, $\epsilon = 0.1$ and $\beta = 0.98$.}
  \label{fig:8}
\end{figure}

Next, Fig.\,\ref{fig:CM_EigenSpectrum_at_beta_0.98_diff_alpha_e_0.1} examines the role of axial variation ($\alpha \neq 0$) of perturbations.
In Fig.\,\ref{fig:CM_ES_e_0.1_beta_0.98_alpha_0}, for the chosen parameter values, the CM is stable for two-dimensional disturbances (i.e., $n \neq 0$ and $\alpha = 0$). However, when three-dimensional disturbances are introduced (i.e., both $n \neq 0$ and $\alpha \neq 0$), surprisingly, the CM becomes  unstable (Fig.\,\ref{fig:CM_ES_e_0.1_beta_0.98_alpha_0.5}).
That is to say, the introduction of an axial variation destabilizes the centre mode. A further increase in $\alpha$ was found to lead to eventual stabilization, however. This implies that, in order to obtain the critical $W\!i$, it is in principle necessary to minimize over both $n$ and $\alpha$.
To this end, 
in Fig.\,\ref{fig:CM_effect_of_alpha}, we show the neutral curves in the $W\!i$-$\alpha$ plane for a fixed $n$, with the latter taken to be the critical value for $\alpha \rightarrow 0$.  The monotonic increase in $W\!i_c$ with increasing $\alpha$, for $\epsilon = 0.1$, appears to be consistent with existence of a Squire's theorem in the limit of $\epsilon \rightarrow 0$ \citep[that is, for rectilinear shearing flows; see][]{BISTAGNINO2007}. 
That this is not the case is shown in Fig.\,\ref{fig:CM_omega_i_vs_alpha_diff_Wi}, where we plot the variation of growth rate as a function of $\alpha$,
for $W\!i$ significantly larger than the threshold for neutral stability, but for a $\beta$ ($= 0.98$) that lies below the rectilinear channel-flow threshold for the chosen $\epsilon$. Even for the relatively low value, $\epsilon = 0.1$, an increase in $\alpha$ is seen to lead to an unstable CM, where no such instability existed for $\alpha = 0$. The initial increase in growth rate, however, is followed by stabilization for sufficient large $\alpha$. Returning to Fig.\,\ref{fig:CM_effect_of_alpha},
for $\epsilon = 1$, the $W\!i$--$\alpha$ neutral curve shows that, while there is an initial destabilization (a decrease in $W\!i$ up until $\alpha \approx 2$, in contrast to $\epsilon = 0.1$), this is followed by a stabilizing effect 
at sufficiently large $\alpha$.
On the whole, both the growth rates and the critical $W\!i$ exhibit a non-monotonic dependence on $\alpha$. Thus, there exist three-dimensional perturbations that are more unstable than two-dimensional ones, in apparent contradiction with Squire's theorem for rectilinear shear flows.
 However, the theorem is not applicable for curvilinear geometries such as the Dean/Taylor-Couette configurations, and thence there is no real contradiction.
The extent of decrease in $W\!i_c$ for $\epsilon = 1$ (from $227.76$ for $\alpha = 0$ to $202.03$ for $\alpha = 1.3$) is, however,  not substantial , and $W\!i_c$ for $\alpha =  0$ serves therefore as a reasonably accurate estimate for the true threshold (obtained by minimizing  over both $n$ and $\alpha$). Thus, for the CM instability, in the ensuing discussion, we present the critical $W\!i_c$ obtained for two-dimensional disturbances.

\begin{figure}
  \centering
  \subfigure[]{\includegraphics[width=0.45\textwidth]{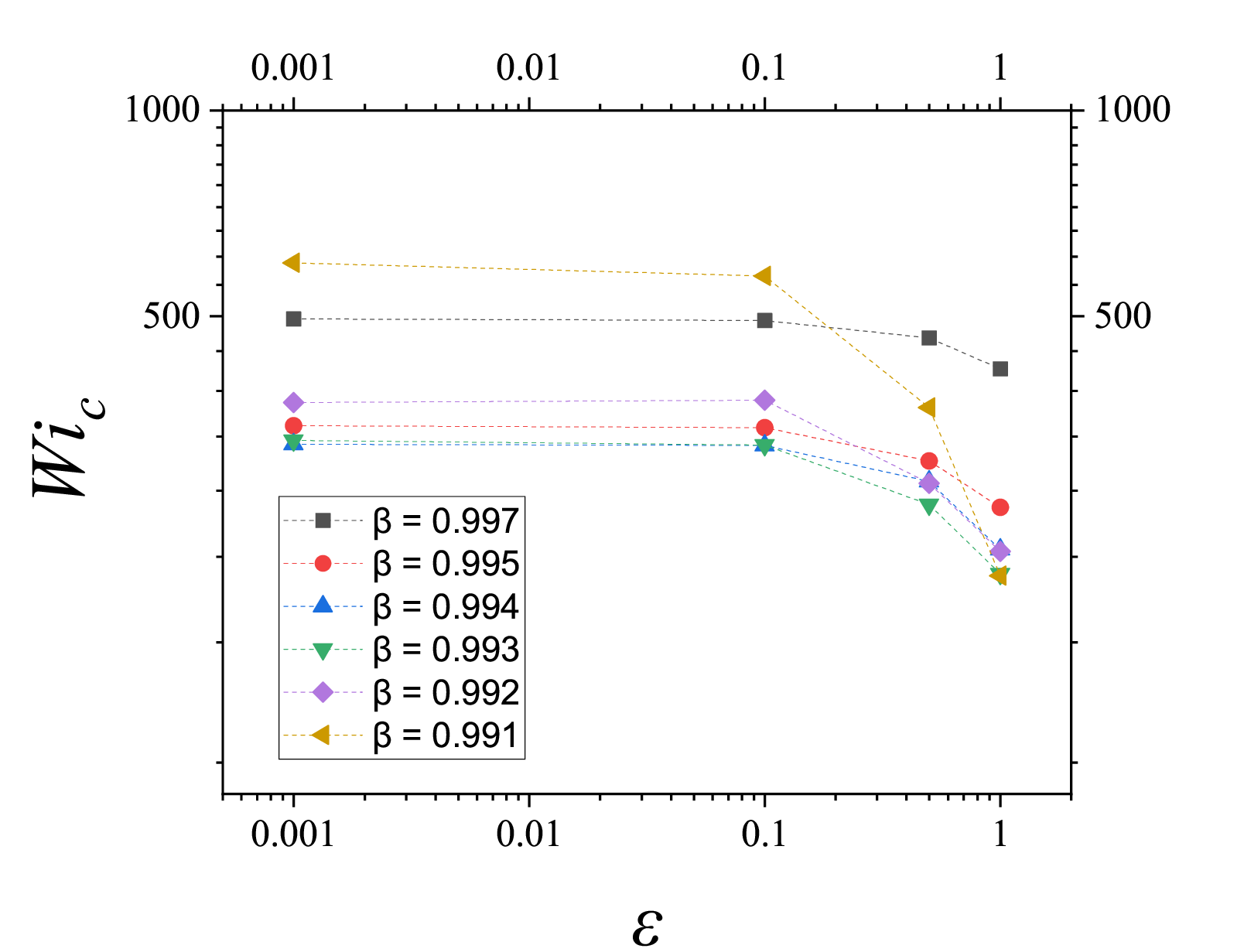}%
  \label{fig:CM_W_c_vs_e}}
  \subfigure[]{\includegraphics[width=0.45\textwidth]{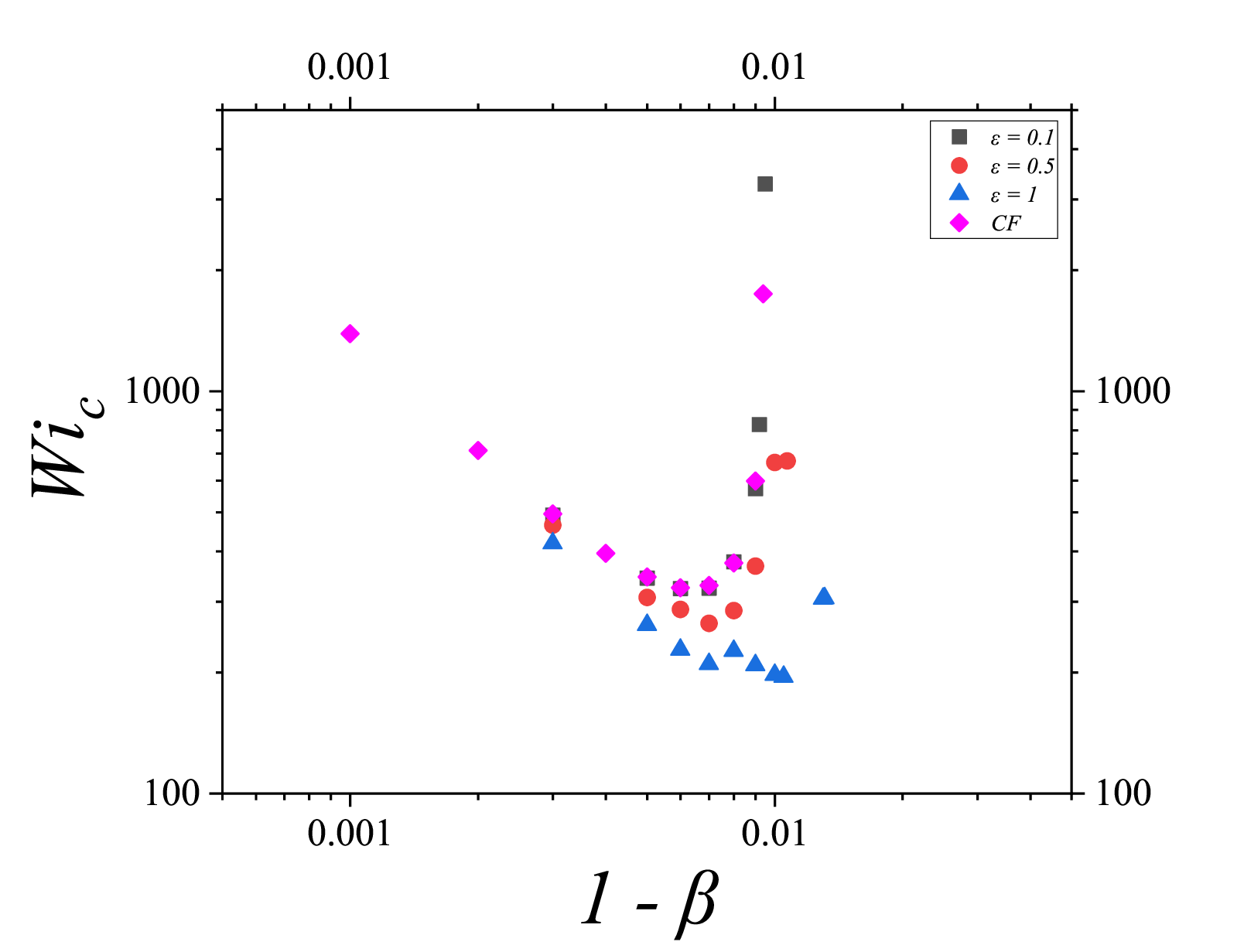}%
  \label{fig:CM_W_c_vs_beta}}
  \caption{The centre-mode instability in Dean flow of an Oldroyd-B fluid at $Re = 0$, $\alpha = 0$: (a) Variation of critical Weissenberg number ($W\!i_c$) with $\epsilon$ at different $\beta$'s; (b) Variation of $W\!i_c$ with  $(1 - \beta)$ at different $\epsilon$. Here `CF' represents the results for plane Poiseuille flow \citep{khalid_creepingflow_2021}.}
  \label{fig:7a}
\end{figure}

Figure\,\ref{fig:CM_W_c_vs_e} shows the reduction in $W\!i_c$ with increasing $\epsilon$, for different $\beta$'s;  representative data, corresponding to this figure, is given in Table\,\ref{W_c_vs_e} for completeness.
 \begin{table}
  \begin{center}
\def~{\hphantom{0}}
  \begin{tabular}{lcc}
  \vspace{5pt}
   $\epsilon$  & $n_c$ & $ W\!i_c $  \\
   0.001 & 1566 & 324.61    \\ 
   0.1 & 16 & 323.38    \\ 
    1 & 3 & 227.76    \\ 
  \end{tabular}
  \caption{Variation of critical Weissenberg number $(W\!i_c)$ with $\epsilon$ at $ \beta = 0.994 $ and $Re = 0$ for the centre mode in Dean flow of an Oldroyd-B fluid.}
  \label{W_c_vs_e}
  \end{center}
\end{table}
Figure\,\ref{fig:CM_W_c_vs_beta} shows the  variation of the critical Weissenberg number $W\!i_c$ with $(1-\beta)$,  for different  $\epsilon$. The CM instability is seen to be restricted to $\beta \geq 0.99$ for $\epsilon \leq 0.1$, consistent with the $\beta$-threshold of 0.9905 for plane Poiseuille flow \citep{khalid_creepingflow_2021}. The instability must clearly be absent for $\beta = 1$, and  we indeed find that $Wi_c$ diverges as $(1-\beta)^{-1}$ for $\beta \rightarrow 1$. Further, the unstable interval of $\beta$'s increases with increasing $\epsilon$, implying that the instability is present over a larger range of polymer concentrations as domain curvature becomes important. Interestingly, the crossing of the different curves in  Fig.\,\ref{fig:CM_W_c_vs_e} shows that, in addition to the increase in the unstable $\beta$-interval, it is the smallest $\beta$'s that become the most unstable with increasing $\epsilon$, in terms of having the smallest $W\!i$-thresholds - this is consistent with both the widening and lowering of the unstable region in Fig.\,\ref{fig:CM_W_c_vs_beta}.

\begin{figure}
  \centering
  \subfigure [$\alpha = 0$]{\includegraphics[width=0.4\textwidth]{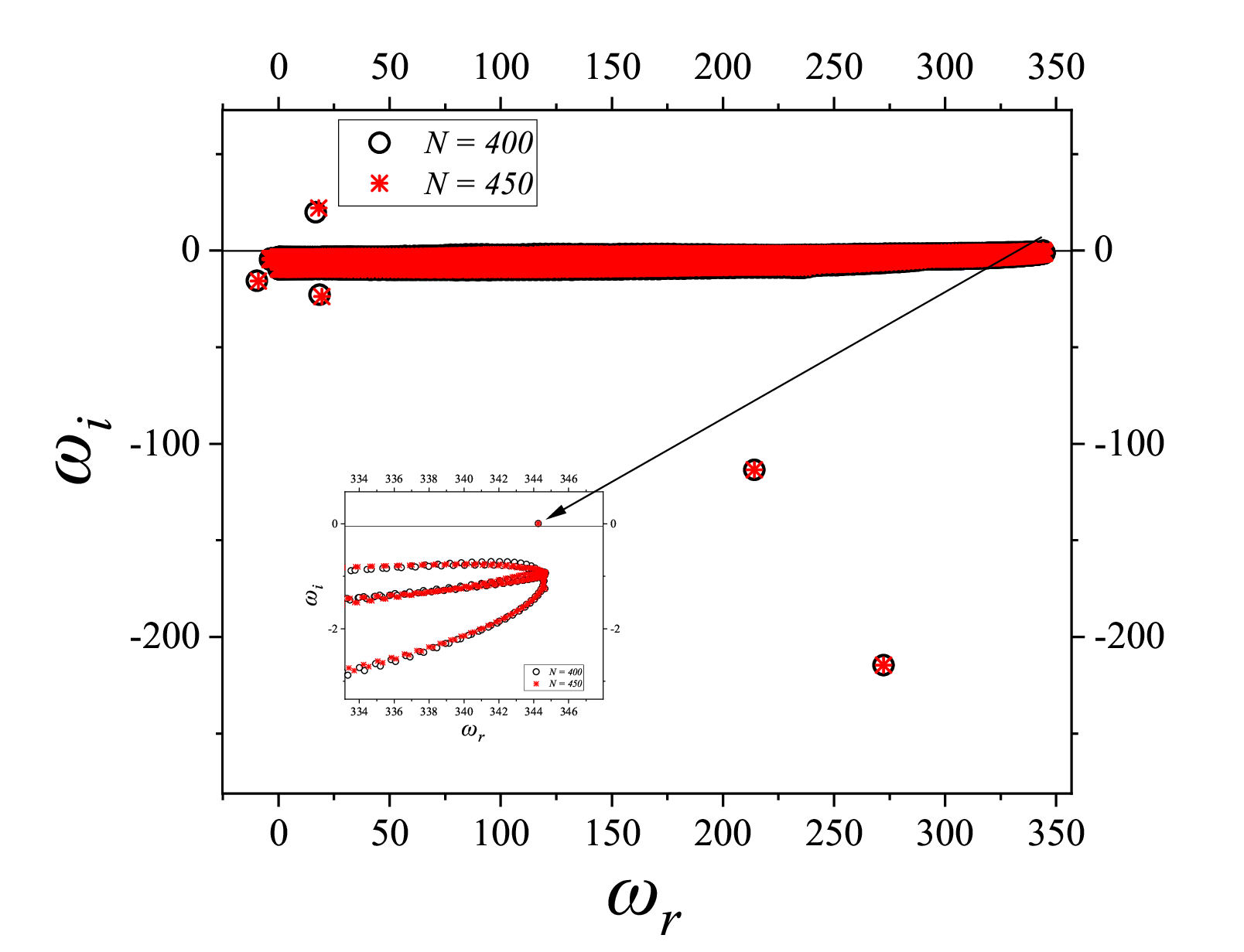}}\label{fig:CM_ES_L_100_e_0.1_beta_0.98_alpha_0}
  \caption{Eigenspectrum showing the presence of unstable centre mode in Dean flow of a FENE-P fluid (with $n = 40$, $\epsilon = 0.1$, $\beta = 0.98$, $L = 100$  and $W\!i = 60$). The insets show the zoomed-in view highlighting the region near the unstable centre mode; the discrete eigenvalues obtained for $N = 400$ and $N = 450$ show good convergence.}
   \label{FENE_P_ES_e_0.1}
\end{figure}   
\subsection {Role of finite extensibility on CM in Dean flow}
\label{subsec:LonCM}

To begin with, in Fig.\,\ref{FENE_P_ES_e_0.1}, we present the elastic spectrum for the FENE-P fluid with  $L = 100$ and $\beta = 0.98$, focusing on two-dimensional disturbances. Interestingly, the CM is unstable here, even at a relatively lower $Wi = 60$. The effect of decreasing $L$ on the CM has been shown to be quite subtle in the case of pressure-driven rectilinear channel flow \citep{KhalidFENEP2025}, in that it could be uniformly stabilising or destabilising, or even non-monotonic; we find a similar trend for Dean flow below. The effect of introducing axial variation was also investigated for $L = 100$ and we again found (data not shown) a slight destabilization  as $\alpha$ is increased up to approximately 1; however, a further increase in $\alpha$ leads to stabilization, similar to the Oldroyd-B case discussed above (Fig.\,\ref{fig:CM_effect_of_alpha}).

\begin{figure}
  \centering
  \subfigure[]{\includegraphics[width=0.45\textwidth]{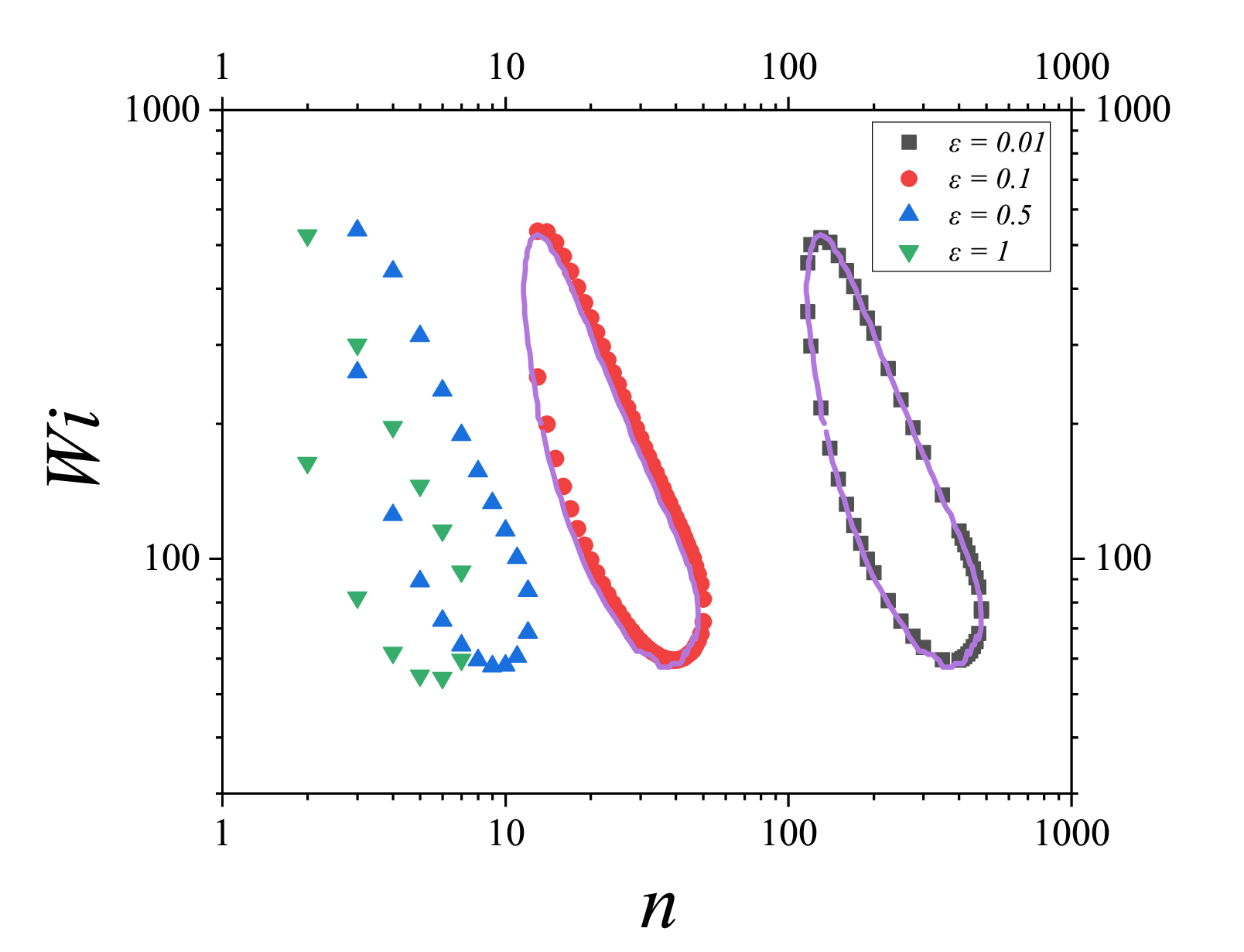}%
  \label{fig:L_100_effect_of_e_on_CM}}
  \subfigure[]{\includegraphics[width=0.45\textwidth]{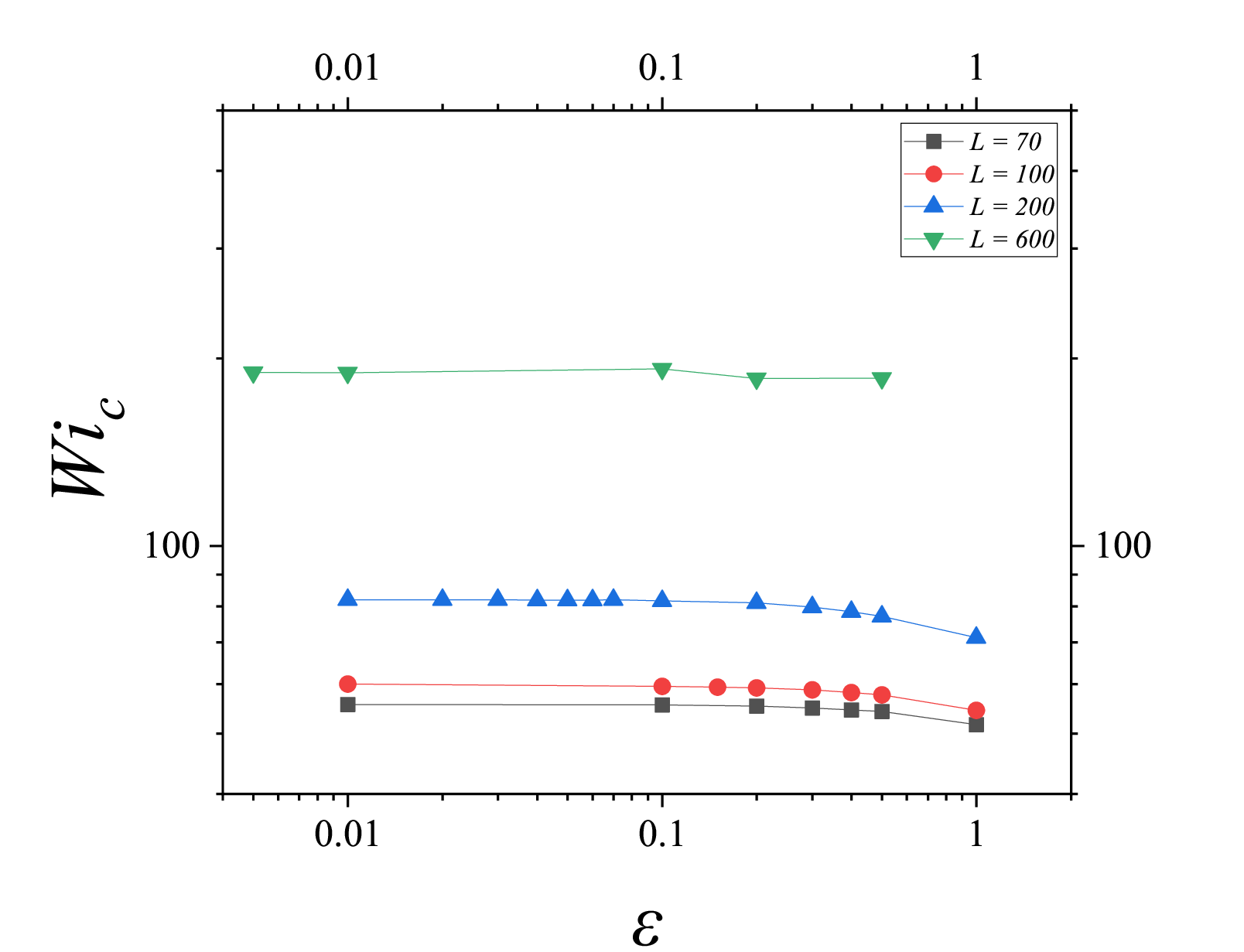}%
  \label{fig:L_affect_on_CM}}
  \caption{Centre-mode instability in Dean flow of a FENE-P fluid: (a) Neutral curves at different $\epsilon$'s for $L = 100$. For $\epsilon = 0.01$ and $0.1$, we also plot  the neutral curve (continuous lines) obtained for pressure-driven flow in a rectilinear channel \citep{KhalidFENEP2025}; (b) Variation of $W\!i_c$ with $\epsilon$ at different $L$'s. Data for both panels correspond to $Re = 0$, $\alpha = 0$, and $\beta = 0.98$.}
  \label{fig:9}
\end{figure}

 \cite{KhalidFENEP2025} first showed, for plane Poiseuille flow, that the neutral curves, for the FENE-P model, form closed loops in the $W\!i$-$k$ plane, as opposed to the open tongues (enclosing the unstable region) seen earlier in Fig.\,\ref{fig:CM_NeutralCurve_at_beta_0.994} for the Oldroyd-B case.  The closing out of the tongues, leading to loops, is a manifestation of the expected stabilizing role of shear thinning, mentioned earlier in Sec.\,\ref{subsec:litreviewHSM}, and which results in a weakening of elastic effects at sufficiently high $W\!i$'s. As shown in Fig.\,\ref{fig:L_100_effect_of_e_on_CM}, in Dean flow too, the neutral curves are closed loops in the $W\!i$-$n$ plane. Importantly, the destabilizing effect of curvature (i.e., increasing $\epsilon$) persists for the FENE-P model as well, as evident from the neutral loops in Fig.\,\ref{fig:L_100_effect_of_e_on_CM} shifting downward when $\epsilon$ is increased to $O(1)$ values.  We also find very good agreement between the loop for $\epsilon$ $= 0.01$ with the one obtained by \cite{KhalidFENEP2025} for pressure-driven channel flow. Furthermore, as shown in Fig.\,\ref{fig:L_affect_on_CM}, the destabilizing effect of $L$ on the CM, originally predicted for pressure-driven channel flow \citep{page2020exact,KhalidFENEP2025}, persists even for Dean flow at finite $\epsilon$. Indeed, an obvious signature of the finite-extensibility-induced destabilization is the fact that we continue to get unstable islands even for $\beta = 0.98$ in Fig.\,\ref{fig:L_100_effect_of_e_on_CM}; the flow is stable for the Oldroyd-B model for this $\beta$.

 In Appendix\,C, we demonstrate that the CM  instability is not governed solely by the ratio $W\!i/L$ (for different $(W\!i, L)$ pairs), although the base state profiles are, as was demonstrated in Sec.\,\ref{subsec:BaseFENEP}.  The eigenfunctions for various field variables corresponding to the CM in Dean flow (for $L = 100$ and $\epsilon = 0.1$) are shown in Appendix\,D by way of comparison with those for the HSM. While the the key features of the CM eigenfunctions are rather robust to changes in $\epsilon$ and $L$, they are qualitatively different from the corresponding HSM eigenfunctions.

\section{Elastic hoop-stress instability in Dean flow}
\label{sec:HSMinDeanflow}
\begin{figure}
  \centering
  \subfigure [$ \epsilon = 0.1  $]{\includegraphics[width=0.45\textwidth]{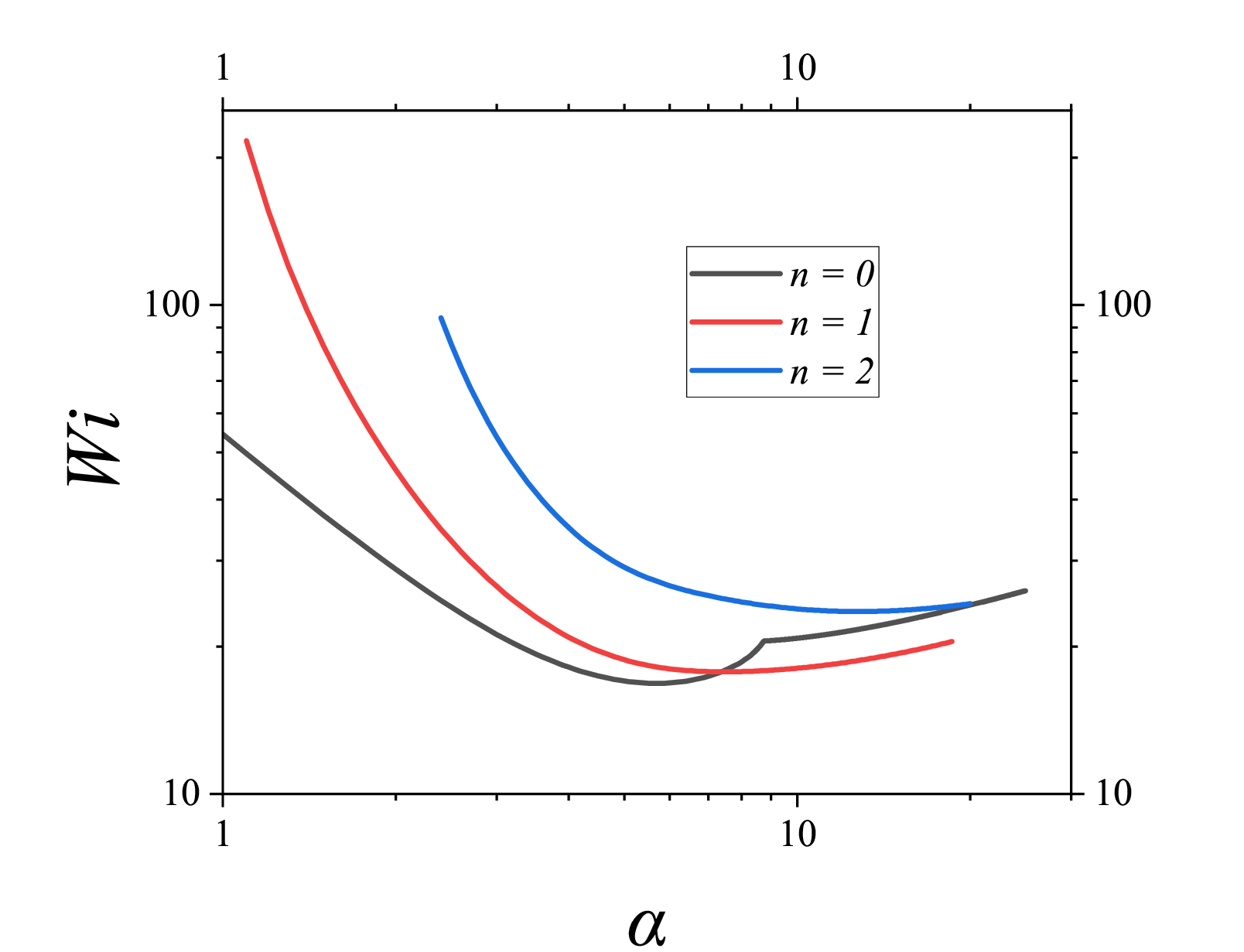}\label{fig:HSM_NeutralCurve_at_e_0.1}}
   \subfigure[$ \epsilon = 1 $]{\includegraphics[width=0.45\textwidth]{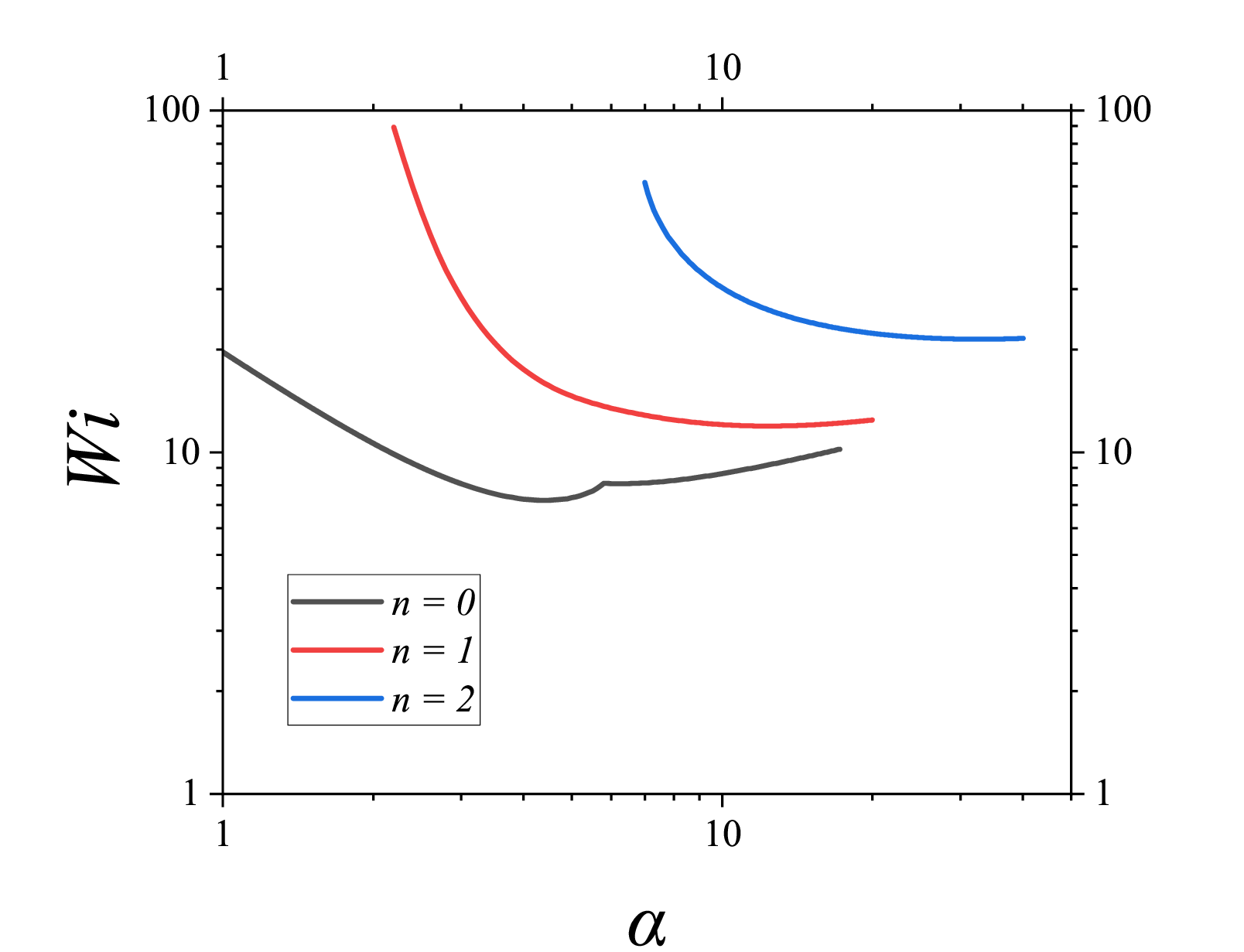}\label{fig:HSM_NeutralCurve_at_e_1}}
  \caption{HSM1 neutral curves for both  axisymmetric and non-axisymmetric disturbances at different $\epsilon$'s. Data for the Oldroyd-B model ($Re = 0$ and $\beta = 0.98$).}
   \label{fig:HSM_NeutralCurves}
\end{figure}

 As mentioned in the Introduction, previous efforts by  \cite{joo_shaqfeh_1991,joo_shaqfeh_1994} 
 have characterized, using the Oldroyd-B model, the hoop-stress-induced elastic instability in Dean flow. Here, we use the more general FENE-P model and examine the role of finite extensibility on the HSM.
 As evident from Figs.\,\ref{Axisymmetric_Stationary_vs_Oscillatory} and \ref{fig:Spectrum_mode_1_and_mode_2} of Sec.\ref{sec:Nature_of_Spectrum}, the Oldroyd-B model predicts both axisymmetric (stationary or propagating, depending on $\alpha$) and  non-axisymmetric (propagating) modes to turn unstable on account of a well-known hoop-stress mediated mechanism. The neutral curves for this case were already seen in  Fig.\,\ref{fig:HSM_NeutralCurves_in_narrowgaplimit}, in the narrow-gap limit, in the context of the validation exercise. The comparison between the axisymmetric and non-axisymmetric thresholds is shown in Figs.\,\ref{fig:HSM_NeutralCurve_at_e_0.1} and \ref{fig:HSM_NeutralCurve_at_e_1}, which plot the neutral curves for $\epsilon = 0.1$ and $1$, respectively.  For both $\epsilon$'s, the axisymmetric mode is the most critical, consistent with the earlier efforts of \cite{joo_shaqfeh_1991,joo_shaqfeh_1994}. A distinct kink is observed in the associated neutral stability curves (at both $\epsilon$'s shown), owing to the bifurcation from stationary to propagating modes.

\begin{figure}
    \centering
    \subfigure[$W\!i_c$ vs $\epsilon$]{
        \includegraphics[width=0.45\textwidth]{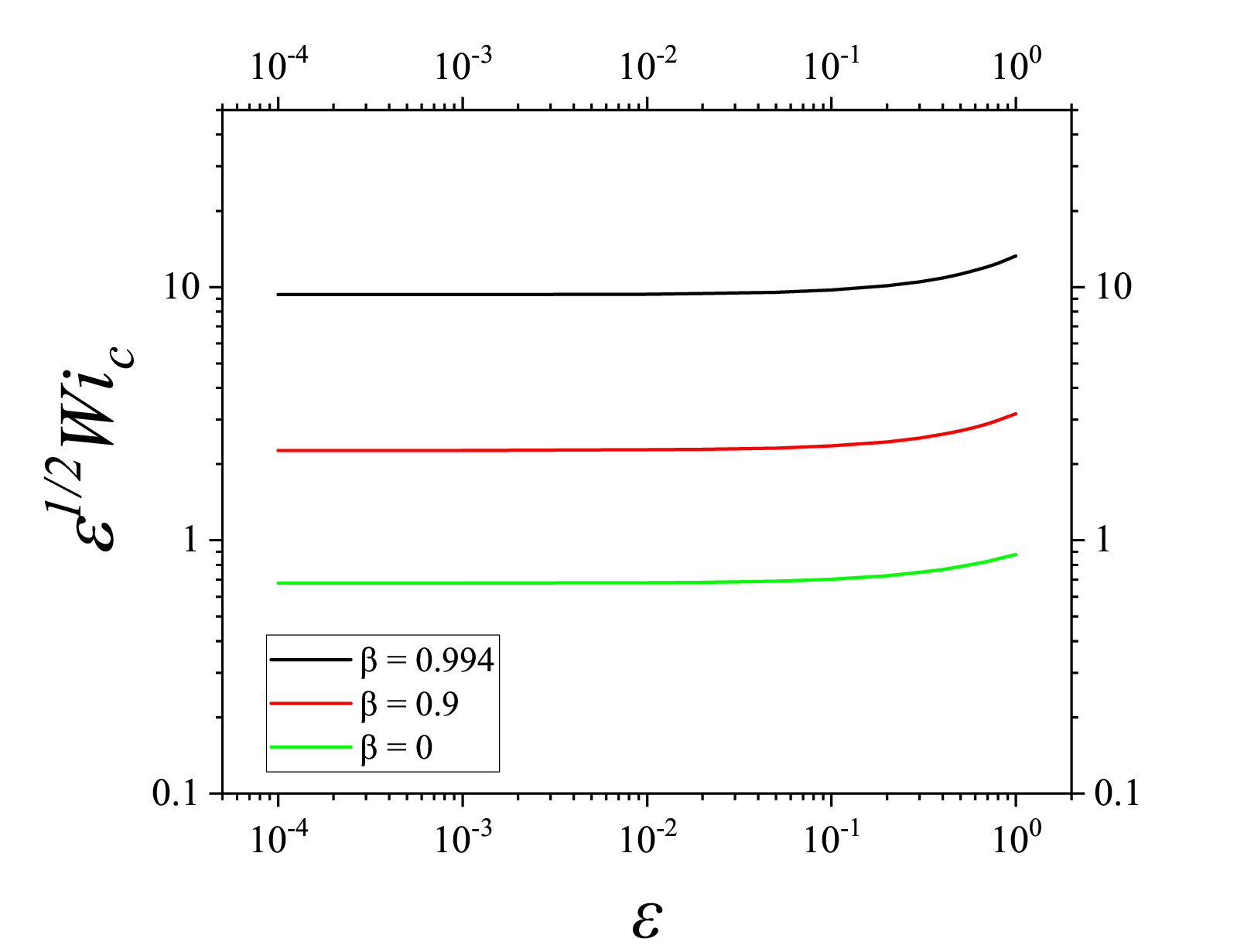} \label{fig:HSM_W_c_vs_e}
    }
    \subfigure[$W\!i_c$ vs $1 - \beta$]{
        \includegraphics[width=0.45\textwidth]{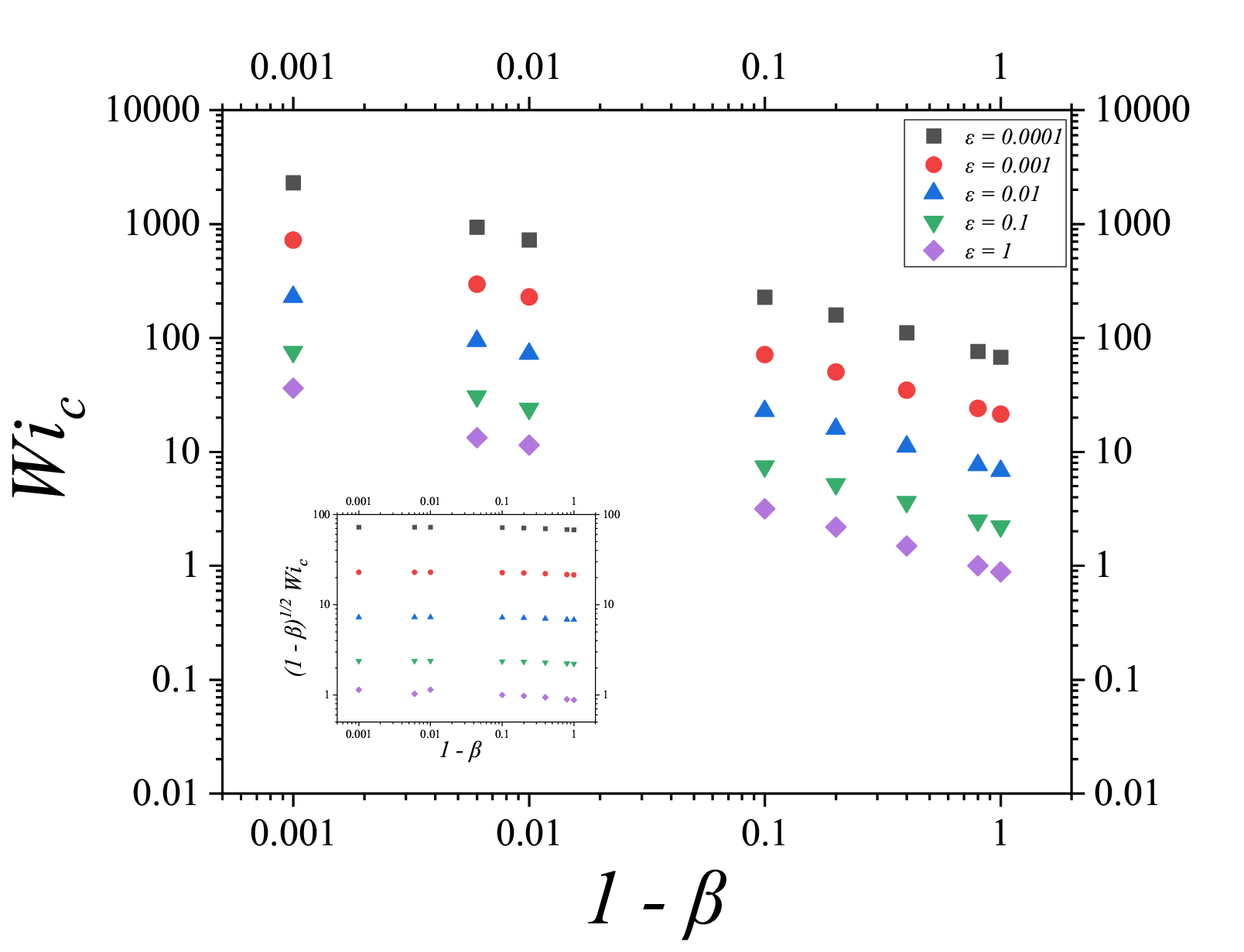}\label{fig:HSM_W_c_vs_beta}
    }
    \caption{Critical Weissenberg number $(W\!i_c)$ for HSM1  in Dean flow  of an Oldroyd-B fluid at $ Re = 0 $ and $ n = 0 $: (a) $ \epsilon^{1/2} W\!i_c $ as a function of $ \epsilon $ for various values of $ \beta $; (b)$W\!i_c $ as a function of $ 1 - \beta $ for different gap width ratios $ \epsilon $; the inset shows $ (1- \beta)^{1/2} W\!i_c $  remains approximately a constant as $\beta$ is varied.}
    \label{HSM_W_c_vs_e_and_beta}
\end{figure}

 Recall from the discussion in Sec.\,\ref{sec:intro} that, for $\epsilon \ll 1$, the critical Weissenberg number ($W\!i_c$) diverges as $\epsilon^{-1/2}$ for the axisymmetric mode.
 Accordingly, in Fig.\,\ref{fig:HSM_W_c_vs_e}, $\epsilon^{1/2} W\!i_c$ is almost independent of $\epsilon$, until $\epsilon \approx 0.1$. This is followed by a mild increase in $\epsilon^{1/2} W\!i_c$ as $\epsilon$ is increased from $0.1$ to $1$.
 Note that, unlike the CM instability, which is present only for $\beta > 0.99$ in plane Poiseuille flow within the Oldroyd-B framework
 \citep[we note, however, that the CM instability has been shown to exist in Kolmogorov flow even for $\beta \rightarrow 0$ by][]{Lewy_Kerswell_2025}, the hoop-stress mode (HSM1) is present even in the absence of solvent ($\beta = 0$), as was indeed already evident from  Fig.\,\ref{fig:HSM_NeutralCurves_in_narrowgaplimit}. However, both instabilities must cease to exist for $\beta \rightarrow 1$.
 The critical $W\!i_c$ for the HSM diverges as $(1-\beta)^{-1/2}$ in this limit -  Fig.\,\ref{fig:HSM_W_c_vs_beta} shows that this scaling continues to hold for any $\beta \in [0, 1)$.
 The scaling can be anticipated from the Pakdel-McKinley argument (Eq.\,\ref{eq:Pakdelcriterion}) as follows. Using ${\cal R} = R_1$, $N_1 \propto \lambda \eta_p (U_m/d)^2$, and the total shear stress $|\tau| \propto \eta U_m/d $ in the PM criterion, we obtain
 \begin{equation}
\frac{\lambda U_m}{R_1} \frac{\lambda \eta_p (U_m/d)^2}{\eta U_m/d } \geq O(1) \, .
 \end{equation}
 Upon using $\epsilon = d/R_1$, and $\eta_p/\eta = (1-\beta)$, the above equation becomes
 \begin{equation}
     \left(\frac{\lambda U_m}{d}\right)^2 \epsilon (1-\beta) \geq O(1)\, ,
  \end{equation}
  which yields $Wi_c \geq  (\epsilon (1-\beta))^{-1/2} $; note that the divergence exponent $-1/2$ applies to 
   both $\epsilon$ and $(1-\beta)$.
   In contrast,  $Wi_c$ for the centre mode obeys an $\epsilon$-independent scaling $Wi_c  \sim (1-\beta)^{-1}$, for $\beta \rightarrow 1$, in the narrow-gap limit  (Fig.\,\ref{fig:CM_W_c_vs_beta}), with this being followed by a rapid increase and divergence at a (slightly lower) threshold $\beta \approx 0.99$.
 
\subsection{Role of finite extensibility on axisymmetric modes}
\label{subsec:Lonaxisym} 

 \begin{figure}
    \centering
    \subfigure[$\epsilon = 0.01$]{
        \includegraphics[width=0.45\textwidth]{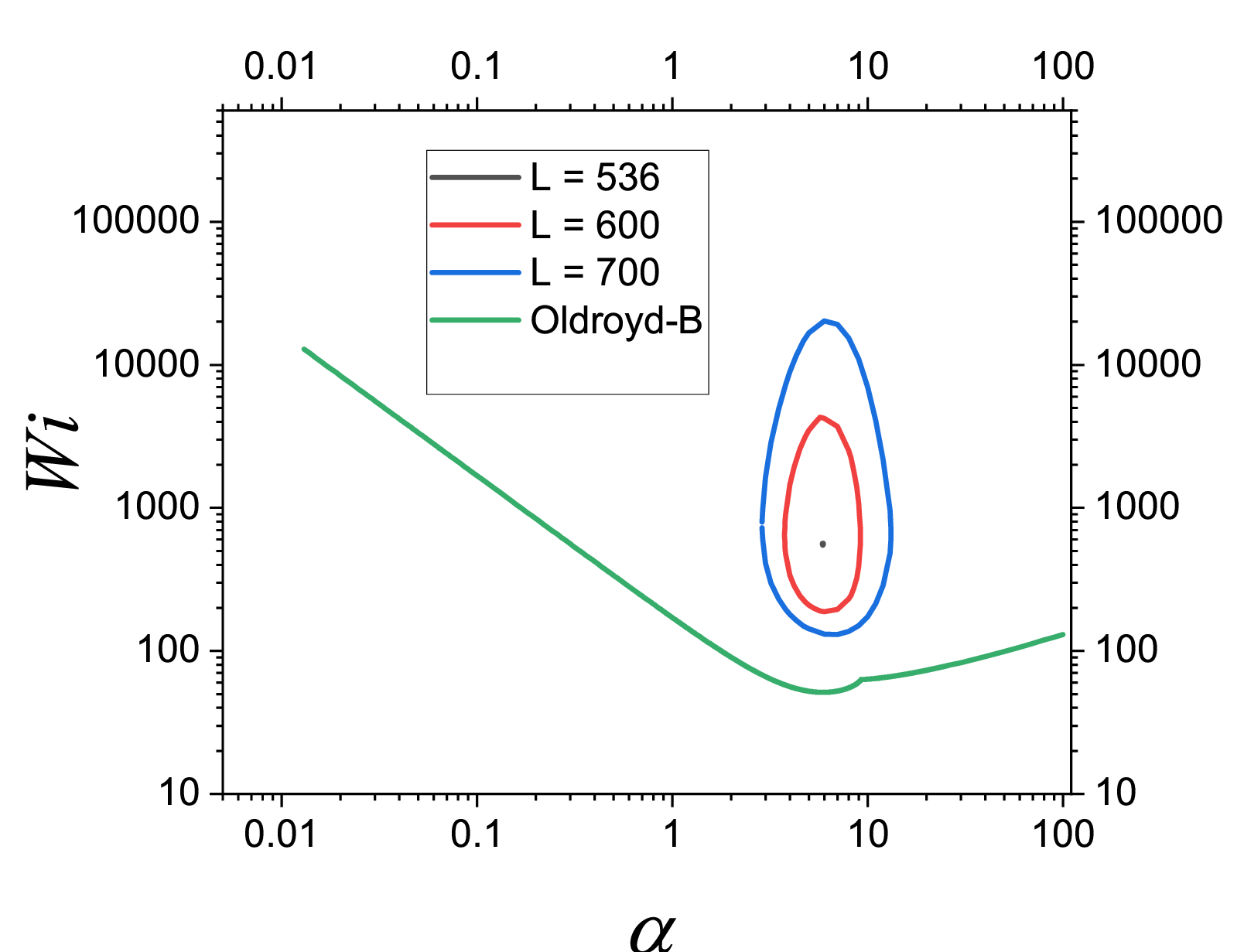}
        \label{fig:subfigures_closed_loops_n_0_e_0.1}
    }
    \subfigure[$\epsilon = 0.1$]{
        \includegraphics[width=0.45\textwidth]{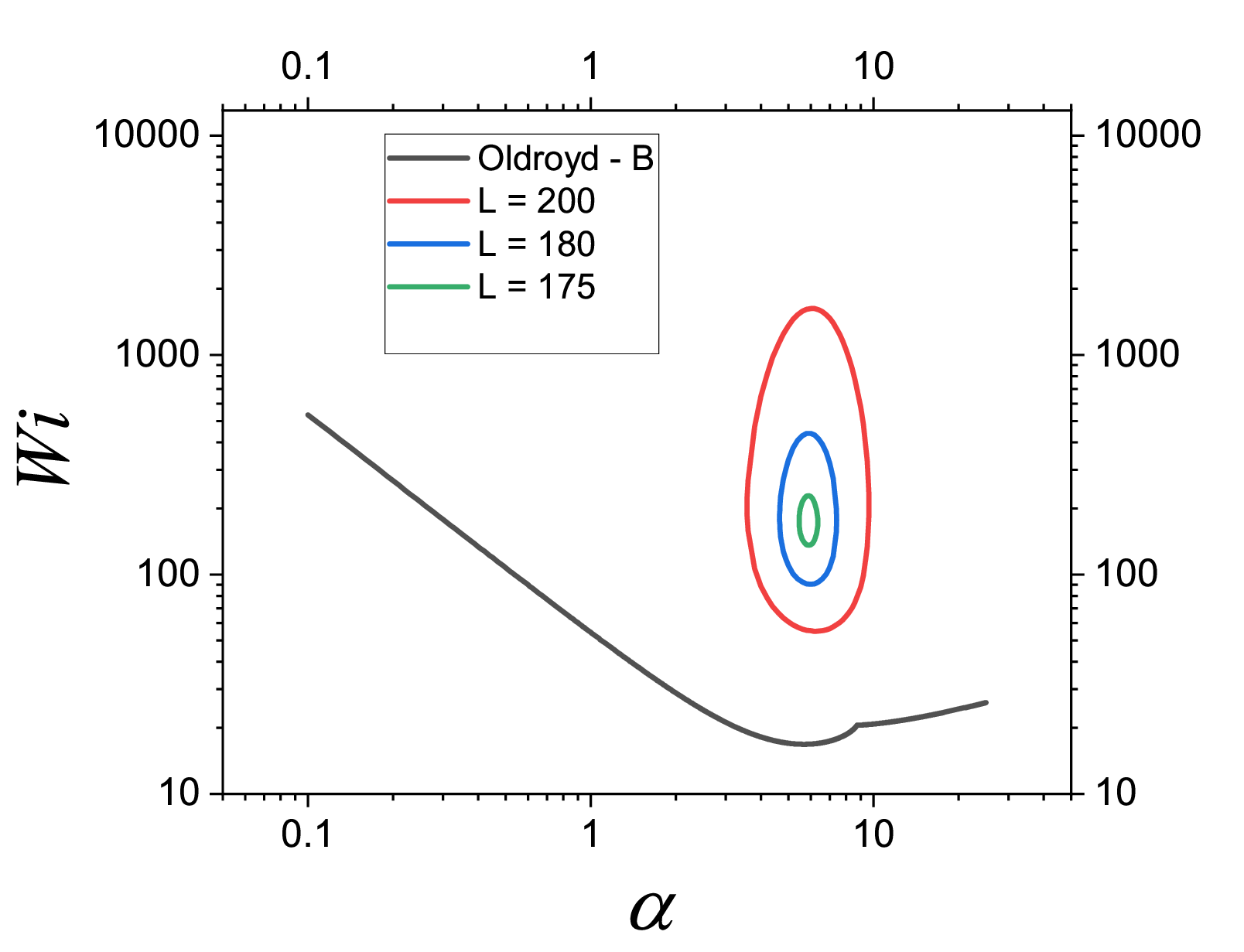}
    }
    \subfigure[$\epsilon = 0.5$]{
        \includegraphics[width=0.45\textwidth]{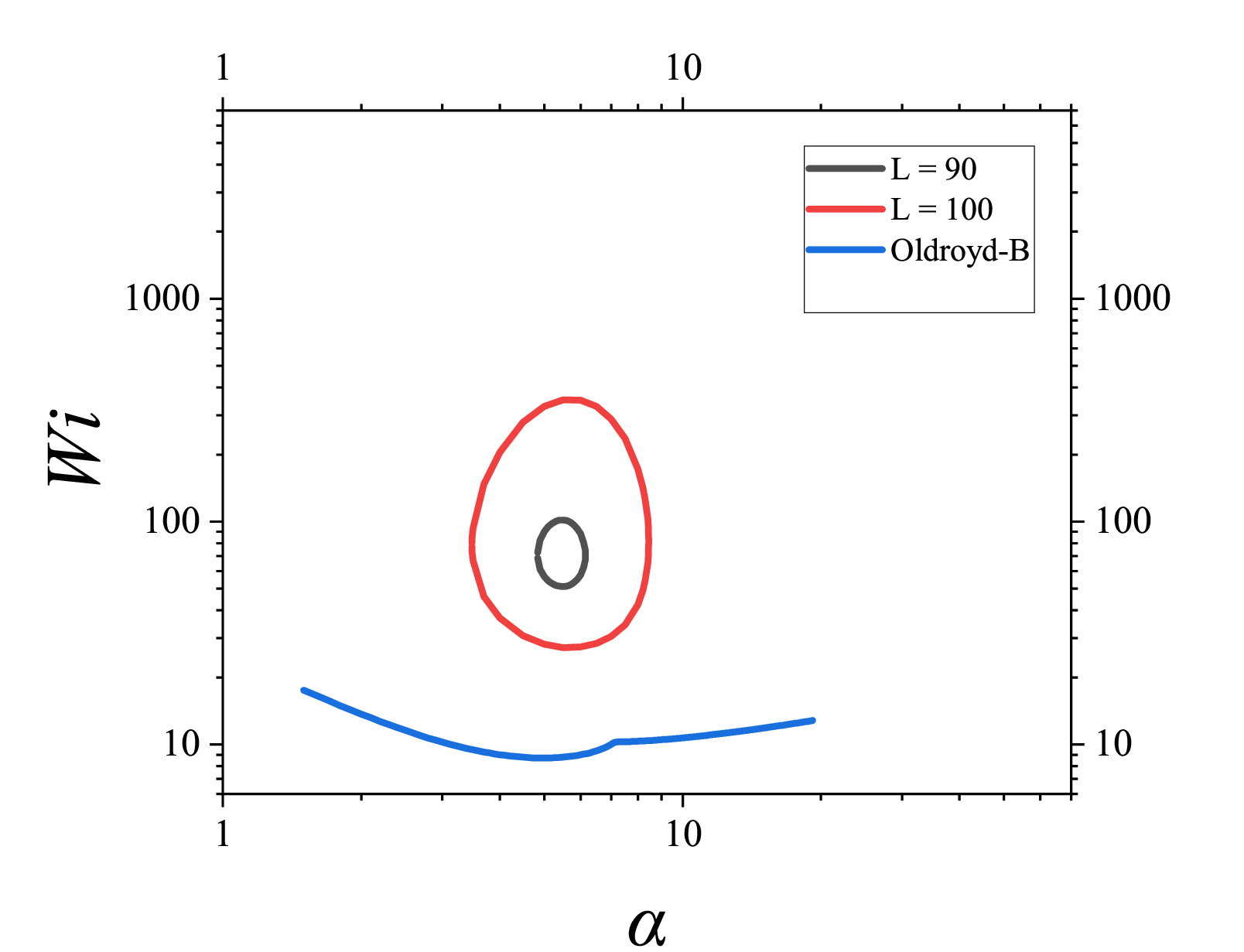}
    }
    \subfigure[$\epsilon = 1$]{
        \includegraphics[width=0.45\textwidth]{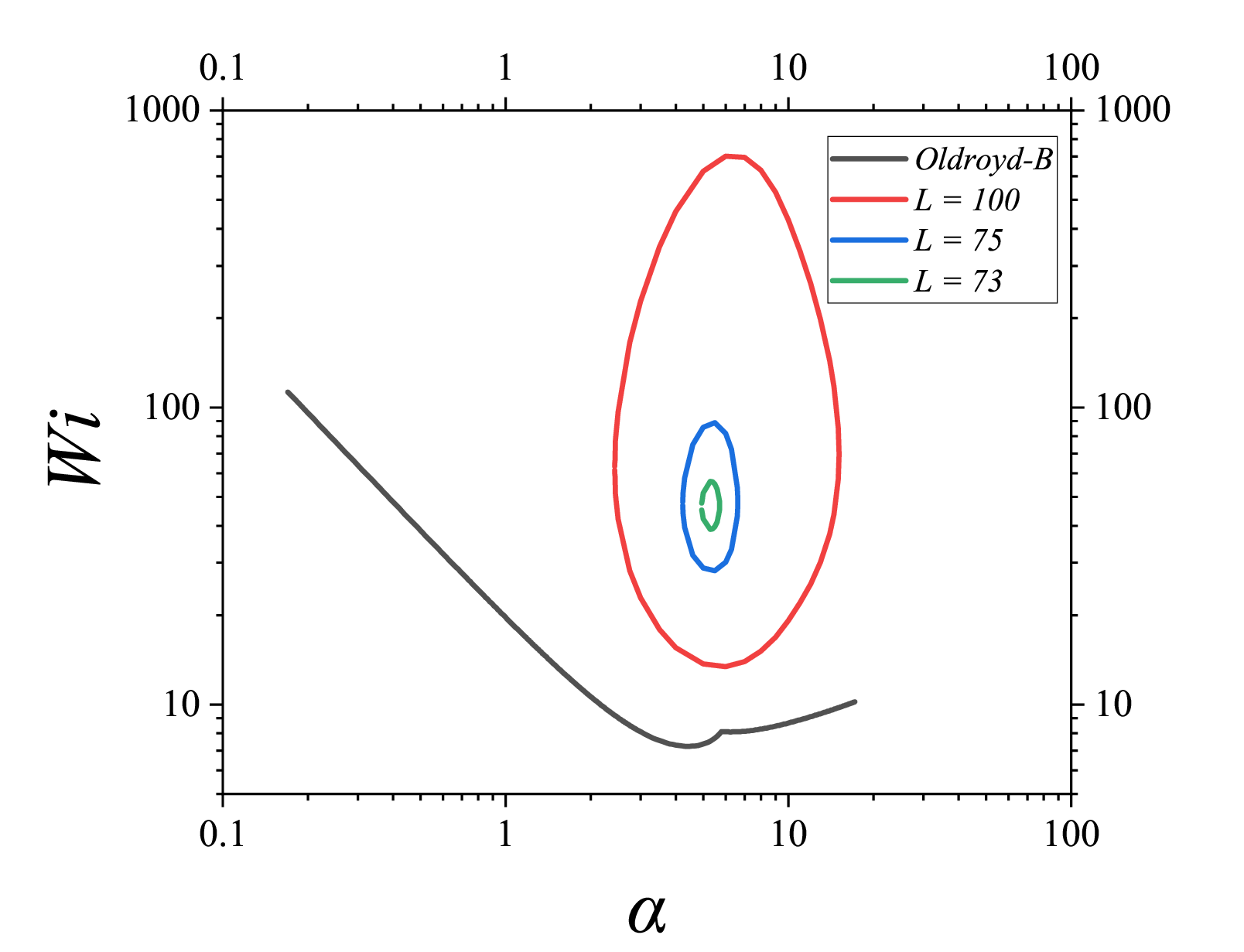}
    }
    \caption{Neutral curves for HSM1 in the $W\!i$--$\alpha$ plane for Dean flow of a FENE-P fluid at $Re = 0$,  $\beta = 0.98 $, $n = 0$.}
    \label{fig:subfigures_closed_loops_n_0}
\end{figure}

 In Fig.\,\ref{fig:subfigures_closed_loops_n_0}, we show the neutral curves  for  HSM1  (the most critical mode in the Oldroyd-B model) in the $W\!i$--$\alpha$ plane for $L$'s ranging from the Oldroyd-B limit down to $100$.
 The eigenspectra corresponding to these neutral curves were presented in  Fig.\,\ref{fig:multi_subfigures_FENE_P_Axisymmetric_same_Wi_diff_L}.
 Owing to the stabilising role of finite extensiblity, the neutral stability curves  for finite $L$ are  again closed loops in the $W\!i$--$\alpha$ plane  for the entire range of gap width ratios examined,  analogous to the neutral loops for CM in Fig.\,\ref{fig:L_100_effect_of_e_on_CM}. Similar closing out of the neutral curves due to shear-thinning-induced weakening of the first normal stress difference has been demonstrated using the Giesekus \citep{Oztekin1994} and FENE-CR \citep{McKinley1995} models for the hoop-stress instability for flow in the cone-and-plate geometry.  Unlike the Oldroyd-B model, which exhibits a kink due to a transition from stationary to propagating modes, for the $L$'s shown in Fig.\,\ref{fig:subfigures_closed_loops_n_0}, the transition to instability occurs via propagating modes for all the $\epsilon$'s considered in this figure.

\begin{figure}
  \centering
  \includegraphics[width= 0.45\textwidth]{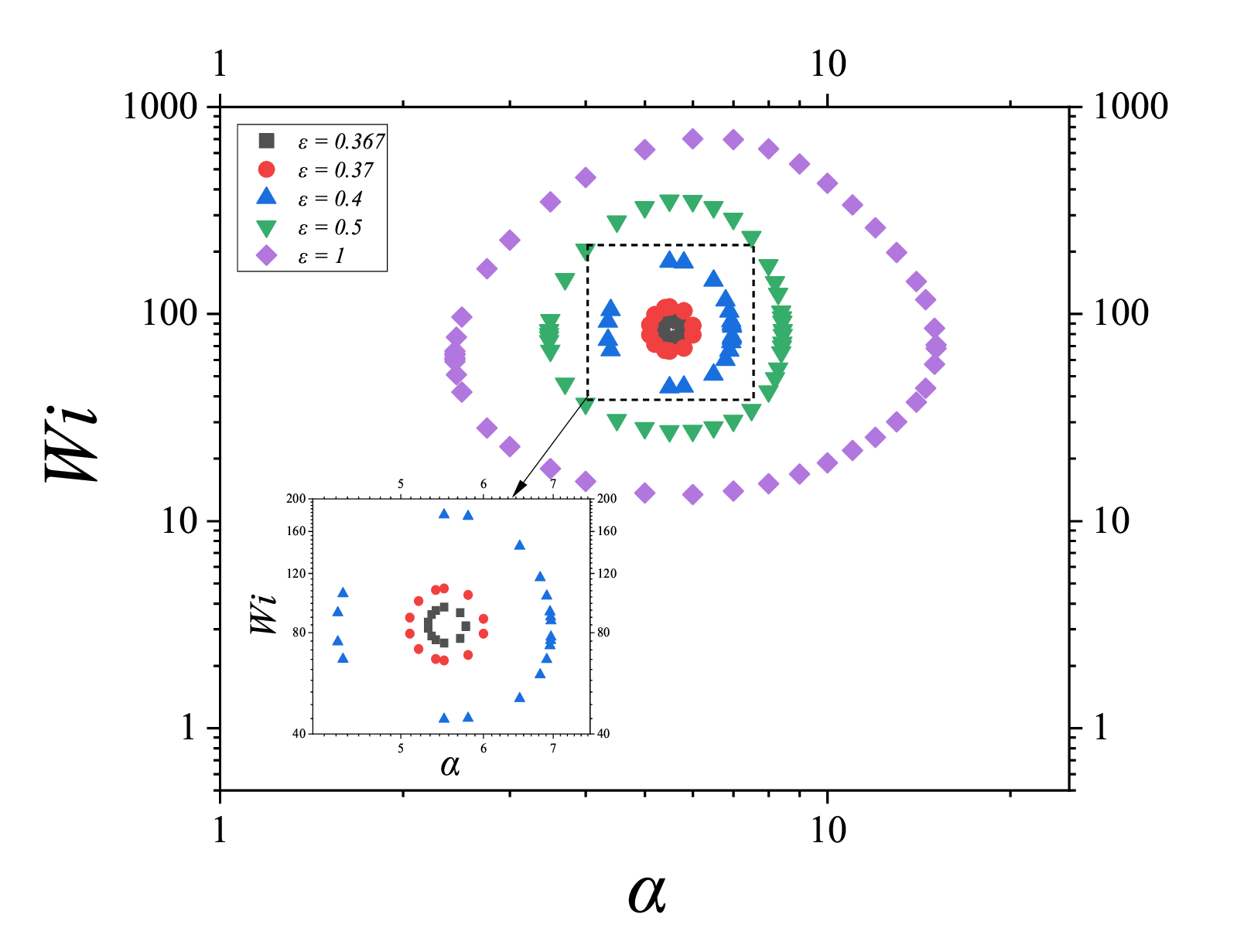}
  \caption{ Neutral curves   for HSM1 at different $\epsilon$'s showing the absence of HSM1 for $\epsilon \lesssim 0.36$. Data for $Re = 0$, $n = 0$, $\beta = 0.98 $ and $L$ = 100. The inset shows an enlarged view of the region $40 < W\!i < 200$.}
  \label{fig:L_100_n_0}
\end{figure}

 For Dean flow considered in this effort, the shear-thinning-induced stabilization, as expected, is strongest for the smallest $\epsilon$ ($ = 0.01$), owing to the high $W\!i$ -- the loop 
 in Fig.\,\ref{fig:subfigures_closed_loops_n_0_e_0.1} 
 already being vanishingly small at $L = 536$. In contrast, for $\epsilon = 0.1$, the loop remains finite in extent, even for $L = 175$. This sensitivity to $\epsilon$ is further highlighted in Fig.\,\ref{fig:L_100_n_0}, where we show neutral stability loops in the $Wi$--$\alpha$ plane for $L = 100$, over a range of $\epsilon$'s.  The size of the unstable region decreases with  decreasing $\epsilon$, for the said $L$, with the HSM being absent for $\epsilon \lesssim 0.36$. For smaller $\epsilon$'s, it is necessary to further examine modes with $n \neq 0$ in order to ascertain the presence of an HSM instability, which we do in the next subsection.

\subsection{Role of finite extensibility on non-axisymmetric modes}
\label{subsec:Lonnonaxy}

\begin{figure}
  \centering
  \subfigure [$ L = 10^6 $]{\includegraphics[width=0.45\textwidth]{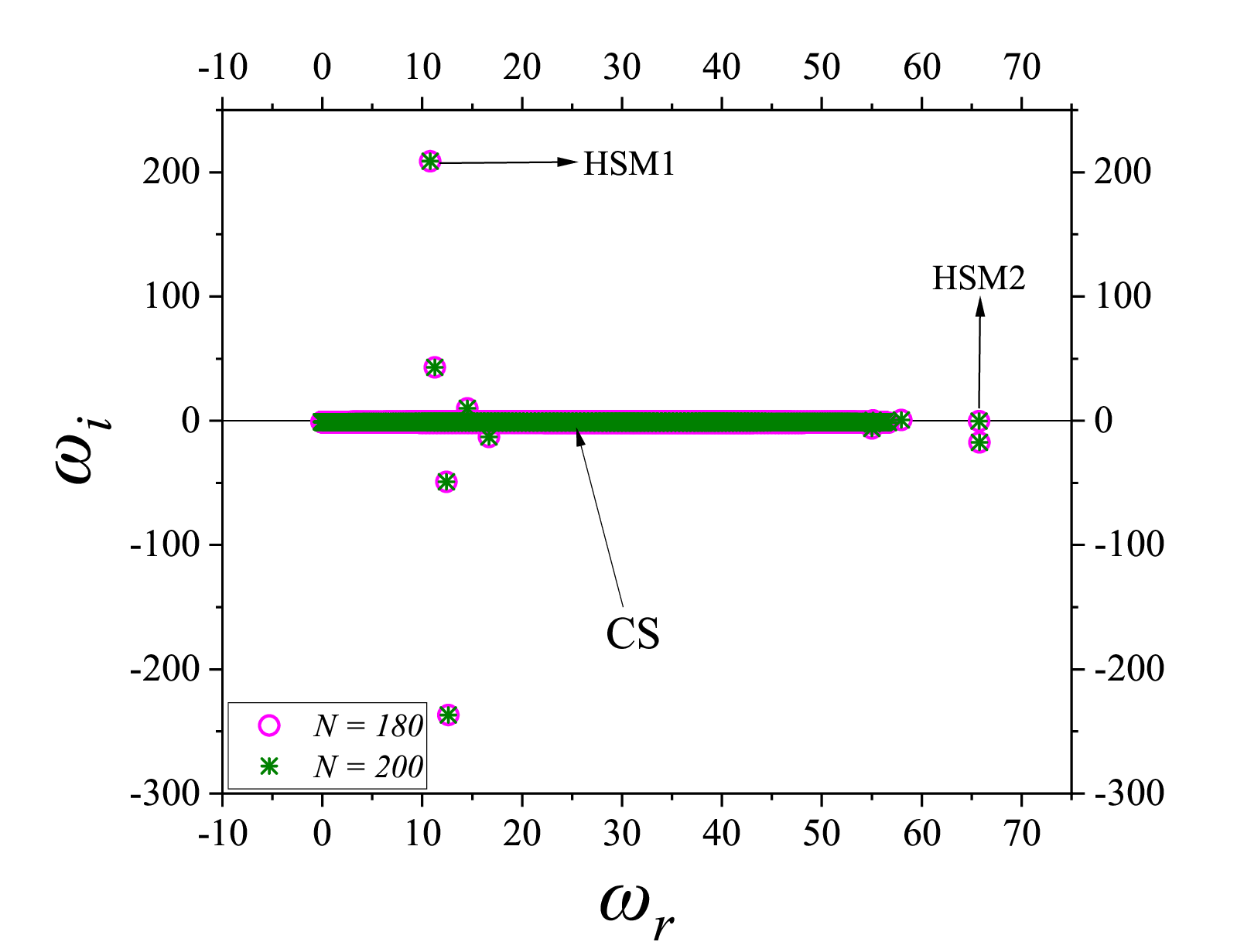}}
  \label{Showing_HSM2_L_1000000}
   \subfigure[$ L = 10^3 $]{\includegraphics[width=0.45\textwidth]{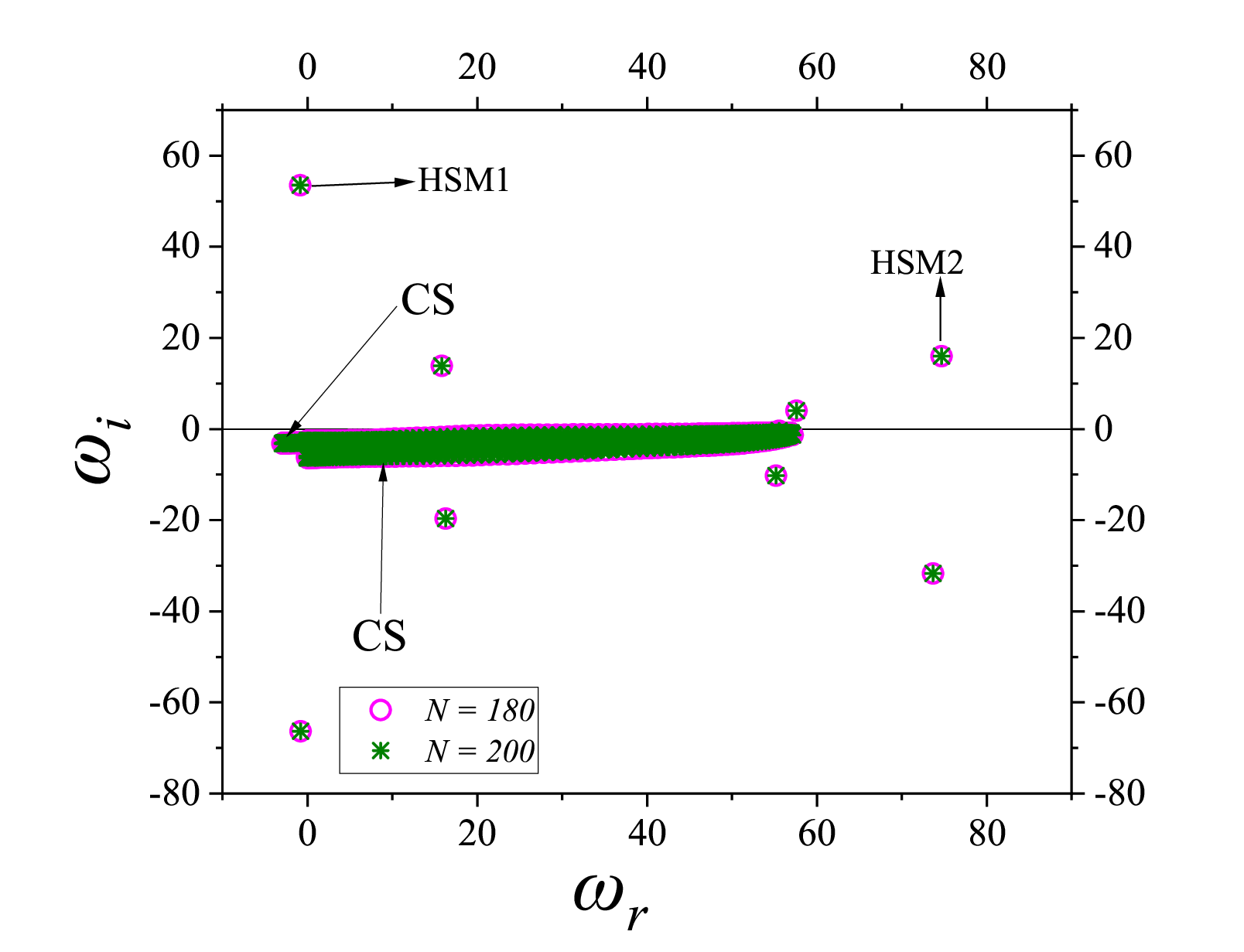}}\label{Showing_HSM2_L_1000}
    \subfigure[$ L = 100 $]{\includegraphics[width=0.45\textwidth]{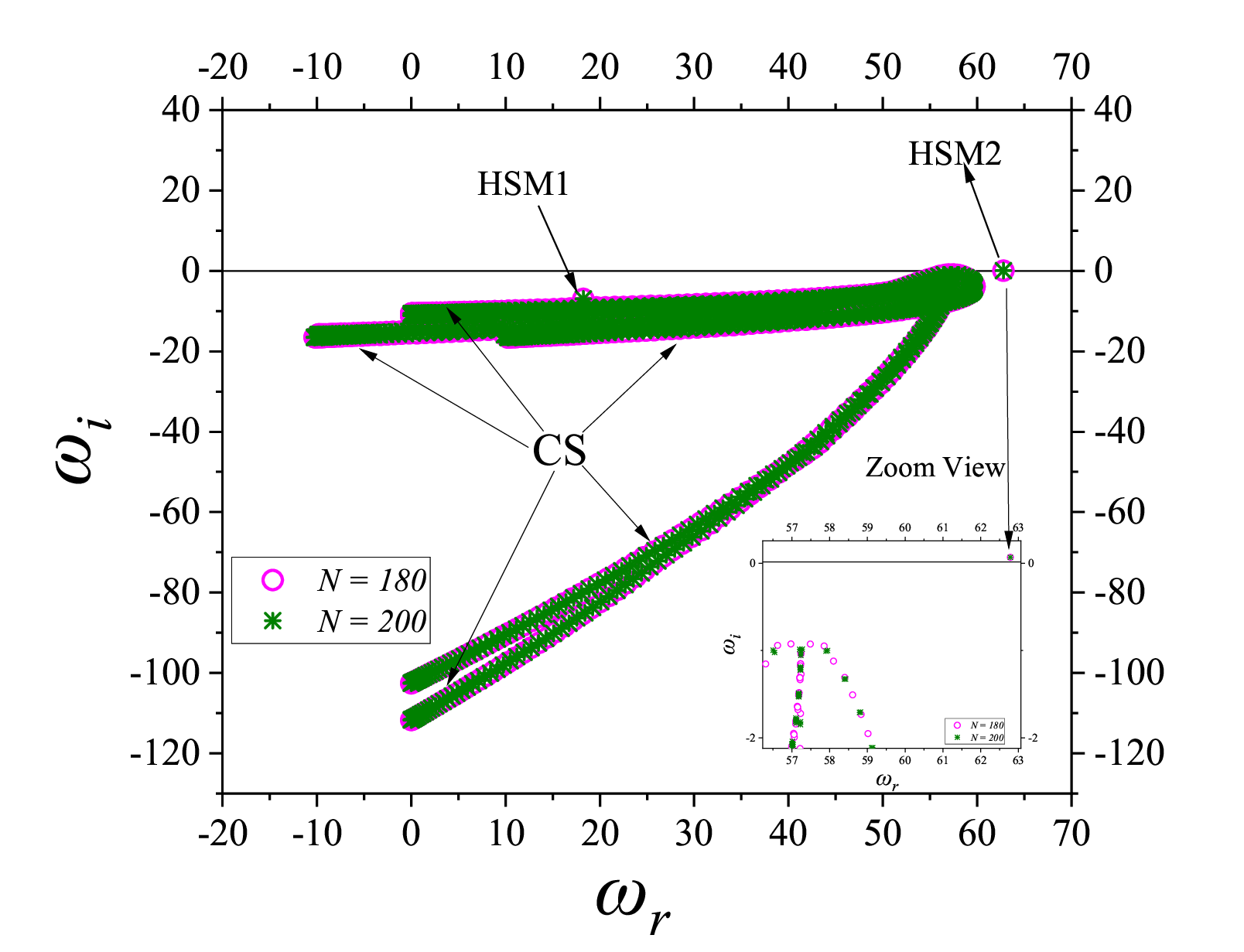}\label{Showing_HSM2_L_100} }
  \caption{Eigenspectra showing HSM1 and 2 in Dean flow of a FENE-P fluid at $Re = 0$, $\alpha = 7$, $\beta = 0.98$, $ \epsilon = 0.1, n = 1 , W\!i = 400 $ at different $L$'s. The continuous spectra are indicated as `CS'.}
   \label{Showing_HSM2_at_diff_L}
\end{figure}

\begin{figure}
  \centering
  \subfigure[]{\includegraphics[width=0.45\textwidth]{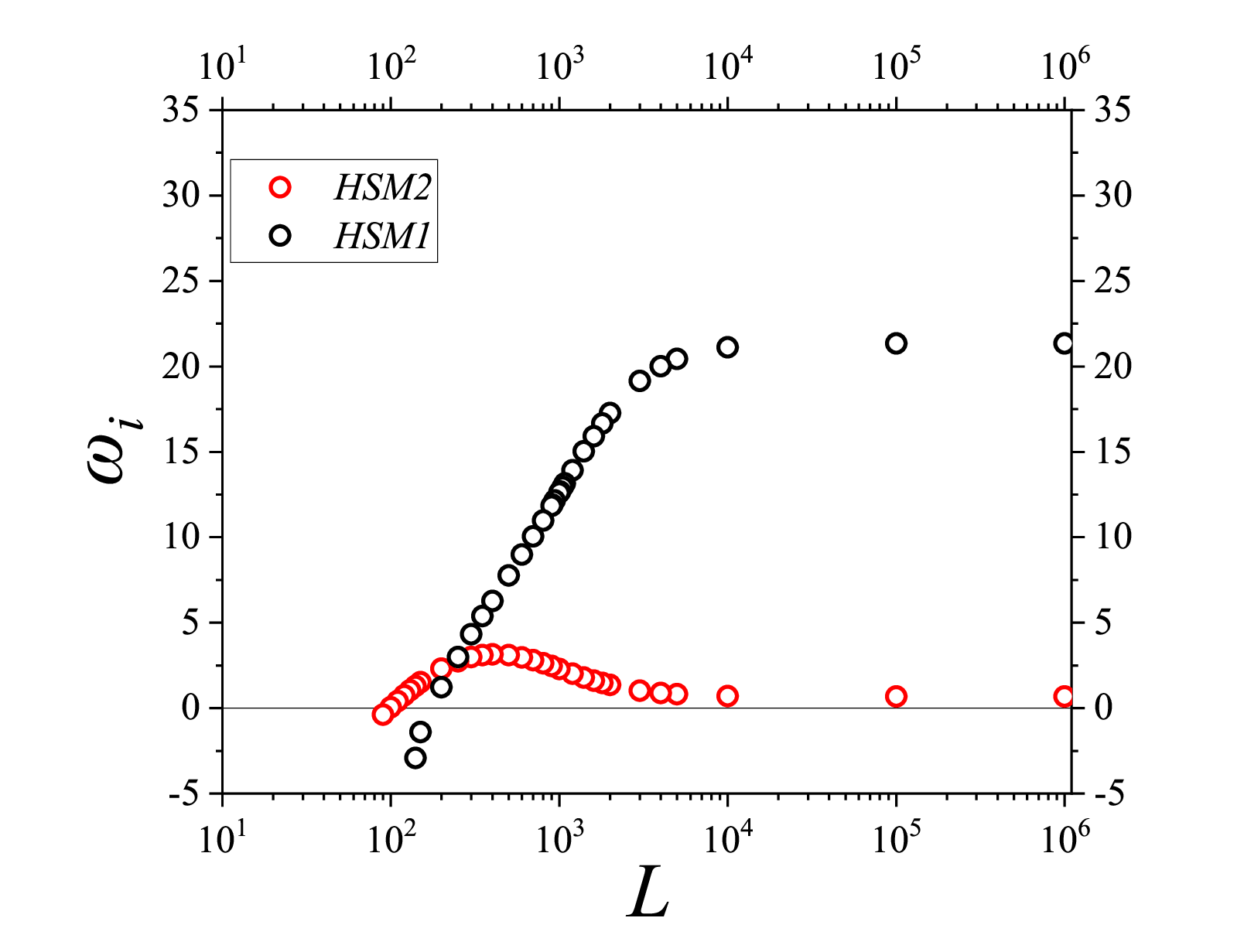}%
  \label{Mode_1_2_track}
  }
  \subfigure[]{\includegraphics[width=0.45\textwidth]{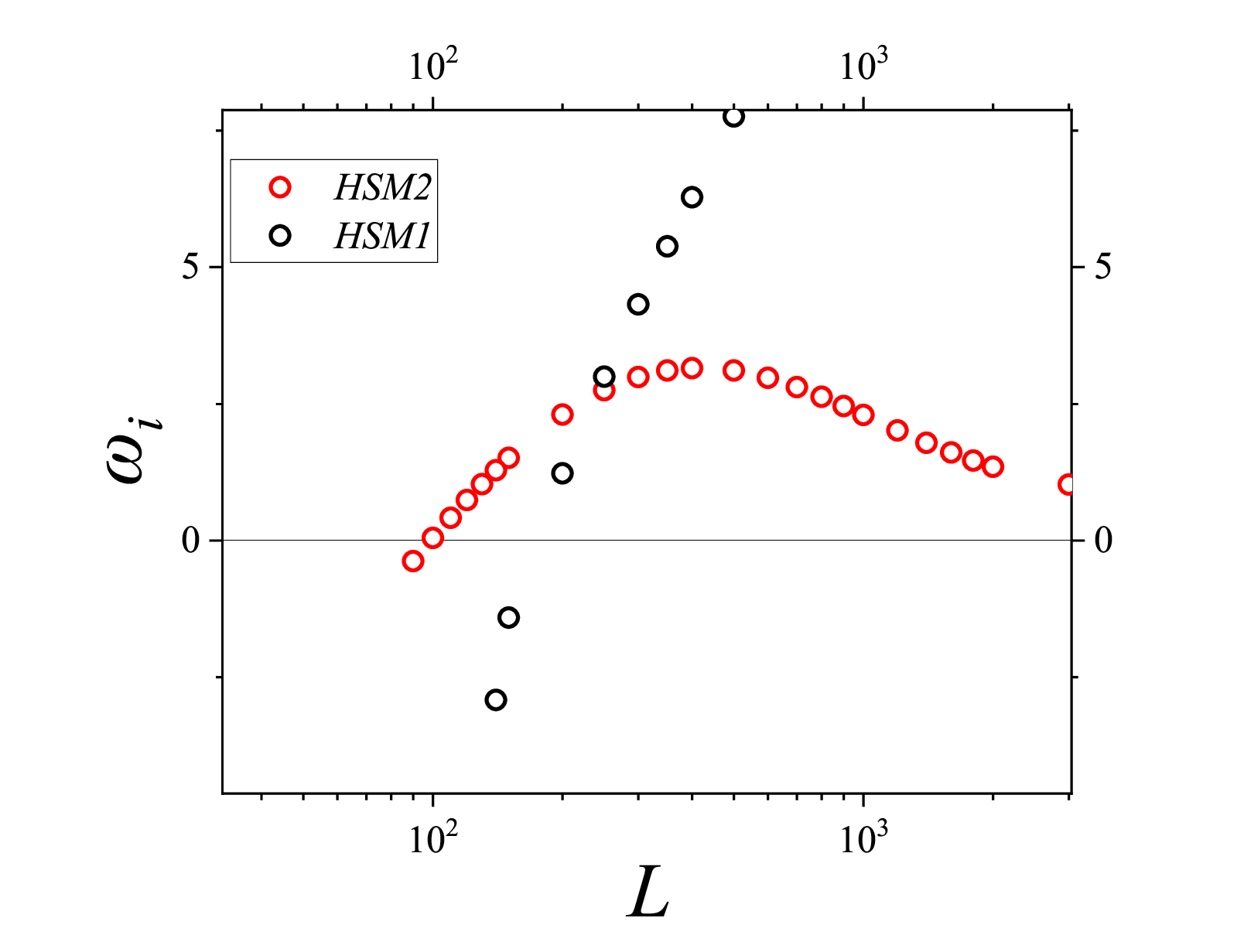}%
  \label{Zoom_in_Mode_1_2_track}
  }
  \caption{Variation of $\omega_i$ with $L$ for the hoop stress modes (HSM1 and HSM2) in Dean flow of a FENE-P fluid for $\epsilon = 0.1$, $\alpha = 7$, $n = 2$, $Re = 0$, $W\!i = 130$, and $\beta = 0.98$: (a) for $10 < L < 10^6$, and (b)  zoomed-in view of (a) highlighting the crossover from HSM1 to HSM2.}
  \label{fig:Tracking_of_Two_Modes_Non_axi_HSM_n_2}
\end{figure}

To begin with, we recall the discussion in Sec.\,\ref{subsec:nonaxy_spectrum} which had identified HSM2, a second discrete mode in the elastic spectrum, besides the dominant one (HSM1) that governed the transition for the Oldroyd-B case. Herein, we examine HSM1 and HSM2 for the FENE-P case in more detail.
Thus, while Fig.\,\ref{Showing_HSM2_L_200} was for $L = 200$, Fig.\,\ref{Showing_HSM2_at_diff_L} examines the non-axisymmetric spectrum ($n = 1$) for $L$ decreasing from $10^6$ (the Oldroyd-B limit for all practical purposes) down to 100.
Decreasing $L$ has a stabilising effect on HSM1, as one might anticipate from the stabilising role of shear thinning on the hoop-stress instability mentioned in our earlier discussion.
Rather intriguingly, for HSM2, decreasing $L$ has an initial destabilizing effect, although an ultimately stabilizing one.  This destabilising influence of finite extensibility on HSM2 is reminiscent of its effect  on the centre mode in plane channel flow \citep{buza_page_kerswell_2022,KhalidFENEP2025}, and in Dean flow (as discussed in Sec.\,\ref{subsec:LonCM}).  
However, HSM2 is not a centre mode, and has more in common with HSM1, in terms of the trends shown by its eigenfunctions - this is demonstrated in  Appendix\,\ref{sec:Appendixeigenfunctions} which compares the eigenfunctions for HSM1, HSM2, and CM. 
Figure\,\ref{fig:Tracking_of_Two_Modes_Non_axi_HSM_n_2} shows the variation of the growth rate ($\omega_i$) with $L$ for both HSM1 and 2 (for $n = 2$), with 
Fig.\,\ref{Zoom_in_Mode_1_2_track} presenting a magnified view of 
the region $50 < L < 10^3$. The non-monotonic variation of $\omega_i$ for HSM2 is evident in Fig.\,\ref{Mode_1_2_track} and is in contrast to the behaviour of HSM1.
This leads to HSM1 being the critical mode for $L > 200$, and HSM2 being critical for $L < 200$. 
Indeed, for $L = 100$, the only unstable mode in the spectrum is HSM2, with the classical hoop-stress mode (HSM1) being completely stabilised.
An identical scenario prevails for $n = 1$ (not shown).

\begin{figure}
    \centering
    \subfigure[Oldroyd-B]{
        \includegraphics[width=0.45\textwidth]{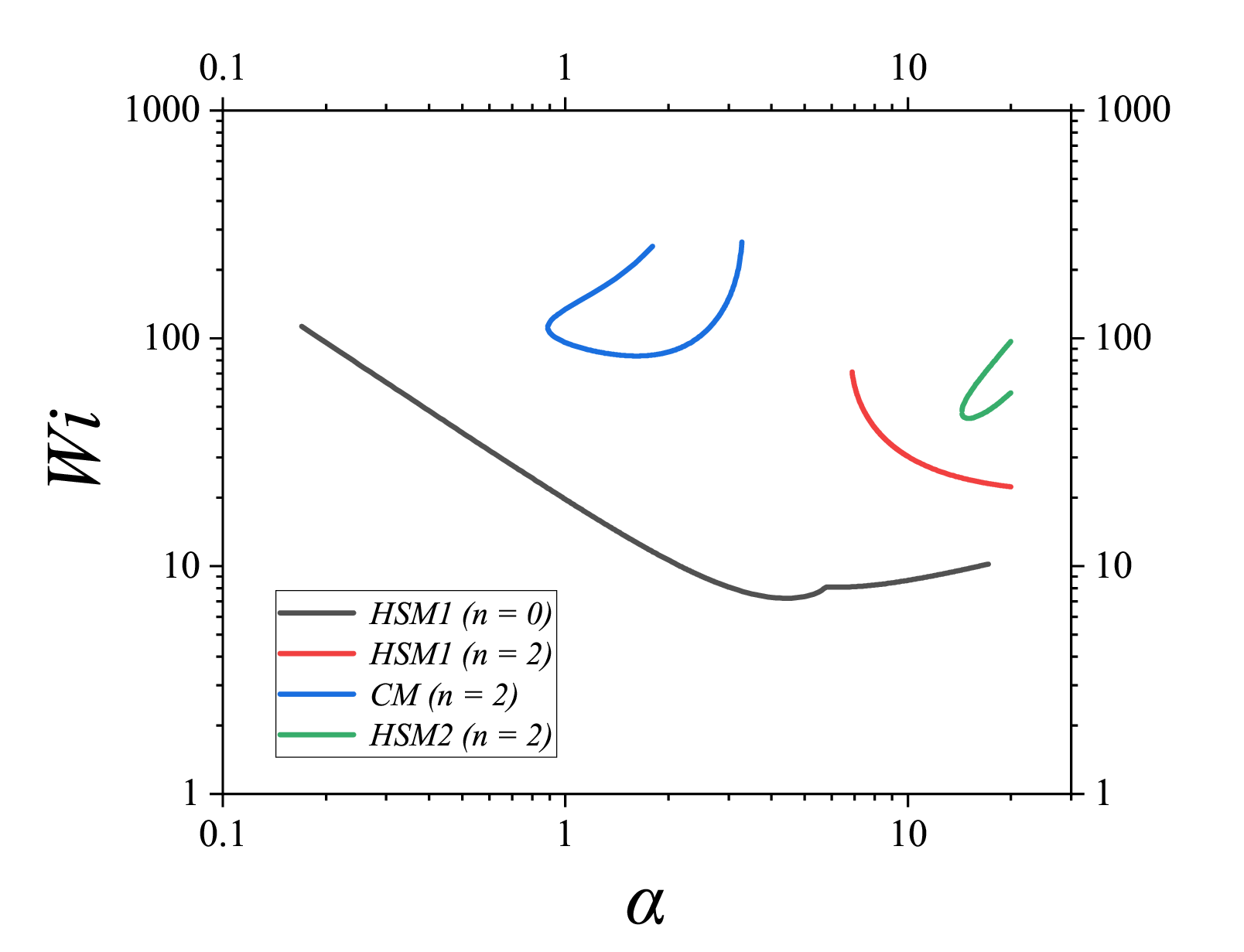} \label{fig:e_1_Old_B}
    }
    \subfigure[$L = 100$]{
        \includegraphics[width=0.45\textwidth]{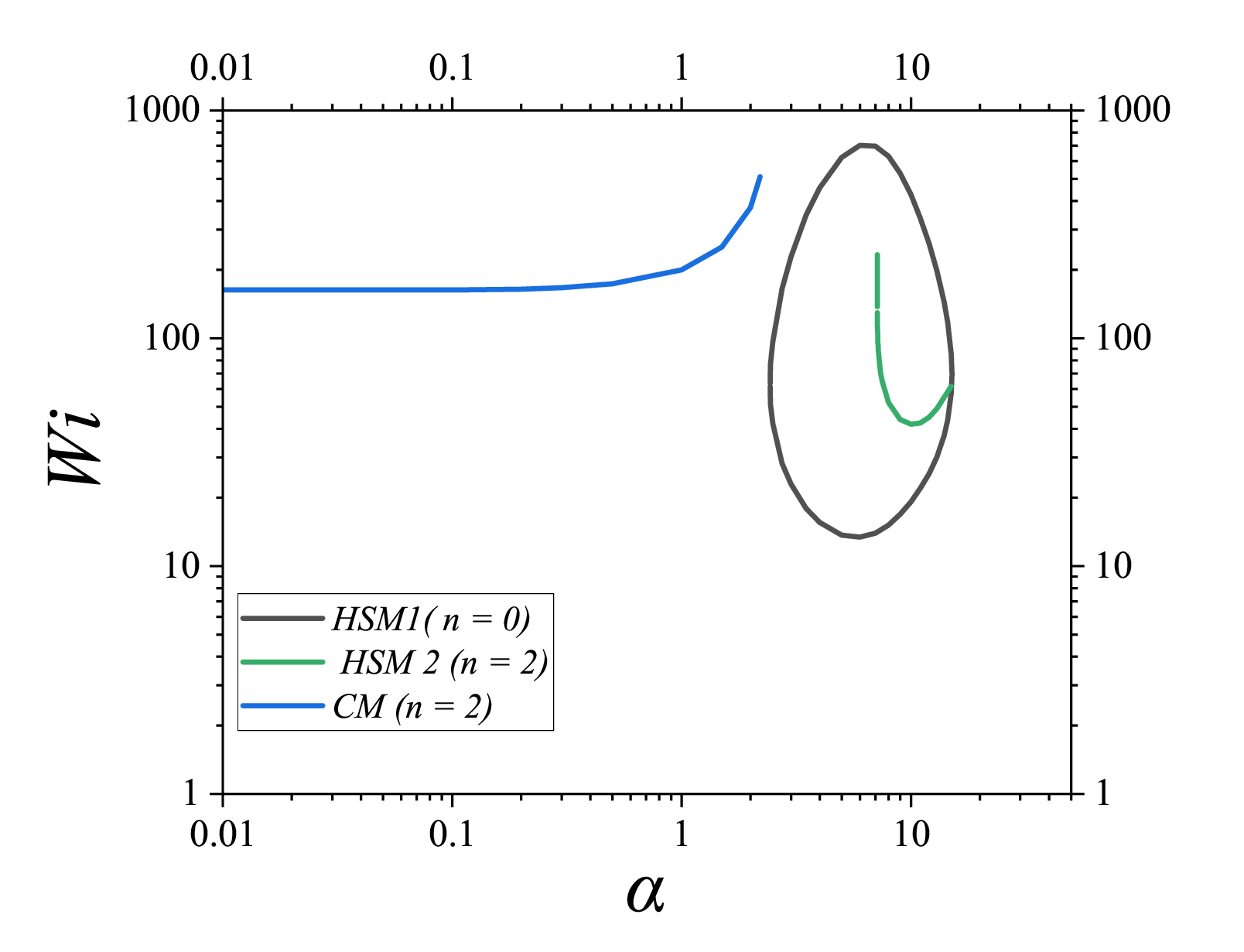}\label{fig:e_1_FENE_P_L_100}
    }
    \caption{Neutral curves for CM and  non-axisymmetric HSM2 for the Oldroyd-B and FENE-P ($L = 100$) models. Data for $n = 2$, $Re = 0$, $\epsilon = 1$,  $\beta = 0.98$. The results for axisymmetric HSM1 are shown for reference.}
    \label{fig:Showing_CM_and_HSM2_in_Oldroyd_B_and_FENE_P_n_2_e_1}
\end{figure}

For the FENE-P case, the destabilizing influence of finite extensibility on both CM and HSM2 might lead one to anticipate that these two modes might be connected in some regions of the parameter space. Figure\,\ref{fig:Showing_CM_and_HSM2_in_Oldroyd_B_and_FENE_P_n_2_e_1} shows that this is not the case.
This figure shows the neutral curves for HSM2 and CM, in the $W\!i$--$\alpha$ plane, for both Oldroyd-B and FENE-P ($L = 100$) fluids, for $\epsilon = 1$.   
For the Oldroyd-B fluid, while the CM is stable for $\alpha = 0$ (when $\beta = 0.98$), the CM is destabilized by nonzero $\alpha$. In addition, there are neutral curves related to HSM1 and 2; in this figure, we do not show the HSM1 neutral curve for $n = 2$. 
The critical $W\!i$ for the $n = 2$ mode in the FENE-P ($L = 100$) model is significantly higher than that for the Oldroyd-B case, due to the stabilising influence of finite extensibility.
The unstable region for  HSM2 (with $n = 2$) is completely disconnected from the CM, both for Oldroyd-B and FENE-P fluids. 
As $\alpha$ is increased for a  fixed $W\!i$ (in Fig.\,\ref{fig:Showing_CM_and_HSM2_in_Oldroyd_B_and_FENE_P_n_2_e_1}), we find that the CM becomes stable and eventually merges into the CS, a trend that is reminiscent of the CM in rectilinear channel and pipe flows \citep{chaudharyetal_2021,khalid2021centermode}.
Similarly, when we tracked HSM2 for a $W\!i$ from the unstable region, the $\omega_r$ for this mode is such that it is beyond the CS range, and hence this mode circumvents the CS and eventually ends up below it, without crossing the same. This shows that the two modes (HSM2 and CM) do not continue to each other, as $\alpha$ is varied for a fixed $W\!i$. 
While the above discussion pertains to HSM2 and CM, it is relevant to
emphasise that for the larger gap-width ratio of $\epsilon = 1$, the HSM1 (with $n = 0$) is the most critical for both the Oldroyd-B and FENE-P models.

\begin{figure}
  \centering
   \subfigure [$\epsilon = 0.1$]{\includegraphics[width=0.4\textwidth]{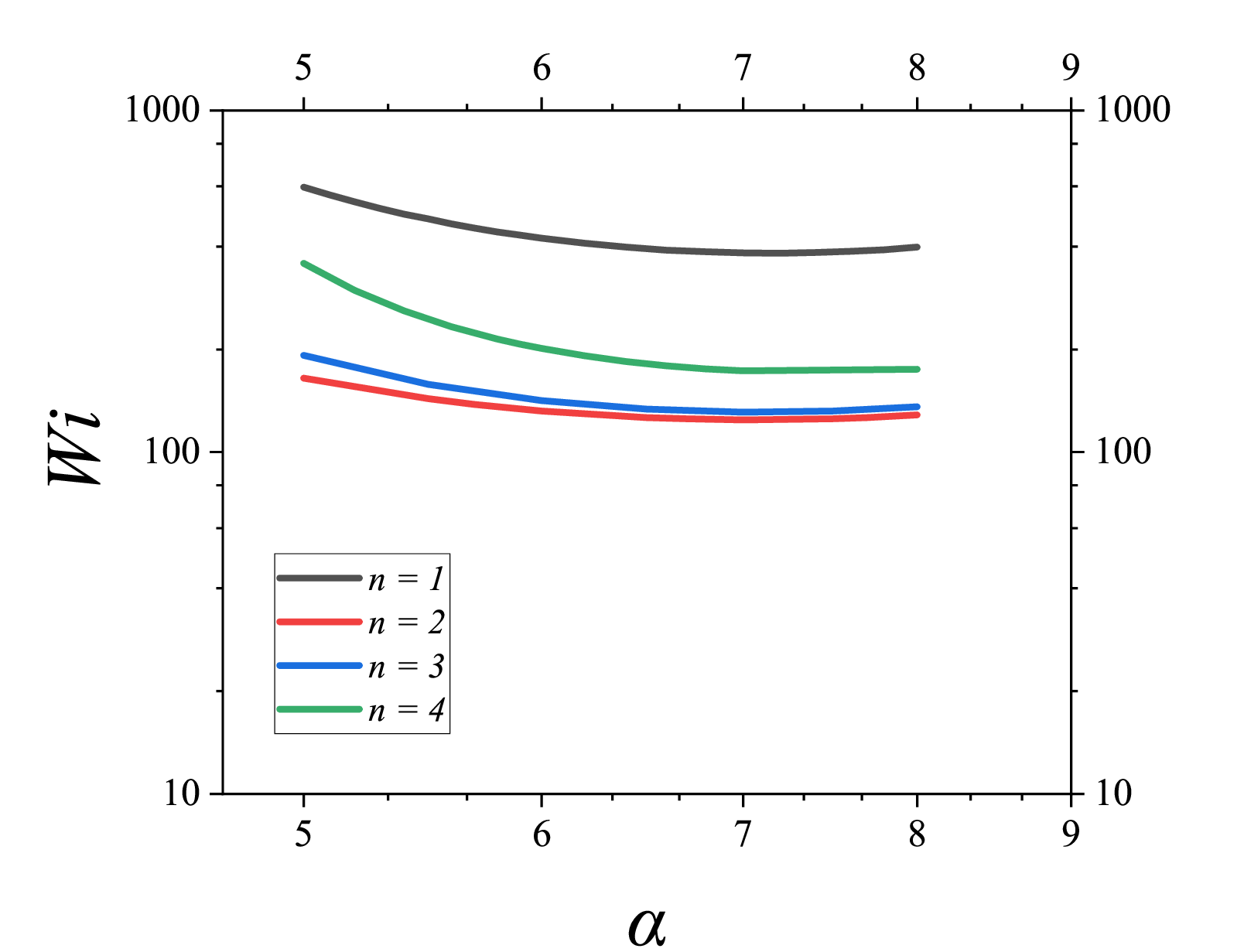}\label{fig:HSM_dean_e_0.01_beta_0.98_diff_L}}
  \subfigure[$\epsilon = 0.2$]{\includegraphics[width=0.4\textwidth]{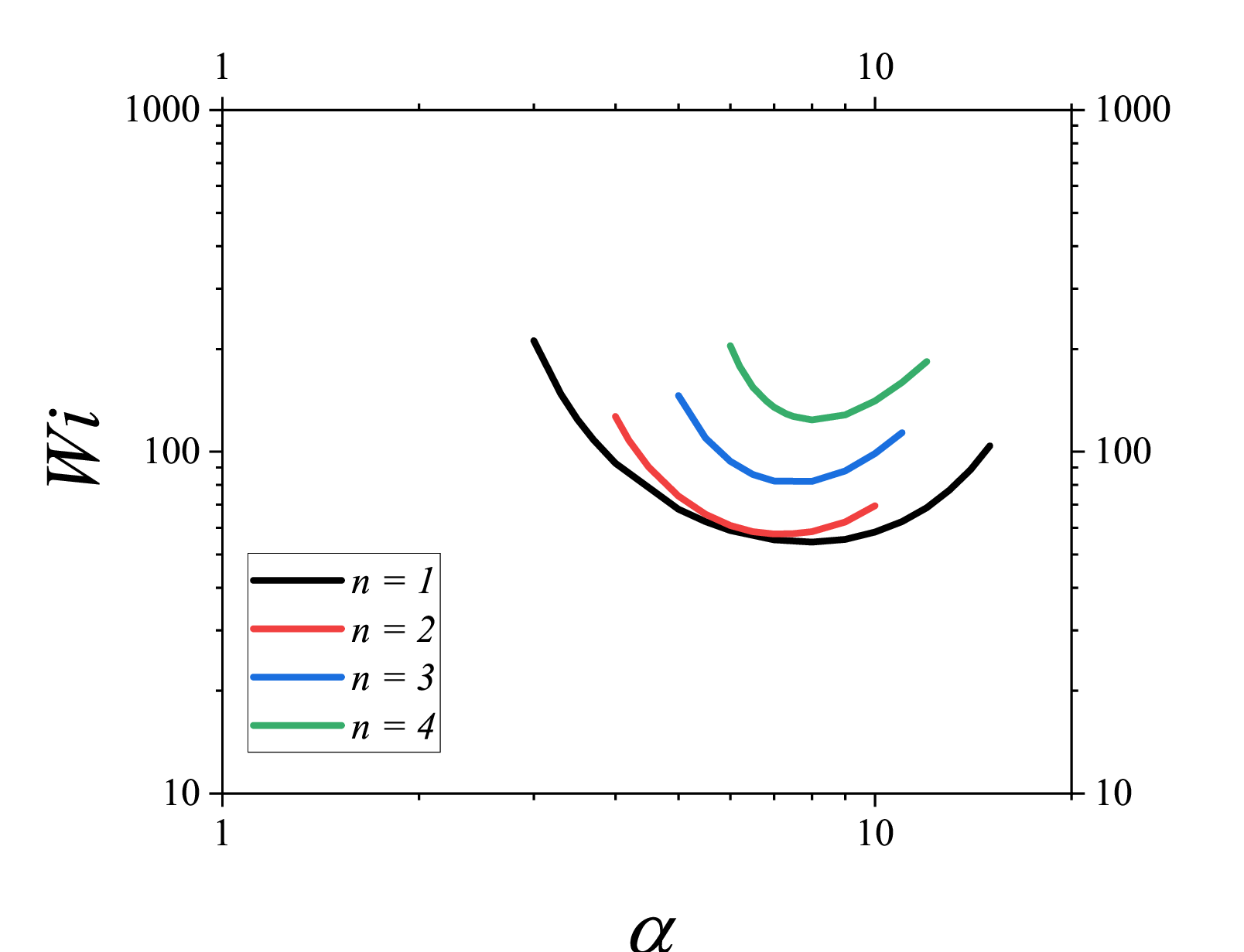}\label{fig:HSM_dean_e_0.2_beta_0.98_diff_L}}
   \subfigure [$\epsilon = 0.5$]{\includegraphics[width=0.4\textwidth]{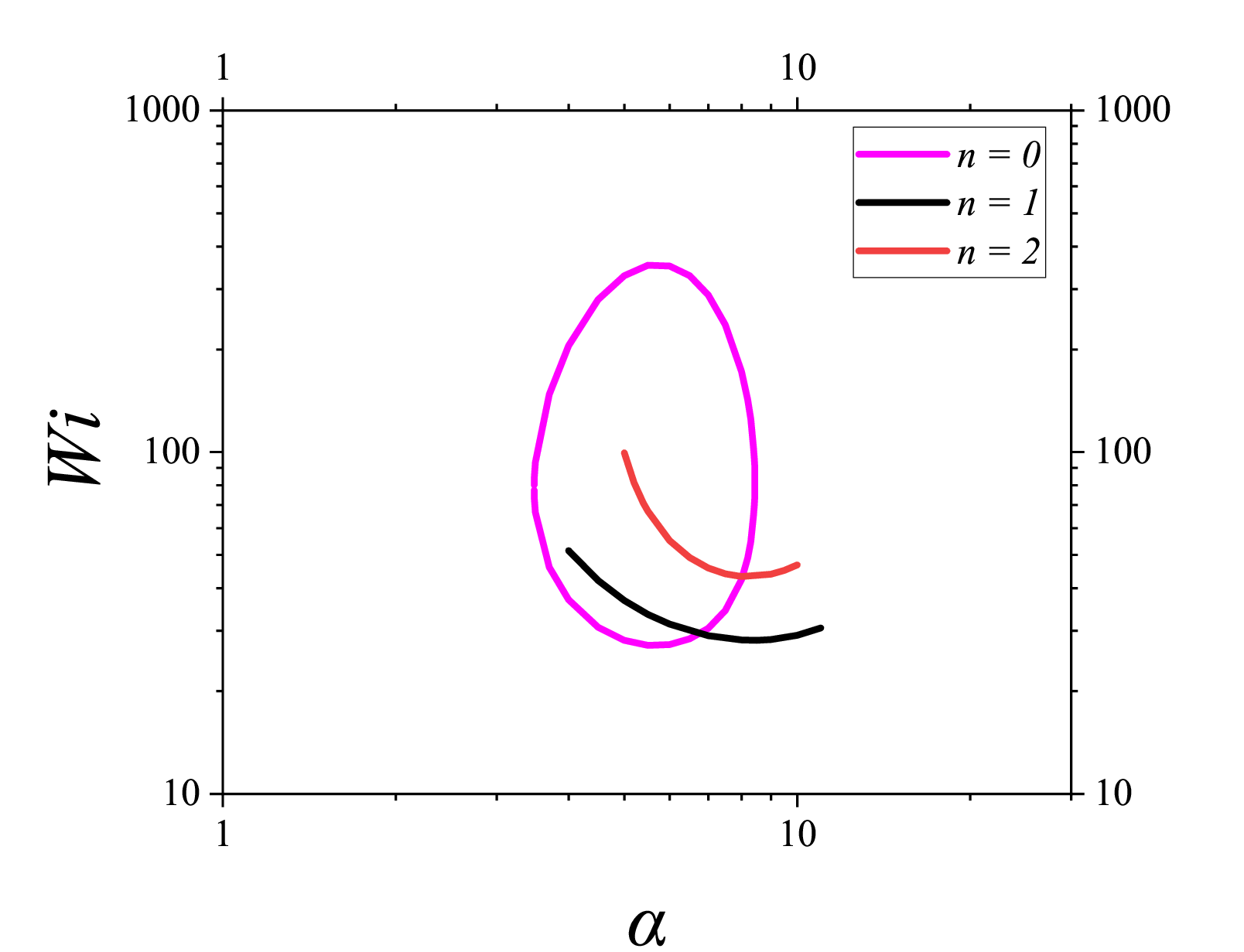}\label{fig:HSM_dean_e_0.5_beta_0.98_diff_L}}
  \subfigure[$\epsilon = 1$]{\includegraphics[width=0.4\textwidth]{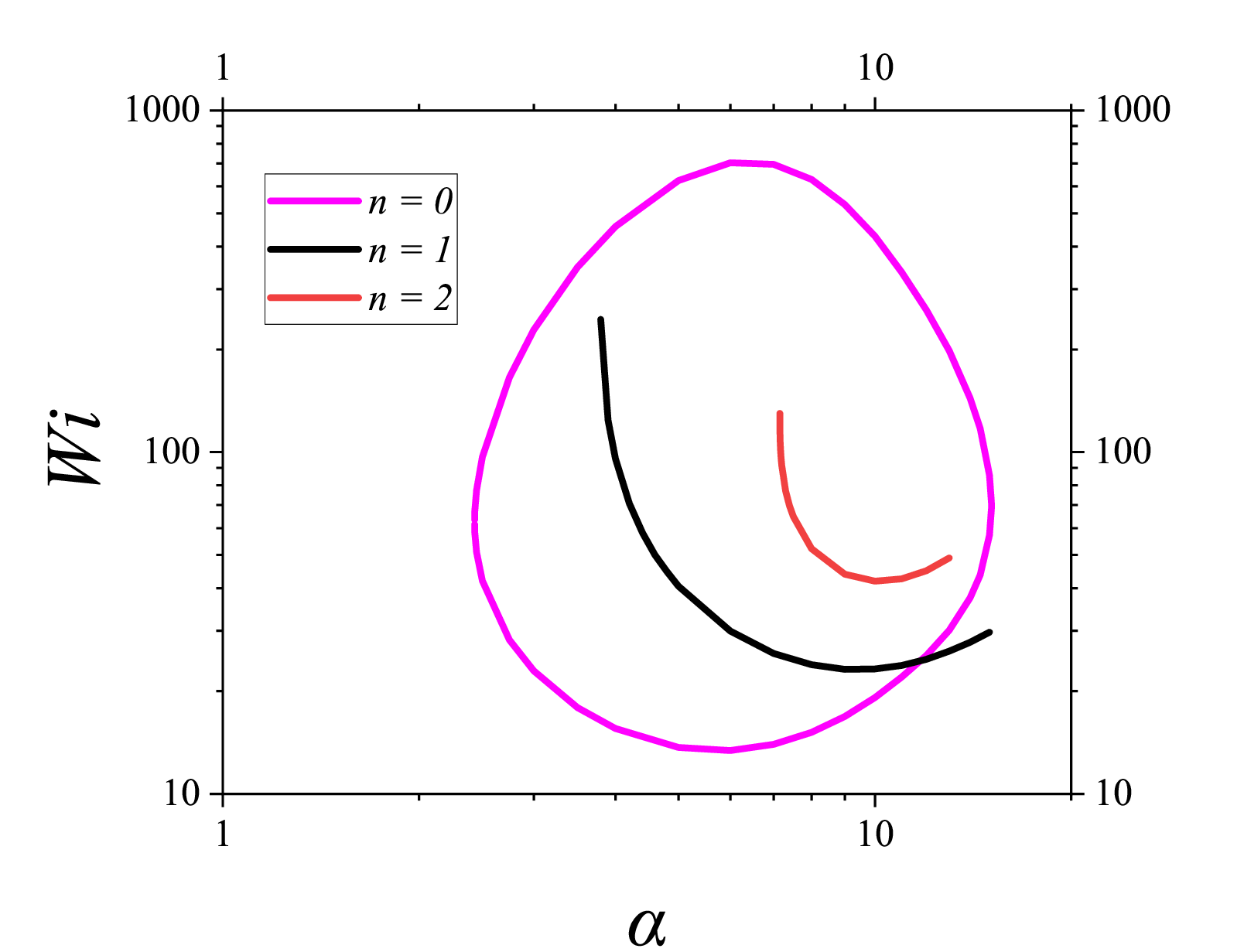}\label{fig:HSM_dean_e_1_beta_0.98_diff_L}}
  \caption{Neutral curves for HSM (1 and 2) at $Re = 0$, $\beta = 0.98$ and $L = 100$. 
While the $n \neq 0$ modes (HSM2) are the most unstable for $\epsilon = 0.1$ and $0.2$, 
and remain unstable even for $\epsilon = 0.5$ and $1$, the $n = 0$ mode (HSM1) becomes 
critical at $\epsilon = 0.5$ and $1$.}
   \label{fig:Fene_P_HSM4}
\end{figure}

As mentioned in Sec.\,\ref{sec:HSMinDeanflow}, for the Oldroyd-B case, the axisymmetric HSM1 is the most unstable mode for any $\epsilon$. However, the trend is very different for the FENE-P case, being sensitive to the choice of $\epsilon$ and $L$.
For $L \sim 100$, the nature of the most unstable mode (viz., axisymmetric vs. non-axisymmetric HSM2)  depends on $\epsilon$, as shown in  Fig.\,\ref{fig:Fene_P_HSM4}.  In the narrow-gap regime ($\epsilon = 0.01 - 0.2$), the axisymmetric modes (both HSM1 and 2) remain stable for $L = 100$, while a non-axisymmetric HSM2 becomes critical. Specifically, at $\epsilon = 0.1$, $n = 2$ is the critical mode (HSM2), whereas at $\epsilon = 0.2$, $n = 1$ emerges as the critical mode (HSM2). In contrast, for the larger values $\epsilon = 0.5$ and $1$, the axisymmetric (HSM1) mode becomes critical. Notably, the range of $\epsilon$'s between $0.1$ and $0.3$ corresponds to gap widths used in the Taylor-Dean experiments of \citep{joo_shaqfeh_1994}, although the serpentine channel experiments of \cite{Groisman2001,Groisman2004} used a higher $\epsilon = 1$; we return to this issue in Sec.\,\ref{sec:concl}.

\begin{figure}
    \centering
    \subfigure[$W\!i_c$ vs. $\epsilon$]{
        \includegraphics[width=0.6\textwidth]{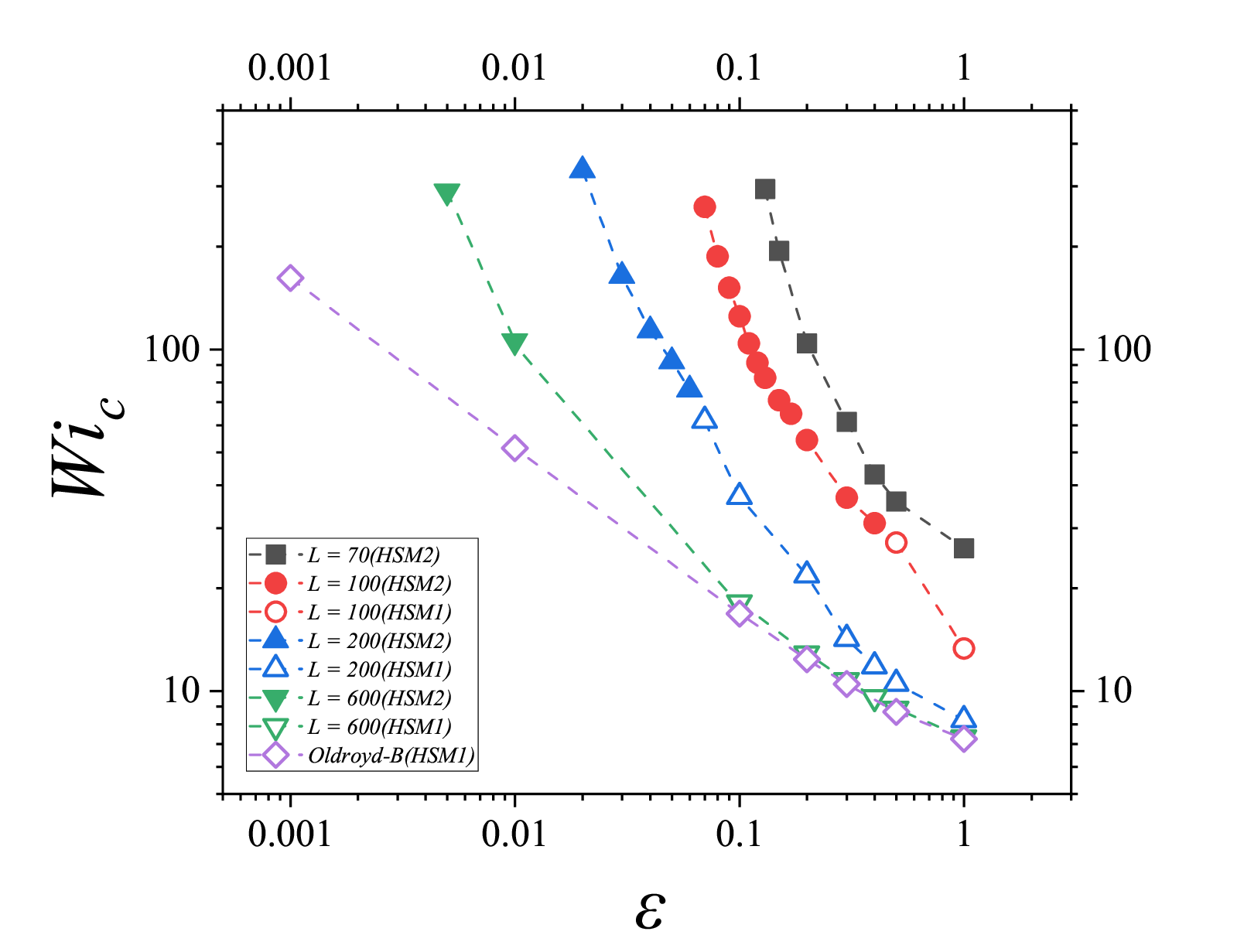} \label{fig:L_affect_on_HSM}
    }
    \subfigure[$n_c$ vs. $\epsilon$]{
        \includegraphics[width=0.45\textwidth]{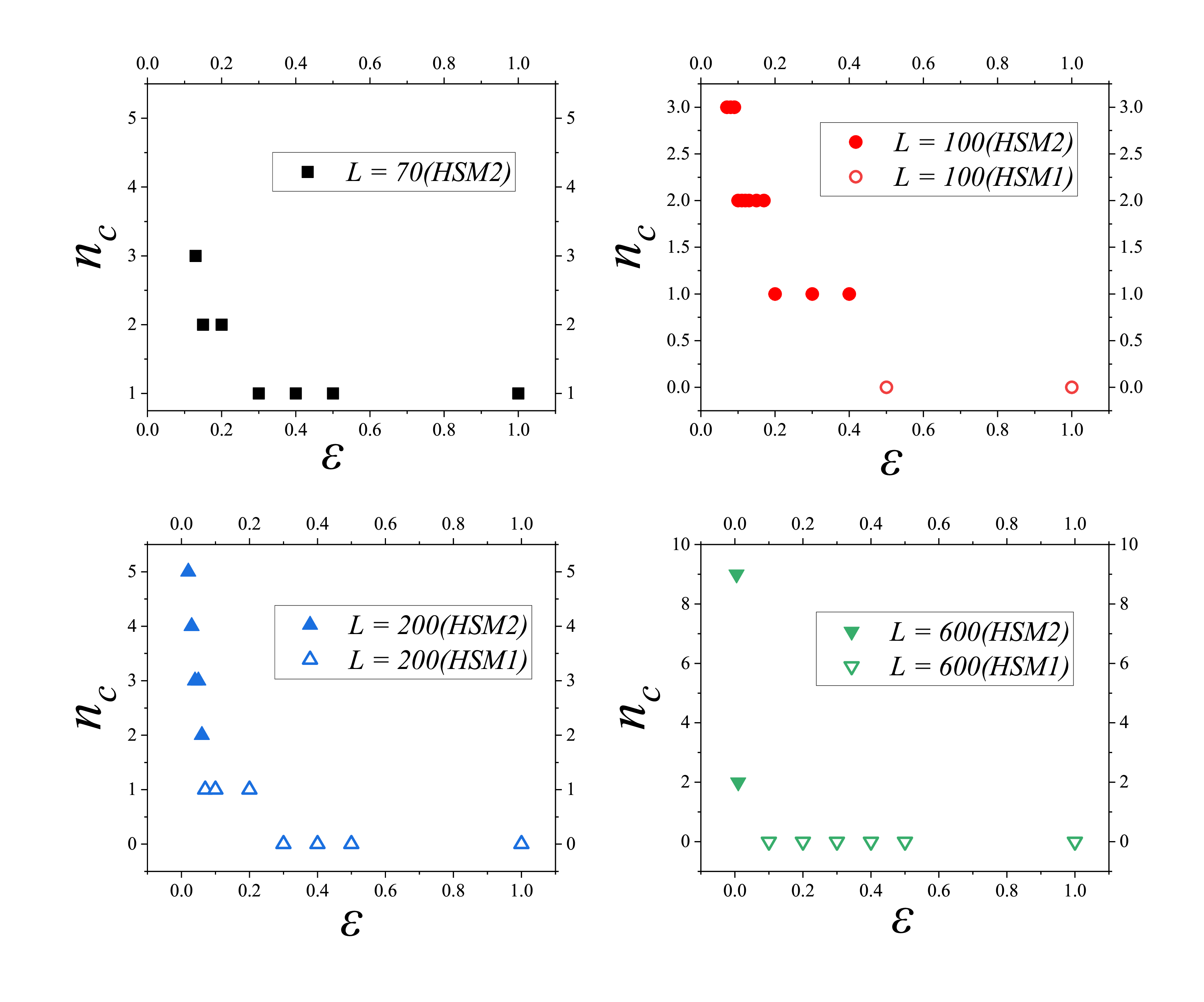}\label{fig:L_affect_on_HSM_n_c_vs_epsilon}
    }
    \subfigure[$\alpha_c$ vs. $\epsilon$]{
        \includegraphics[width=0.45\textwidth]{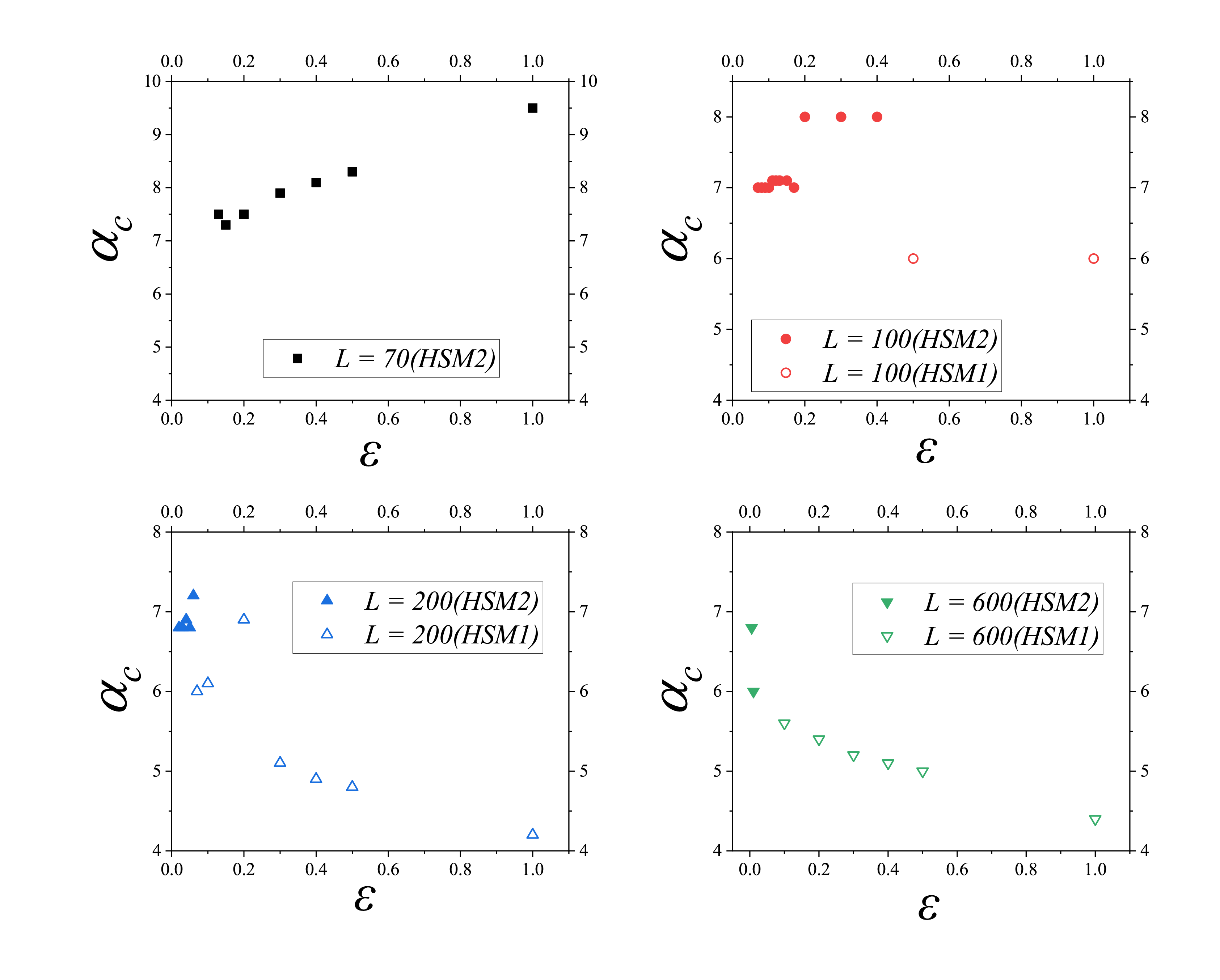}\label{fig:L_affect_on_HSM_alpha_c_vs_epsilon}
    }
    \caption{(a) $W\!i_c$ vs. $\epsilon$ for HSM (1 and 2) in Dean flow at different $L$'s illustrating the stabilising influence of finite extensibility. Here, filled symbols of a given shape denote HSM2 and the corresponding open ones denote HSM1; different symbol shapes corresponds to different $L$'s. Panels (b) and (c) show the corresponding $n_c$ and $\alpha_c$ variation with $\epsilon$. Data for $Re = 0$ and $\beta = 0.98$.}
    \label{fig:L_affect_on_HSM1_and_HSM2}
\end{figure}

For the Oldroyd-B model, the critical Weissenberg number ($W\!i_c$) was seen to scale as $\epsilon^{-1/2}$ for $\epsilon \ll 1$.  In contrast, and
as demonstrated in Figs.\,\ref{fig:subfigures_closed_loops_n_0} and \ref{fig:Fene_P_HSM4}, one expects the HSM instability to be fully suppressed for $\epsilon \rightarrow 0$, as $L$ is decreased. This is illustrated in Fig.\,\ref{fig:L_affect_on_HSM}, where the critical $W\!i_c$ is seen to diverge faster than $\epsilon^{-1/2}$ for all finite $L$. Note that the data points in Fig.\,\ref{fig:L_affect_on_HSM} for $W\!i_c$ (at different $\epsilon$'s or $L$'s) could correspond to either HSM1 ($n = 0$), or HSM1 ($n \neq 0$), or HSM2 ($n \neq 0$). Importantly, as $L$ is decreased to realistic values $\sim 100$, the HSM1 is fully suppressed in Dean flow, with a suggestion of divergence at a finite $\epsilon$.
This result is in qualitative agreement with what one obtains at a scaling level by incorporating shear thinning in the Pakdel-McKinley criterion \citep[see Fig.\,15 of][]{CastilloSanchez2022}. 
Figure\,\ref{fig:L_affect_on_HSM_n_c_vs_epsilon} shows the critical azimuthal wavenumber $n_c$ as function of $\epsilon$ at different $L$'s; the notation used is similar to 
Fig.\,\ref{fig:L_affect_on_HSM}, in that open symbols pertain to HSM1, while filled symbols to HSM2.
While in the Oldroyd-B limit, $n = 0$ always remains critical for all $\epsilon$'s,  the critical azimuthal wavenumber for lower $L$'s changes with $\epsilon$, and there does not appear to be a simple trend.  Figure\,\ref{fig:L_affect_on_HSM_alpha_c_vs_epsilon} shows the corresponding critical axial wavenumber $\alpha_c$. The abrupt variation of $\alpha_c$ with $\epsilon$ in this figure is on account of the transition from HSM1 to HSM2, and also due to changes in $n_c$.

\begin{figure}
  \centering
  \includegraphics[width= 0.45\textwidth]{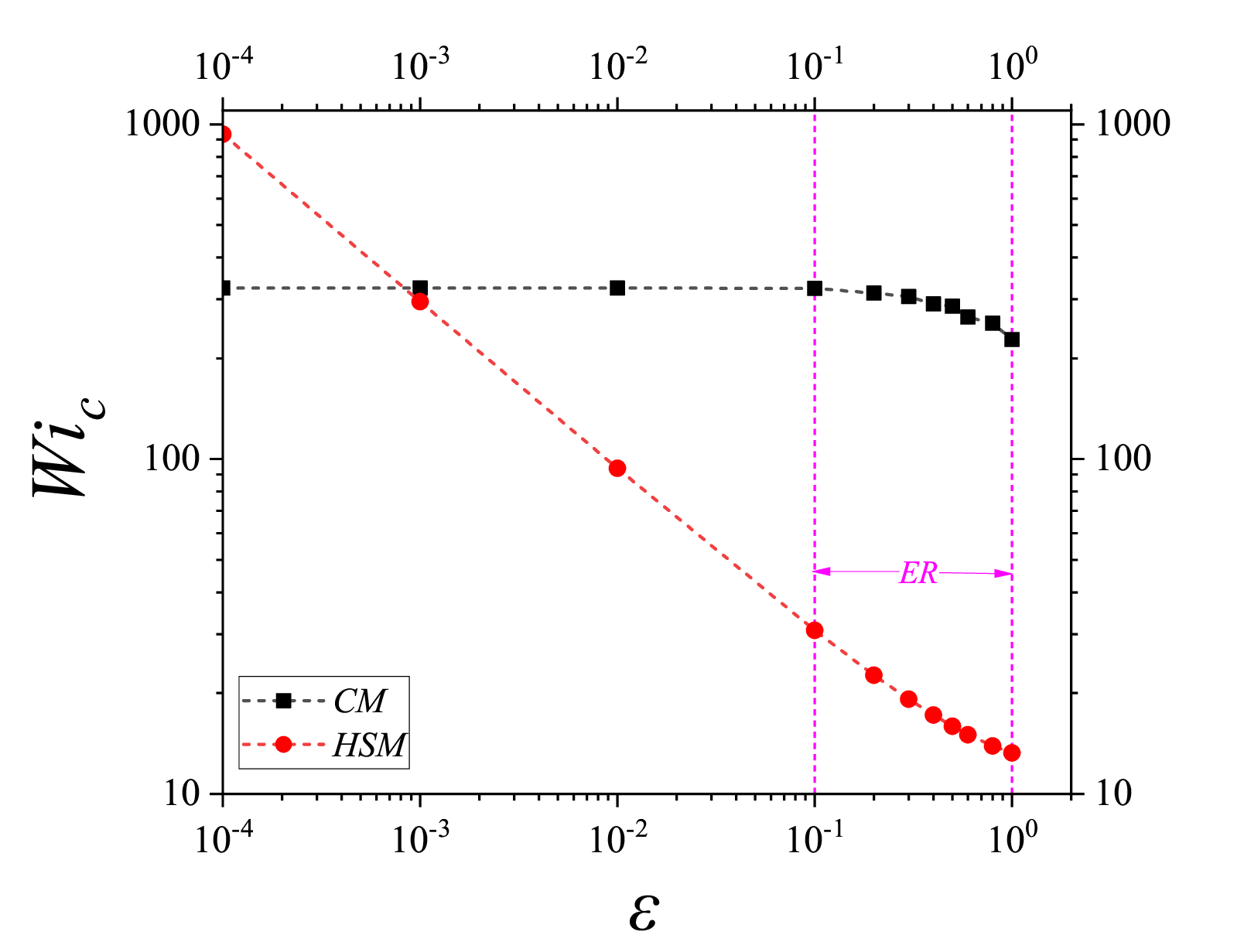}
  \caption{ For the Oldroyd-B fluid, the CM becomes the critical mode only for $\epsilon < 10^{-3}$; for $\epsilon > 10^{-3}$, HSM is more unstable. Data showing the
  variation of $W\!i_c$ with $\epsilon$ for $Re = 0$ and $\beta = 0.994$. The range indicated by `ER' denotes the $\epsilon$'s typically used in experiments involving Taylor-Couette and Taylor-Dean configurations.}
  \label{fig:W_c_vs_e_HSM_vs_CM}
\end{figure}

\begin{figure}
\centering
\subfigure[$ L = 600$]{\includegraphics[width=0.45\textwidth]{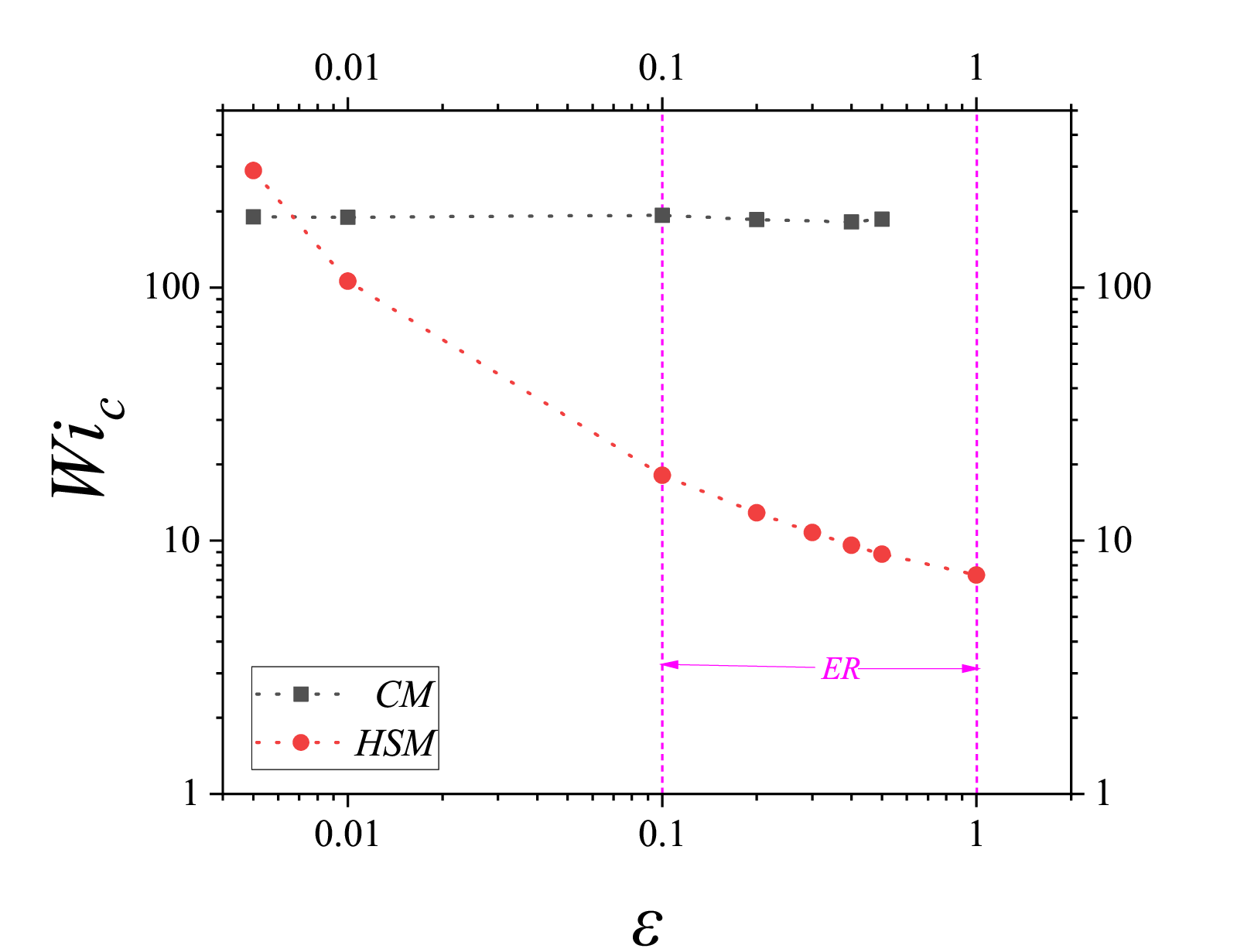}\label{fig:CM__L_600}}
  \,\, 
  \subfigure [$ L = 200$]{\includegraphics[width=0.45\textwidth]{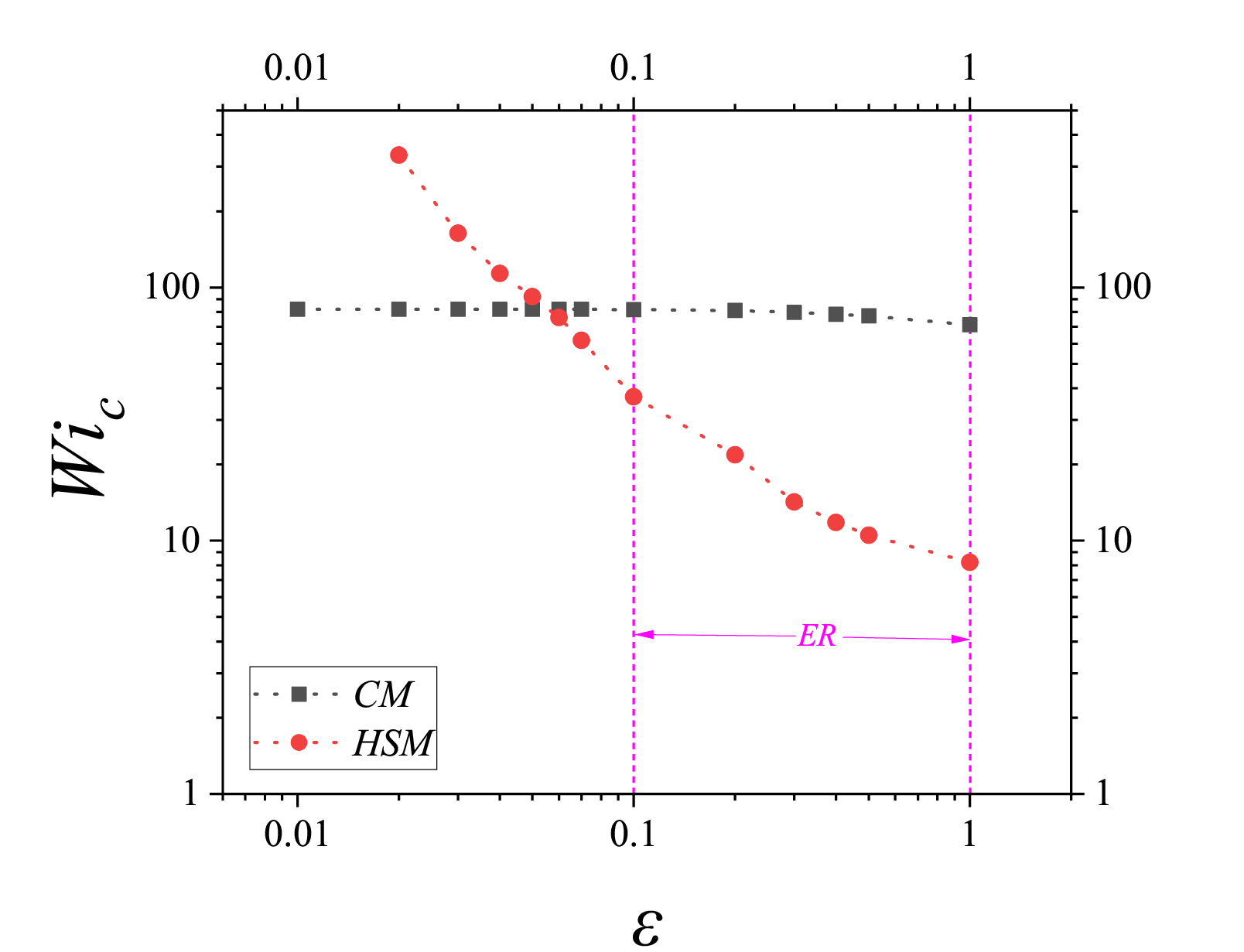}}\label{fig:CM__L_200}
  \,\, 
  \subfigure[$ L = 100$]{\includegraphics[width=0.45\textwidth]{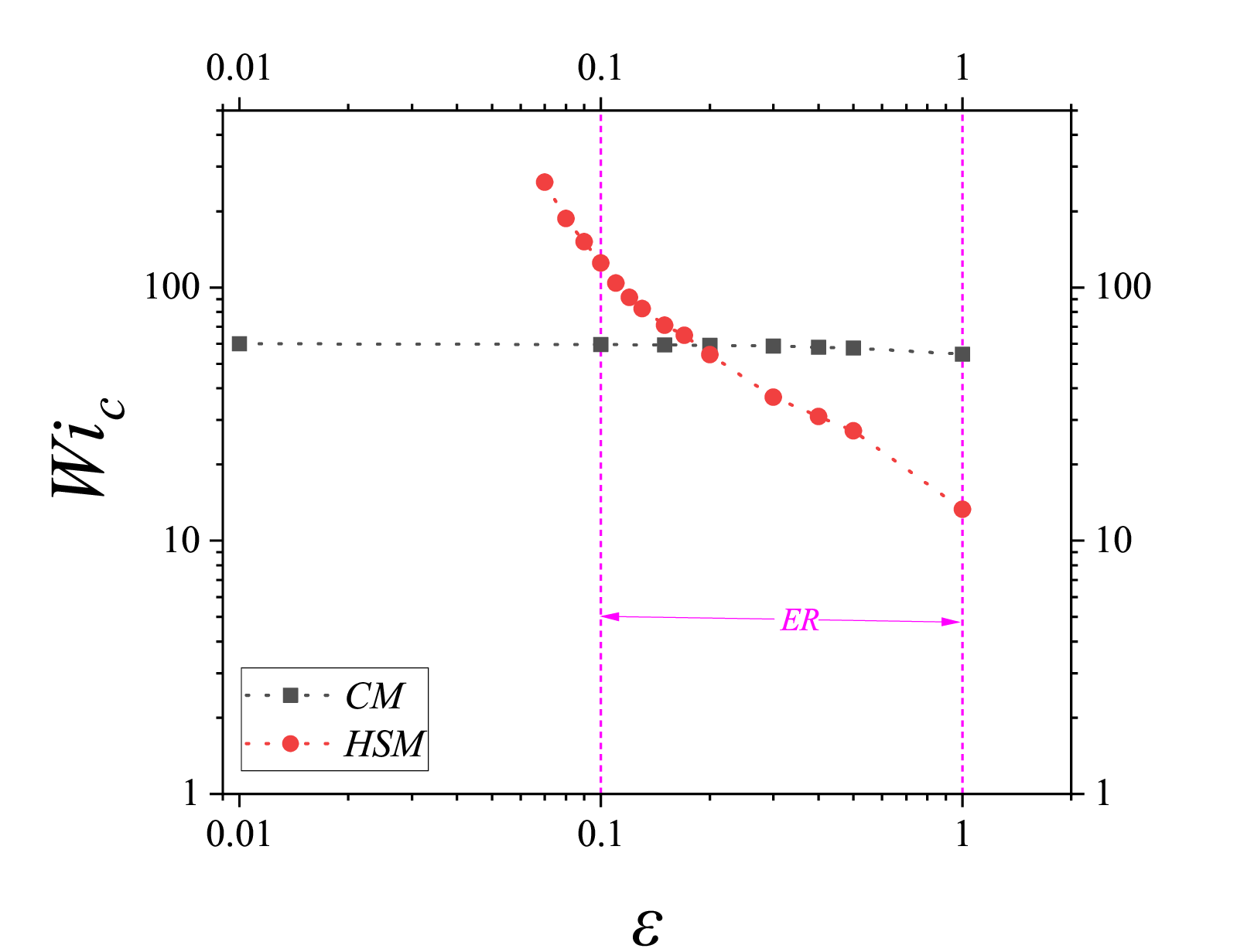}
  \label{fig:CM_L_100}}
   \centering
  \subfigure [$ L = 70$]{\includegraphics[width=0.45\textwidth]{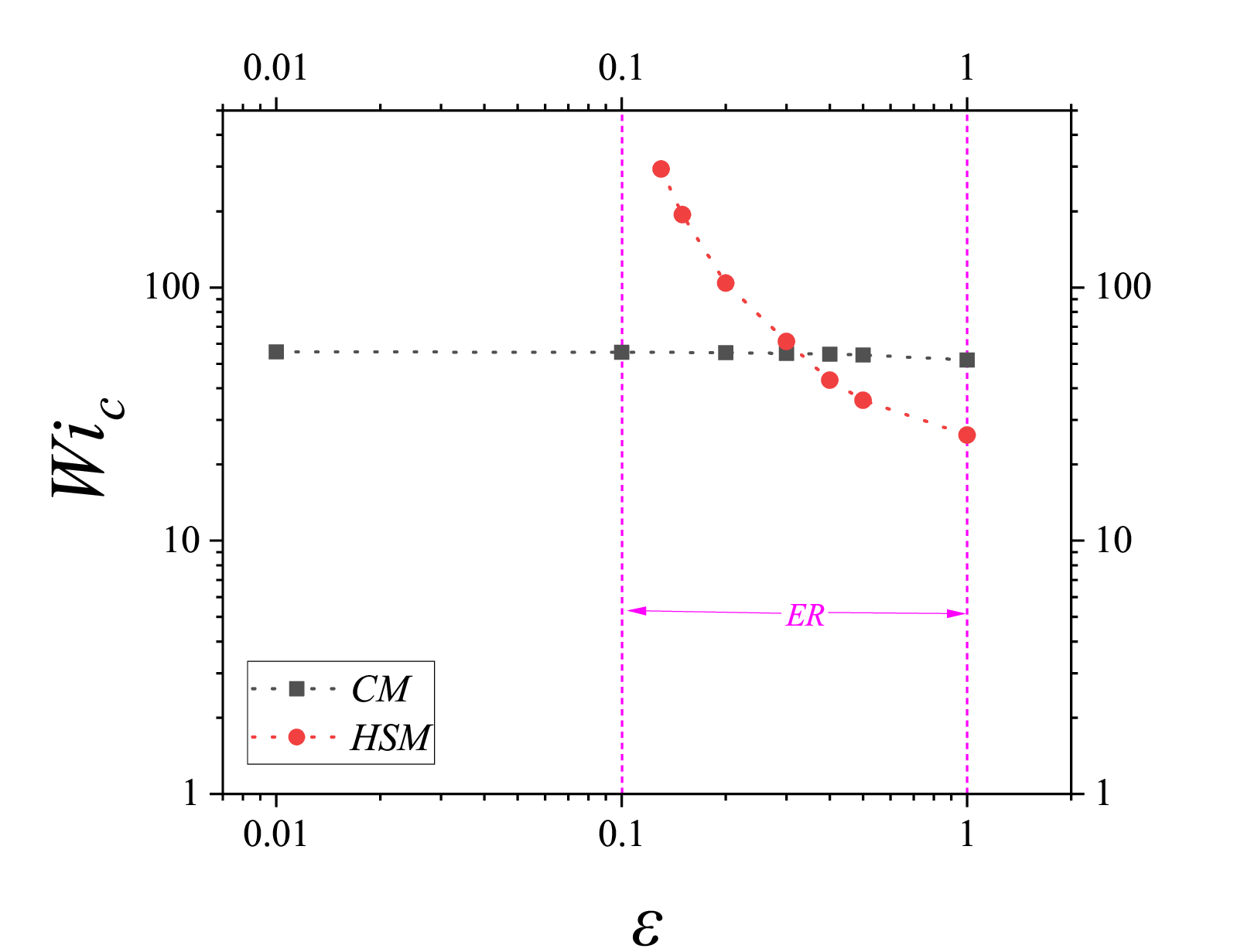}}\label{fig:CM_L_70}
  \caption{For Dean flow of FENE-P fluids, as $L$ is decreased, the CM remains more critical compared to HSM at increasingly higher $\epsilon$'s. Data showing $W\!i_c$ vs $\epsilon$ at $\beta = 0.98$, $Re = 0$. The range indicated by `ER' denotes the $\epsilon$'s typically used in experiments.}
   \label{fig:Fene_P_Dean_HSM_vs CM}
  \end{figure}  
\section{Conclusions}
\label{sec:concl}
Elastic instabilities in the absence of fluid inertia have traditionally been associated with curvilinear shear flows \citep{Pakdel_McKinley}, driven by hoop stresses induced by streamline curvature. Recently, \cite{khalid_creepingflow_2021} identified a novel centre-mode (CM) elastic instability in rectilinear viscoelastic channel flow, driven by a mechanism that is not reliant on streamline curvature, and analogues of which have subsequently been found in other rectilinear shearing flows \citep{Mamta2023,Yadav2024,Lewy_Kerswell_2025}.  In this study, for the first time, we have shown the presence of both hoop-stress (HSM) and centre-mode (CM) instabilities in the same shearing flow, the  Dean flow configuration, using both Oldroyd-B and FENE-P models. While Dean flow approaches plane Poiseuille flow in the narrow-gap limit (with $\epsilon \rightarrow 0$),  thence guaranteeing the presence of a CM instability in the narrow-gap limit,  as argued below, the CM turns out to be the most unstable mode for realistic values of $L \sim 100$ in the FENE-P model even for experimentally relevant $\epsilon$'s ($\approx 0.1 - 0.3$). Thus,  the present study demonstrates the importance of two qualitatively distinct elastic instabilities in the same curvilinear flow configuration, and thence, the possibility of two different routes to an eventual (elastic) turbulent regime.

It is useful to compare the critical $W\!i_c$ for each of these modes in order to determine which one is more critical, and as to how  the criticality  is influenced by $\epsilon$ and $L$.  On account of the existence of the CM instability only in the dilute limit, in the inertialess regime, the ensuing discussion applies only for $\beta > 0.95$; although, as discussed below, nonlinear stability considerations will likely render the CM significant over a wider range of $\beta$ \citep{buza_beneitez_page_kerswell_2022}. Figure\,\ref{fig:W_c_vs_e_HSM_vs_CM} shows that, 
within the Oldroyd-B framework ($L \rightarrow \infty$), the axisymmetric HSM is the more unstable mode across all experimentally reasonable gap width ratios, viz., $\epsilon \geq 0.1$. As a result, although the range of $\beta$'s over which the CM is unstable is slightly larger than that for pressure-driven channel flow, this is of little consequence within the Oldroyd-B framework.
  
Figure\,\ref{fig:Fene_P_Dean_HSM_vs CM} shows the HSM and CM thresholds as a function of $\epsilon$, for different $L$'s, for $\beta = 0.98$. Figure\,\ref{fig:CM__L_600} suggests that the CM instability is not relevant for $L = 600$; for the polyacrylamide solutions routinely used in experiments, $L \sim 600$ would imply a molecular weight of $\gtrsim 10^8$ g/mol.  However, the highest molecular weight used hitherto in experiments probing elastic instabilities is $1.8 \times 10^7$ g/mol \citep[e.g., see][]{Neelamegam2013}.
 The CM starts to become relevant for $L = 200$, and is certainly the only one present for $\epsilon \leq 0.1$ and  $L < 100$ \citep[as appropriate for the polyacrylamide solutions used in][with a molecular weight of $5 \times 10^6$g/mol]{choueiri2021experimental}. For the latter range of $L$, as shown in Fig.\,\ref{fig:CM_L_100}, the CM is the most critical mode for  $\epsilon = 0.1$–$0.3$ corresponding to the Taylor-Dean experiments of \cite{joo_shaqfeh_1994}, although the HSM will likely be dominant in the serpentine channel experiments of \cite{Groisman2004} with $\epsilon = 1$. To a good approximation, the critical $W\!i_c$ is obtained by setting $\alpha = 0$, implying that the  critical mode is nearly two-dimensional.
   It is worth adding that, 
although the focus of this work is on Dean flow, most of the results are expected to carry over to the Taylor-Dean geometry as well, in light of a similar demonstration for CM in Couette-Poiseuille flows in a rectangular channel \citep{Yadav2024}.  Thus, in experiments involving the Dean or Taylor-Dean configurations, for realistic values of $\epsilon$ and $L$, the CM instability will be relevant for sufficiently dilute solutions such that $\beta > 0.95$, while the HSM is expected be the dominant mode for $\beta < 0.95$; at least within the linear stability framework. Crucially, the demonstration by \cite{buza_beneitez_page_kerswell_2022}, of finite-amplitude travelling waves borne out of the elastic CM instability in rectilinear channel flow being strongly subcritical and existing at $W\!i \lesssim 50$ and for $\beta$ down to $0.7$,  suggests that the centre mode could be relevant even in parameter regimes
where it is linearly stable in Dean flow.

Finally, even in parameter regimes where the CM is critical, it is tempting to speculate that the primary CM instability will essentially act to modify and augment the streamline curvature in Dean flow, which could then be susceptible to a secondary instability, driven by the hoop-stress mechanism, involving spanwise perturbations. While a similar scenario has been earlier proposed by Morozov and coworkers \citep{morozov_saarloos2005,Morozov2005,morozov_saarloos2019} for rectilinear viscoelastic flows in parametric regimes devoid of any linear instability, the above proposal is actually closer in spirit to the three-dimensional secondary instability in (Newtonian) boundary-layer flows, as demonstrated by \cite{Herbert1988}. In the latter context, the Tollmien-Schlichting (TS) mode underlies the primary instability, which then modifies the base flow. This spatially periodic perturbed base state, 
 comprising a superposition of the original boundary-layer flow and a finite amplitude TS mode,
 subsequently becomes unstable to three-dimensional perturbations. 
  
 The demonstration of an analogous secondary instability, for the Dean flow configuration, in regimes where the CM controls the primary transition, will be an interesting avenue for future investigation.

\backsection[Declaration of interests]{The authors report no conflict of interest.
}

\backsection[Author contributions]{V.S and G.S proposed and designed the research. P.S.D.S.P.T and V.S derived the equations and developed the code. P.S.D.S.P.T carried out the simulations.  All authors contributed equally to analysing data, reaching conclusions, and writing the paper.}

\backsection[Author ORCIDs]{V.\,Shankar, https://orcid.org/0000-0003-0233-7494, 

G.\,Subramanian, https://orcid.org/0000-0003-4314-3602}

\appendix
\section{Linearized stability equations for the Oldroyd-B model}
\label{app:Old_b_Linearized_eq}
The linearized continuity, momentum, and constitutive equations for the Oldroyd-B model are given in this Appendix. Here, and in what follows, $D \equiv \frac{d}{dy}$.
\noindent Continuity equation:
\begin{equation}
\frac{d \Tilde v _r}{dy} + \frac{\epsilon}{1 + \epsilon y} \Tilde v _r   + \frac{i n \epsilon}{1 + \epsilon y} \Tilde v _\theta + i \alpha \Tilde v_z  = 0 ,
\label{Normal mode form continuity}
\end{equation}
\noindent $r$-momentum equation:
\begin{equation}
G \Tilde v _r + G_1 \Tilde v _\theta = -D \Tilde p + \beta M \Tilde v _r  + M_1 \Tilde \tau _{rr} + M_2 \Tilde \tau _{r\theta} + i\alpha \Tilde \tau _{rz} - \left(\frac{\epsilon}{1 + \epsilon y}\right) \Tilde \tau _{\theta \theta} ,
\label{Normal mode form R Momemtum}
\end{equation}
\noindent $\theta$-momentum equation:
\begin{equation}
G \Tilde v _\theta + G_2 \Tilde v _r = - M_2 \Tilde p + \beta M \Tilde v _\theta  + (M_1 + \frac{\epsilon}{1 + \epsilon y})\Tilde \tau _{r\theta} + M_2 \Tilde \tau _{\theta \theta} + i\alpha \Tilde \tau _{\theta z} ,  
\label{Normal mode form theta Momemtum}
\end{equation}
\noindent $z$-momentum equation:
\begin{equation}
G \Tilde v _z = - i \alpha \Tilde p + \beta M \Tilde v _z  + M_1 \Tilde \tau _{rz} + M_2 \Tilde \tau _{\theta z} + i\alpha \Tilde \tau _{z z} ,  
\label{Normal mode form Z Momemtum}
\end{equation}
\noindent $\tau_{rr}$ equation:
\begin{equation}
   \zeta \Tilde \tau _{rr} = 2(1 - \beta) \frac{d \Tilde v _r}{dy} + 2 W\!i M_2 \Tilde v _r , 
   \label{Normal mode form RR stress}
\end{equation}
\noindent $\tau_{r\theta}$ equation:
\begin{equation}
 \zeta \Tilde \tau _{r\theta} = W\!i \left(U' - \frac{\epsilon}{1 + \epsilon y} U \right) \Tilde \tau _{rr} 
 + P v_r + \left[(1-\beta)(D -  \frac{\epsilon}{1 + \epsilon y}) + M_2 \overline\tau_{r \theta}\right] \Tilde v_\theta ,
   \label{Normal mode form Rtheta stress}
\end{equation}
\noindent $\tau_{rz}$ equation:
\begin{equation}
    \zeta \Tilde \tau _{rz} = \left[(1-\beta)D + W\!i M_2 \overline\tau_{r \theta}\right]\Tilde v_z  + (1-\beta) i \alpha \Tilde v_r ,
      \label{Normal mode form RZ stress}
\end{equation}
\noindent $\tau_{\theta \theta}$ equation:
\begin{equation}
    \zeta \Tilde \tau _{\theta \theta} = 2 W\!i \left(U' - \frac{\epsilon}{1 + \epsilon y} U \right) \Tilde \tau _{r\theta} + \left[2(1-\beta) \frac{\epsilon}{1 + \epsilon y} - W\!i \left(\overline\tau_{\theta \theta}' - \frac{2 \epsilon}{1 + \epsilon y} \overline\tau_{\theta \theta}\right)\right] \Tilde v_r + F \Tilde v_\theta ,
      \label{Normal mode form thetatheta stress}
\end{equation}
\noindent $\tau_{\theta z}$ equation:
\begin{equation}
    \zeta \Tilde \tau _{\theta z} =  W\!i \left(U' - \frac{\epsilon}{1 + \epsilon y} U \right) \Tilde \tau _{rz} + E \Tilde v_z +  (1-\beta) i \alpha \Tilde v_\theta ,
      \label{Normal mode form thetaZ stress}
\end{equation}
\noindent $\tau_{zz}$ equation:
\begin{equation}
     \zeta \Tilde \tau _{zz} = 2(1- \beta) i \alpha \Tilde v_z ,
       \label{Normal mode form ZZ stress}
\end{equation}
where 
$ \zeta =   \left[1 - i \omega + W\!i\left(\frac{ i n \epsilon}{1 + \epsilon y}\right) U \right]$,

$G = Re \left(\frac{i n \epsilon}{1 + \epsilon y} U -  \frac{i \omega}{W\!i}\right) $,

$M =  \left( D^2 + \frac{\epsilon}{1 + \epsilon y} D - \frac{\epsilon^2}{(1 + \epsilon y)^2} - \frac{n^2 \epsilon^2}{(1 + \epsilon y)^2} - \alpha^2 \right)$,

$ M_1 = \left(D + \frac{\epsilon}{1 + \epsilon y}\right) , M_2 = \left(\frac{i n \epsilon}{1 + \epsilon y}\right), G_1 =  - 2 Re \frac{\epsilon}{1 + \epsilon y} + \beta \frac{2 i n \epsilon^2}{(1 + \epsilon y)^2}  $,

$ G_2 =   Re \frac{\epsilon}{1 + \epsilon y} - \beta \frac{2 i n \epsilon^2}{(1 + \epsilon y)^2} + Re U'  $,

$E = \left[ W\!i\left(\overline\tau_{r \theta} D + M_2 \overline\tau_{\theta \theta} \right) +  (1 - \beta) M_2 \right], F = \left[ 2 W\!i\left(\overline\tau_{r \theta} D + M_2 \overline\tau_{\theta \theta} -\frac{\epsilon}{1 + \epsilon y} \overline\tau_{r \theta} \right) +  2 (1 - \beta) M_2 \right]$,

$ P = \left[ W\!i\left(\overline\tau_{r \theta} D + M_2 \overline\tau_{\theta \theta} + \frac{\epsilon}{1 + \epsilon y} \overline\tau_{r \theta} - \overline\tau_{r \theta}' \right) +  (1 - \beta) M_2 \right] $,  

 No-slip boundary conditions $\Tilde v_r = 0$, $\Tilde v_\theta = 0$ and $\Tilde v_z = 0$ are applicable at both $y = 0$ and $y = 1$.

\section{Linearized stability equations for the FENE-P model}
\label{app:FENE_P_Linearized_eq}
The linearized continuity, momentum, and  constitutive equations for the FENE-P model are given in this Appendix.

\noindent Continuity equation:
\begin{equation}
    \left(D + \frac{\epsilon}{1 + \epsilon y}\right) \Tilde{v}_r + \frac{i n \epsilon}{1 + \epsilon y} \Tilde{v}_\theta + i \alpha \Tilde{v}_z = 0
    \label{FENE_P_Normal_mode_form_continuity}
\end{equation}

\noindent $r$-momentum Equation:
\begin{equation}
\begin{aligned}
   Re \left(-\frac{i \omega}{W\!i} + \frac{ \overline{v}_\theta i n \epsilon}{1 + \epsilon y}\right)\Tilde{v}_r 
    - \frac{2 Re \epsilon \overline{v}_\theta}{1 + \epsilon y}\Tilde{v}_\theta 
    &=  -\frac{\partial \Tilde{p}}{\partial y} + \beta \left( \left(\frac{\partial^2}{\partial y^2} - \frac{\epsilon^2}{(1 + \epsilon y)^2} + \frac{\epsilon}{1 + \epsilon y}\frac{\partial}{\partial y} \right.\right. \\
    &\quad \left.\left. - \frac{n^2 \epsilon^2}{(1 + \epsilon y)^2} - \alpha^2 \right)\Tilde{v}_r - \frac{2 \epsilon^2 i n}{(1 + \epsilon y)^2} \Tilde{v}_\theta \right) \\
    &\quad + (1 - \beta)\left(\frac{\partial \Tilde{\tau}_{rr}}{\partial y} + \frac{\epsilon}{1 + \epsilon y}\Tilde{\tau}_{rr} + \frac{i n \epsilon}{1 + \epsilon y}\Tilde{\tau}_{\theta r} \right. \\
    &\quad \left. + i \alpha \Tilde{\tau}_{zr} - \frac{\epsilon}{1 + \epsilon y} \Tilde{\tau}_{\theta \theta}\right)
\end{aligned}
\end{equation}

\noindent $\theta$-momentum Equation:
\begin{equation}
\begin{aligned}
    Re \left(\frac{\partial \overline{v}_\theta}{\partial y} + \frac{ \overline{v}_\theta \epsilon}{1 + \epsilon y}\right)\Tilde{v}_r 
    + Re \left(\frac{\overline{v}_\theta i n \epsilon}{1 + \epsilon y} - \frac{i \omega}{W\!i}\right)\Tilde{v}_\theta 
    &= -\frac{i n \epsilon}{1 + \epsilon y}\Tilde{p}  \\
    &\quad + \beta \left(\left(\frac{\partial^2}{\partial y^2} - \frac{\epsilon^2}{(1 + \epsilon y)^2} + \frac{\epsilon}{1 + \epsilon y }\frac{\partial}{\partial y} \right.\right. \\
    &\quad \left.\left. - \frac{n^2 \epsilon^2}{(1 + \epsilon y)^2} - \alpha^2 \right)\Tilde{v}_\theta + \frac{2 \epsilon^2 i n}{(1 + \epsilon y)^2}\Tilde{v}_r \right) \\
    &\quad + (1 - \beta)\left(\frac{\partial \Tilde{\tau}_{\theta r}}{\partial y} + \frac{\epsilon}{1 + \epsilon y}\Tilde{\tau}_{\theta r} \right. \\
    &\quad \left. + \frac{i n \epsilon}{1 + \epsilon y}\Tilde{\tau}_{\theta \theta} + i \alpha \Tilde{\tau}_{z \theta}\right)
\end{aligned}
\end{equation}

\noindent $z$-momentum Equation:
\begin{equation}
\begin{aligned}
    Re \left(-\frac{i \omega}{W\!i} + \frac{\overline{v}_\theta i n \epsilon}{1 + \epsilon y}\right)\Tilde{v}_z 
    &= -i \alpha \Tilde{p} + \beta \left(\frac{\partial^2 \Tilde{v}_z}{\partial y^2} - \frac{\epsilon^2}{(1 + \epsilon y)^2}\Tilde{v}_z 
    + \frac{\epsilon}{1 + \epsilon y}\frac{\partial \Tilde{v}_z}{\partial y} - \frac{n^2 \epsilon^2}{(1 + \epsilon y)^2}\Tilde{v}_z 
    - \alpha^2 \Tilde{v}_z\right) \\
    &\quad + (1 - \beta)\left(\frac{\partial \Tilde{\tau}_{rz}}{\partial y} + \frac{\epsilon}{1 + \epsilon y}\Tilde{\tau}_{rz} + \frac{i n \epsilon}{1 + \epsilon y}\Tilde{\tau}_{\theta z} + i \alpha \Tilde{\tau}_{zz}\right)
    \label{FENE_P_Normal_mode_form_Z_Momemtum}
\end{aligned}
\end{equation}

\noindent $C_{rr}$ equation:
\begin{equation}
    \left( \overline{C}_{rr}' - 2 \overline{C}_{rr} D - \frac{2 \overline{C}_{r \theta} i n \epsilon}{1 + \epsilon y} \right) \Tilde{v}_r 
    + \left( \frac{\overline{v}_{\theta} i n \epsilon}{1 + \epsilon y} - \frac{i \omega}{W\!i} \right) \Tilde{C}_{rr} 
    = -\Tilde{\tau}_{rr}
\end{equation}

\noindent $C_{r \theta}$ equation:
\begin{equation}
\begin{aligned}
    \left( \overline{C}_{r\theta}' - \frac{i n \epsilon}{1 + \epsilon y} \overline{C}_{\theta \theta} \right) \Tilde{v}_r
    + \left( \frac{\epsilon \overline{C}_{rr}}{1 + \epsilon y} - \overline{C}_{rr} \frac{\partial}{\partial y} \right) \Tilde{v}_\theta
    + i \alpha \overline{C}_{r \theta} \Tilde{v}_z & \\
    + \left( \frac{\overline{v}_\theta \epsilon}{1 + \epsilon y} -  \overline{v}_\theta' \right) \Tilde{C}_{rr}
    + \left( \frac{\overline{v}_\theta \epsilon i n}{1 + \epsilon y} - \frac{i \omega}{W\!i} \right) \Tilde{C}_{r \theta} 
    &= -\Tilde{\tau}_{r \theta}
\end{aligned}
\end{equation}

\noindent $C_{rz}$ equation:
\begin{equation}
    -i \alpha \overline{C}_{zz} \Tilde{v}_r 
    - \left( \overline{C}_{rr} \frac{\partial}{\partial y} + \frac{\overline{C}_{r \theta} i n \epsilon}{1 + \epsilon y} \right)\Tilde{v}_z 
    + \left(\frac{\overline{v}_\theta i n \epsilon}{1 + \epsilon y} - \frac{i \omega}{W\!i} \right)\Tilde{C}_{rz} 
    = - \Tilde{\tau}_{rz}
\end{equation}

\noindent $C_{\theta \theta}$ equation:
\begin{equation}
\begin{aligned}
    \left(  \overline{C}_{\theta \theta}' - \frac{2 \epsilon }{1 + \epsilon y} \overline{C}_{\theta \theta} \right)\Tilde{v}_r 
    + \left( \frac{2 \epsilon}{1 + \epsilon y} \overline{C}_{r \theta} - \frac{2 i n \epsilon}{1 + \epsilon y}\overline{C}_{\theta \theta} - 2 \overline{C}_{r \theta} \frac{\partial}{\partial y}\right) \Tilde{v}_\theta &\\ 
    + 2\left(\frac{\epsilon}{1 + \epsilon y}\overline{v}_\theta - \overline{v}_\theta' \right)\Tilde{C}_{r \theta} 
    + \left(\frac{\overline{v}_\theta i n  \epsilon}{1 + \epsilon y} - \frac{i \omega}{W\!i} \right)\Tilde{C}_{\theta \theta} 
    &= - \Tilde{\tau}_{\theta \theta}
\end{aligned}
\end{equation}

\noindent $C_{\theta z}$ equation:
\begin{equation}
    -i \alpha \overline{C}_{zz} \Tilde{v}_\theta 
    - \left(\overline{C}_{\theta r} \frac{\partial}{\partial y} + \frac{i n \epsilon}{1 + \epsilon y} \overline{C}_{\theta \theta}\right)\Tilde{v}_z 
    + \left(\frac{\epsilon}{1 + \epsilon y} \overline{v}_\theta - \frac{\partial \overline{v}_\theta}{\partial y}\right)\Tilde{C}_{rz} 
    + \left(\frac{ i n \epsilon}{1 + \epsilon y} \overline{v}_\theta - \frac{i \omega}{W\!i}\right)\Tilde{C}_{\theta z} 
    = - \Tilde{\tau}_{\theta z}
\end{equation}

\noindent $C_{zz}$ Equation:
\begin{equation}
    \frac{\partial \overline{C}_{zz}}{\partial y}\Tilde{v}_r 
    - 2 i \alpha \overline{C}_{zz} \Tilde{v}_z 
    + \left(\frac{i n \epsilon}{1 + \epsilon y} \overline{v}_\theta - \frac{i \omega}{W\!i} \right) \Tilde{C}_{zz} 
    = - \Tilde{\tau}_{zz}
\end{equation}

\section{Eigenspectra for viscoelastic Dean flow}
\label{sec:Appendixspectra}
\begin{figure}
  \centering
  \subfigure [$ \alpha = 2 $]{\includegraphics[width=0.45\textwidth]{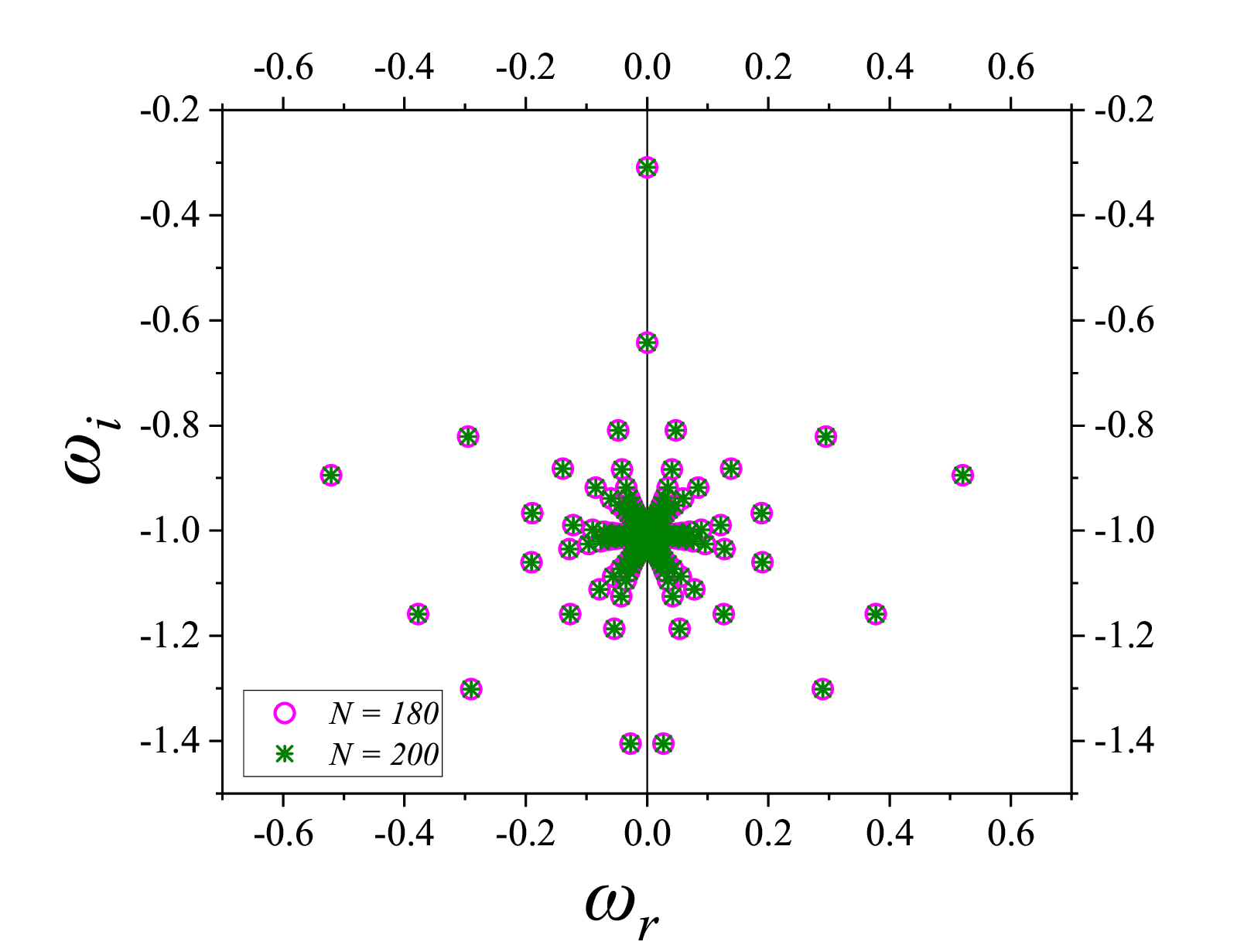}}\label{fig:Axisymm_HSM_dean_e_0.1_beta_0.98_alpha_2}
   \subfigure [$ \alpha = 7 $]{\includegraphics[width=0.45\textwidth]{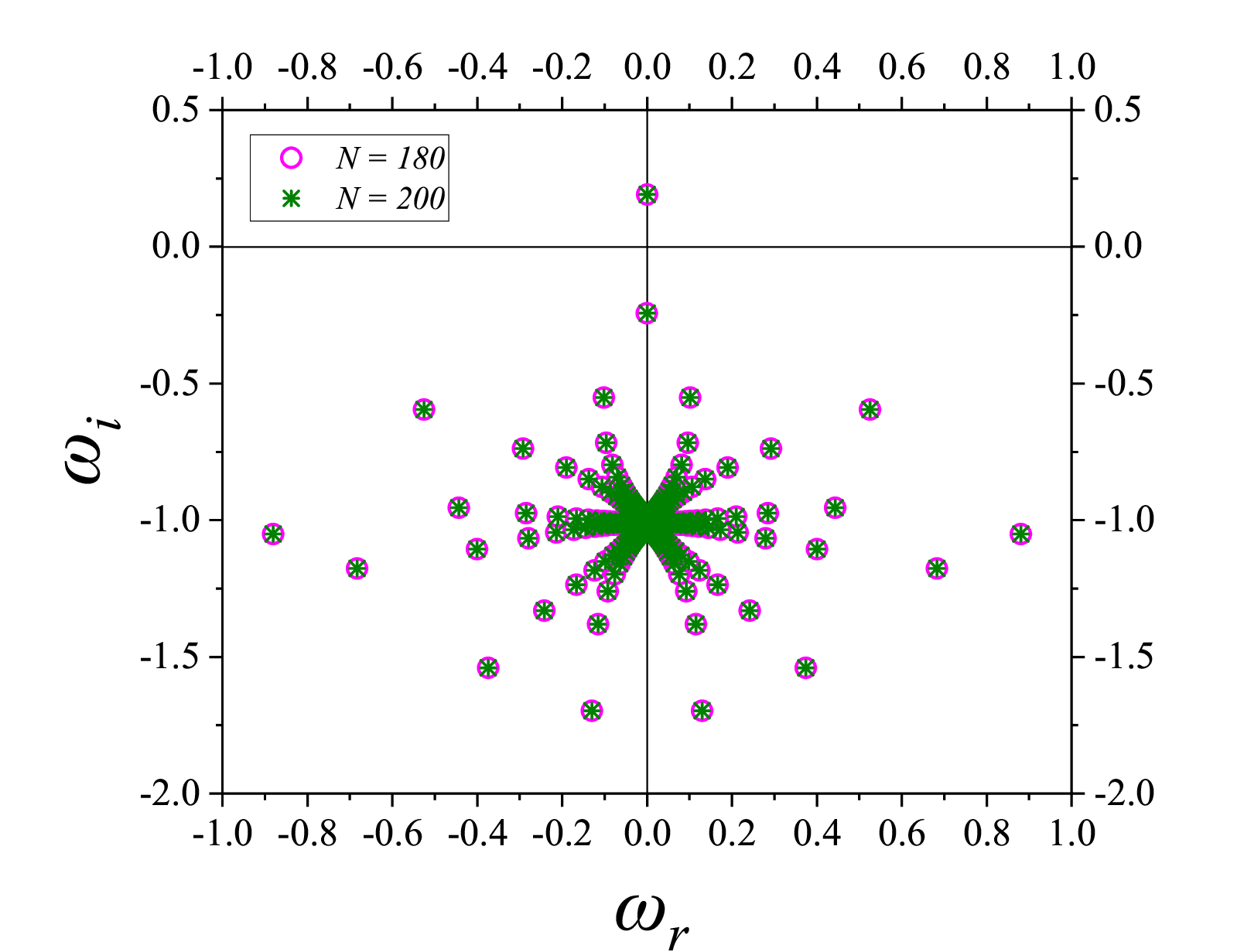}}\label{fig:Axisymm_HSM_dean_e_0.1_beta_0.98_alpha_7}
  \subfigure[$ \alpha = 8 $]{\includegraphics[width=0.45\textwidth]{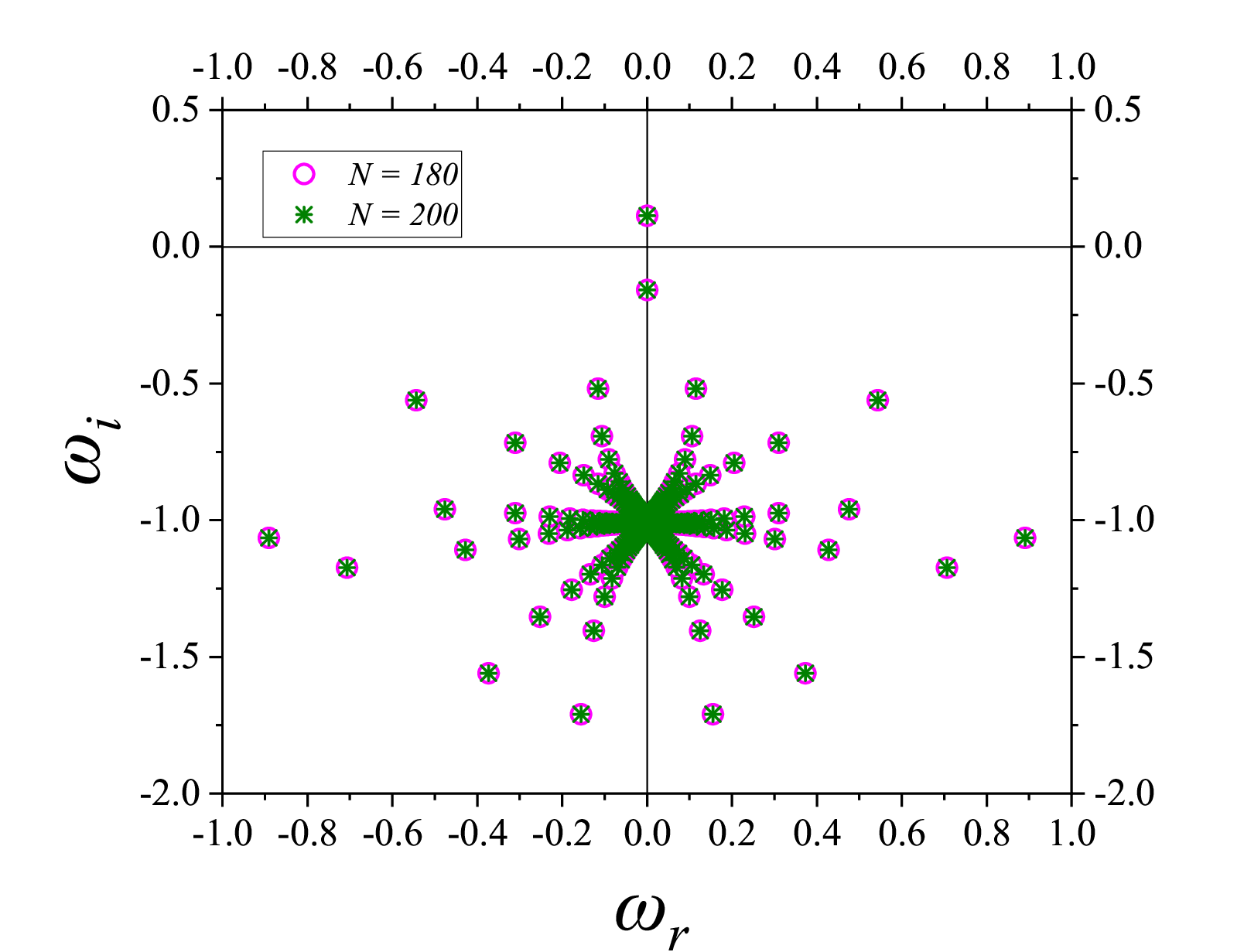}}\label{fig:Axisymm_HSM_dean_e_0.1_beta_0.98_alpha_8}
   \subfigure [$ \alpha = 8.6 $]{\includegraphics[width=0.45\textwidth]{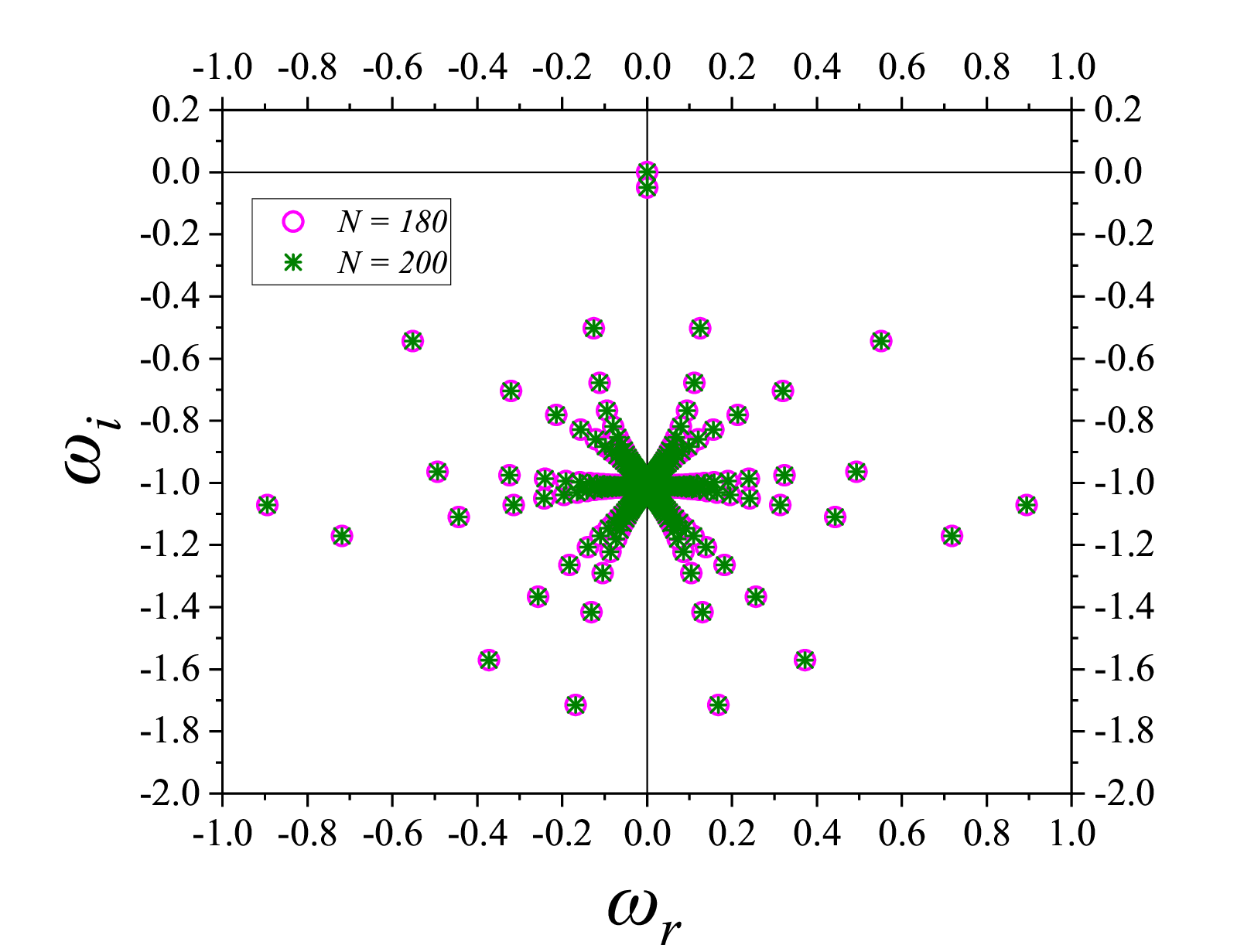}}\label{fig:Axisymm_HSM_dean_e_0.1_beta_0.98_alpha_8.6}
  \subfigure[$ \alpha = 8.7 $]{\includegraphics[width=0.45\textwidth]{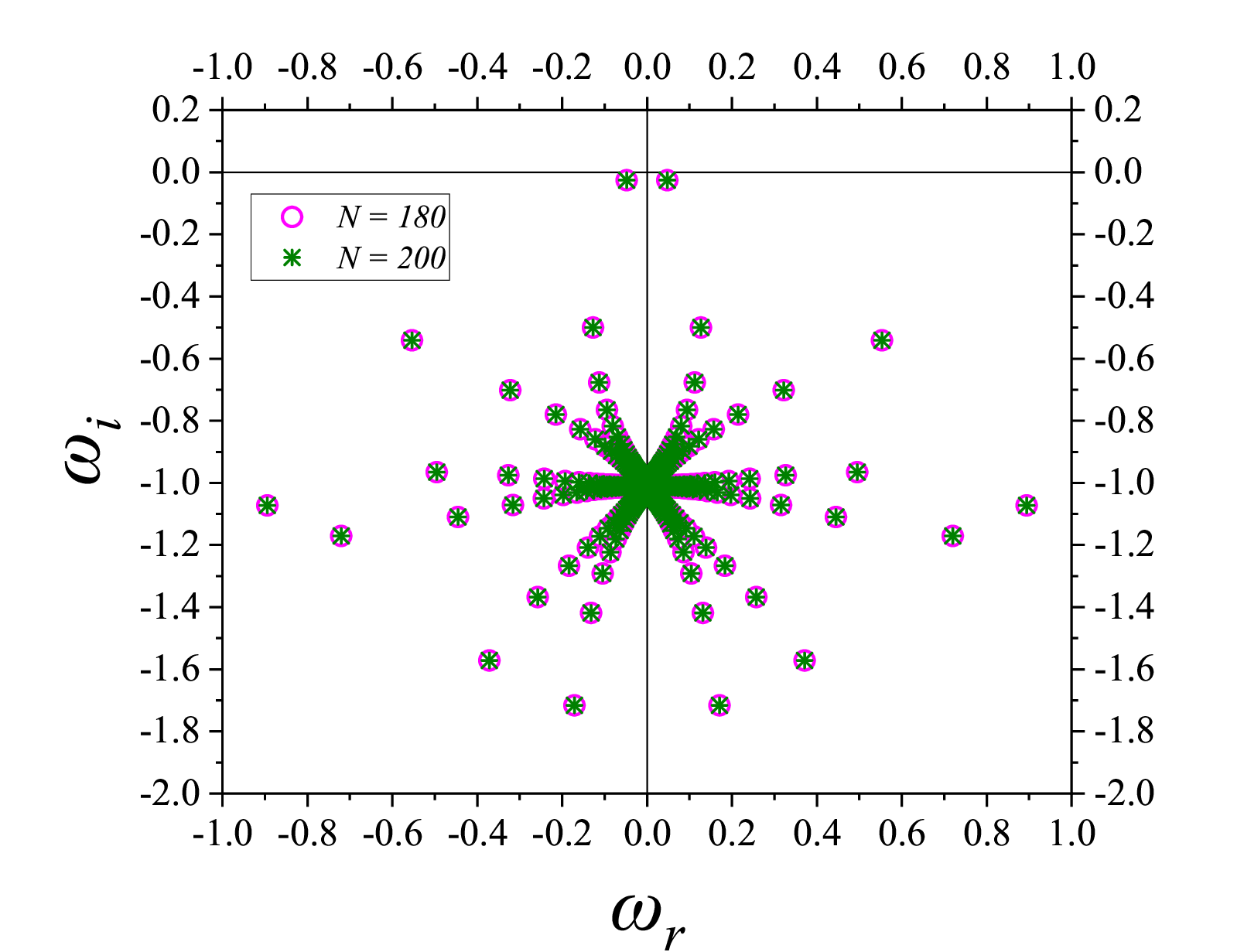}}\label{fig:Axisymm_HSM_dean_e_0.1_beta_0.98_alpha_8.7}
   \subfigure[$ \alpha = 10 $]{\includegraphics[width=0.45\textwidth]{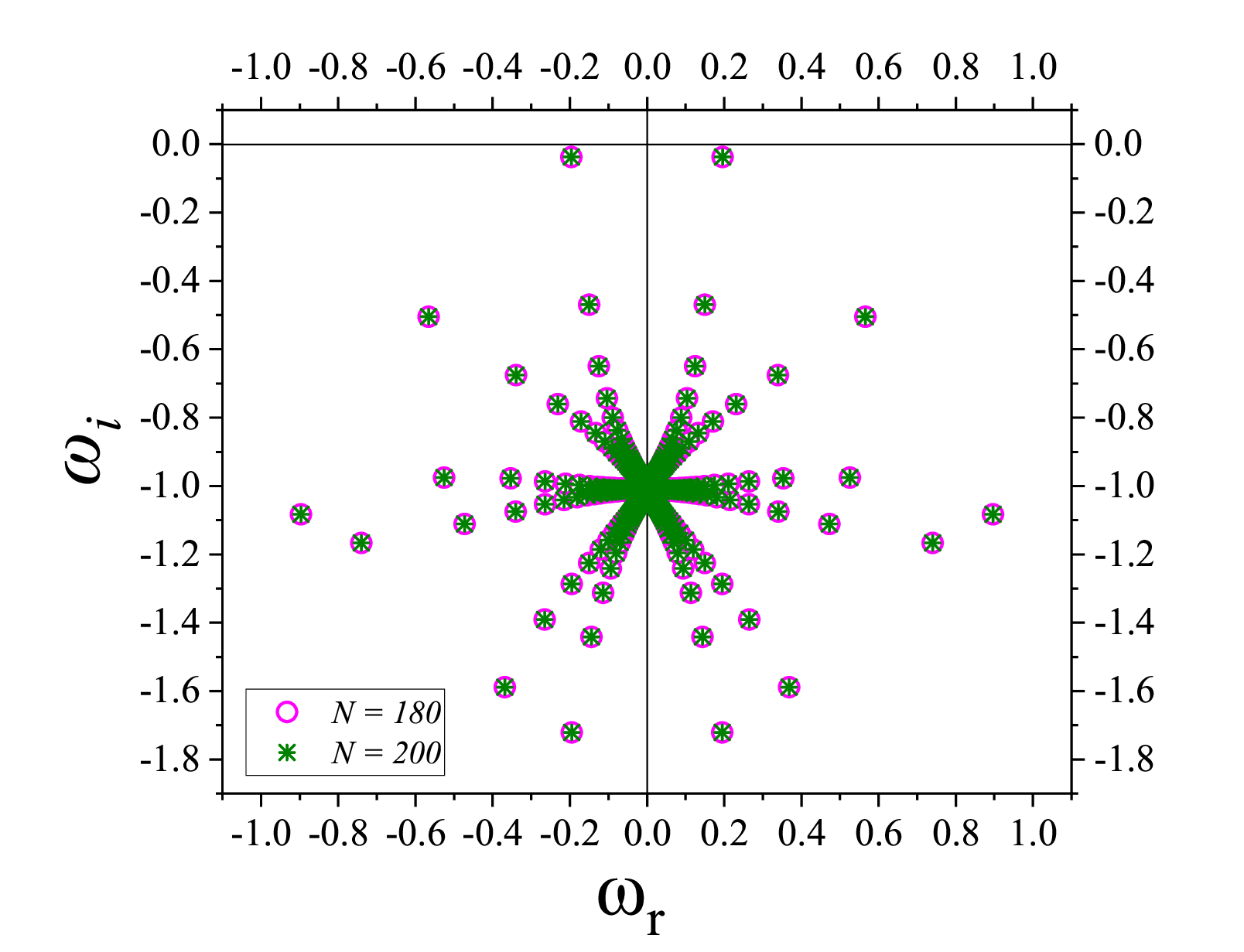}}\label{fig:Axisymm_HSM_dean_e_0.1_beta_0.98_alpha_10}
  \caption{Axisymmetric ($n = 0$) eigenspectra for different $\alpha$ at $Wi = 20$, $\epsilon = 0.1$, $\beta = 0.98$ for Dean flow of an Oldroyd-B fluid demonstrating the bifurcation of stationary HSM into two propagating modes.}
\label{fig:multi_subfigures_Old_b_Axisymmetric_same_Wi_diif-alpha}
\end{figure}

In Fig.\,\ref{fig:multi_subfigures_Old_b_Axisymmetric_same_Wi_diif-alpha}, we present the axisymmetric eigenspectra for the Oldroyd-B fluid at different axial wavenumbers $\alpha$, for a fixed $W\!i$.
As mentioned in Sec.\,\ref{subsec:spectrum_axisym}, for $\beta = 0.98$, the two point CS are nearly overlapping  (with $\omega_i = -1$) in this figure. Twelve branches of discrete modes are seen to spread out from the (point) CS, and $\omega_r \neq 0$ for these modes. At lower $\alpha$'s, we also find  two least stable modes which are  stationary  (i.e. $\omega_r = 0$), with the mode having the higher growth rate turning unstable for $\alpha = 7$. As $\alpha$ is increased further, the aforementioned stationary modes coalesce in a bifurcation, giving rise to symmetrically placed propagating modes (of the form $\pm \omega_r + \omega_i$), with no stationary discrete modes present in the  spectrum.

\begin{figure}
  \centering
   \subfigure [$ W\!i = 1.6 $]{\includegraphics[width=0.45\textwidth]{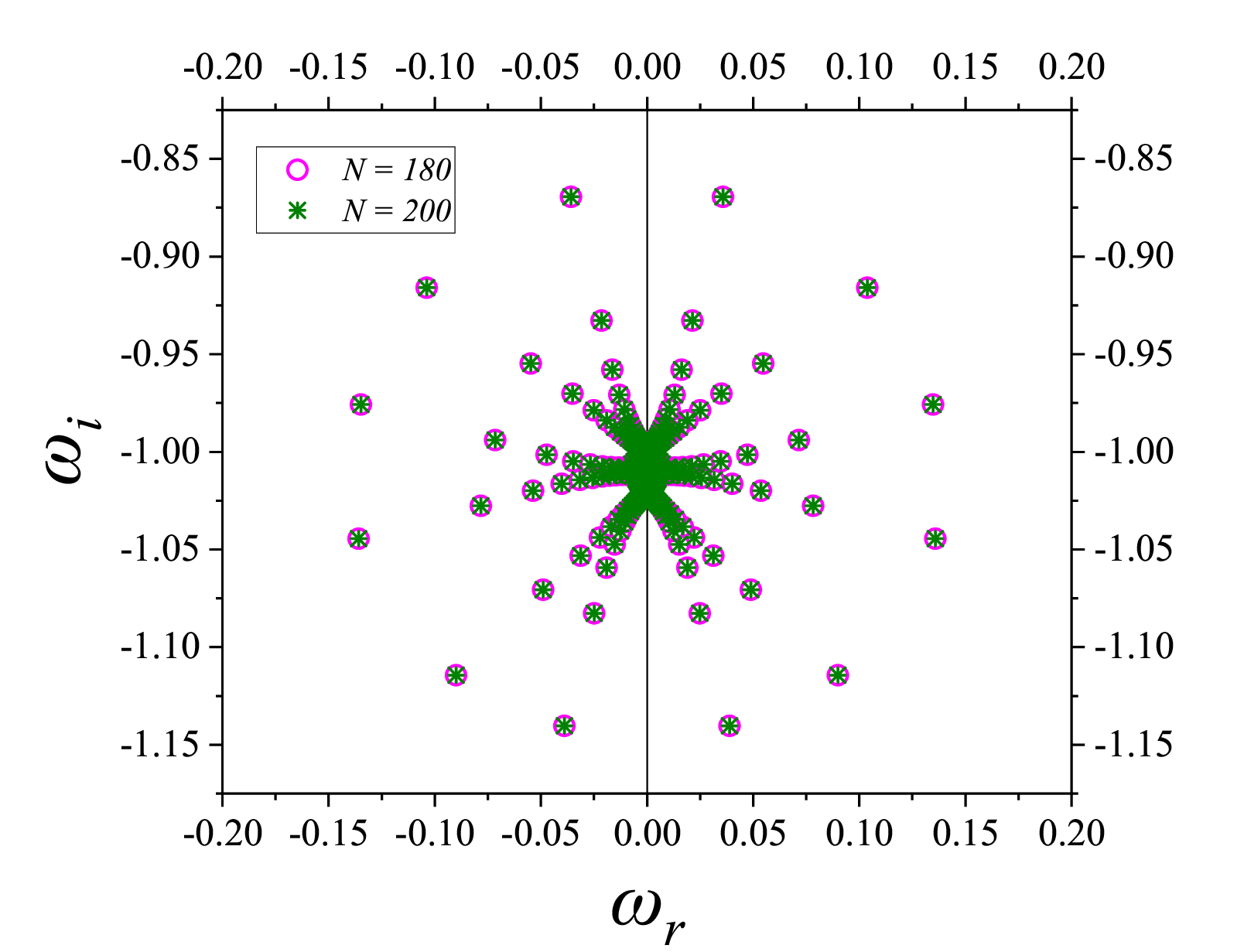}}\label{fig:Axisymm_HSM_dean_e_0.1_beta_0.98_Wi_1.6}
  \subfigure[$ W\!i = 13 $]{\includegraphics[width=0.45\textwidth]{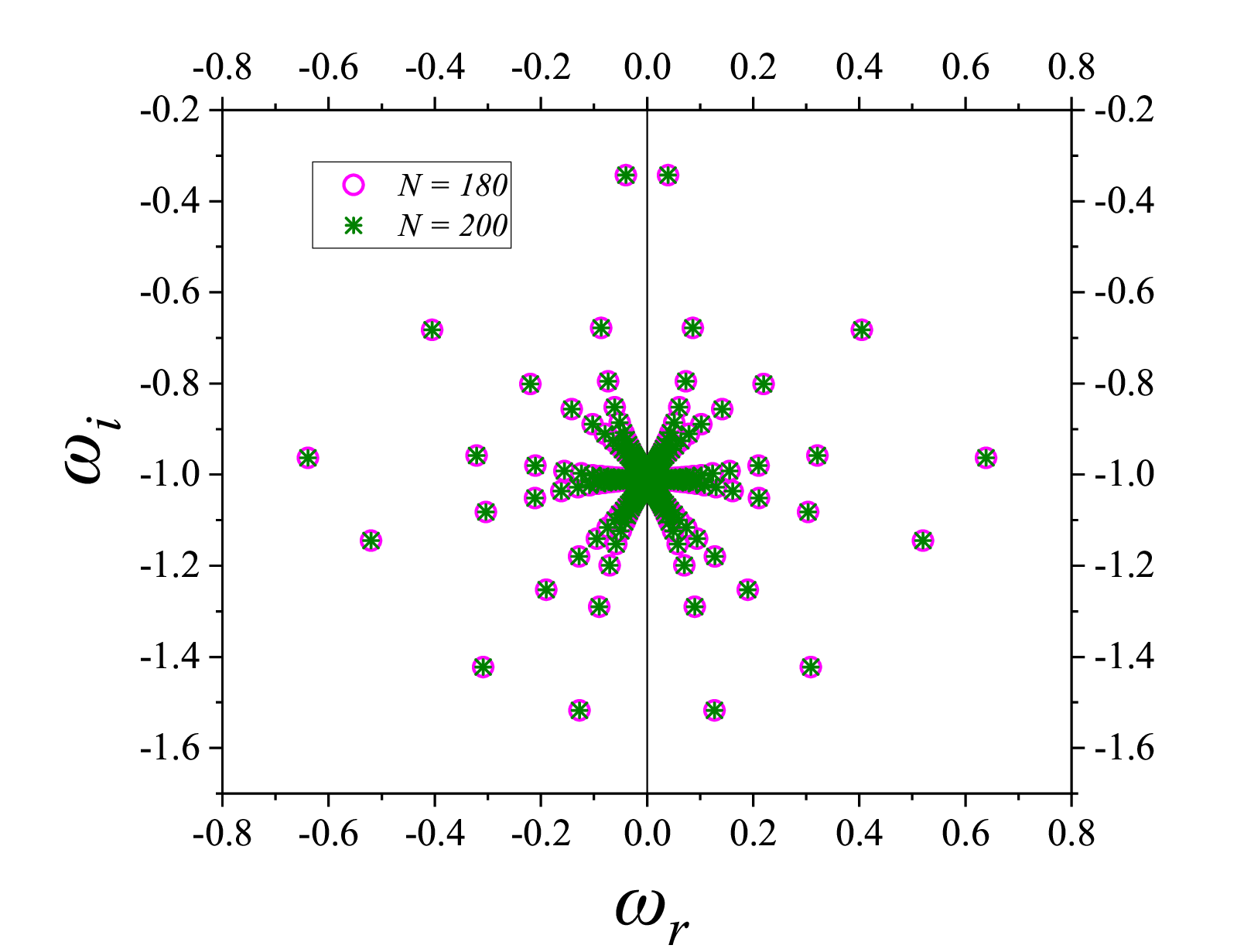}}\label{fig:Axisymm_HSM_dean_e_0.1_beta_0.98_Wi_13}
   \subfigure [$ W\!i = 13.5 $]{\includegraphics[width=0.45\textwidth]{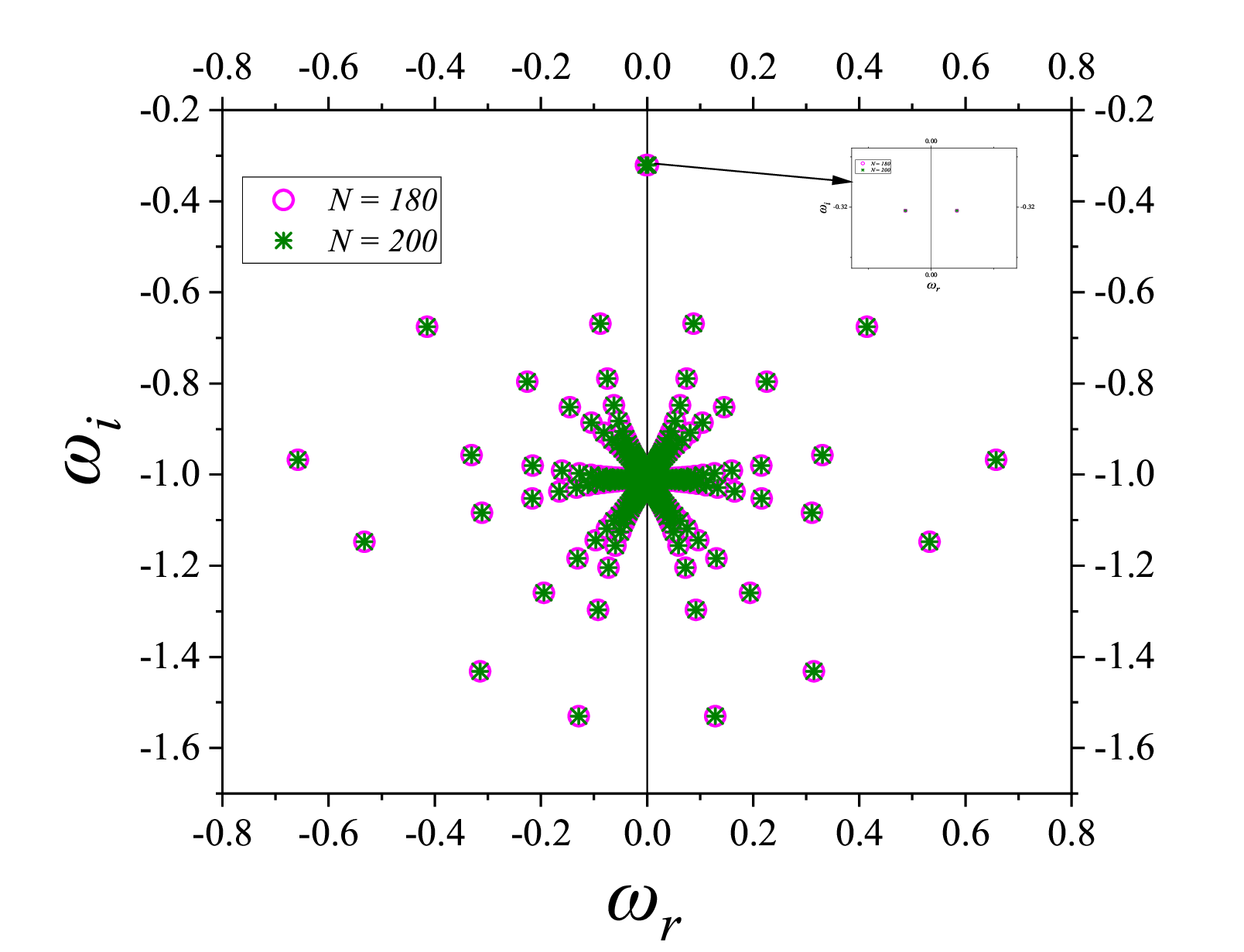}}\label{fig:Axisymm_HSM_dean_e_0.1_beta_0.98_Wi_13_alpha_7}
  \subfigure[$ W\!i = 20 $]{\includegraphics[width=0.45\textwidth]{Axisymmetric_Oldroyd_B_ES_e_0.1_alpha_7_Wi_20_n_0_beta_0.98.eps}}\label{fig:Axisymm_HSM_dean_e_0.1_beta_0.98_Wi_20}
  \caption{Axisymmetric ($n = 0$) eigenspectra for $\epsilon = 0.1$, $\alpha = 7$, $\beta = 0.98$ at different $W\!i$'s for Dean flow of an Oldroyd-B.}
\label{fig:multi_subfigures_Old_b_Axisymmetric_same_Wi_diff_Wi}
\end{figure}

This bifurcation can happen either in the stable ($\omega_i < 0$) or unstable ($\omega_i > 0$) regions of the parameter space; for the $W\!i$ chosen in  Fig.\,\ref{fig:multi_subfigures_Old_b_Axisymmetric_same_Wi_diif-alpha}, the bifurcation occurs prior to the instability. 
For the scaled equations obtained in the narrow-gap limit ($\epsilon \rightarrow 0$, $\epsilon^{1/2} W\!i \sim O(1)$), the most unstable mode appears to always remain stationary. 
Figure\,\ref{fig:multi_subfigures_Old_b_Axisymmetric_same_Wi_diff_Wi} depicts the structure of the spectrum for varying $W\!i$, and for $\alpha$ fixed at $7$. In this case, the spectrum evolves in the opposite direction (relative to Fig.\,\ref{fig:multi_subfigures_Old_b_Axisymmetric_same_Wi_diif-alpha}) as $W\!i$ is increased from $1.6$ to $13$; in that, we start off with the pair of least stable modes being propagating ones, which move towards each other with increasing $W\!i$, coalescing at $Wi \approx 13.52$.  As $Wi$ is increased further, two stationary modes bifurcate out of the coalescence point, with
one of the modes becoming unstable, while the other becomes more stable, viz., they travel in the opposite directions along the $\omega_r = 0$ axis.  Although not shown here, we have verified that, for $\epsilon = 0.1$ and $\beta = 0.98$, the least stable modes are stationary at all $Wi$ for $\alpha < \alpha_{\text{critical}} \approx 4$.
In contrast, for $\alpha > \alpha_{\text{critical}} = 4$, the least stable modes are propagating at lower $Wi$ (e.g., $Wi = 1.6$), but eventually become stationary at higher $Wi$.

\begin{figure}
  \centering
   \subfigure [$ W\!i = 0.016 $]{\includegraphics[width=0.45\textwidth]{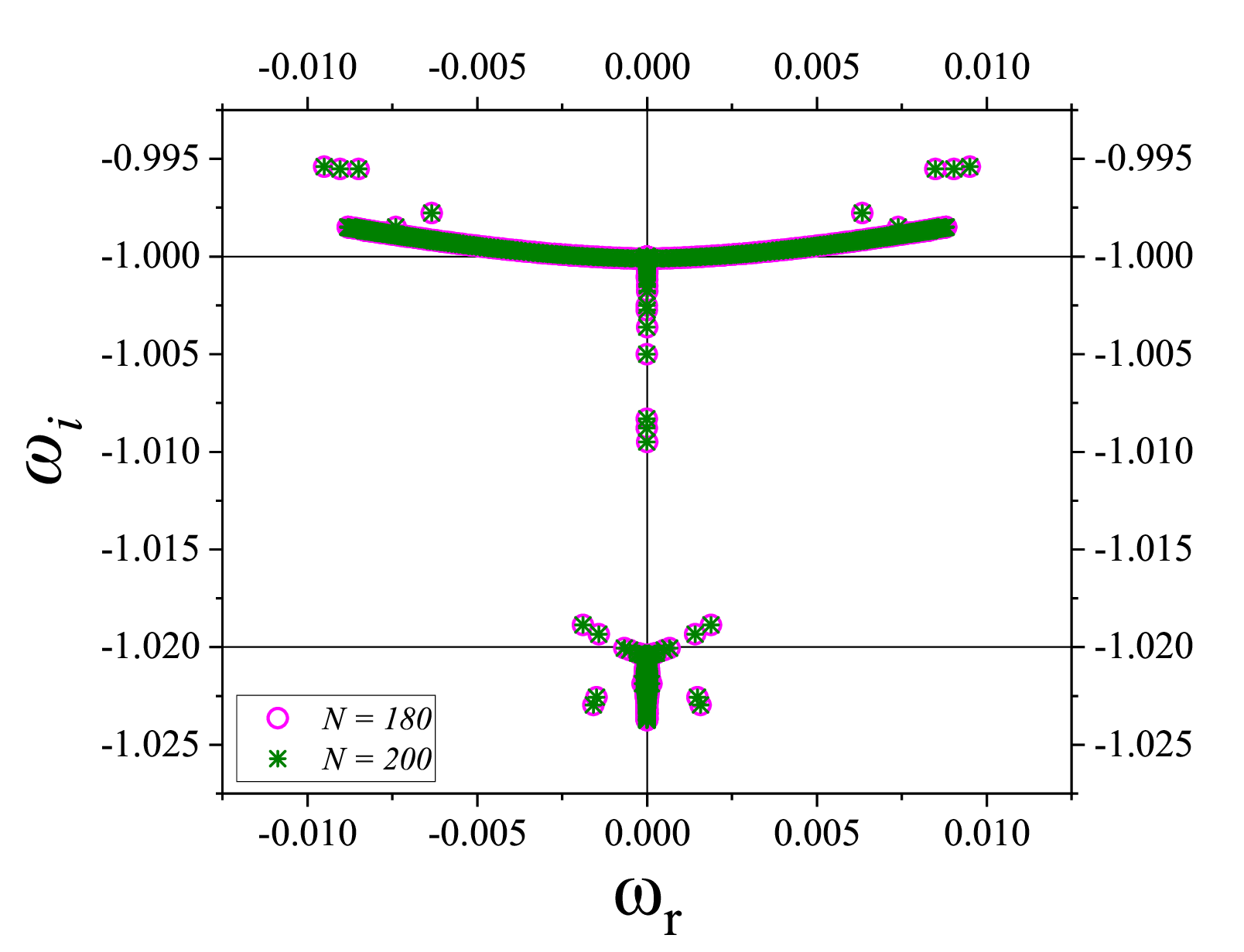}\label{fig:Axisymm_HSM_dean_e_0.1_beta_0.98_1}}
  \subfigure[$ W\!i = 0.16 $]{\includegraphics[width=0.45\textwidth]{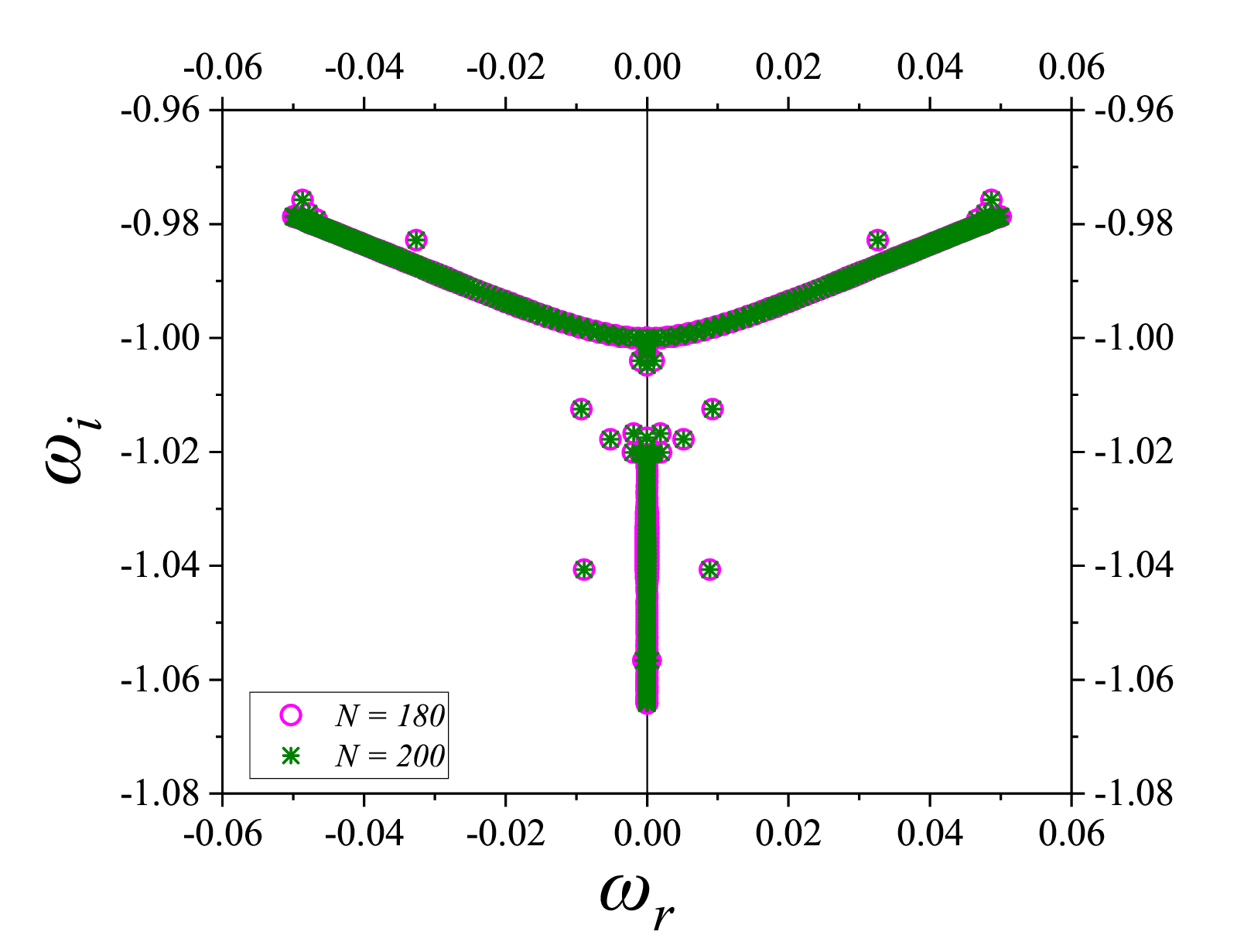}\label{fig:Axisymm_HSM_dean_e_0.1_beta_0.98_Wi_0.16}}
  \subfigure[$ W\!i = 0.8 $]{\includegraphics[width=0.45\textwidth]{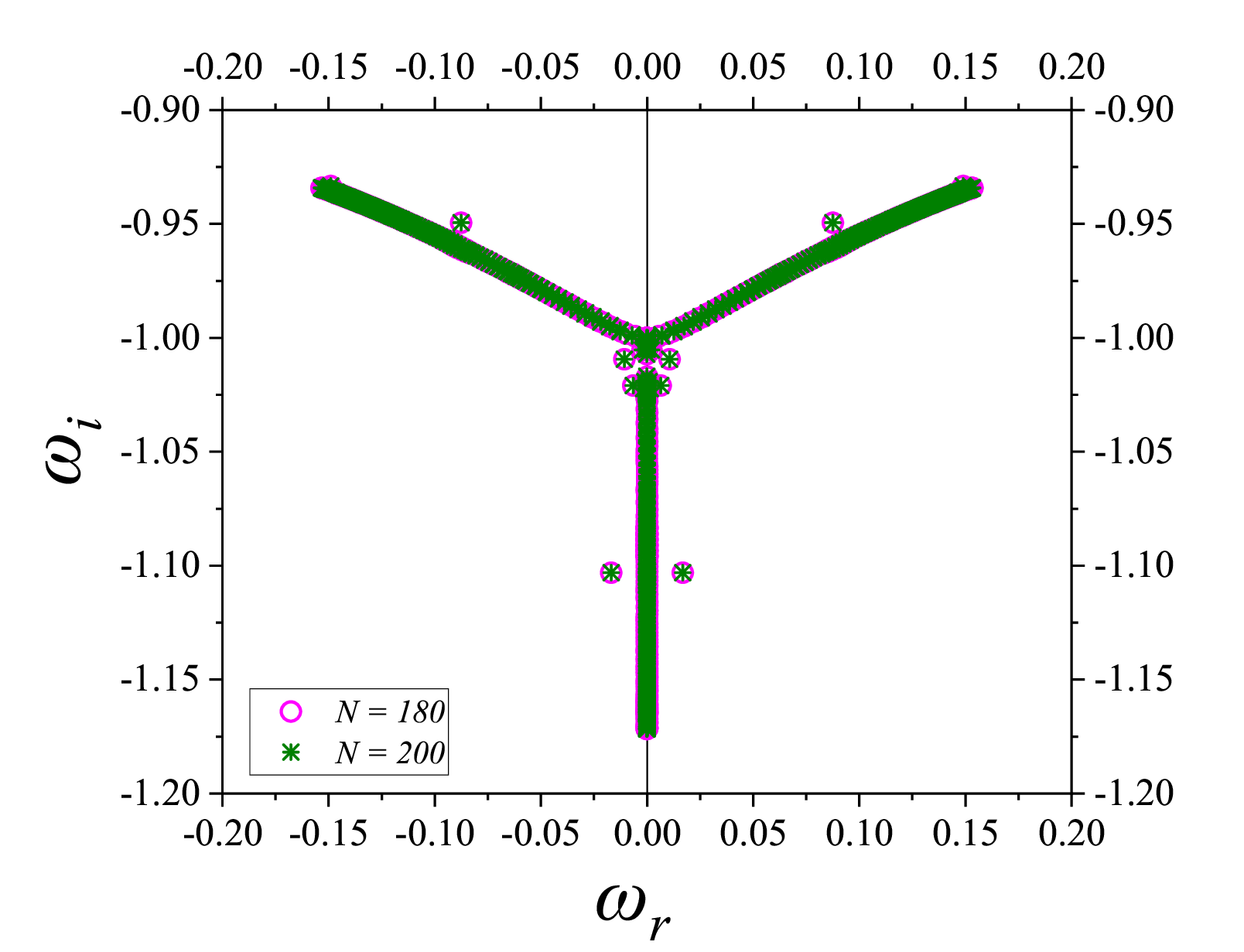}\label{fig:Axisymm_HSM_dean_e_0.1_beta_0.98_Wi_0.8}}
   \subfigure [$ W\!i = 1.6 $]{\includegraphics[width=0.45\textwidth]{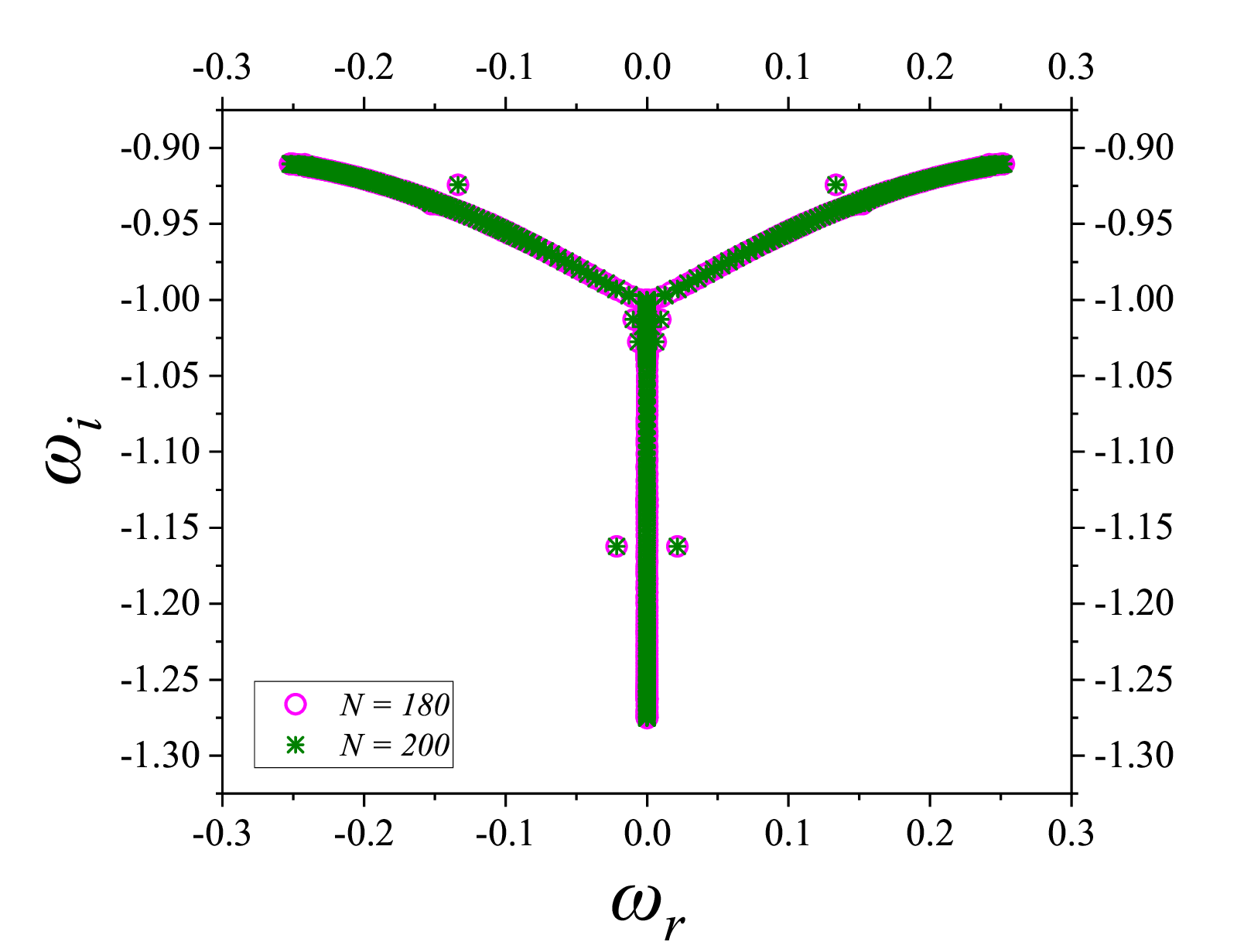}\label{fig:Axisymm_HSM_dean_e_0.1_beta_0.98_L_100_Wi_1.6}}
    \subfigure [$ W\!i = 8 $]{\includegraphics[width=0.45\textwidth]{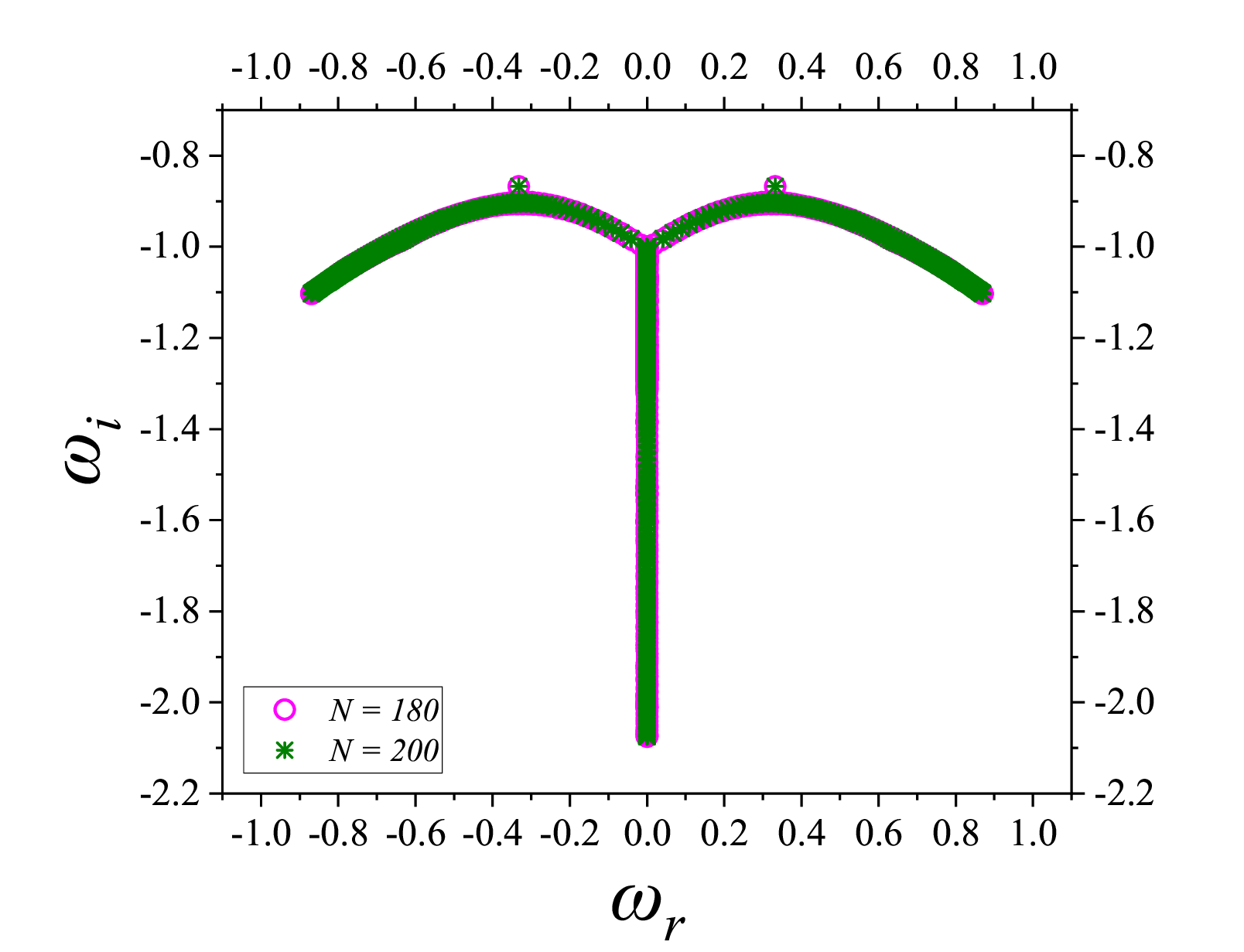}\label{fig:Axisymm_HSM_dean_e_0.1_beta_0.98_Wi_8}}
  \subfigure[$ W\!i = 20 $]{\includegraphics[width=0.45\textwidth]{Axisymmetric_FENE_P_L_100_ES_e_0.1_alpha_7_Wi_20_n_0_beta_0.98.eps}\label{fig:Axisymm_HSM_dean_e_0.1_beta_0.98_Wi_20_L_100}}
  \caption{Axisymmetric ($n = 0$) eigenspectra  at different $W\!i$'s for Dean flow. Data for $\epsilon = 0.1$, $\alpha = 7$, $\beta = 0.98$ , $L = 100$.}
\label{fig:Showing_Stabilization_effect_of-Wi_Fene_p_Axisymmetric_L_100_diff_Wi}
\end{figure}

The effect of increasing $W\!i/L$ on the spectra is shown in Fig.\,\ref{fig:Showing_Stabilization_effect_of-Wi_Fene_p_Axisymmetric_L_100_diff_Wi}; this may be achieved by increasing $W\!i$ for a fixed $L$ (as in the figure), or by decreasing $L$ for a fixed $W\!i$. Here, the two symmetrically placed wing-like structures as well as the vertical line-like segments on the imaginary axis (as discussed in Sec.\,\ref{sec:Nature_of_Spectrum}) are numerical approximations of the CS in Dean flow \citep{Mohanty_etal_CS}.
\begin{figure}
    \centering
    \subfigure[$W\!i = 200$ and $L = 100$]{
        \includegraphics[width=0.45\textwidth]{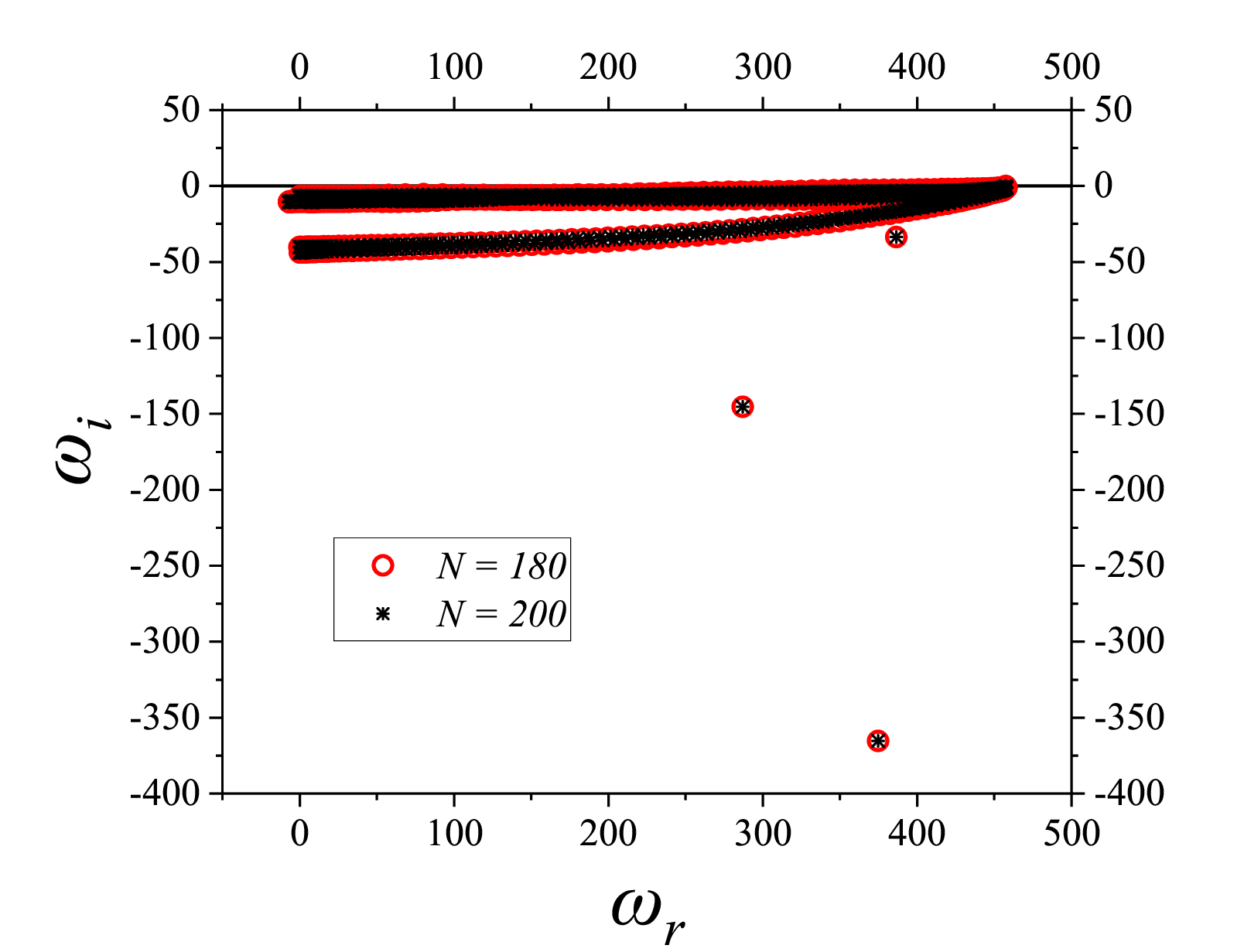} \label{fig:CM_W_200_L_100_eigen_spectrum}
    }
    \subfigure[$W\!i = 200$ and $L = 100$]{
        \includegraphics[width=0.45\textwidth]{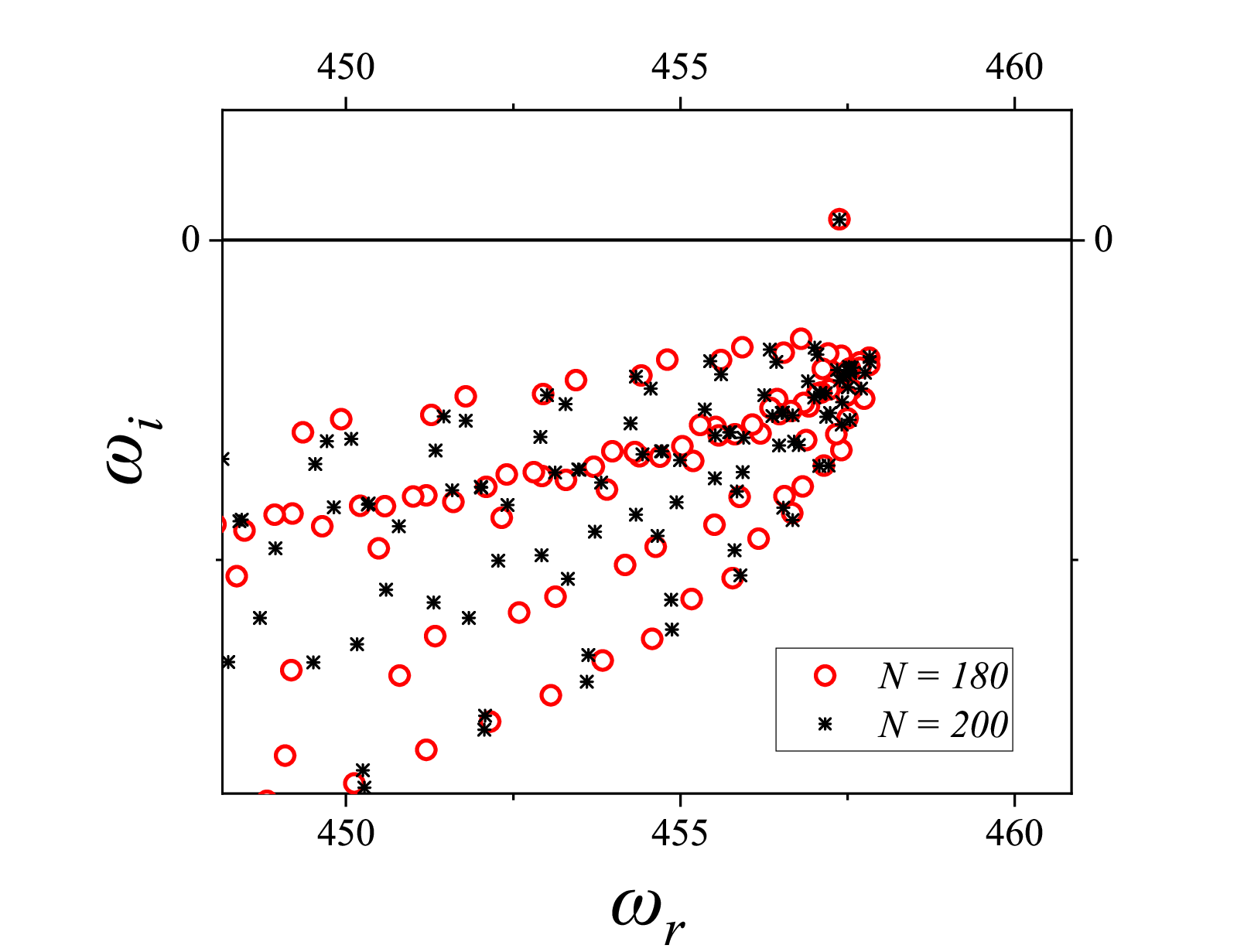} \label{fig:Zoom_in_CM_W_200_L_100_eigen_spectrum}
    
    }
    \subfigure[$W\!i = 400$ and $L = 200$]{
        \includegraphics[width=0.45\textwidth]{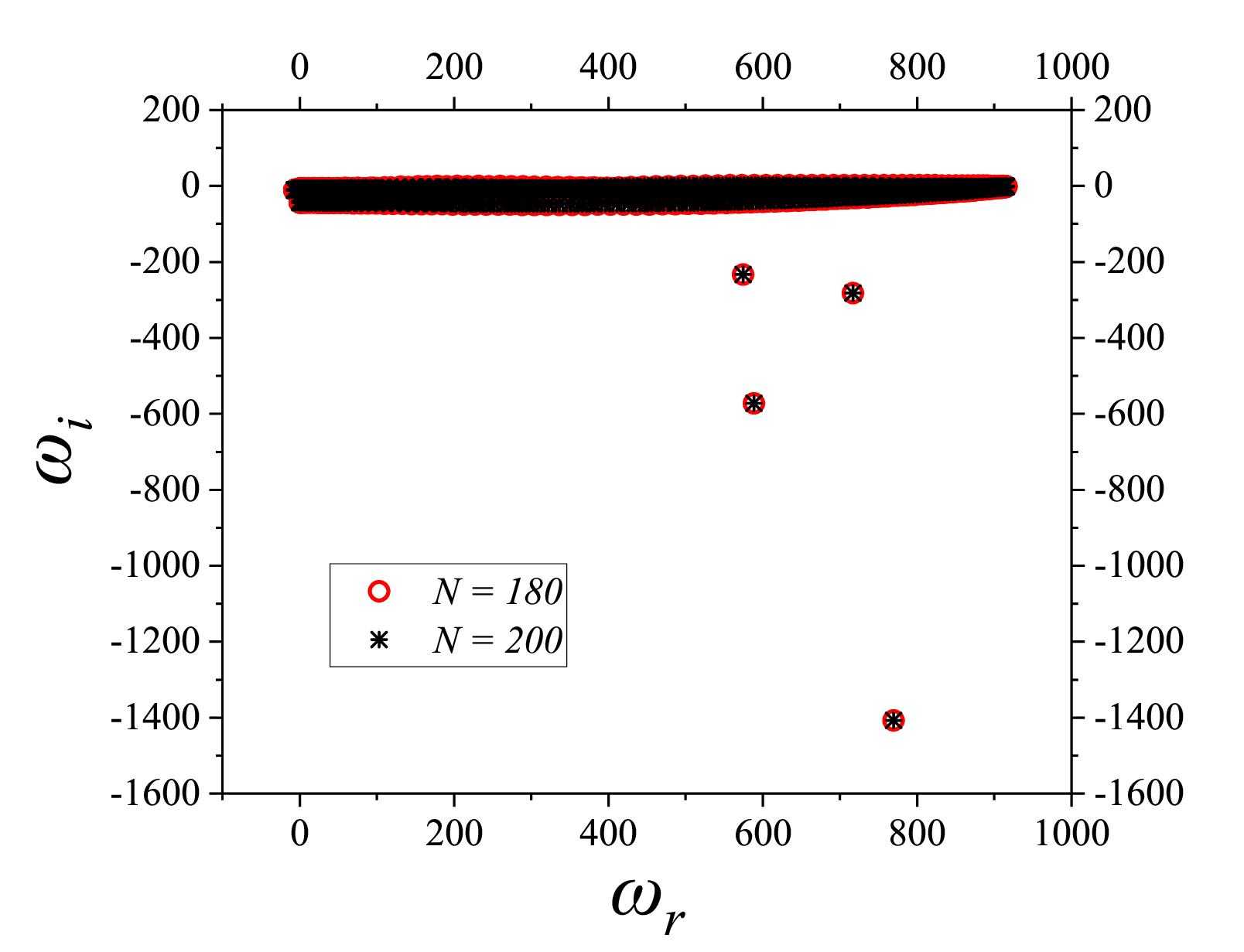}\label{fig:CM_W_400_L_200_eigen_spectrum}
    }
    \subfigure[$W\!i = 400$ and $L = 200$]{
        \includegraphics[width=0.45\textwidth]{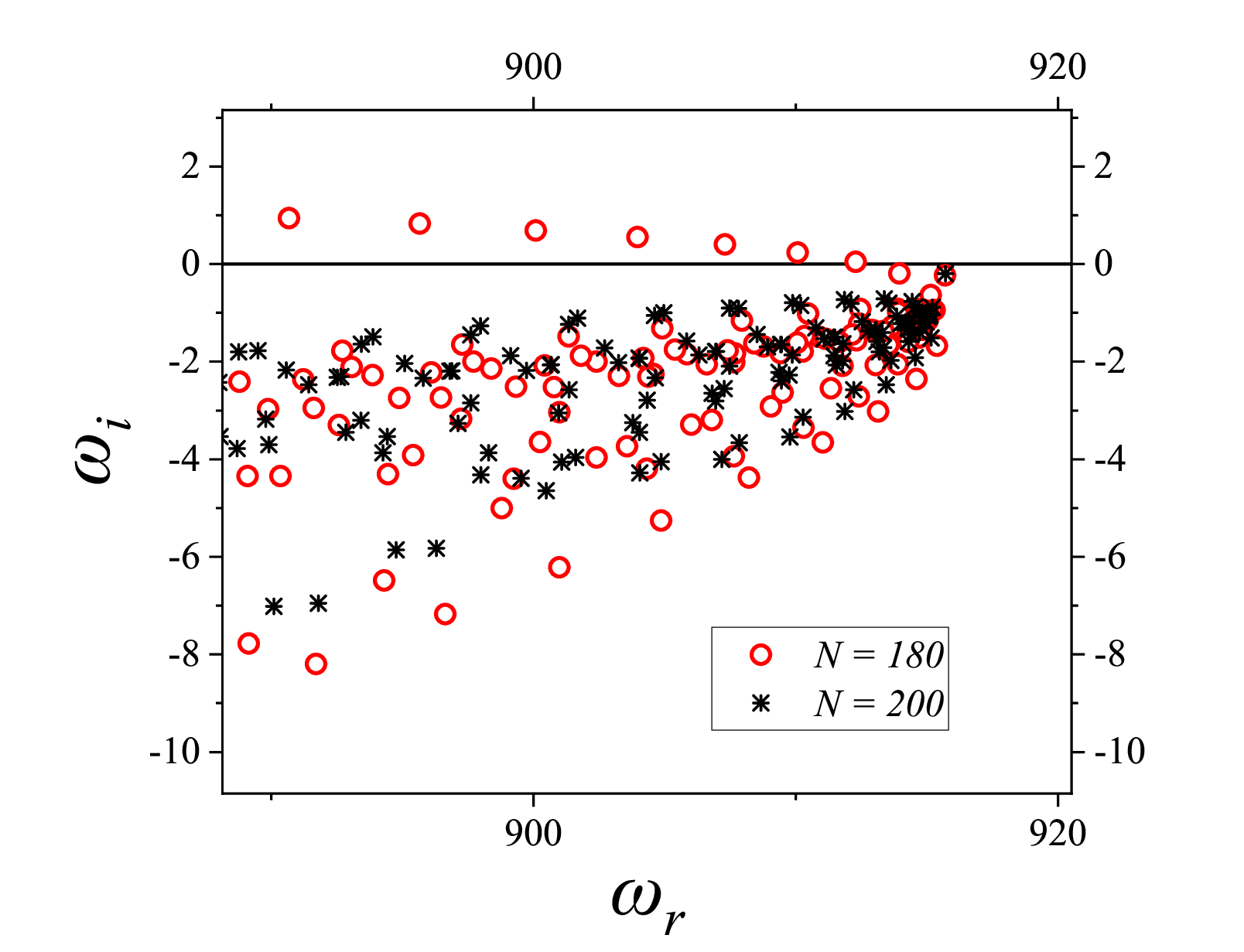}\label{fig:Zoom_in_CM_W_400_L_200_eigen_spectrum}
    }
    \caption{Eigenspectra demonstrating the centre mode in Dean flow of a FENE-P fluid at $Re = 0$, $\alpha = 0.1$, $\beta = 0.98$, $\epsilon = 0.1$, and $n = 16$, for  $W\!i/L = 2$ but for two ($W\!i$, $L$) pairs. (a) Full eigenspectrum for the given parameters, and (b) zoomed-in view of (a). Similarly, (c) full eigenspectrum for another set of $W\!i$ and $L$, and (d) zoomed-in view of (c). The eigenspectra obtained for $N = 180$ and $200$ show good convergence.}
    \label{fig:CM_same_W_by_L_2_Eigen_spectrum}
\end{figure}
The spectrum approaches a purely continuous one as $W\!i$ is increased from Fig.\,\ref{fig:Axisymm_HSM_dean_e_0.1_beta_0.98_1} - \ref{fig:Axisymm_HSM_dean_e_0.1_beta_0.98_Wi_20_L_100}, with the wings in particular assuming their full form for the largest $W\!i/L$.

Figures\,\ref{fig:CM_W_200_L_100_eigen_spectrum} and \ref{fig:CM_W_400_L_200_eigen_spectrum} depict the spectra for the same $W\!i/L$, albeit at different choices of $W\!i$ and $L$;  Figs.\, \ref{fig:Zoom_in_CM_W_200_L_100_eigen_spectrum} and \ref{fig:Zoom_in_CM_W_400_L_200_eigen_spectrum} show their corresponding magnified views. 
A data collapse of the base-state velocity profiles was demonstrated for different pairs of ($W\!i,L$) for a fixed $W\!i/L$ ratio for Dean flow for fixed $\epsilon$ and $\beta$ (see Fig.\,\ref{fig:Deandiff_W_by_L_same_beta_0.98}). 
However, Fig.\,\ref{fig:CM_same_W_by_L_2_Eigen_spectrum} shows that the eigenspectrum is independently influenced by $W\!i$ and $L$, and not just by the $W\!i/L$ alone. This can be attributed to the two- or three-dimensional nature of the disturbance flow field; the original data collapse of \cite{Yamani_McKinley2023,Tej2024} is expected to hold only for viscometric flows.

\section{Eigenfunctions for viscoelastic Dean flow}
\label{sec:Appendixeigenfunctions}

\begin{figure}
    \centering
    \subfigure[$\tilde{v}_{r}$]{
        \includegraphics[width=0.45\textwidth]{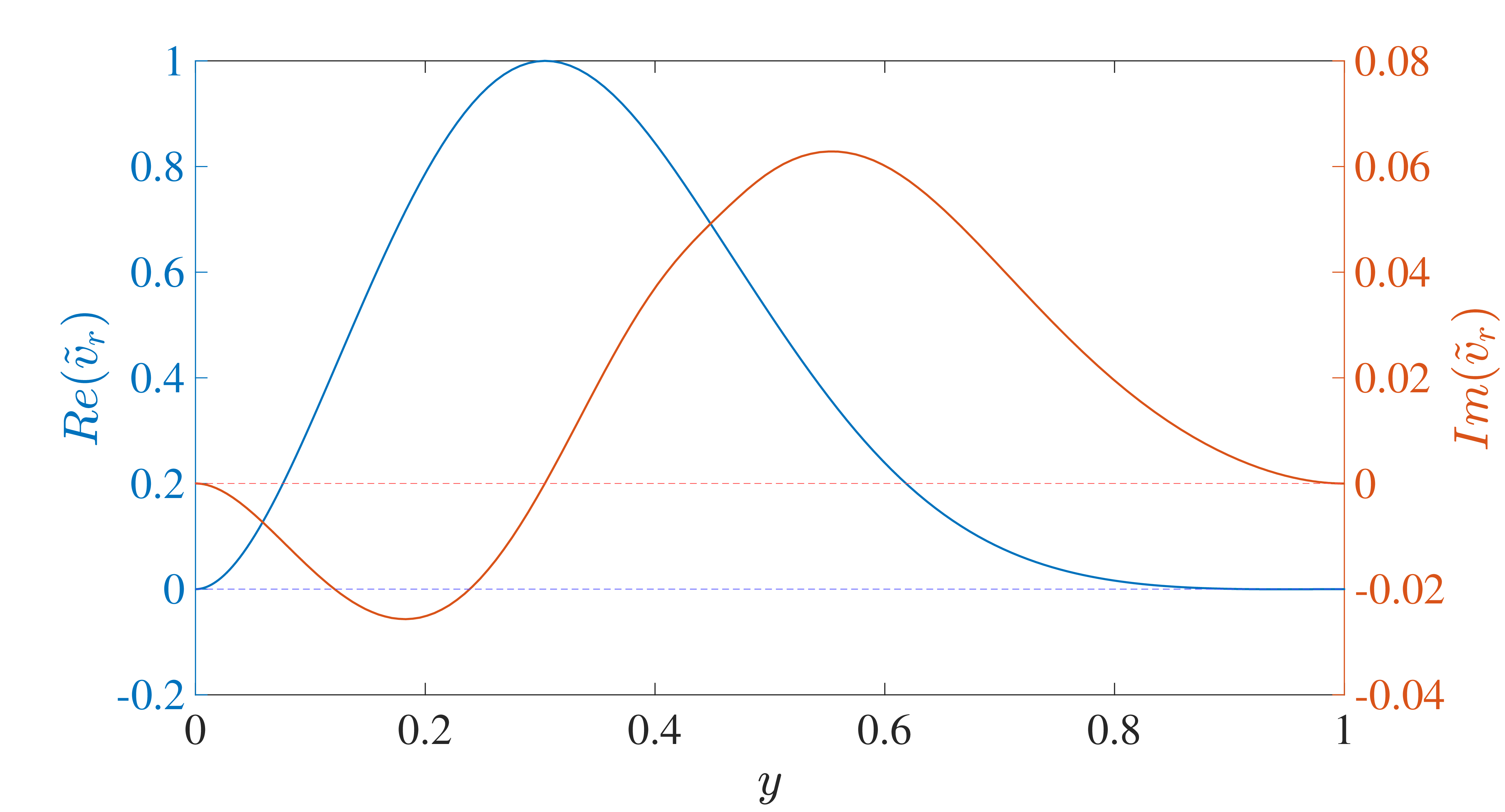} \label{fig:HSM2_u_r}
    }
    \subfigure[$\tilde{v}_{\theta}$]{
        \includegraphics[width=0.45\textwidth]{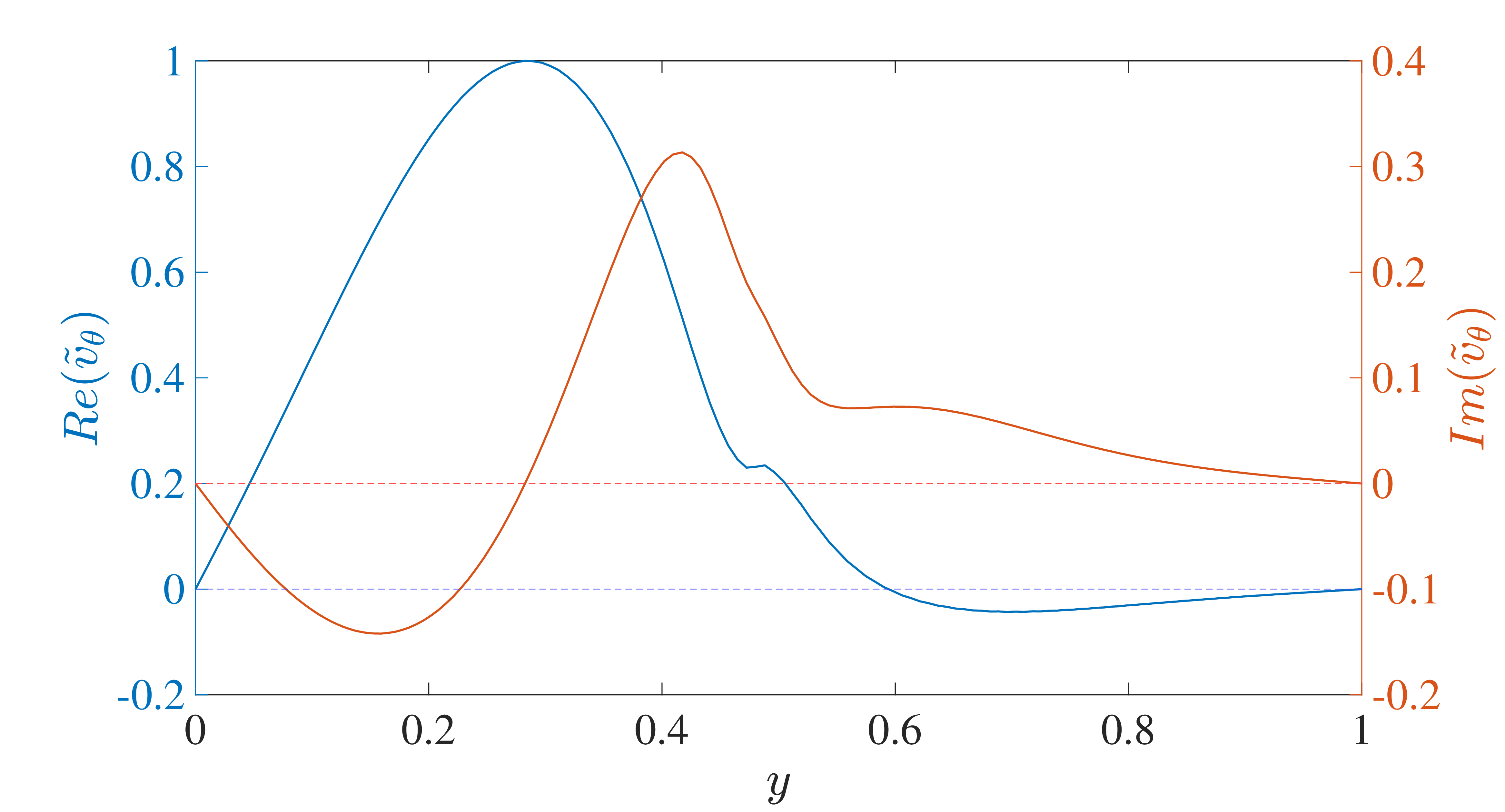} \label{fig:HSM2_u_0}
    
    }
    \subfigure[$\tilde{v}_{z}$]{
        \includegraphics[width=0.45\textwidth]{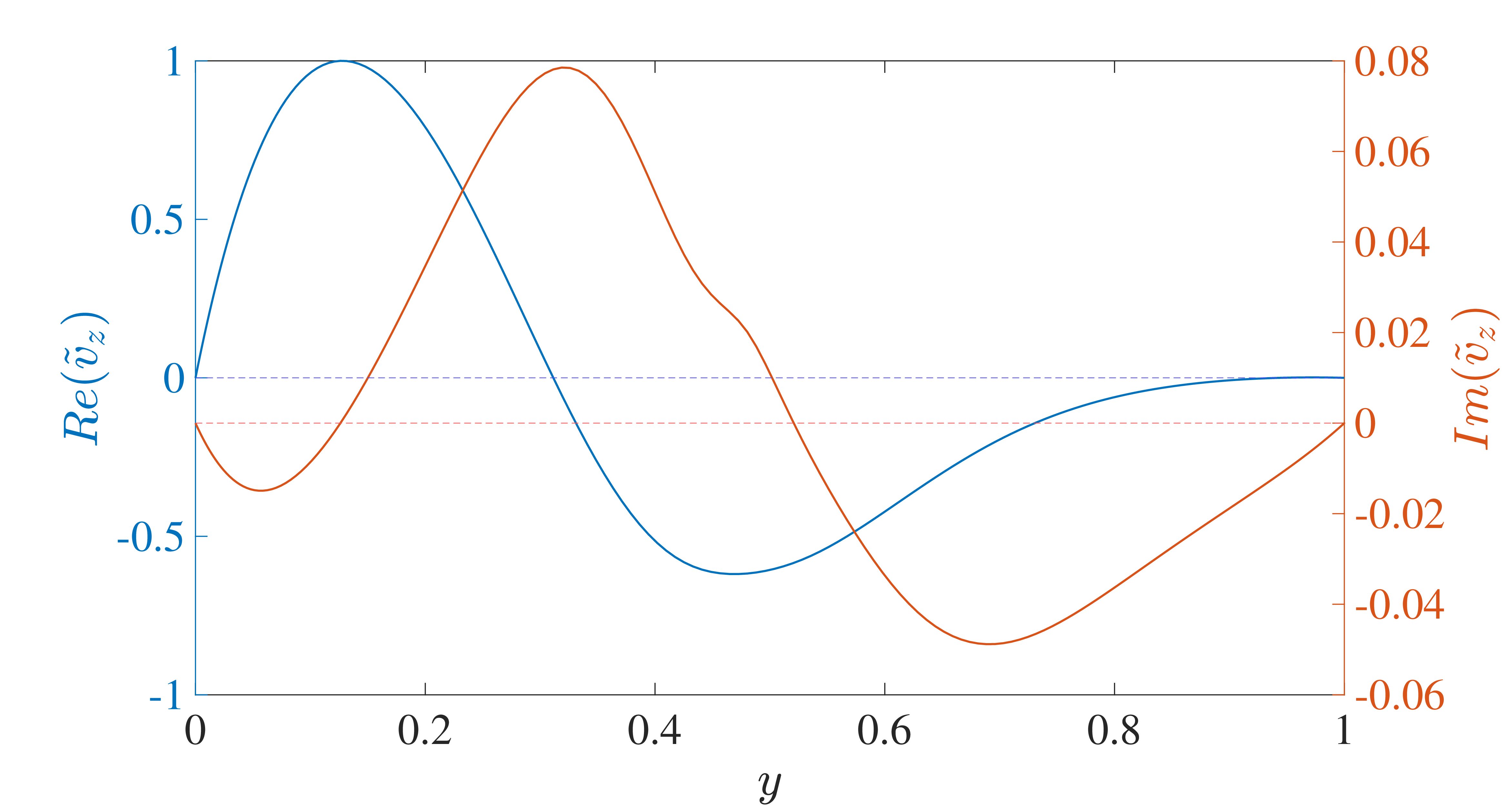}\label{fig:HSM2_u_z}
    }
    \subfigure[$\tilde{\tau}_{rr}$]{
        \includegraphics[width=0.45\textwidth]{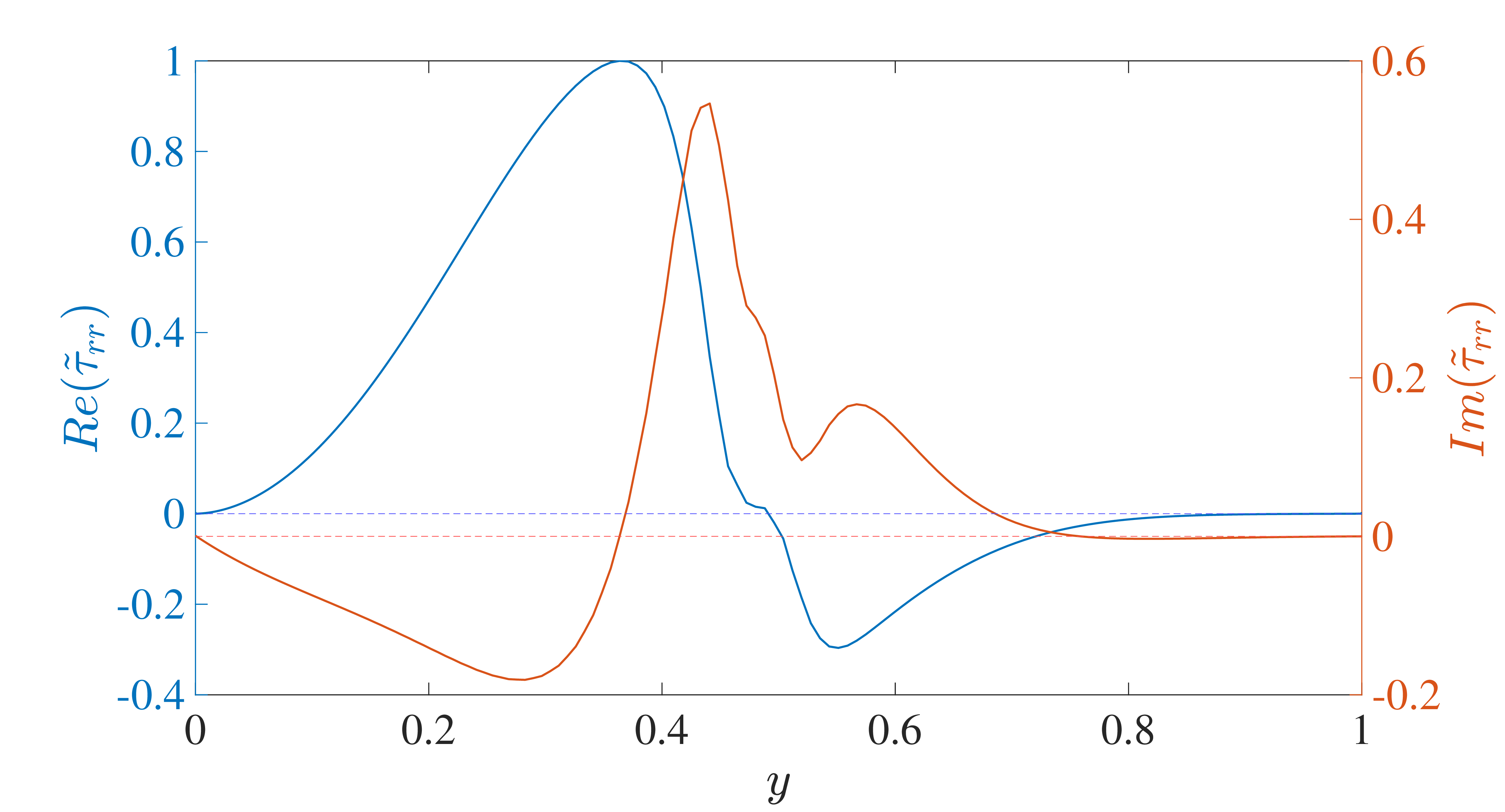}\label{fig:HSM2_Tau_rr}
    }
    \subfigure[$\tilde{\tau}_{r\theta}$]{
        \includegraphics[width=0.45\textwidth]{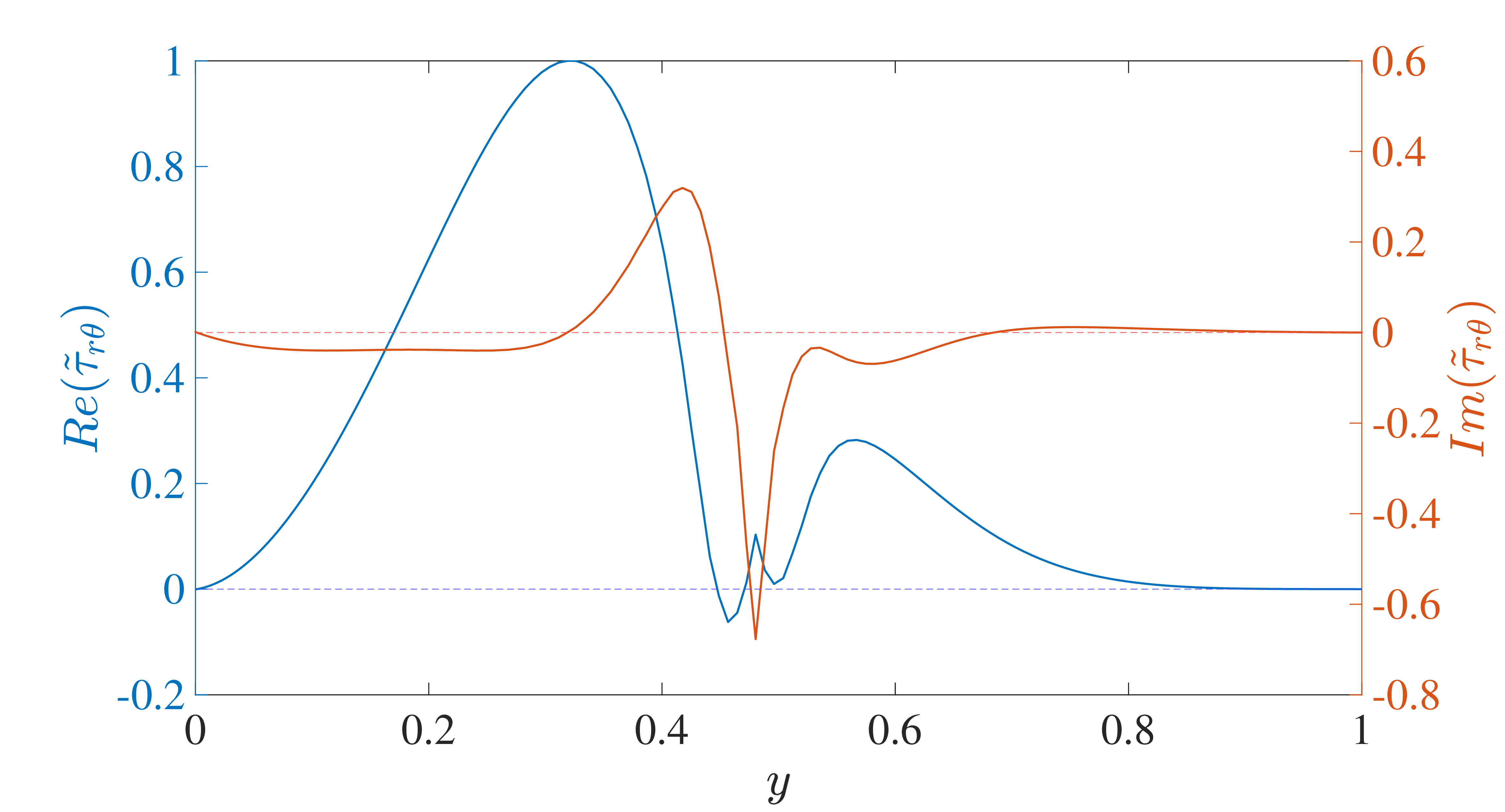}\label{fig:HSM2_Tau_r0}
    }
    \subfigure[$\tilde{\tau}_{rz}$]{
        \includegraphics[width=0.45\textwidth]{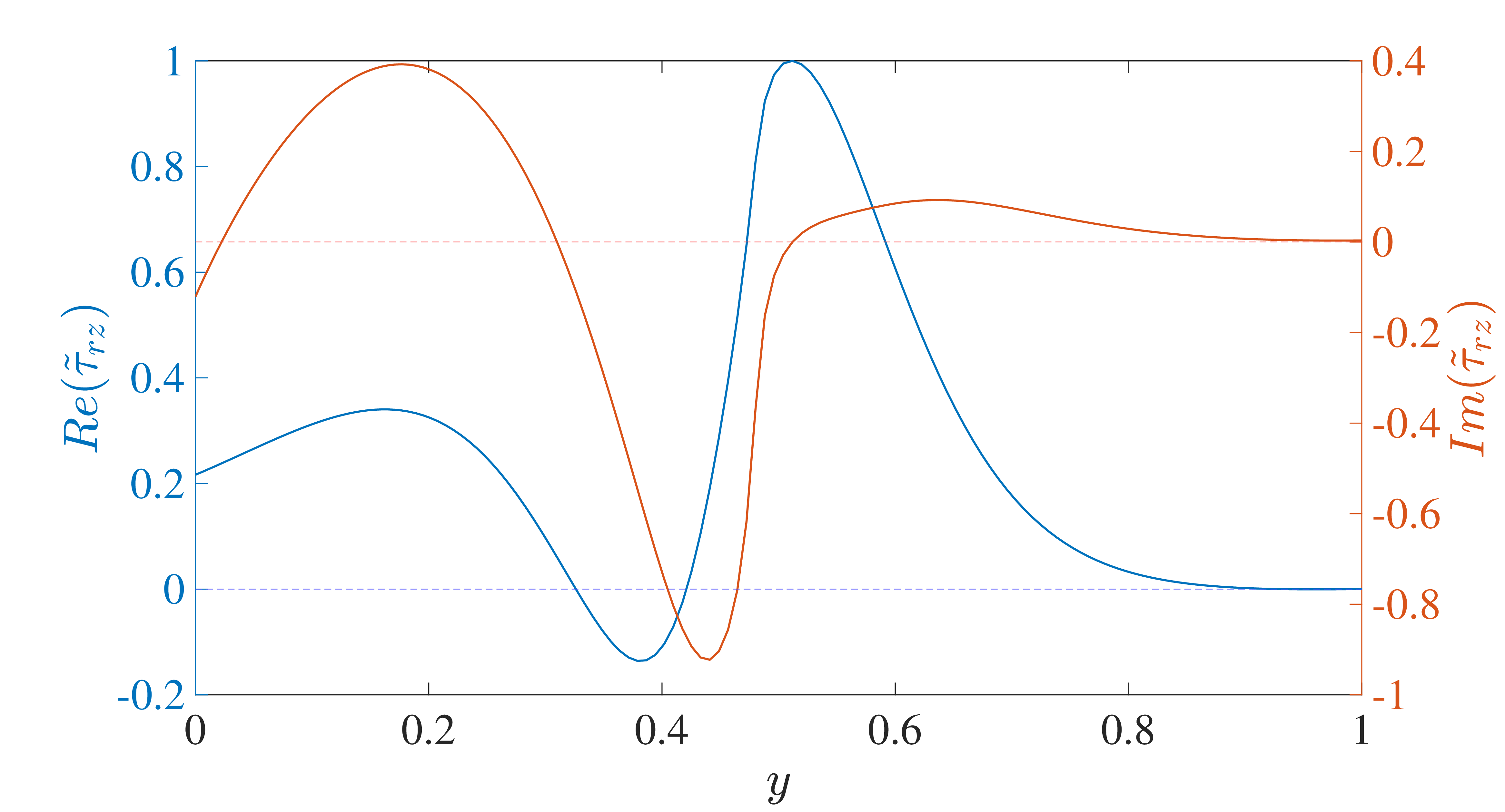}\label{fig:HSM2_Tau_rz}
    }
    \subfigure[$\tilde{\tau}_{\theta \theta}$]{
        \includegraphics[width=0.45\textwidth]{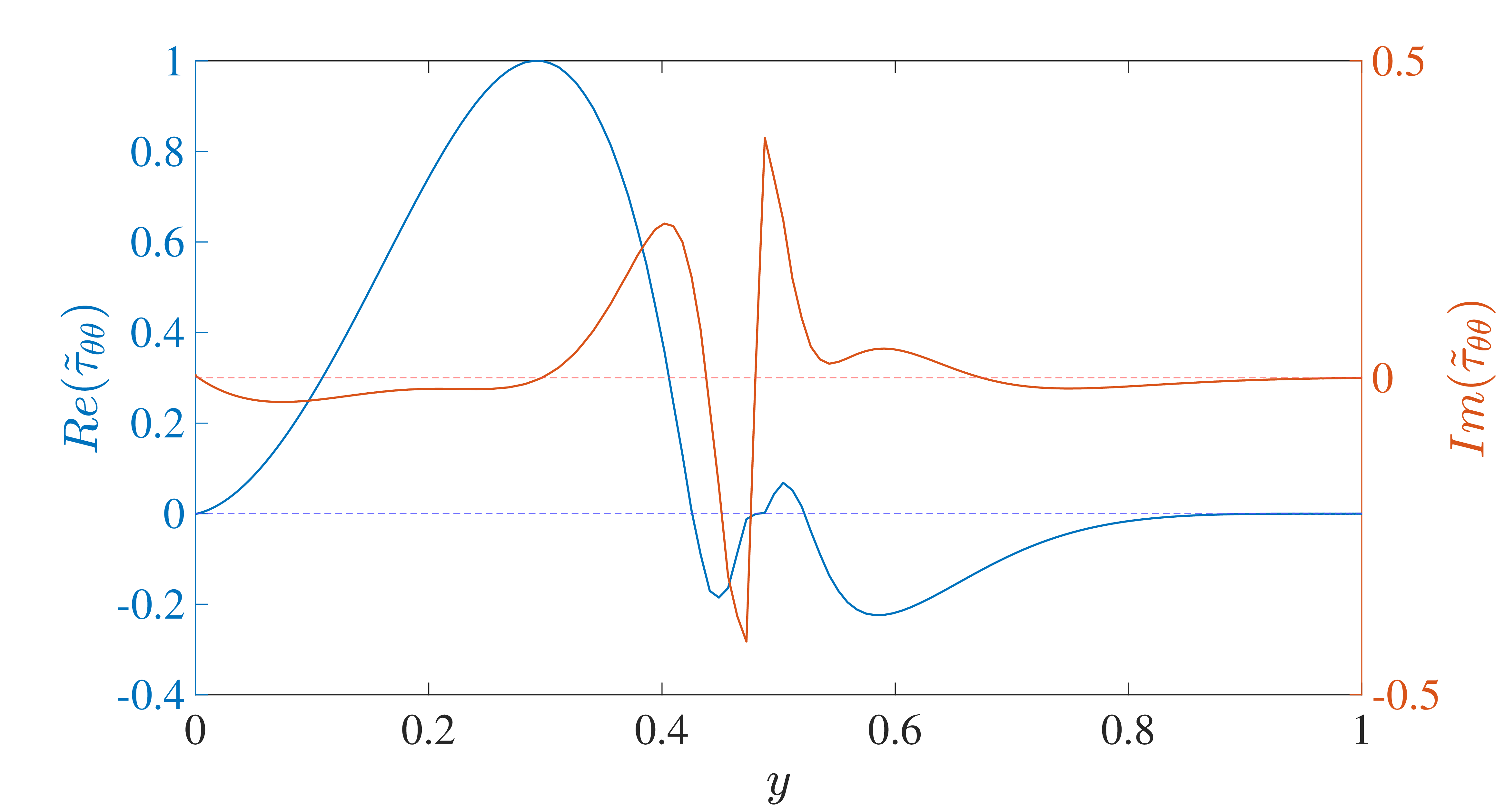}\label{fig:HSM2_Tau_00}
    }
    \subfigure[$\tilde{\tau}_{\theta z}$]{
        \includegraphics[width=0.45\textwidth]{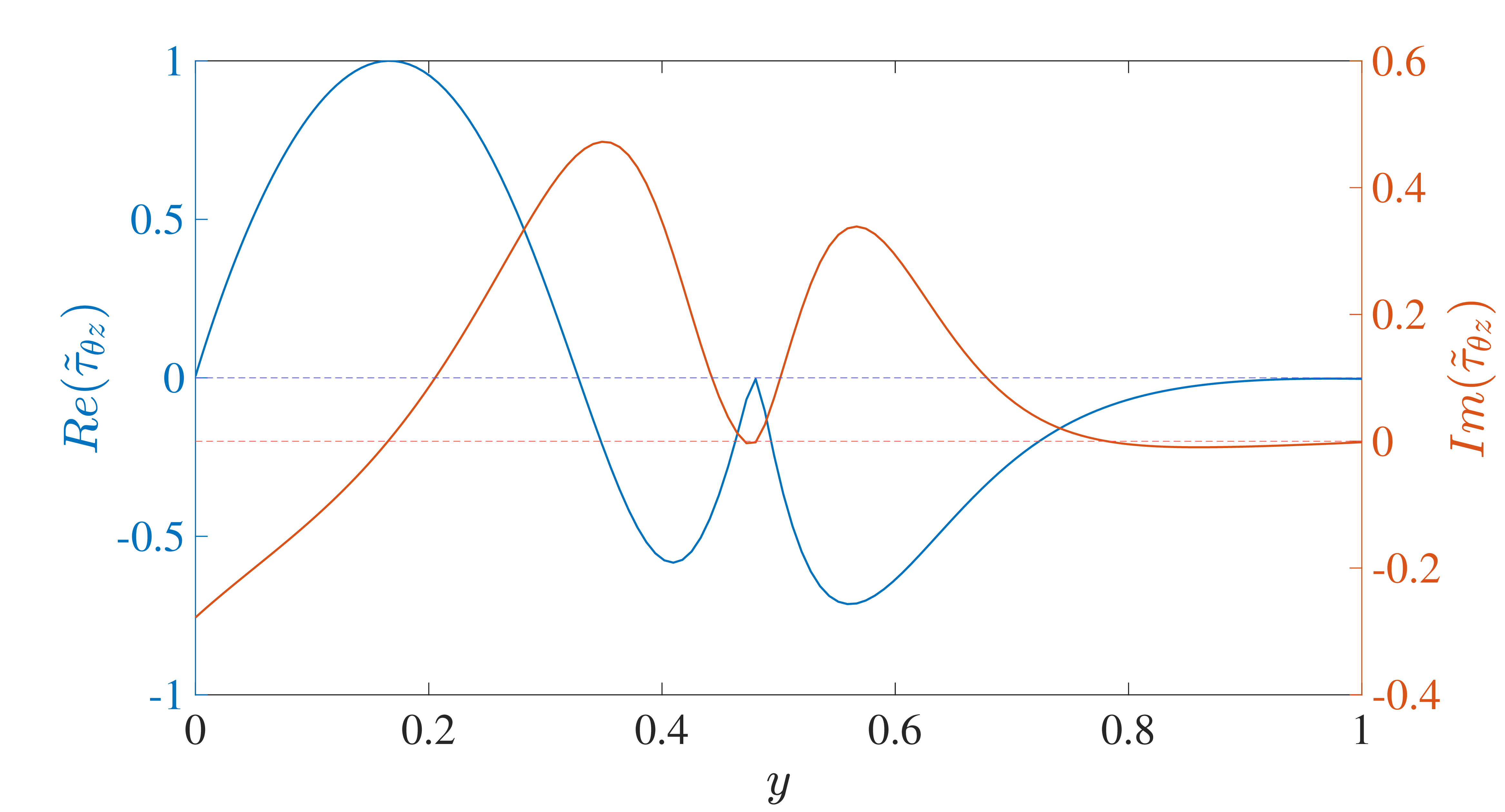}\label{fig:HSM2_Tau_0z}
    }
     \subfigure[$\tilde{\tau}_{zz}$]{
        \includegraphics[width=0.45\textwidth]{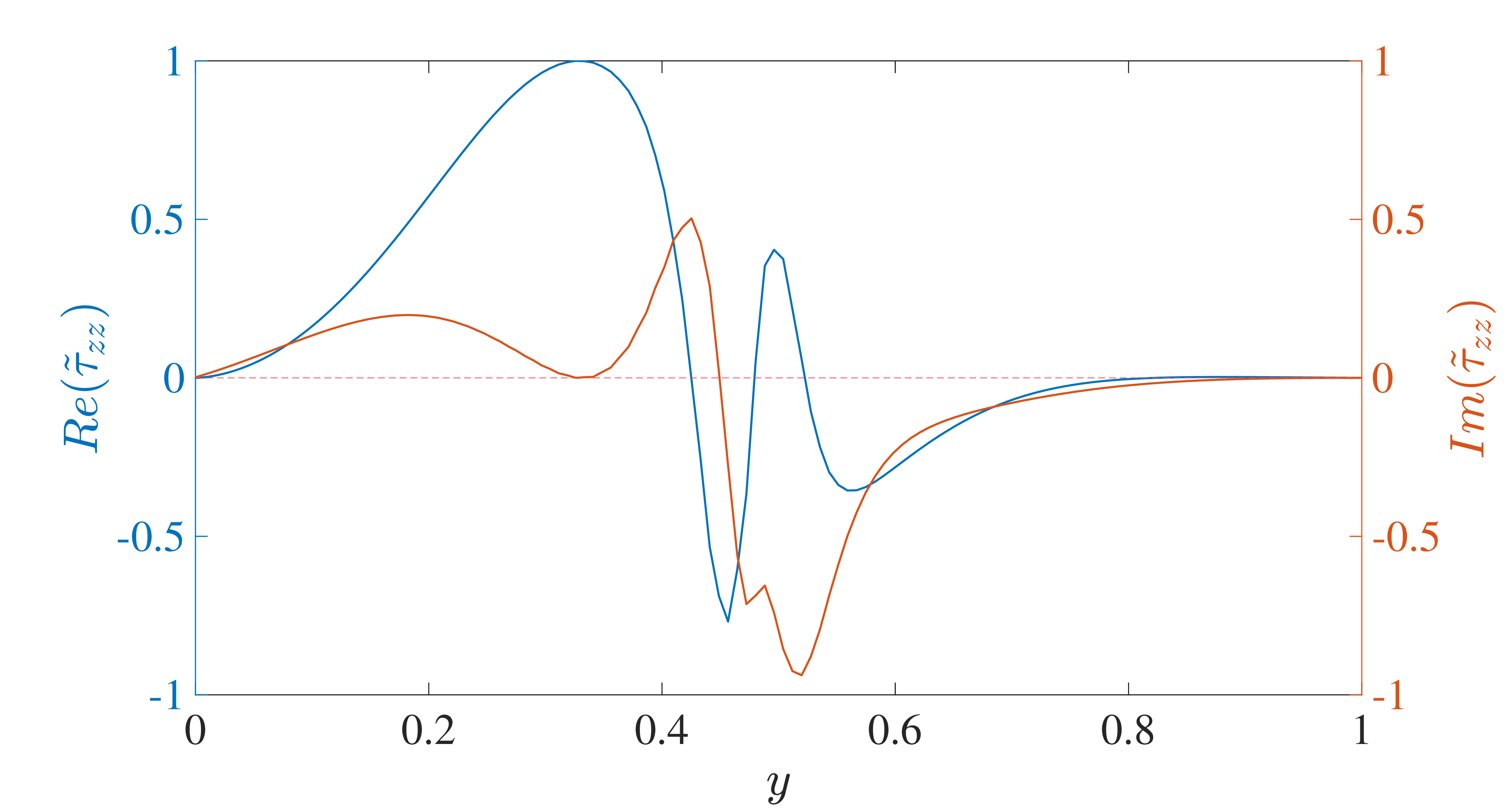}\label{fig:HSM2_Tau_zz}
    }
    \caption{Eigenfunctions for the HSM2 mode 
$\left(\omega = 62.786065 + 0.06752153\,i\right)$ in Dean flow of a FENE-P fluid at 
$Re = 0$, $\alpha = 7$, $\beta = 0.98$, $\epsilon = 0.1$, $L = 100$, $n = 1$ and $W\!i = 400$. The corresponding HSM2 eigenvalue was identified in the 
eigenspectrum shown previously in Fig.\,\ref{Showing_HSM2_L_100}.  The scales on the left and right $y$-axes are different, and dotted lines indicate the origins of the respective $y$-axes.
}
    \label{fig:HSM2_L_100_Eigen_function}
\end{figure}

\begin{figure}
    \centering
    \subfigure[$\tilde{v}_{r}$]{
        \includegraphics[width=0.45\textwidth]{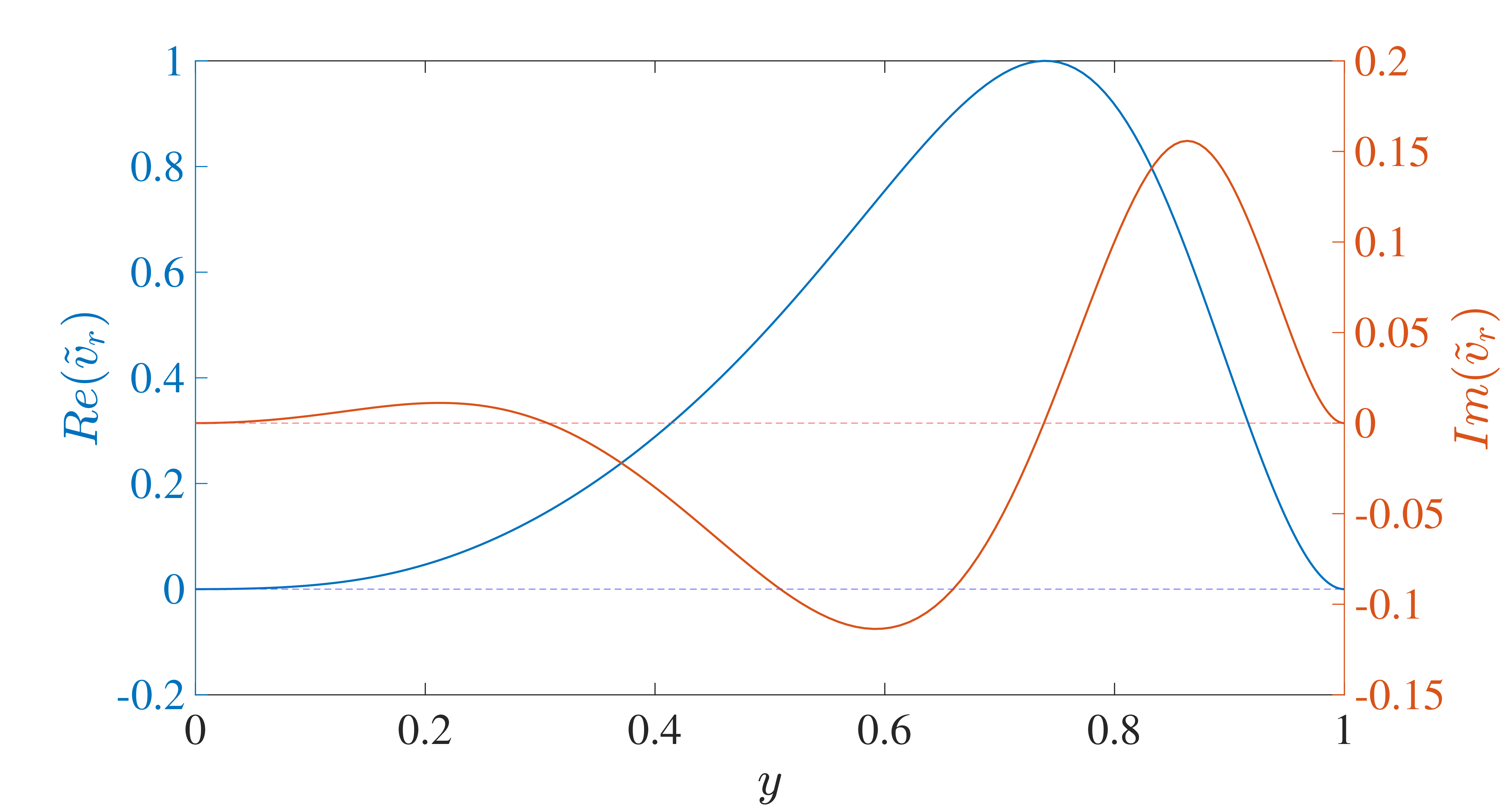} \label{fig:HSM1_u_r}
    }
    \subfigure[$\tilde{v}_{\theta}$]{
        \includegraphics[width=0.45\textwidth]{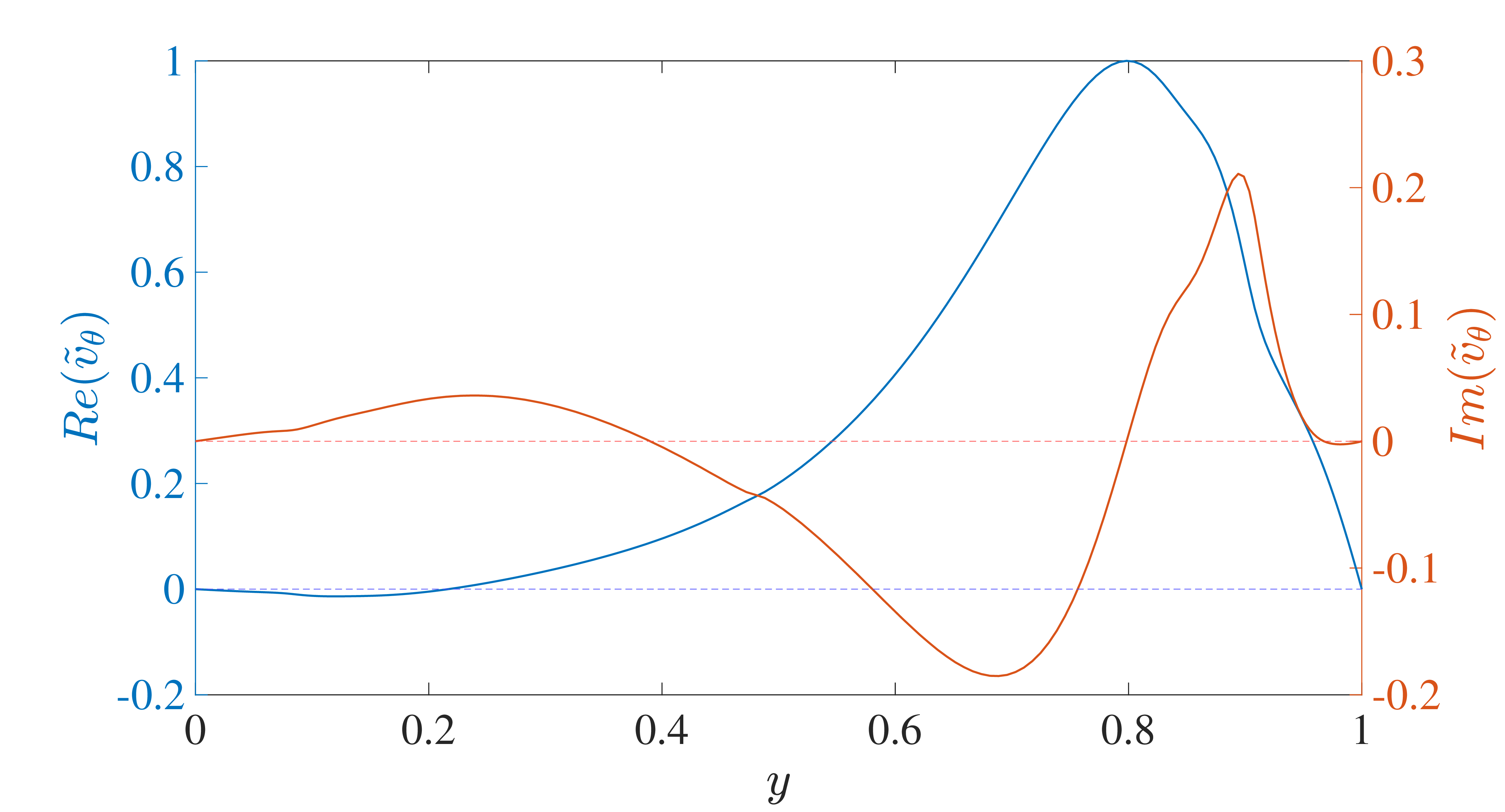} \label{fig:HSM1_u_0}
    
    }
    \subfigure[$\tilde{v}_{z}$]{
        \includegraphics[width=0.45\textwidth]{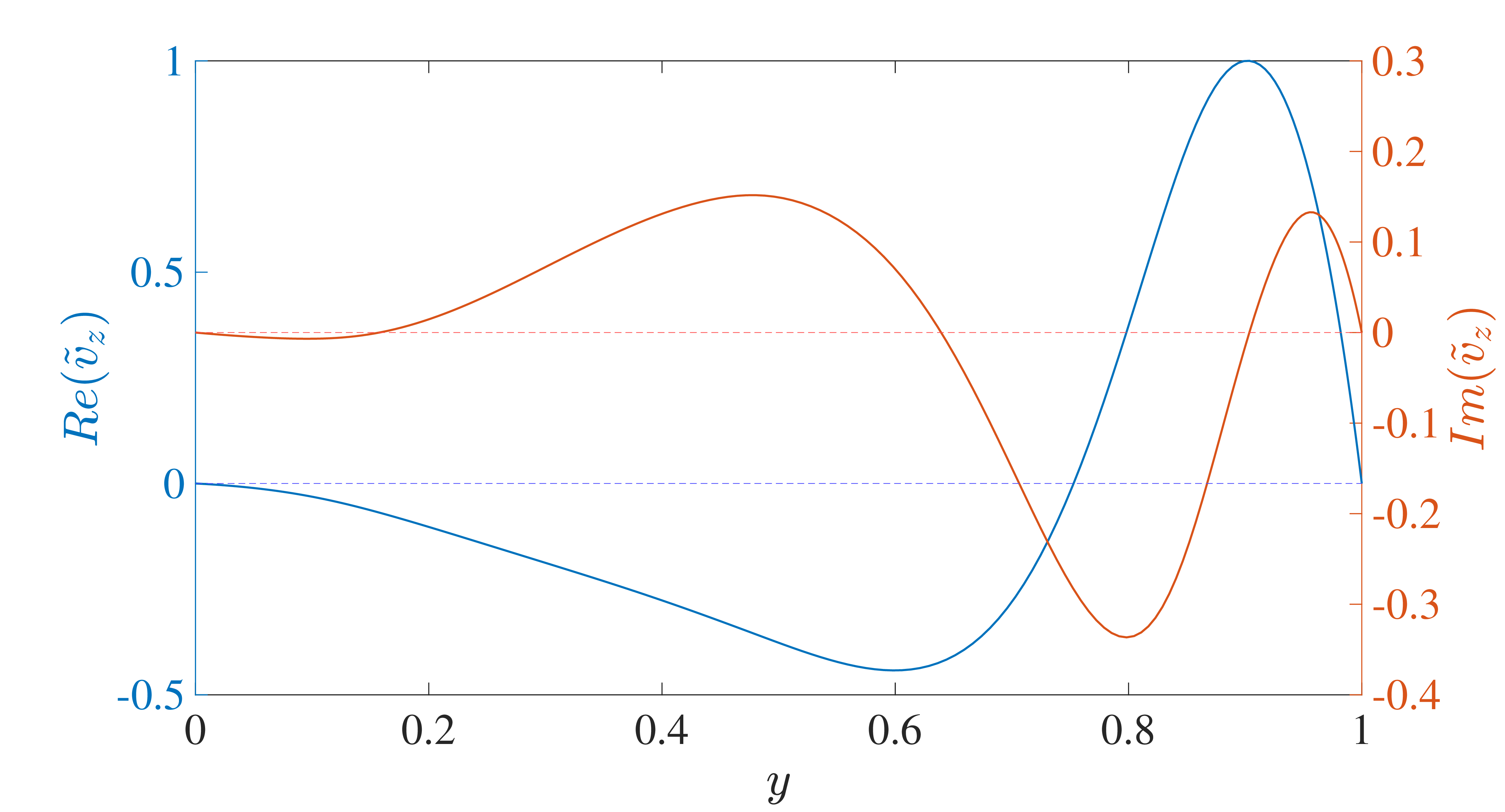}\label{fig:HSM1_u_z}
    }
    \subfigure[$\tilde{\tau}_{rr}$]{
        \includegraphics[width=0.45\textwidth]{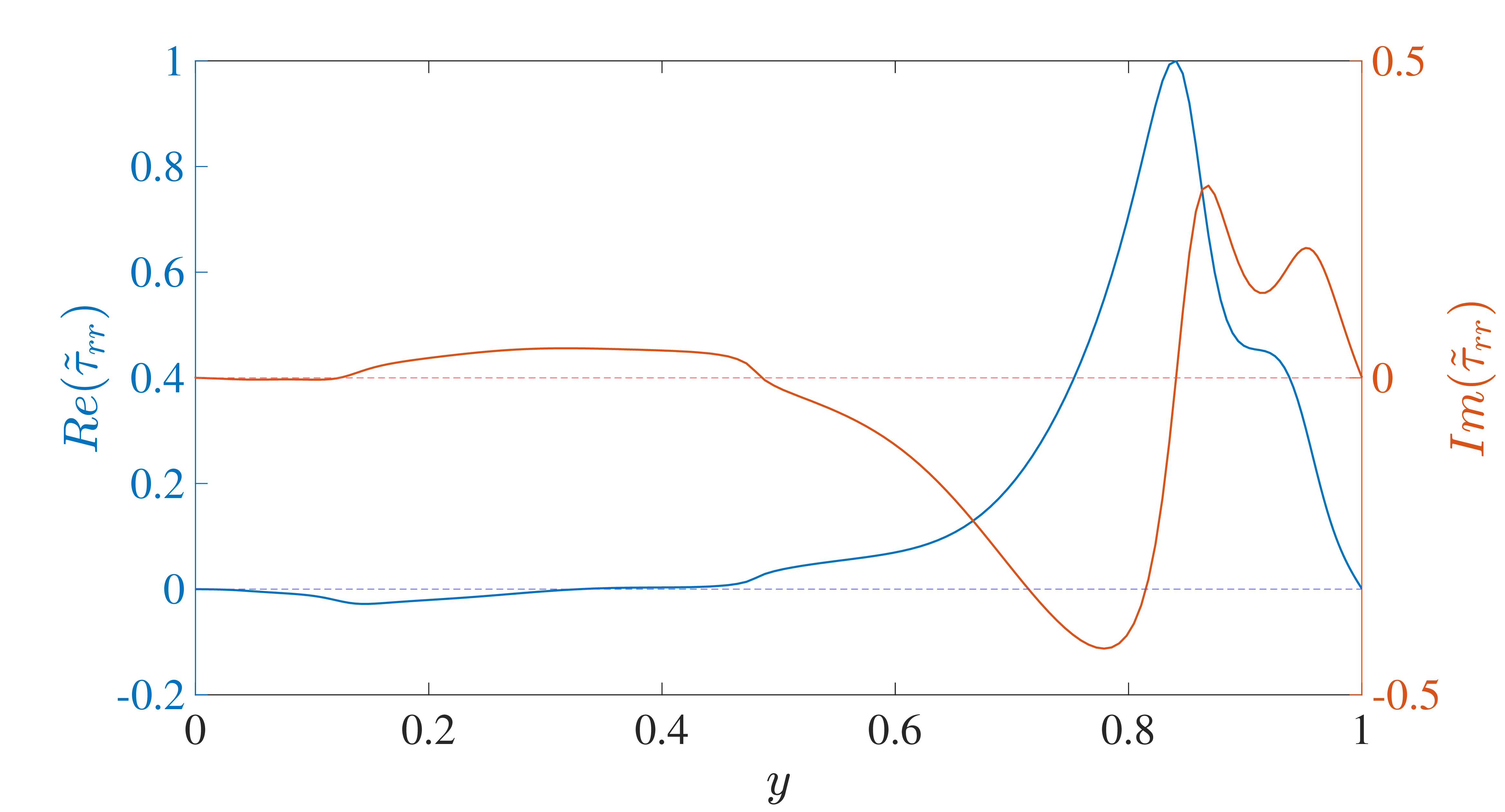}\label{fig:HSM1_Tau_rr}
    }
    \subfigure[$\tilde{\tau}_{r\theta}$]{
        \includegraphics[width=0.45\textwidth]{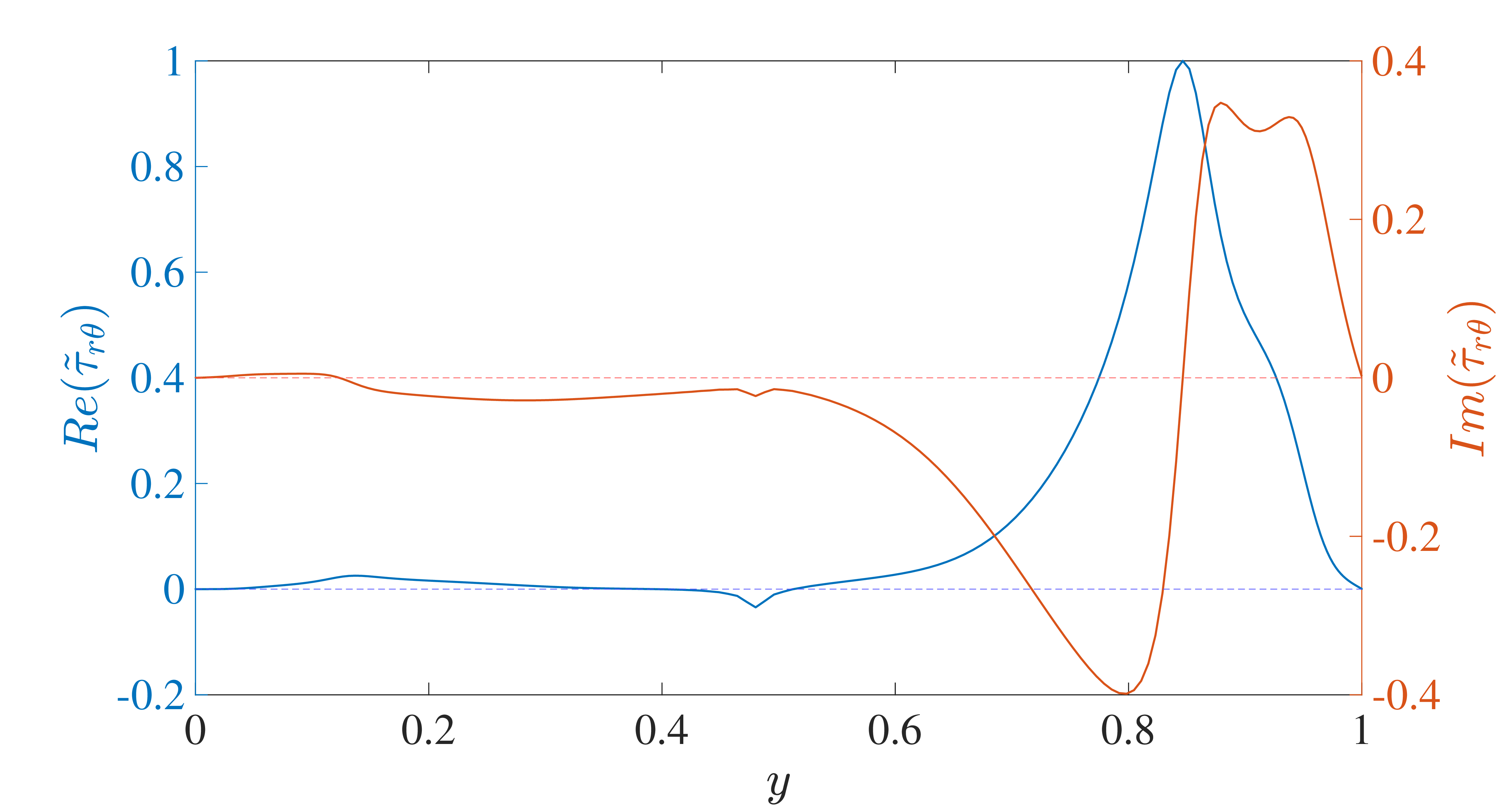}\label{fig:HSM1_Tau_r0}
    }
    \subfigure[$\tilde{\tau}_{rz}$]{
        \includegraphics[width=0.45\textwidth]{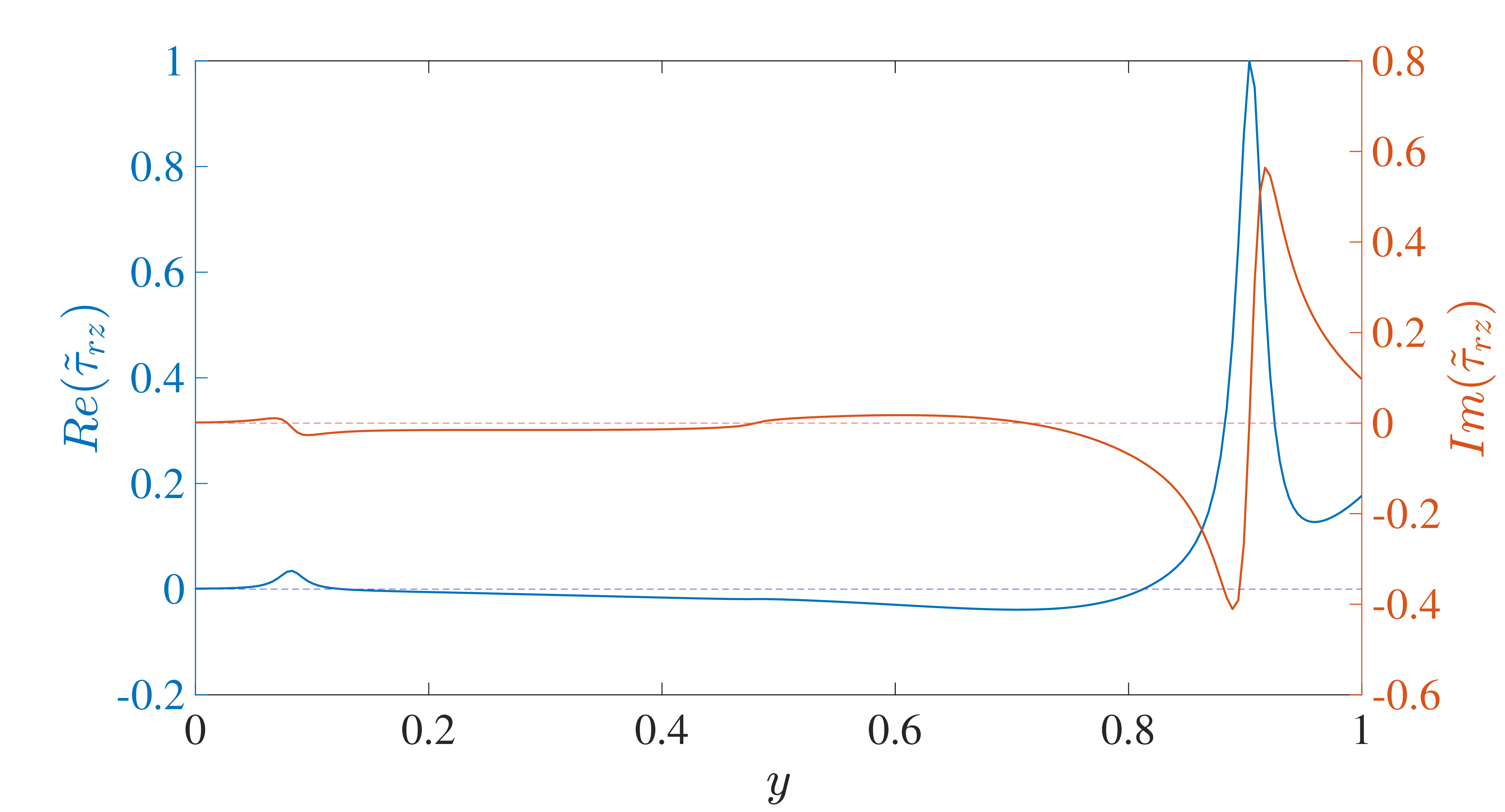}\label{fig:HSM1_Tau_rz}
    }
    \subfigure[$\tilde{\tau}_{\theta \theta}$]{
        \includegraphics[width=0.45\textwidth]{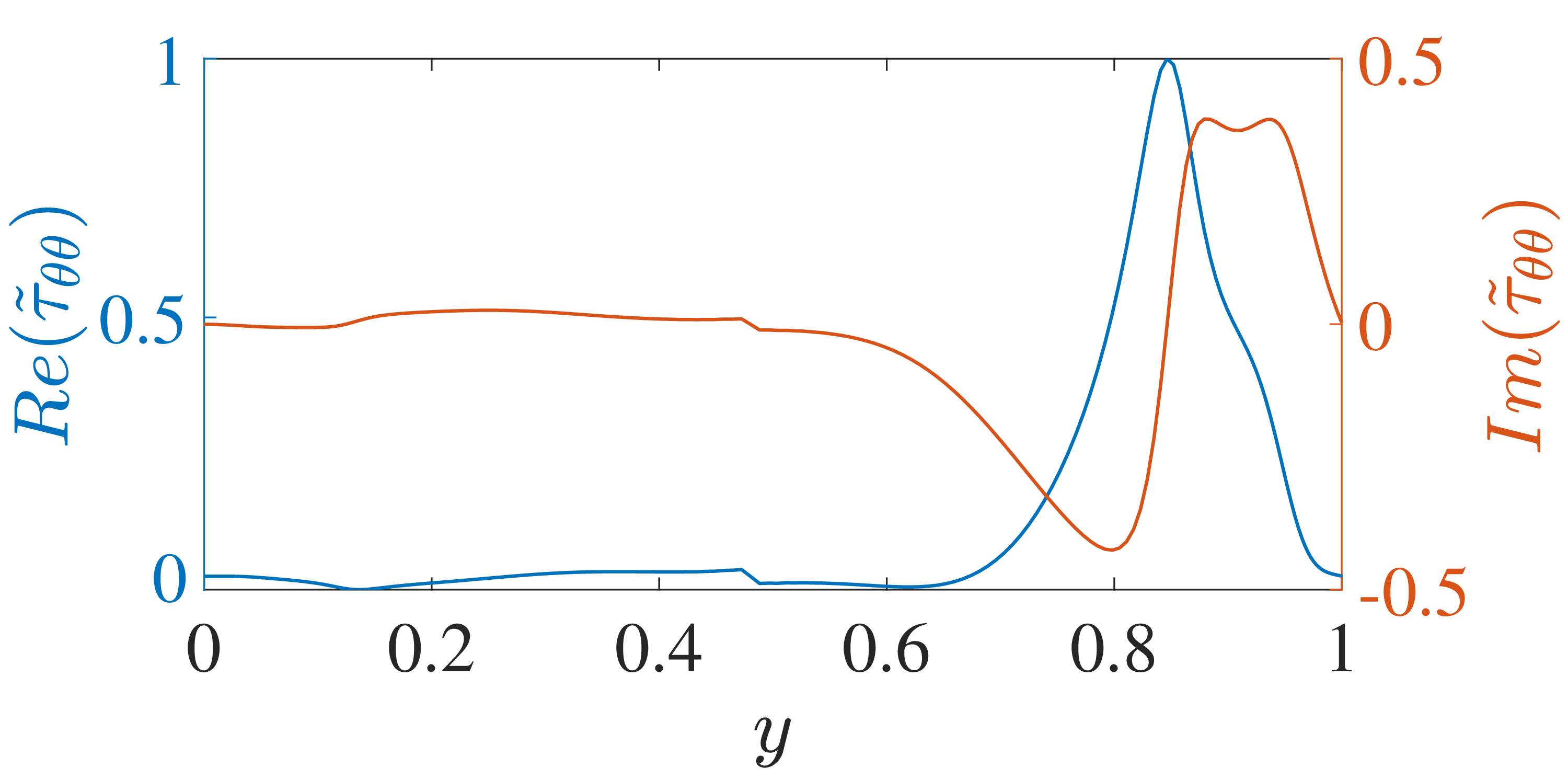}\label{fig:HSM1_Tau_00}
    }
    \subfigure[$\tilde{\tau}_{\theta z}$]{
        \includegraphics[width=0.45\textwidth]{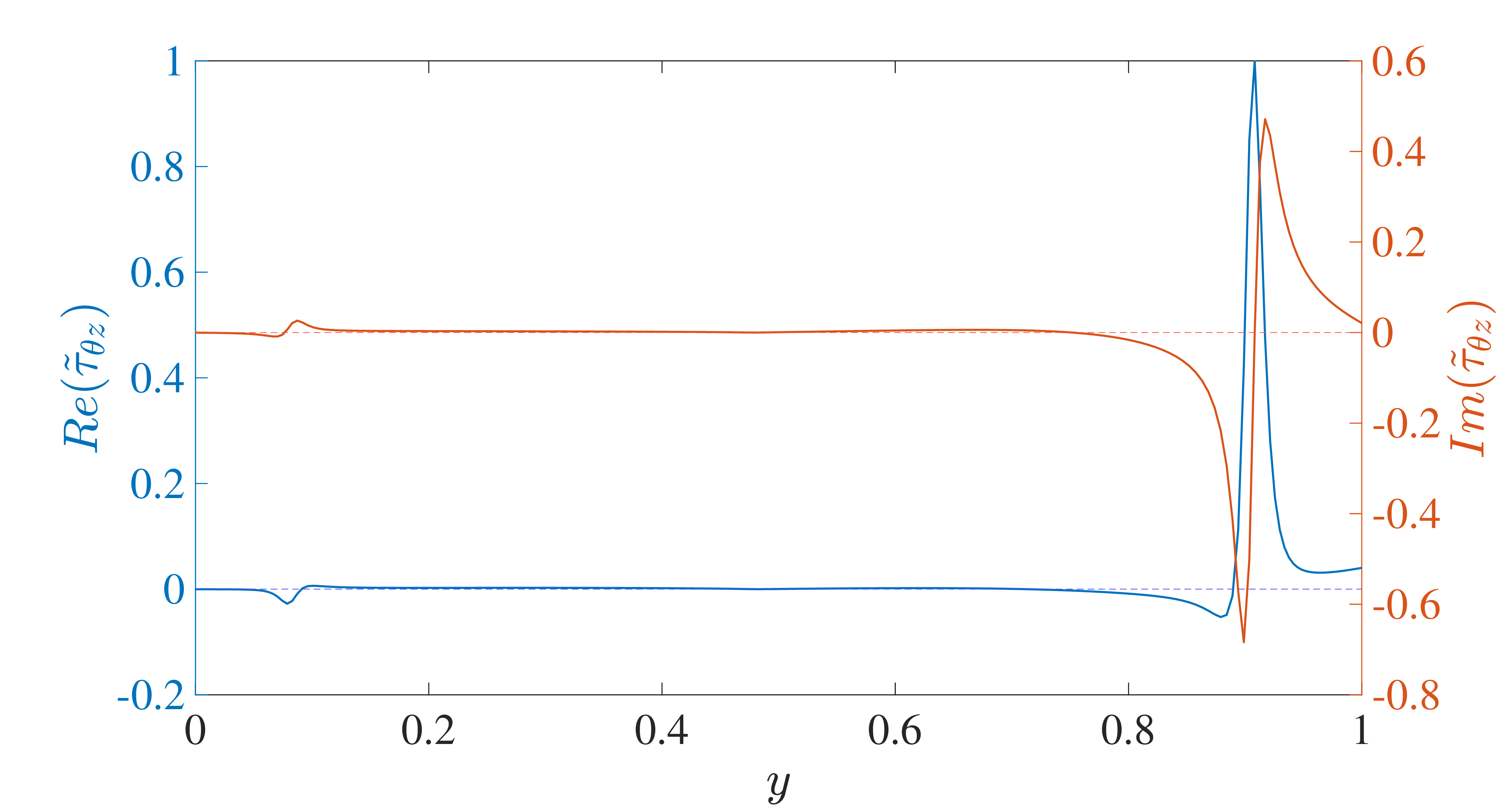}\label{fig:HSM1_Tau_0z}
    }
     \subfigure[$\tilde{\tau}_{zz}$]{
        \includegraphics[width=0.45\textwidth]{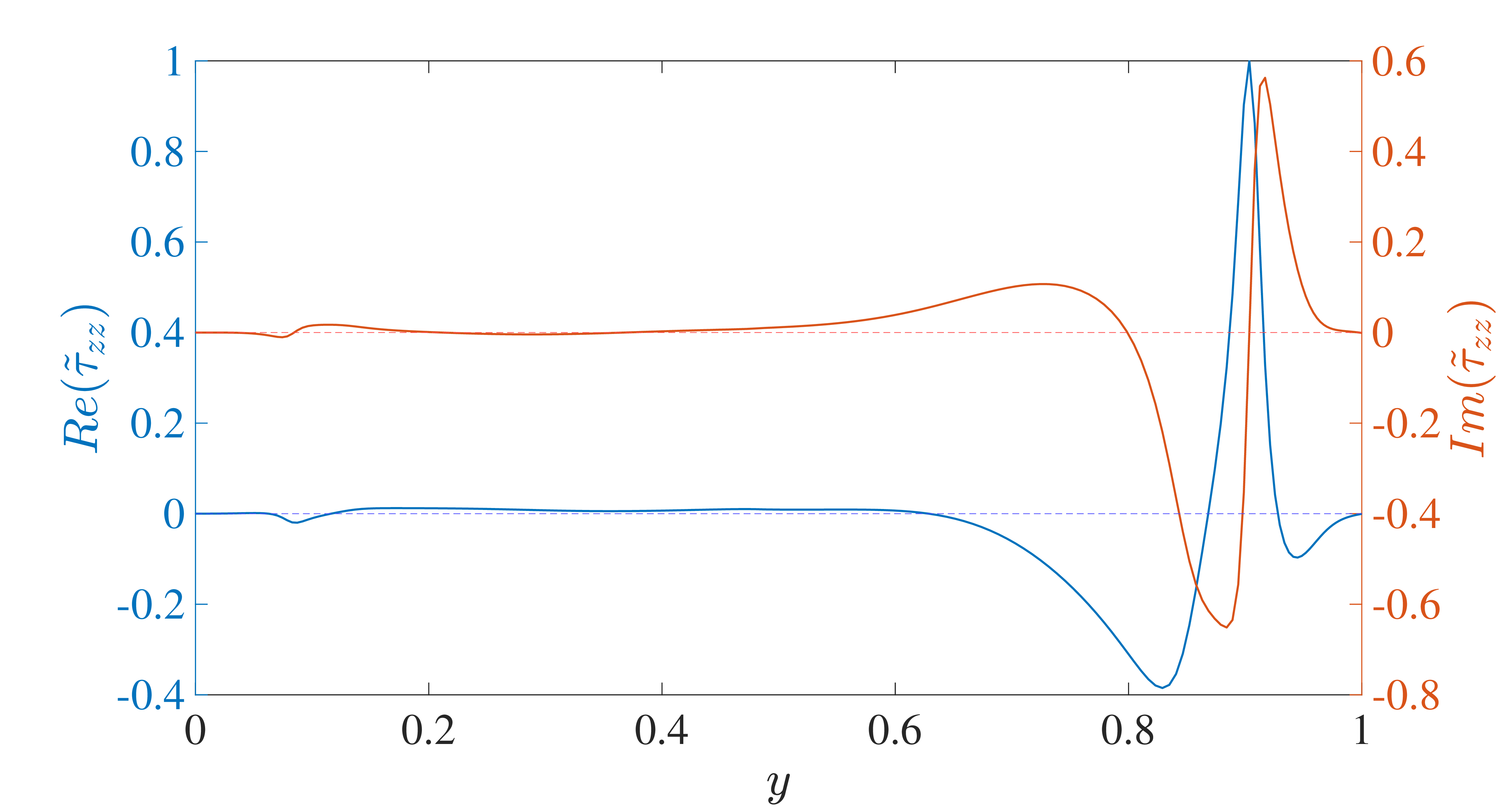}\label{fig:HSM1_Tau_zz}
    }
    \caption{Eigenfunctions for the HSM1 mode $\left(\omega = 18.2424473 - 6.952907\,i\right)$  in Dean flow of a FENE-P fluid at $Re = 0$, $\alpha = 7$, $\beta = 0.98$, $\epsilon = 0.1$, $L = 100$ and $n = 1$, for  $W\!i = 400$. The corresponding HSM1 eigenvalue was identified in the 
eigenspectrum shown previously in Fig.\,\ref{Showing_HSM2_L_100}.  The scales on the left and right $y$-axes are different, and dotted lines indicate the origins of the respective $y$-axes.}
    \label{fig:HSM1_L_100_Eigen_function}
\end{figure}

In this Appendix, we present the eigenfunctions for the centre mode (CM) and hoop stress modes (HSM1 and HSM2). As discussed earlier, for the eigenspectrum in Fig.\,\ref{FENE_P_ES_e_0.1}, the CM instability observed in our work for Dean flow is analogous to the instability reported by \cite{KhalidFENEP2025} for channel flow. 
The eigenfunctions for the CM show strong peaks in $\tilde\tau_{\theta\theta}$ and $\tilde\tau_{zz}$ near the location of the base-state maximum, a signature that is prominent even in rectilinear channel flow. The eigenfunctions of HSM1 and HSM2, presented in Figs.\,\ref{fig:HSM2_L_100_Eigen_function} and \ref{fig:HSM1_L_100_Eigen_function}, in contrast, are qualitatively different from those for  CM. In fact, the eigenfunctions for HSM2 are closer to those  of HSM1, thereby reinforcing that HSM2 is not a continuation of CM.

\begin{figure}
    \centering
    \subfigure[$\tilde{v}_{r}$]{
        \includegraphics[width=0.45\textwidth]{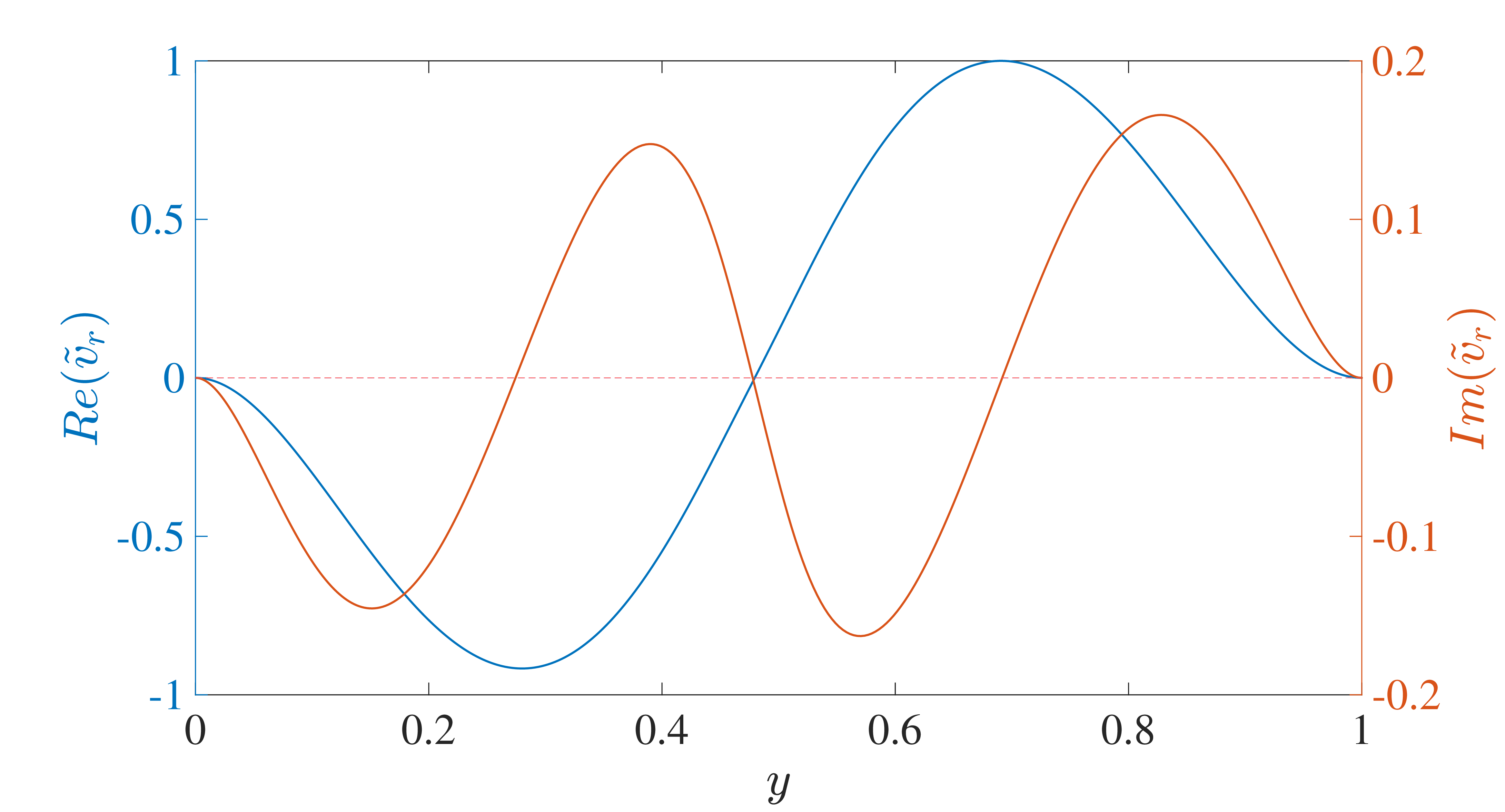} \label{fig:2D_CM_u_r}
    }
    \subfigure[$\tilde{v}_{\theta}$]{
        \includegraphics[width=0.45\textwidth]{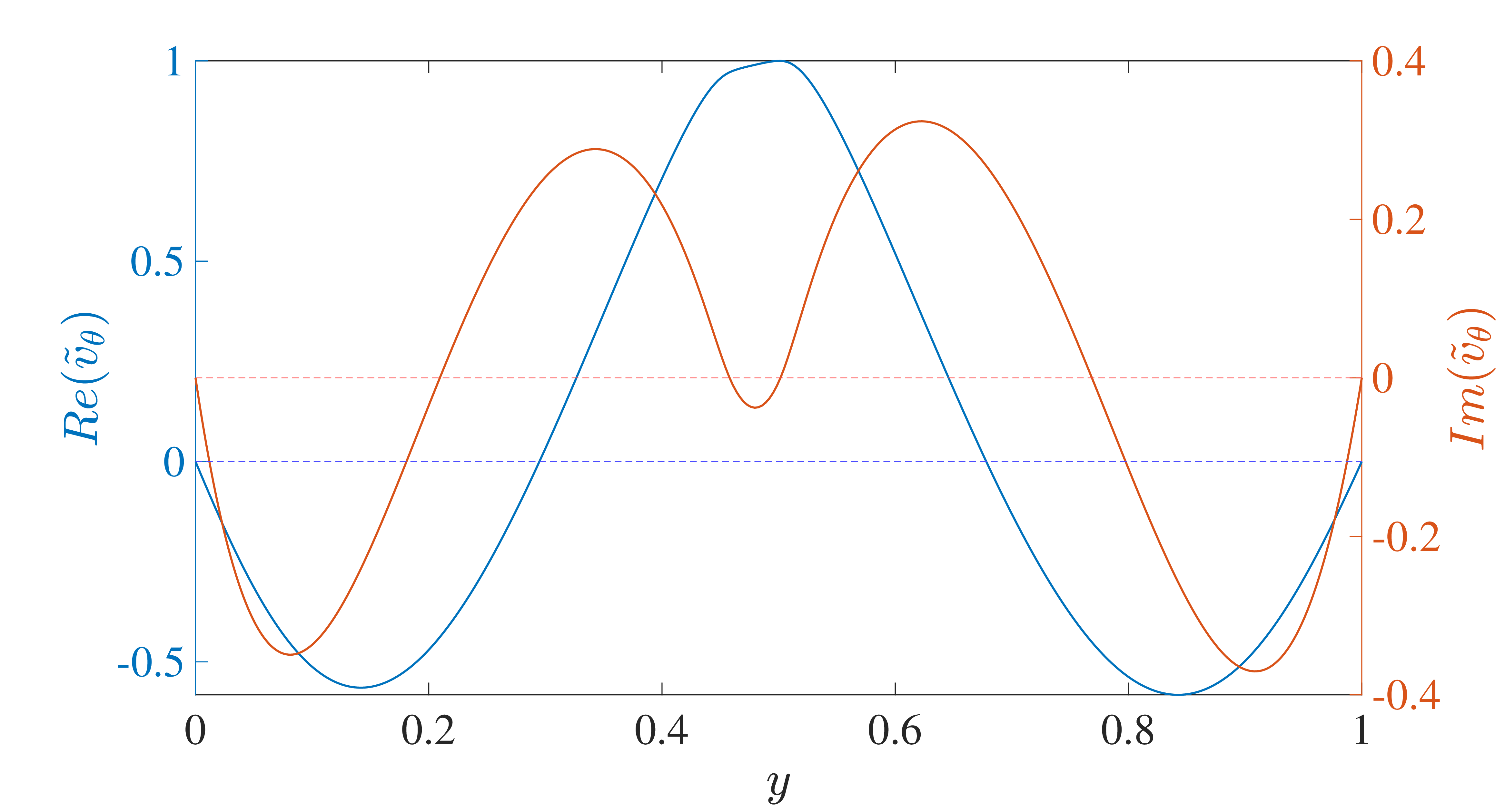} \label{fig:2D_CM_u_0}
    
    }
    \subfigure[$\tilde{\tau}_{rr}$]{
        \includegraphics[width=0.45\textwidth]{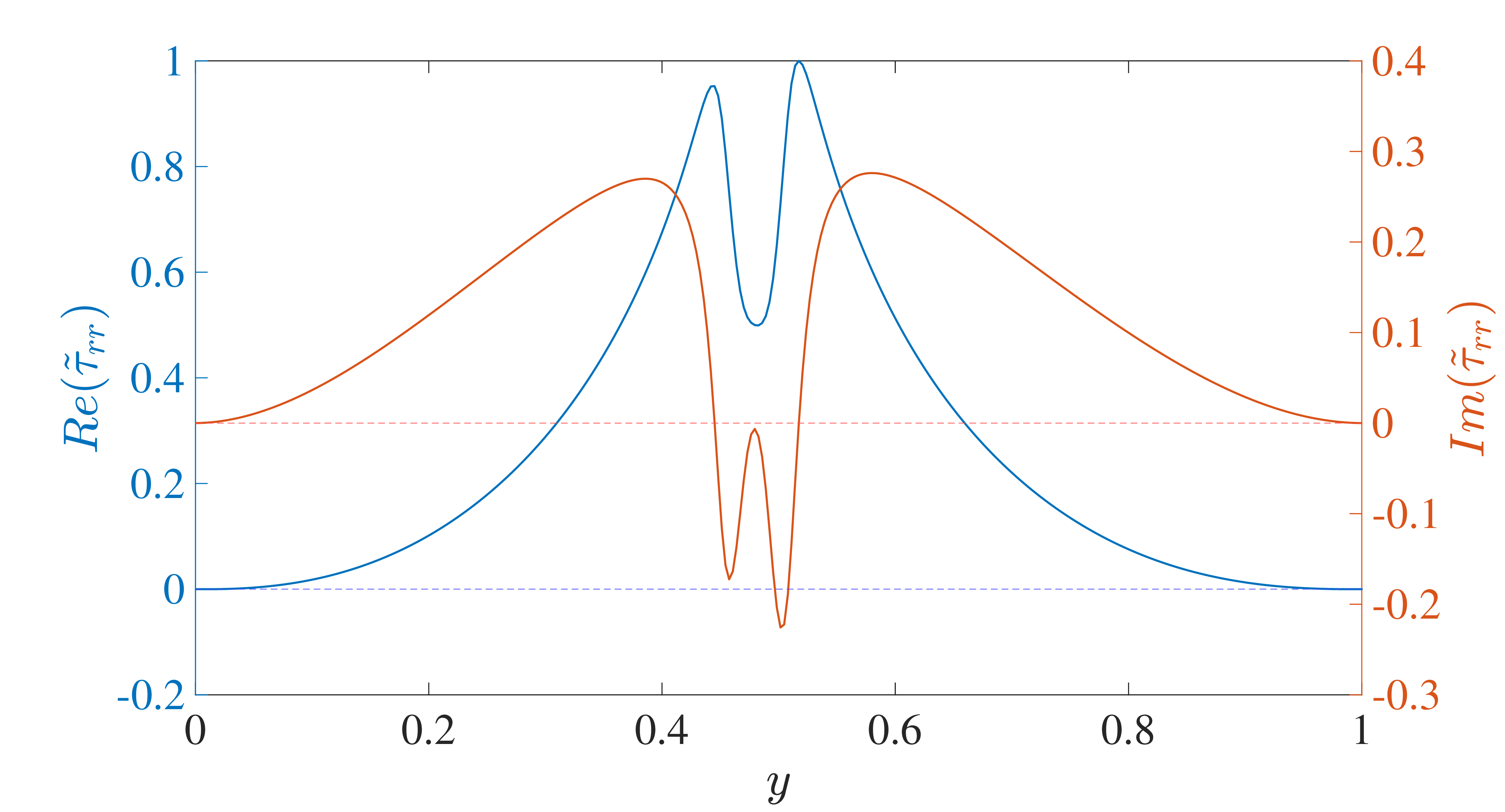}\label{fig:2D_CM_Tau_rr}
    }
    \subfigure[$\tilde{\tau}_{r\theta}$]{
        \includegraphics[width=0.45\textwidth]{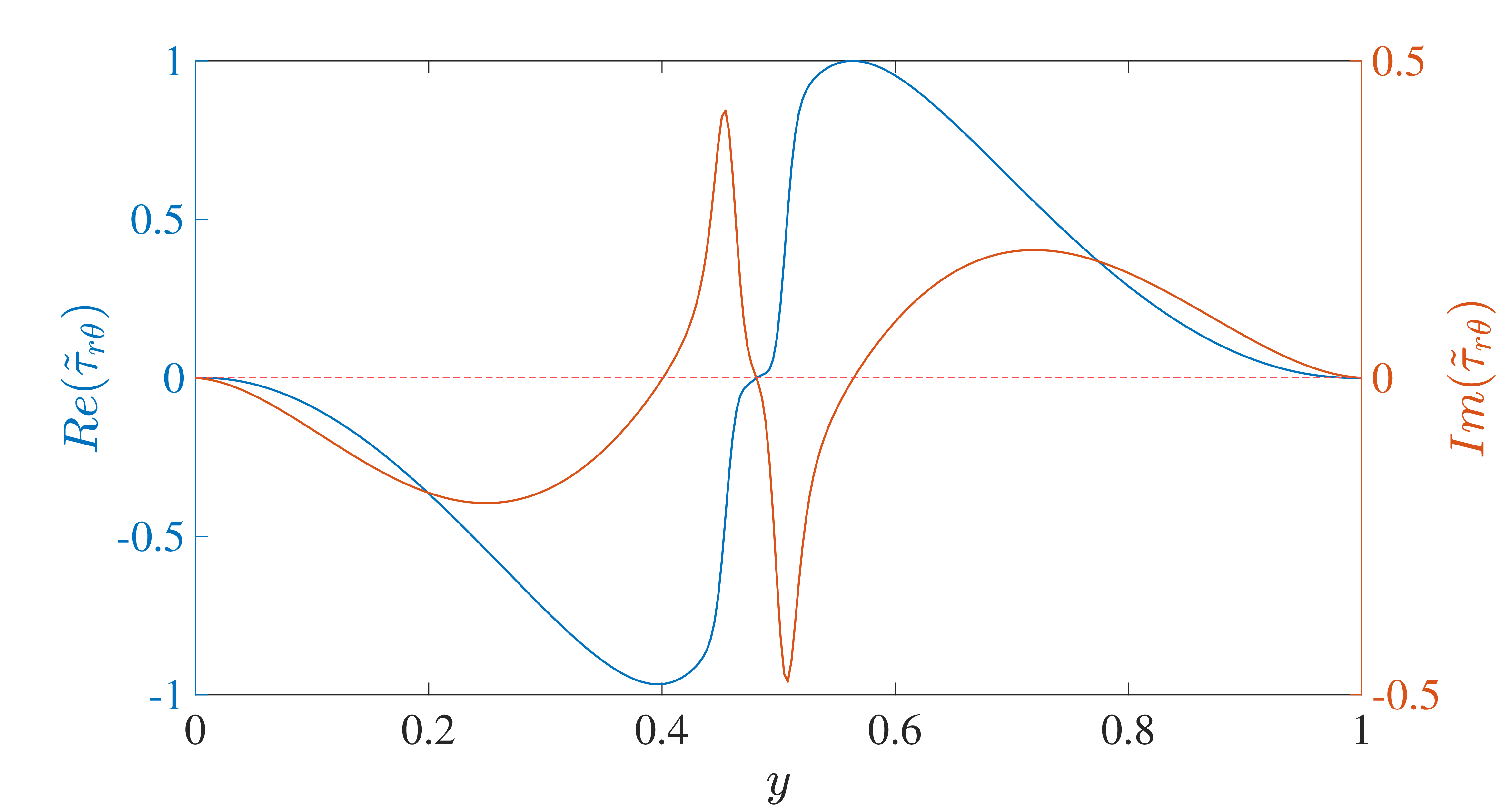}\label{fig:2D_CM_Tau_r0}
    }
    \subfigure[$\tilde{\tau}_{\theta \theta}$]{
        \includegraphics[width=0.45\textwidth]{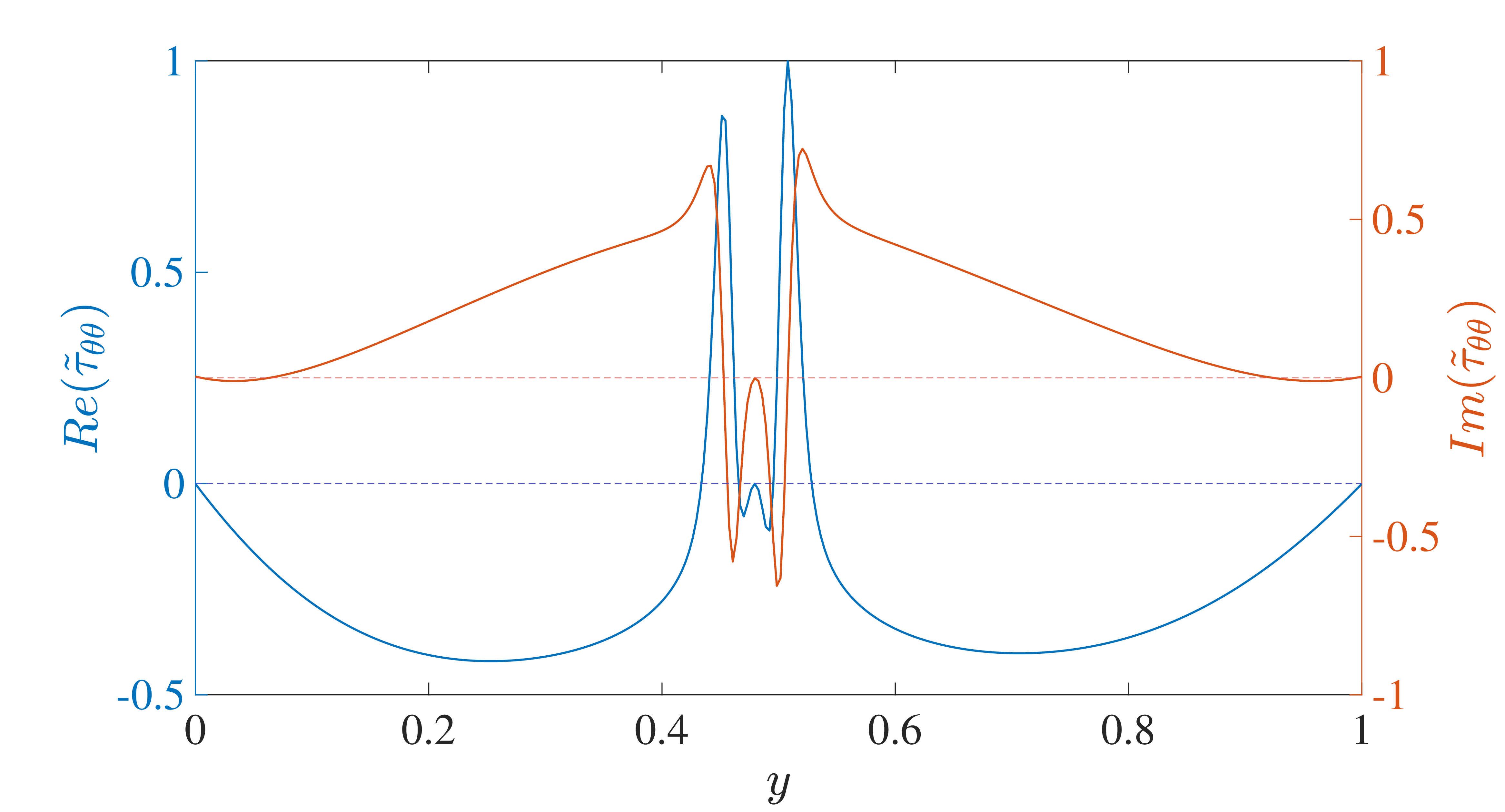}\label{fig:2D_CM_Tau_00}
    }
    \subfigure[$\tilde{\tau}_{zz}$]{
        \includegraphics[width=0.45\textwidth]{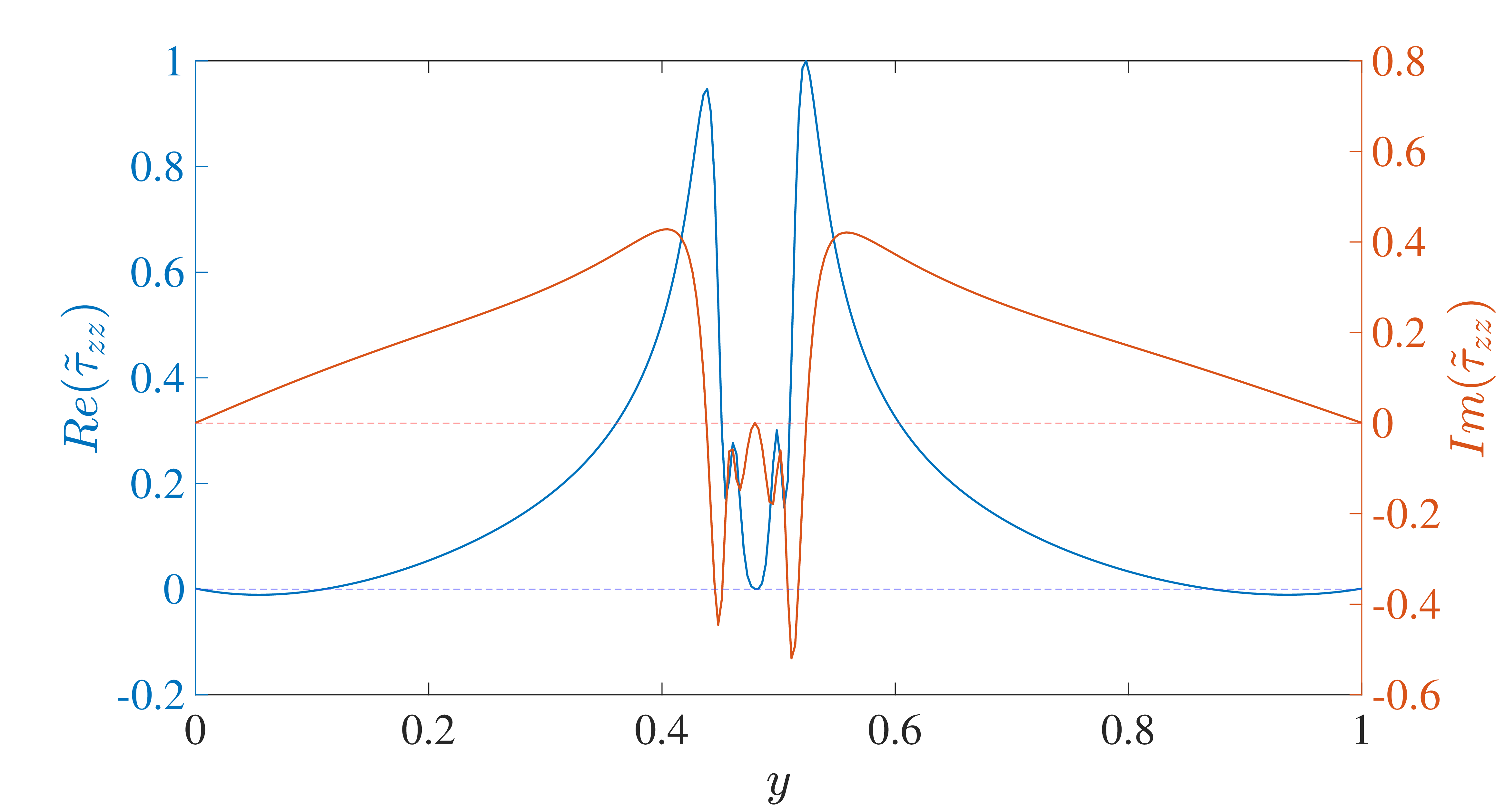}\label{fig:2D_CM_Tau_zz}
    }
    \caption{Eigenfunctions for the centre mode $\left(\omega = 344.258 + 0.00630277\,i\right)$ in Dean flow of a FENE-P fluid at $Re = 0$, $\alpha = 0$, $\beta = 0.98$, $\epsilon = 0.1$, $L = 100$ and $n = 40$, for  $W\!i = 60$. The corresponding CM eigenvalue was identified in the 
eigenspectrum shown previously in Fig.\,\ref{FENE_P_ES_e_0.1}. The scales on the left and right $y$-axes are different, and dotted lines indicate the origins of the respective $y$-axes.}
    \label{fig:2D_CM_e_0.1_L_100_Eigen_function}
\end{figure}

\bibliographystyle{jfm}
\bibliography{References}
\end{document}